\documentclass[epj]{svjour}
\usepackage{graphics,feynmf,epsfig,amssymb}
\newcommand{\SU}{SU}

\newcommand{\U}{U}

\newcommand{\Slash}[1]{\mbox{\it #1\hspace{-0.6em}\slash}}

\newlength{\ziffer}

\newlength{\vorzeichen}
\newcommand{\Plus}{\settowidth{\vorzeichen}{$-$}\hspace*{\vorzeichen}}

\newcommand{\TeV}{\,\mbox{Te\kern-0.2exV}}
\newcommand{\GeV}{\,\mbox{Ge\kern-0.2exV}}
\newcommand{\mGeV}{\,\mathrm{Ge\kern-0.2exV}}
\newcommand{\MeV}{\,\mbox{Me\kern-0.2exV}}
\newcommand{\keV}{\,\mbox{ke\kern-0.2exV}}
\newcommand{\eV}{\,\mbox{e\kern-0.2exV}}

\newcommand{\m}{\,\mbox{m}}
\newcommand{\cm}{\,\mbox{cm}}
\newcommand{\mm}{\,\mbox{mm}}
\newcommand{\um}{\,\mbox{$\mathrm\mu$m}}

\newcommand{\ps}{\,\mbox{ps}}
\newcommand{\hz}{\,\mbox{Hz}}
\newcommand{\Hz}{\hz}
\newcommand{\kHz}{\,\mbox{kHz}}

\newcommand{\ipb}{\,\mbox{pb}^{-1}}
\newcommand{\ifb}{\,\mbox{fb}^{-1}}
\newcommand{\pb}{\,\mbox{pb}}

\newcommand{\T}{\,\mbox{T}}

\newcommand{\bea}{\pagebreak[3]\begin{samepage}\begin{eqnarray}}
\newcommand{\eea}{\end{eqnarray}\end{samepage}\pagebreak[3]}
\newcommand{\beq}{\begin{equation}}
\newcommand{\eeq}{\end{equation}}

\newcommand{\met}{\,\slash\!\!\!\!{E_T}}
\newcommand{\mpt}{\,\slash\!\!\!{\vec{p}_T}}

\newcommand{\abb}{Fig.~\ref}
\newcommand{\fig}{\abb}
\newcommand{\tab}{Tab.~\ref}

\newcommand{\eq}[1]{Eq.~(\ref{#1})}

\newcommand{\geant}{{\sc Geant}}
\newcommand{\pythia}{{\sc Pythia}}
\newcommand{\comphep}{{\sc CompHEP}}
\newcommand{\alpgen}{{\sc Alpgen}}
\newcommand{\mcatnlo}{{\sc Mc@nlo}}
\newcommand{\mcfm}{{\sc Mcfm}}
\newcommand{\singletop}{{\sc SingleTop}}
\newcommand{\madgraph}{{\sc MadGraph}}
\newcommand{\toprex}{{\sc TopRex}}

\newcommand{\herwig}{{\sc Herwig}}
\newcommand{\madevent}{{\sc Madevent}}

 \newlength{\howlong}

\newcommand{\refblue}{Lyons:1988rp}
 
\newcommand{\refpdg}{Nakamura:2010zzi}
\newcommand{\refcdfsecvtx}{Affolder:2001wd}
\newcommand{\refcdfdet}{Abe:1988me,Blair:1996kx,Acosta:2004yw}
\newcommand{\refdzerodet}{Abachi:1993em,Abazov:2005pn}
\newcommand{\subsubsubsection}[1]{\paragraph{#1}}
\begin{document}
\title{Properties of the Top Quark}
%\subsubtitle{Do you have a subtitle?\\ If so, write it here}
\author{Daniel Wicke\inst{1}\inst{2}% etc
}                     % Do not remove
\institute{%
 Inst. f. Physik, Johannes Gutenberg-Universit\"at, Staudinger Weg 7, D-55099 Mainz, Germany \and
 Bergische Universit\"at, Gau{\ss}str. 20, D-42097 Wuppertal, Germany}
\date{Received: date / Revised version: date  (CVS $Revision: 1.1.2.41 $)}
\abstract{
The top quark was discoverd at the CDF and D0 experiments in 1995. As the
partner of the bottom quark its properties within the Standard Model are fully
defined. Only the mass is a free parameter. The measurement of the top quark
mass and the verification of the expected properties have been an important
topic of experimental top quark physics since. In this review the recent
results on top quark properties obtained by the Tevatron experiments CDF and
D0 are summarised. At the advent of the LHC special emphasis is given to the
basic measurement methods and the dominating systematic uncertainties. 
\PACS{
      {14.65.Ha}{Top quarks}  
% \and      {14.65.Jk}{Other quarks (e.g., 4th generations)} % tprime
% \and      {14.80.Fd}{Other charged Higgs bosons}           % charged Higgs
% \and      {14.70.Pw}{Other gauge bosons}                   % Zprime
     } % end of PACS codes
} %end of abstract
\maketitle
\tableofcontents 
%\newpage

\section{Introduction}
\label{sec:Intro}
The aim of particle physics is the understanding of
elementary particles and their interactions. 
%that drive our world 
% that govern our world
The current theory of elementary particle
physics, the Standard Model, contains twelve different types of fermions 
which (neglecting gravity) interact through the gauge bosons of three
forces~\cite{Glashow:1961ez,Goldstone:1962es,PhysRevLett19_1264,Fritzsch:1973pi}.  
In addition a scalar particle, the Higgs boson, is
needed for theoretical consistency~\cite{Higgs:1964ia}.
These few building blocks explain all 
experimental results found in the context of particle physics, so far.

Nevertheless, it is believed that the Standard Model is only an approximation
to a more  complete theory. First of all the fourth known
force, gravity, has withstood all attempts to be included until now.
Furthermore, the Standard Model describes several features of the elementary particles 
like the existence of three families of fermions or the quantisation of
charges, but does not explain these properties from underlying principles.  
Finally, the lightness of the Higgs boson needed to explain the symmetry breaking
is difficult to maintain in the presence of expected corrections from gravity
at high scales. This is the so called hierarchy problem.

In addition astrophysical results indicate that the universe consists only to
a very small fraction of matter described by the Standard Model. Large
fractions of dark energy and dark matter are needed to describe the
observations~\cite{\refpdg,deBernardis:2000gy,Spergel:2003cb,Astier:2005qq,Spergel:2006hy}. %,Komatsu:2008hk}.  
 Both do not have any correspondence in the Standard Model. 
Also the very small asymmetry between matter and anti-matter
that results in the observed universe built of matter (and not of anti-matter) cannot be
explained until now.

It is thus an important task of experimental particle physics to test the
predictions of the Standard Model to the best possible accuracy and to search
for deviations pointing to necessary extensions or modifications of our current
theoretical understanding.

The top quark was predicted to exist by the Standard Model as the partner of
the bottom quark.  It was first observed in 1995 by the Tevatron experiments
CDF and D\O~\cite{Abe:1995hr,Abachi:1995iq}
and was the last of the quarks to
be discovered.  As the partner of the bottom quark the top quark is expected
to have quantum numbers identical to that of the other known up-type
quarks. Only the mass is a free parameter. We now know that it is
more than 30 times heavier than the next heaviest quark, the bottom quark.  

Thus, within the Standard Model all production and decay properties are fully
defined. Having the complete set of quarks further allows to verify
constraints that the Standard Model puts on the sum of all quarks or
particles. This alone is reason enough to experimentally study the top quark
properties. The high value of the top quark mass and its closeness to the
electroweak scale has inspired people to speculate that the top quark
could have a special role in the electroweak symmetry breaking, see e.g.~\cite{Chivukula:1998wd,Dobrescu:1997nm}.
Confirming the expected properties of the top quark  experimentally
establishes the  top quark as we expect it to be. Any deviation from the
expectations gives hints to new physics that may help to solve the outstanding
questions. 

Since the last review on top quark physics in this journal~\cite{Quadt:2007jk} the luminosity at the
Tevatron has been increased by more than an order of magnitude. Now measurements of
the top quark mass are no longer limited by statistical but by systematic
uncertainties. Even discussions about the knowledge of the implicitly used
theoretical mass definition become relevant. Moreover measurements of additional
top quark properties now reach a viable precision.
In this review the recent results on top quark properties
obtained by the Tevatron experiments CDF and
{D\O} are summarised. 
At the advent of the LHC special emphasis is given to the basic measurement
methods and the dominating systematic uncertainties.
Other reviews with different emphases are available in the 
literature~\cite{Beneke:2000hk,Wagner:2005jh,Kehoe:2007px,%
Demina:2008zz,%
Pleier:2008ig,%
Bernreuther:2008ju,%
Incandela:2009pf,%
Deliot:2010ey,%
Heinson:2010xh}.

After a short introduction to the Standard Model and the
experimental environment in the remainder of this section,
Chapter~\ref{section:Masses} describes the current status of top quark mass
measurements. Then measurements of interaction properties  are
described in Chapter~\ref{section:interaction}.  
Finally, Chapter~\ref{sect:BSM} deals with
analyses that consider hypothetical particles beyond the Standard Model
in the observed events.

%%% Local Variables: 
%%% mode: latex
%%% TeX-master: "EPJC_TopProperties"
%%% End: 

\subsection{Theory}
\label{sec:Theory}
The physics of elementary particles is
described by the Standard Model of particle physics. 
It describes three of the four known forces that act on elementary particles,
the electromagnetic and the weak force are unified in the 
GSW-theory~\cite{Glashow:1961ez,Goldstone:1962es,PhysRevLett19_1264},  
the strong force is described by quantum chromodynamics
(QCD)~\cite{Fritzsch:1973pi}. So far the influence of gravitation could not be
unified with the other three forces in a consistent quantum theory. Due to its
weakness it is usually safe to neglect its influence in the context of
particle physics.

In the following a short description of the Standard Model shall be given to
set the stage for later descriptions. For this the natural units where
$\hbar=c=1$ will be used. 

\subsubsection{The Standard Model}
\subsubsubsection{Lagrangian}
The Standard Model is a quantum field theory with local gauge symmetry under 
the group $\SU(3)\times\SU(2)\times\U(1)$. Its Lagrangian contains fields
corresponding to three types of particles: Gauge or vector bosons, fermions
and scalars. Gauge boson are described by vector fields, $A_\mu$, scalars by
complex fields, $\phi$. Fermions can be described by Weyl spinors, $\psi$,
with left and right handed helicity.
Using Einstein's summation convention the Lagrangian can be written as
\bea\label{eq:lagrangian}
{\cal L}&=&-\frac{1}{4}F^A_{\mu\nu}F^{A\,\mu\nu}
         +i\overline{\psi}_\alpha\Slash{D}\;\psi_\alpha
         +(D_{\mu}\phi^a) (D^{\mu}\phi^a)\\
       &&+y_{\alpha\beta}^a\overline{\psi}_\alpha\psi_\beta\phi^a
         +V(\phi)
         +(\mbox{ghost- and gauge terms}).\nonumber
        \label{Lagrangedichte}
\eea
Here capital Latin letters run over the gauge bosons, lower case Latin letters
over the scalar fields and Greek letters $\alpha$, $\beta$ index the fermions
of the standard model. $\mu$ and $\nu$ are Dirac indices.
The field tensor is defined as
\beq
 F_{\mu\nu}^{A}=\partial_{\mu}A^A_{\nu}
                    -\partial_{\nu}A^A_{\mu}
                    -g C_{ABC}A^B_{\mu}A^C_{\nu}\quad,
        \label{Feldstaerketensor}
\eeq
where $C_{ABC}$ are the structure constants of the gauge group. The covariant
derivative is defined from the gauge symmetry to be
\beq
D_\mu=\partial_\mu - i \frac{g_{(B)}}{2}\; t^B A^B_\mu\quad.
\label{kovarianteAbleitung}
\eeq
Here  $g_{(B)}$ is the coupling strength for the gauge boson $A^B$, i.e. $g$,
$g'$ or $g_s$. $t^B$~are the generators of the gauge symmetry in the
representation that corresponds to the particle field on which the derivative acts. 
\beq
\Slash{D}=\sigma^\mu D_\mu\quad\mbox{with}\quad\sigma^\mu=(1,\pm \vec\sigma)\quad,
\eeq 
where   $\vec\sigma$ is the vector of Pauli matrices. The positive sign applies
to fermions of right handed helicity, the negative for left handed ones.

The Yukawa couplings, $y_{\alpha\beta}^a$, are free parameters that may be
non-zero only when the combination of the fermions $\alpha$ and $\beta$ with
the scalars is gauge invariant. $V(\phi)$ is a
quartic form.

\subsubsubsection{Particle Content}
The particle content of the Standard Model is specified by defining the
representations of the gauge symmetry for each of the fields contained.

The Weyl spinors describing the fermions transform according to fundamental
representations for each of the subgroups or may be invariant under a
subgroup. Quarks transform under the three dimensional representation of the
$SU(3)$ group,~${\bf 3}$. All left handed spinors transform under $SU(2)$
according to the two dimensional representation,~${\bf 2}$, while the right
handed spinors are  singlets, i.e. invariant under  $SU(2)$ rotations. This
reflects the left handed nature of weak interactions.
The transformation properties under $U(1)$ are specified by the
hypercharge. In the Standard Model all representations are repeated three
times and build the three generation of fermions.

In \tab{TabFDarst} the representations of the fermions of the Standard Model
are summarised. The representations of the non-Abelian gauge groups
 $SU(3)$ and $SU(2)$ are specified by their dimension (in bold-face), the 
hypercharge corresponding to the $U(1)$ group is given as index.
\begin{table}
   \caption{Fermions of the Standard Model and their representations\label{TabFDarst}.}
\begin{center}
\begin{tabular}{|l|ccc|cc|}
\hline
Gene- & \multicolumn{3}{c|}{Quarks} &  \multicolumn{2}{c|}{Leptons} \\
ration&$\!{\bf(3,2)}_{\frac{1}{6}}\!$&$\!{\bf(3,1)}_{\frac{2}{3}}\!$&$\!{\bf(3,1)}_{-\frac{1}{3}}\!$&$\!{\bf(1,2)}_{-\frac{1}{2}}\!$&$\!{\bf(1,1)}_{-1}\!$\\[1mm]
\hline
 1st & $\displaystyle {u \choose d}_L $ & $u_R$ & $d_R$ & $\displaystyle {\nu_e \choose e}_L $ & $e_R$        \\
 2nd & $\displaystyle {c \choose s}_L $ & $c_R$ & $s_R$ & $\displaystyle {\nu_{\mu} \choose \mu}_L $& $\mu_R$ \\
 3rd & $\displaystyle {t \choose b}_L $ & $t_R$ & $b_R$ & $\displaystyle {\nu_{\tau} \choose \tau}_L $&$\tau_R$\\
\hline
\end{tabular}
\end{center}
\end{table}
\begin{table}
   \caption{Gauge bosons of the Standard Model and their representations\label{TabEBDarst}.}
\begin{center}
\begin{tabular}{|c|c|c|}
\hline
   Symbol      & Representation   &  Coupling strength \\
\hline
     g         &${\bf(8,1)}_0 $&  $g_s$\\
$(W^1,W^2,W^3)$&${\bf(1,3)}_0 $&  $g=g_2$\\
    $B$        &${\bf(1,1)}_0 $&  $g'=\sqrt{\frac{3}{5}}g_1$\\
\hline
\end{tabular}
\end{center}
\end{table}
For a few years it is known from the measurement of neutrino
oscillations~\cite{Fukuda:1998mi,Ahmad:2001an,Ahmad:2002jz,\refpdg} 
that also the neutrinos have mass and thus  right handed neutrinos should be
added to \tab{TabFDarst}. The necessary extensions of the Standard Model
are not unique and thus they are usually not considered part of the Standard Model.
In the context of top quark physics neutrino oscillations do not play any role and can thus
be ignored for the purpose of this review.

The gauge bosons of a quantum field theory need to transform according to the
adjoint representation of their sub-group. Thus we get eight gauge bosons for
the $SU(3)$ symmetry, the gluons, three for $SU(2)$ and one for the $U(1)$
symmetry, c.f.~\tab{TabEBDarst}.
In addition to the bosons and  fermions described the Standard Model contains
a complex scalar  doublet, the Higgs doublet $\Phi$, which transforms according to
${\bf(1,2)}_\frac{1}{2}$. It is needed for symmetry breaking.

\subsubsubsection{Symmetry breaking: Higgs mechanism}
In nature the symmetry of the Standard Model is broken. The symmetry
breaking is implemented by the Higgs mechanism~\cite{Higgs:1964ia},
which assumes that the scalar isospin doublet $\Phi$ has a vacuum
expectation value. This is achieved by proper choice of parameters in  the most
general potential, $V(\phi)$, for the scalar field in~\eq{eq:lagrangian}.
According to the symmetry this can be chosen to exist in
the lower component of the doublet:
\beq
   \Phi =  {\phi_1 \choose \phi_2 }
   \qquad\mbox{with}\qquad
   \langle \Phi \rangle = {0 \choose \langle\phi_2\rangle}\quad.
\eeq
By expanding the complex scalar field around the vacuum expectation value 
first four real scalar fields are specified. Three of these can 
behave like longitudinal  components of the $SU(2)$ gauge bosons,
$W^i$. Usually the theory is thus written in terms of three massive vector
bosons $W^+$, $Z$, $W^-$, a massless vector boson, the photon $A$, and the
fourth 
real scalar field,  the Higgs boson~$H$:
\begin{eqnarray}
 W^{\pm} & = & \frac{1}{\sqrt{2}}(W^1\mp iW^2) \nonumber\\
 Z       & = & \frac{g' B - g  W^3}{\sqrt{g'^2+g^2}} 
               \;=\; \sin \theta_W \, B - \cos \theta_W \, W^3\nonumber\\
 A       & = & \frac{g  B + g' W^3}{\sqrt{g'^2+g^2}}
               \;=\; \cos \theta_W \, B + \sin \theta_W \, W^3\quad.
\end{eqnarray}
Here $B$ is the gauge boson of the (hypercharge) $U(1)$ symmetry
and the Weinberg angle  $\theta_W$ is defined by the ratio of coupling
constants
\begin{equation}        
        \tan\left(\theta_W\right):=\frac{g'}{g}\quad.
\end{equation}
After this rewriting the theory remains $SU(3)\times U(1)$ invariant. The
$U(1)$ symmetry now corresponds to the electrical charge. So the Standard
Model parts are quantum chromodynamics (QCD) and  quantum
electrodynamics with their $SU(3)$ and $U(1)$ symmetries, respectively.

The Higgs mechanism not only yields massive vector bosons, it is also responsible
for the masses of the fermions. 
For the specified particle content, the Yukawa terms in \eq{eq:lagrangian} may
be non-zero for left-handed quark or lepton doublets paired with the
corresponding right-handed quark and lepton 
$SU(2)$-singlets. For the first generation
these terms are
\beq
y_{ee} \overline{(\nu_e,e)}_L \Phi e_R + %\mbox{h.c}\quad
y_{dd} \overline{(u,d)}_L \Phi d_R+ % \mbox{h.c}\quad
y_{uu} \overline{(u,d)}_L i\sigma^2\Phi u_R+ \mbox{h.c}\quad\mbox{.}
\eeq
Corresponding terms can be written not only for the other generations, but
in general also for fermion pairs between different generations. 
Unitary rotations in the three dimensional space of  generations are commonly used to
redefine the lepton and quark fields such that Yukawa couplings occur only
between particles of the same generation. This provides the mass eigenstates
of the quark and lepton fields. 
These rotations cancel in most terms of the Lagrangian. The only observable
remainder of this rotation is the Cabibbo-Kobayashi-Maskawa (CKM) matrix~\cite{Cabibbo:1963yz,Kobayashi:1973fv,\refpdg}
which occurs in the coupling of the $W^\pm$ bosons to quarks:
\beq
V_\mathrm{CKM}=\left(
\begin{array}{ccc}
V_{ud} & V_{us} & V_{ub} \\
V_{cd} & V_{cs} & V_{cb} \\
V_{td} & V_{ts} & V_{tb} \\
\end{array}
\right)
\eeq
At the same time this is the only process in the Standard Model that connects
the different generations. Numerically this unitary matrix has diagonal
entries close to unity and off-diagonal entries 
that are around $0.2$ between the
first and second generation, around $0.04$ between the second and third
generation and even smaller for the transition of the first to the third generation~\cite{\refpdg}.

\subsubsection{Perturbation Theory}
Predictions of the Standard Model for high energy reactions are so far generally
performed in perturbation theory. The reactions are described as one or more
point-like interactions between otherwise free particles. 
The allowed reactions can be read off the Lagrangian and are usually
represented by Feynman diagrams. For example 
the Yukawa term $y \bar{\psi}\phi\psi$
yields an interaction vertex of the strength $y$ with two fermions $\psi$ and
the scalar field $\phi$. % and is represented by two Fermion lines and one s. 

Each Feynman dia\-gram serves simultaneously as
a diagrammatic description of the reaction and as a short hand notation for the
corresponding computation of the transition amplitude. The quantum mechanical amplitude of a given process that
transforms a set of initial state particles to a set of final state particles
is given by the sum of all possible diagrams with the corresponding initial and
final state particles as external lines. 

Diagrams with few interactions usually give the largest contributions and
higher order corrections are suppressed by factors of the additional coupling
strengths. Thus calculations are usually performed in a fixed order of the
coupling constant(s). 
In some cases, however, kinematic enhancements of logarithmic type 
may compensate the suppression by additional powers of the coupling
constant. Notably these occur in cases of collinear or soft gluon radiation and
for top quark pair production near threshold. 
In these cases resummation of  the leading
(or next to leading) logarithms to all order of the coupling are performed.
 
Diagrams of higher orders generally involve loops which require to integrate over all possible momenta of
the internal lines. Naively, such integrals diverge.  It is necessary to
renormalise the theory  in order to obtain finite predictions. There are
several possible schemes to perform this renormalisation. Most commonly the so
called $\overline{\mathrm{MS}}$ scheme is used, which itself depends on a
continuous parameter the renormalisation scale $\mu$. The dependence of
results on the choice of this parameter is often used as a measure for theoretical
uncertainties of a prediction.

Perturbation theory described so far deals with the particles of the Standard
Model named above, i.e. with quarks, leptons, gauge bosons and the Higgs
boson. In nature, however, quarks have not been observed as free particles,
rather they are confined in bound states of colour neutral hadrons. 
The dynamics of quarks and gluons inside hadrons cannot be described by
perturbation theory.
% as the assumption of (interaction of .

To describe the collisions of hadrons with perturbation theory 
the (soft) physics that governs the behaviour of quarks and gluons in the
hadron needs to be factorised from the hard process in the collisions.  
The partons inside an incoming hadron are considered as a number of ``free''
partons that may enter the hard interaction. The distribution of partons
inside the incoming hadron is taken from parton distribution functions (PDFs)
that are derived from experiments. With this the cross-section, $\sigma (p\bar{p}\rightarrow X; s)$, to produce
final state particles $X$ can be
described in terms of cross-sections $\hat{\sigma}$ of incoming quarks and
gluons to produce $X$: 
\bea
&& \sigma (p\bar{p}\rightarrow X; s) = \nonumber\\
&& \sum_{i, j=g,u,\bar{u},d,\bar{d},...} 
 \!\!\int\! \mathrm{d} x_1 \mathrm{d} x_2 \,\,
  f_{i} (x_1) \bar{f}_{j} (x_2) \,\,
  \hat{\sigma}(ij \rightarrow X; \hat{s})\qquad
  \label{eq:factorization}  
\eea
Here $s$ is the centre-of-mass energy squared of the incoming hadrons. At the Tevatron
Run\,II  $\sqrt{s}=1.96\TeV$. The colliding partons carry momentum fractions
$x_1$ and $x_2$, and $\hat s=x_1 x_2 s$ is their centre-of-mass energy squared.
$f_i$~and $\bar{f}_j$ are the parton distribution functions 
for the partons $i$ and $j$,
i.e. the probabilities to find this parton in the proton and anti-proton,
respectively. Formally, the parton density functions $f$, $\bar f$
and the partonic cross-section, $\hat\sigma$, depend on the factorisation
scale, $\mu_F$, which specifies which effects are included in the PDFs and
which remain to be described by the hard matrix element. They also depend on
the renormalisation scale, $\mu$. These dependencies are not written out in \eq{eq:factorization}.

\subsubsection{Top Quark Production in Hadron Collisions}

\begin{figure*}
  \centering
  \setlength{\unitlength}{1mm}
\null\hfill
  \begin{fmffile}{mf_doubletop}
   \begin{minipage}[t]{31mm}
    \begin{fmfgraph*}(31,22)
      \fmfstraight
      \fmfleft{q,qb}
      \fmfright{t,tb}
      \fmf{fermion}{q,v,qb}
      \fmf{gluon,lab=$g$}{v,vv}
%      \fmfdot{v}
%      \fmfdot{vv}
      \fmf{fermion}{tb,vv,t}
      \fmflabel{$q$}{q}
      \fmflabel{$\bar{q}$}{qb}
      \fmflabel{$t$}{t}
      \fmflabel{$\bar{t}$}{tb}
    \end{fmfgraph*}
   \end{minipage}%\\\vspace*{10mm}
\hfill\null\hfill
   \begin{minipage}[t]{25mm}
    \begin{fmfgraph*}(22,14)
      \fmfstraight
      \fmfleft{g1,g2}
      \fmfright{t,tb}
      \fmf{gluon}{g1,v,g2}
      \fmf{gluon,lab=$g$}{v,vv}
%      \fmfdot{v}
%      \fmfdot{vv}
      \fmf{fermion}{tb,vv,t}
      \fmflabel{$g$}{g1}
      \fmflabel{$g$}{g2}
      \fmflabel{$t$}{t}
      \fmflabel{$\bar{t}$}{tb}
    \end{fmfgraph*}
   \end{minipage}\hfill
   \begin{minipage}[t]{25mm}
    \begin{fmfgraph*}(25,14)
      \fmfstraight
      \fmfleft{g1,g2}
      \fmfright{t,tb}
      \fmf{gluon}{g1,v1}
      \fmf{gluon}{v2,g2}
%      \fmfdot{v}
%      \fmfdot{vv}
      \fmf{fermion}{tb,v2}
      \fmf{fermion,tension=0}{v2,v1}
      \fmf{fermion}{v1,t}
      \fmflabel{$g$}{g1}
      \fmflabel{$g$}{g2}
      \fmflabel{$t$}{t}
      \fmflabel{$\bar{t}$}{tb}
    \end{fmfgraph*}
   \end{minipage}\hfill
   \begin{minipage}[t]{25mm}
    \begin{fmfgraph*}(25,14)
      \fmfstraight
      \fmfleft{g1,g2}
      \fmfright{tb,t}
      \fmf{gluon}{g1,v1}
      \fmf{gluon}{v2,g2}
%      \fmfdot{v}
%      \fmfdot{vv}
      \fmf{phantom}{tb,v1}
      \fmf{phantom}{v2,t}
      \fmffreeze
      \fmf{fermion}{tb,x1}
      \fmf{plain}{x1,v2}
      \fmf{fermion}{v2,v1}
      \fmf{plain}{v1,x2}
      \fmf{fermion}{x2,t}
      \fmflabel{$g$}{g1}
      \fmflabel{$g$}{g2}
      \fmflabel{$t$}{t}
      \fmflabel{$\bar{t}$}{tb}
   \end{fmfgraph*}
   \end{minipage}\hfill\null
\end{fmffile}\vspace*{5mm}
  \caption{Born level Feynman diagrams contributing to top quark pair
    production. The quark annihilation (leftmost diagram) is dominating
    top pair production at the Tevatron. The gluon fusion processes (three
    right diagrams) contribute about $15\%$, only.}
  \label{fig:feyn-doubletop}
\end{figure*}
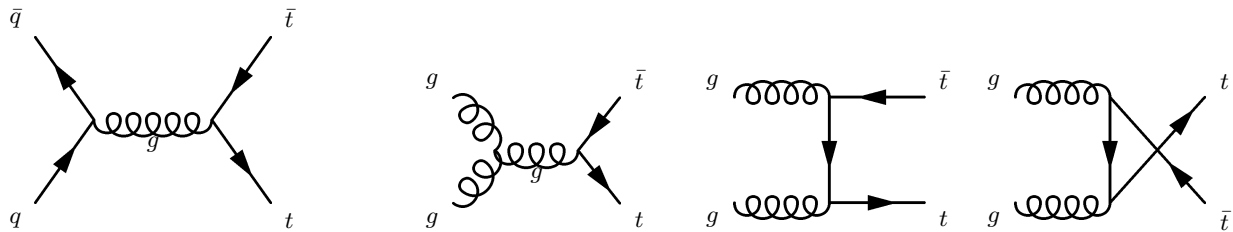

In hadron collisions, like the proton-antiproton collisions at the Tevatron, top
quarks can be produced singly or in pairs. The pair production occurs
dominantly via the
strong interaction. In leading order the quark anti-quark annihilation and gluon
fusion processes shown in \fig{fig:feyn-doubletop} contribute. In higher
orders also quark-gluon processes exist. The relative contribution of
these diagrams depend on the parton distribution functions. At the Tevatron
with a centre-of-mass energy of $1.96\TeV$ next-to-leading order predictions
lead to an expectation of $85\%$ contribution from $q\bar q$ annihilation and $15\%$ from gluon
fusion.
The total cross-section of top quark pair production 
has been computed in perturbation theory using various
approximations~\cite{Kidonakis:2003qe,Cacciari:2003fi,Cacciari:2008zb,Moch:2008qy,Moch:2008ai,Kidonakis:2008mu}.  
Its value has a significant dependence on the top quark mass, which near the
world average value is about $-0.2\pb/\GeV$. For a top quark
(pole) mass of $175\GeV$ Moch
and Uwer~\cite{Moch:2008ai} find $\sigma_{t\bar t}=6.90^{+0.46}_{-0.64} \pb$,
based on the CTEQ6.6~\cite{Pumplin:2002vw} PDF. The uncertainty includes
uncertainties of the PDF and the scale uncertainty.
% echo "7.80984-0.243547*4+0.425424*0.01*16-6.01643*0.0001*4*4*4+7.29114*0.000001*4*4*4*4"|bc
% echo "7.53093-0.235746*4+0.4212856*0.01*16-5.84596*0.0001*4*4*4+7.08203*0.000001*4*4*4*4"|bc
% echo "7.36172-0.22749*4+0.39309*0.01*16-5.5075*0.0001*4*4*4+6.6412*0.000001*4*4*4*4"|bc

\begin{figure}
  \centering
  \setlength{\unitlength}{1mm}
  \vspace*{5mm}
  \begin{fmffile}{mf_singletop}
\null\hfill
   \begin{minipage}[t]{35mm}
    \begin{fmfgraph*}(31,22)
      \fmfstraight
      \fmfleft{q1,q2}
      \fmfright{t,b}
      \fmf{fermion}{q1,v,q2}
      \fmf{photon,lab=$W$}{v,vv}
%      \fmfdot{v}
%      \fmfdot{vv}
      \fmf{fermion}{b,vv,t}
      \fmflabel{$q$}{q1}
      \fmflabel{$\bar{q}$}{q2}
      \fmflabel{$t$}{t}
      \fmflabel{$\bar{b}$}{b}
    \end{fmfgraph*}
   \end{minipage}\hfill
   \begin{minipage}[t]{35mm}
    \begin{fmfgraph*}(35,22)
      \fmfstraight
      \fmfleft{g,q}
      \fmfright{b,t,p,q2}
      \fmf{fermion}{q,v,q2}
      \fmf{photon,lab=$W$,label.side=right,tension=0.5}{v,v2}
      \fmf{gluon,tension=2.5}{g,v3}
      \fmf{fermion}{b,v3,v2,t}
%      \fmfdot{v}
%      \fmfdot{vv}
      \fmflabel{$q$}{q}
      \fmflabel{$q'$}{q2}
      \fmflabel{$g$}{g}
      \fmflabel{$t$}{t}
      \fmflabel{$\bar{b}$}{b}
    \end{fmfgraph*}
   \end{minipage}\hfill\null
\end{fmffile}\vspace*{5mm}
  \caption{Leading order Feynman diagrams contributing to single top quark 
    production
    at the Tevatron. According to the structure of the diagrams the left
    process is called $s$-channel and the right $t$-channel production.
}
  \label{fig:feyn-singletop}
\end{figure}
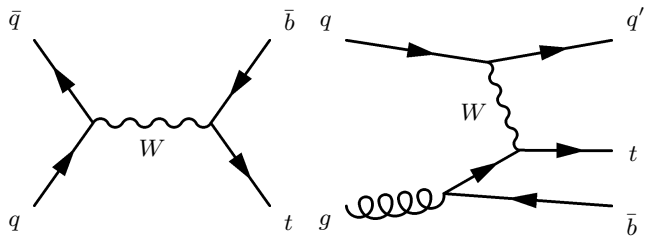

Single top quark production can only take place via the weak interaction. The
leading processes are  quark annihilation through a $W$ boson, also called
$s$-channel, the quark gluon fusion with a $W$ boson in a $t$-channel,
c.f.~\fig{fig:feyn-singletop}, and production of single top quarks 
in association with a (close to) on-shell $W$-boson. Charge conjugate
diagrams apply for anti-top quark production.
At the Tevatron Run\,II the cross-section to produce a single top or anti-top quark
is $3.4\pm 0.22\pb$. The $s$- and $t$-channel contribute a little less than $1/3$ and
$2/3$, respectively. The associated production 
contributes less than $10\%$~\cite{Harris:2002md,Sullivan:2004ie,Kidonakis:2006bu}.
These numbers were derived assuming $m_t=175\GeV$ and using the MRST2004~\cite{Martin:2004ir} PDFs.

After production top quarks decay very rapidly through the weak interaction
into a $W$ boson and a $b$ quark. In the Standard Model contributions from
decays to light quarks are suppressed due to the smallness of the corresponding
entries of the CKM matrix. The expected decay width of about $1.34\GeV$
corresponds to a lifetime of the order of $5\cdot 10^{-25}\,\mathrm{s}$~\cite{\refpdg}.
Thus the top quark decays before it can couple to light quarks and 
form hadrons. The lifetime of $t\bar t$ bound states, toponium, is too small, 
$\Gamma_{t\bar t} \sim 2\,\Gamma_{t}$, to allow for a proper
definition of a bound state as 
already pointed out in the early 1980s~\cite{kuehn1981}.

\begin{figure}[b]
  \centering
  \includegraphics[width=0.8\linewidth,clip,trim=0mm 0mm 0mm 15mm]{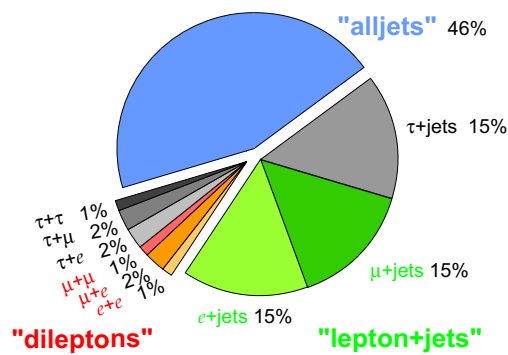}  
  \caption{Top quark pair branching fractions~\cite{d0topfigures}.}
  \label{fig:topbranching}
\end{figure}
The decay modes are defined purely by the $W$ boson
decays. $W$ bosons may decay to quarks, i.e. hadronically, or leptonically. 
For top quark pairs this yields three basic decay modes: the all hadronic
channel, the semileptonic channel and the dilepton channel. 
The all hadronic channel has a branching fraction of $46\%$. Each of the three
charged Standard Model leptons contributes $15\%$ in the semileptonic
channel. The dileptonic decays have a total branching fraction of $9\%$. See
\fig{fig:topbranching} for a graphical representation.
Decays that involve $\tau$ leptons are usually 
not considered in analyses of the semileptonic and dileptonic decay modes, because $\tau$
leptons are difficult to identify. The analyses, however, include the
events in which the tau decayed to an electron or muon.

Following this experimental nomenclature the semileptonic and
dileptonic channels are considered to include only electrons and muons. 
If an analysis  considers identified tau leptons this fact shall be explicitly stated.
For single top quark analyses so far only the leptonic $W$ boson decays (to electrons
and muons) are considered.

%%% Local Variables: 
%%% mode: latex
%%% TeX-master: "EPJC_TopProperties"
%%% End: 

\subsection{Experiments}
\label{sec:Experiments}
% $Id: Experiments.tex,v 1.17.2.8 2011-03-22 17:46:58 wicke Exp $ 

Up to recently 
only one collider provides centre-of-mass energies sufficiently high
to produce top quarks: the Tevatron at the Fermi National Accelerator
Laboratory (FNAL) near Chicago, IL, USA. At the two collision points 
typical general purpose experiments of present collider physics are
positioned, CDF and D\O. Each consists of a cylindrical part that covers
particles produced at large angles to the beam and two end-caps that detect
particles at smaller angles. Close to the beam tracking and vertex
reconstruction components are placed, followed to the outside by calorimetry
and muon detection systems. 
Some more details of the Tevatron and the two detectors shall be described
below. 

\subsubsection{The Tevatron}

The Tevatron collides beams of protons and antiprotons.
The protons and antiprotons are produced and
pre-accelerated in a series of smaller machines and then filled into the
Tevatron to circulate in opposite directions.
In a first phase of operation between 1992 and 1996 the beams were accelerated
to $900\GeV$. This phase is commonly called Run\,I. 
After an upgrade the Run\,II started in 2001. The upgrade enables the Tevatron
to accelerate the beams to a final 
energy of $980\GeV$ to yield a centre-of-mass energy of
$1.96\TeV$.  Also the peak
luminosity was gradually improved and now reaches a factor of 
approximately 10 over the 
Run\,I performance~\cite{tev-rookie-books,tev-2}.
%The beams collide at two interaction points
%where the detectors of the CDF and D\O\ collaboration record the events.

In Run\,I an integrated luminosity of about $160\ipb$ was delivered to both
experiments. This was sufficient to discover the top quark in
proton-antiproton collisions with a centre-of-mass energy of
$1.8\TeV$~\cite{Abe:1995hr,Abachi:1995iq}.  With the increased centre-of-mass
energy a $40\%$ increased cross-section was achieved for top quark pair
production.  At the time of writing an
integrated luminosity of more than $6\ifb$ was delivered to both experiments.
Preliminary results using up to $3.6\ifb$ have been made public by the experiments.

It is currently foreseen to continue running the Tevatron at least until the end
of 2010 and an extension into 2011 is being discussed. The total integrated
luminosity is expected to increase by about $2.5\ifb$ for each year of running.
Thus until the end of the Tevatron program the total luminosity will more than double
compared to what has been analysed so far~\cite{tev-plans}.

\subsubsection{The CDF Detector}
The Collider Detector at Fermilab (CDF)~\cite{\refcdfdet} is one of two
detectors recording collisions at the Tevatron. 
The vertexing and  tracking  components, the  calorimetry
and muon detection systems as used in the Run\,II measurements
shall be described shortly in turn.

The tracking system of CDF is placed in a $1.4\T$ solenoidal magnetic field.
The heart of the CDF tracking for Run\,II consists of three separate silicon
detectors. The innermost (L00) consists of a single layer of single sided sensors at
a radius of about $1.5\cm$.  The upgraded silicon vertex detector (SVX II) consists
of five cylindrical double-sided layers that along the beam axis reach to $\pm45\cm$  from the centre of the
detector at radii between $2.5\cm$ and $10.6\cm$~\cite{Sill:2000zz,Bardi:2001uv}.
Finally the intermediate silicon layer (ISL) consists of 
double-sided sensors in one central layer at $23\cm$ and in two layers at
$20\cm$ and $29\cm$ for the forward regions~\cite{Affolder:2000tj}.
The tracking system is completed by wire drift chambers with an outer radius
of $1.32\m$. Their length of $3.2\m$ allows to cover the central region of
$\left|\eta\right|<1$.  The tracking systems provides full coverage and thus
precise tracking for $\left|\eta\right|\le2.0$  to measure the momentum of
charged particles, to find the primary vertex of the collision as well as to find
possible secondary vertices from long lived particles like $b$ quark hadrons.

Outside the tracking system a time of flight system (TOF) based on
scintillator bars is positioned. This allows particle identification and is
used in identifying $b$ hadrons~\cite{Acosta:2004kc}.  The solenoid that
provides the magnetic field for the tracking system is placed  between the
TOF and the calorimetry.
\begin{figure}[t]
    \centering
\includegraphics[width=0.77\linewidth]{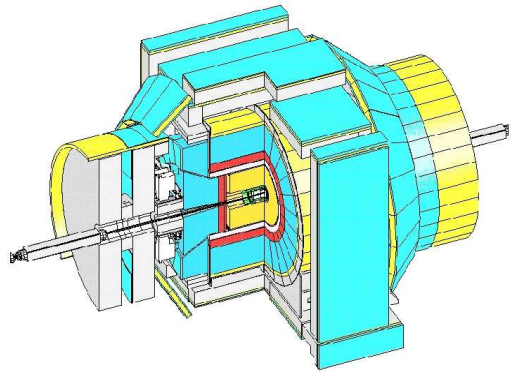}
\caption{Schema of the CDF detector at  Run\,II~\cite{Sill:2000zz}.}
\label{fig:cdf-det}
\end{figure}

The central electromagnetic calorimeter is composed of alternating layers of
lead and scintillator. It covers the pseudo-rapidity range of
$\left|\eta\right|<3.5$. To detect more
forwardly produced particles layers of lead and proportional chambers are
installed at pseudo-rapidities $\left|\eta\right|<4.2$.
% For more
% forwardly produced particles lead and proportional chambers are used to
% cover pseudo-rapidities to $\left|\eta\right|<4.2$.
% The achieved resolution is $\sigma_E/E\sim 10/\sqrt{E/\GeV}\oplus 7\%$.
The hadronic calorimeter uses lead as absorber material. The central region
and the end cap wall use scintillator as active material. The forward
regions also use gas proportional chambers.  In total the calorimeter covers
pseudo-rapidities of $\left|\eta\right|<4.2$.  The calorimeter allows energy
measurements and identification of electrons, photons, jets and missing
transverse momentum.

The calorimeter is surrounded by several systems of drift chambers to
identify muons. The `central muon system' is placed at a radius of
$347\cm$. Behind an additional 3.3 interaction lengths of $60\cm$ steel of
the return joke the central muon upgrade system is located. Both systems
cover pseudo-rapidities of $\left|\eta\right|<0.6$. The third system, the
central muon extension, extends this coverage at
$0.6<\left|\eta\right|<1.0$. So called barrel muon chambers extend the
coverage for $1.0<\left|\eta\right|<1.5$.  Track segments in these
components are used to identify muons.

CDF uses a three-level trigger system. The level 1 system has a pipeline for
$42$ beam-crossings. After the level 1 trigger the event rate is approximately
$10\kHz$. At level 2 trigger processors analyse a substantial fraction of
the event and further reduce the rate to $200\Hz$. The 3rd level consists of
a cluster of computers which perform an optimised event reconstruction. With
this information the event rate is reduced to about $40\Hz$. The events
selected by level 3 are stored permanently to tape.

\subsubsection{The D\O\ Detector}

The D\O\ (Dzero) detector~\cite{\refdzerodet} is the second of the two
detectors recording collisions at the Tevatron. 
It has been significantly upgraded to adapt to the reduced bunch distance and
increased luminosity of the Tevatron Run\,II. The upgraded D\O\ detector shall
be shortly described here.

The tracking system at the centre of D\O\ has been fully replaced since Run\,I
and now consists of a silicon  micro-strip tracker (SMT) and a
scintillating-fibre tracker within a $2\T$ solenoidal magnet. The SMT consists
of a barrel with four layers of single and double sided silicon micro-strip
detectors with a total length of about 70\cm. 
The barrel is separated  into six subsections
along the beam-pipe. Each subsection is capped with a
disk of silicon detectors (F-disks). Three additional F-disks are
placed on each side further outside of the barrel. Larger disks (H-disks) are
placed at distances of $100\cm$ and $121\cm$ from the beam pipe.
The SMT barrel provides excellent $r$-$\phi$-information, the disks
provide $r$-$z$ as well as $r$-$\phi$ measurements. 
During a shutdown period in 2006 the
D\O\ silicon system was extended by adding an additional layer of sensors
directly on top of the beam pipe. This Layer-0 significantly improves the ability of
vertex reconstruction. D\O\  measurements are usually performed separately for
data taken before and after the installation of Layer-0. The two run periods
are commonly denoted as Run\,IIa and Run\,IIb.
Outside the SMT the central fibre tracker (CFT) is placed.  It consists of
scintillating fibres mounted on eight concentric support cylinders at radial
distances between $20$ and $52\cm$ covering $\left|\eta\right|$ to about $1.7$.
At each distance one  layer of fibres is oriented along the beam axis 
and a second is mounted with stereo angles of of $\pm 3^\circ$.

\begin{figure}[b]
  \centering
  \includegraphics[width=0.9\linewidth]{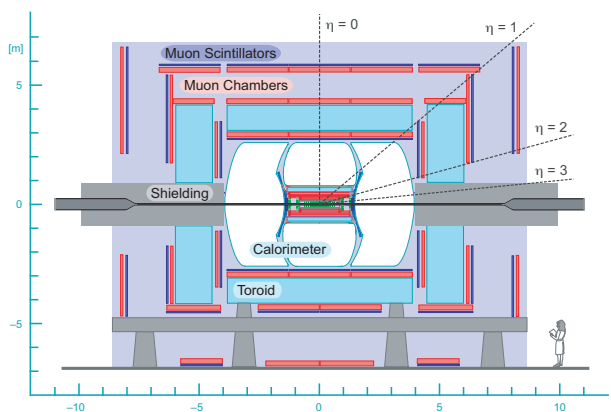}
  \caption{Schematic view of the Run\,II D\O\ detector~\cite{d0-det-scheme}.}
\label{fig:d0schema}
\end{figure}

To the outside of the CFT the solenoidal magnet is placed that produces a
$2\T$ magnetic field for the tracking components.
It is followed by the calorimetry that consists of a pre-shower detector that
is placed in front of the cryostat and the sampling calorimeter based primarily
on uranium/liquid-argon  inside the cryostat. The end-caps that cover the
forward regions have a similar structure, i.e. a pre-shower detector and a
calorimeter cryostat. 
The calorimeters consists of three regions. 
The innermost is the electromagnetic calorimeter which uses thin ($3-4\mm$) plates of
depleted uranium as absorber material. It is followed by the fine hadronic 
calorimeter with $6\mm$ thick plates of uranium-niobium. The outermost
subsection, the coarse hadronic calorimeters, uses thick (about $47\mm$) 
absorber plates of  copper (in the central) and
stainless steel (in the end-caps).
The active medium in all three regions is liquid argon.
Additionally, inter-cryostat-detectors (ICD) are mounted between the central and
the forward cryostat to improve on the incomplete coverage of the calorimeters at
$0.8<\left|\eta\right|<1.4$. 

The D\O\ muon systems outside the calorimeter consist of tree layers of
drift chambers, one inside and two outside a toroidal magnetic
field. The central muon chambers cover $\left| \eta\right|\lesssim1.0$ with
proportional drift chambers and have a magnetic field of $1.9\T$. The
forward muon chambers use mini drift chambers and a toroidal field of $2.0\T$.
They extend the coverage to $ \left|\eta\right|\approx2.0$.

Also D\O\ uses a trigger system with three stages. 
The first level consists of a set of hardware trigger elements that provide a
trigger accept rate of $2\kHz$. In the second level hardware engines and
embedded microprocessors provide information to a global processor 
that considers individual detector information as well as correlations. It reduces
the rate by a factor of about $2$. The third level consists of a farm of computers
that reduces the rate to $50\Hz$ based on a limited event reconstruction.
The accepted events are stored to tape for offline analysis.

%%% Local Variables: 
%%% mode: latex
%%% TeX-master: "EPJC_TopProperties"
%%% End: 

\subsection{Basic Event Selection}
The selection of events in general, in particular the selection
of top quark events, is based on the reconstruction of a number of
different objects: The primary vertex of the collision, electrons, muons, jets
and transverse missing energy. In addition, to reduce the background, often methods to
identify jets from $b$-quarks are applied. 
In the following these objects
shall be shortly described in turn.

The reconstruction of the primary vertex is determined by assigning well
measured tracks to a common origin in the interaction region. The primary
vertex is constructed event-by-event and is used as reference for some of the
following objects.

Electrons are reconstructed by a combination of tracking and calorimeter
information. Quality cuts on the tracks typically include a $p_T$ threshold of
the order of $10\GeV$ and the matching energy deposit in the calorimeters
should be well contained in the electromagnetic subsection. In the absence of
bremsstrahlung the energy deposit is expected to
have a small radius in the $\eta$-$\phi$ plane. Sophisticated algorithms take
into account bremsstrahlung photons. Discriminating observables include the
relative sizes of the energy deposit in the hadronic and the electromagnetic
calorimeter. D\O\ uses a fixed ratio of the electromagnetic to the total 
energy $f_\mathrm{em}=E_\mathrm{em}/E<0.9$~\cite{Abazov:2007kg}
while CDF uses an energy dependent cut on $E_\mathrm{em}/E_\mathrm{had}\le
0.055+0.00045\GeV^{-1} E$~\cite{Abulencia:2006kv}.

Muons are identified by the presence of signals in the dedicated muon
chambers that can be matched to a track found in the tracking system. 
For this purpose CDF and D\O\ extrapolate isolated tracks with standardised
quality cuts through the calorimeter out to the muon chambers to find matching
track segments. 

For both lepton objects different quality classes are defined, named
``loose'', ``medium'', ``tight'', etc, with increasingly stricter
requirements on the isolation of the track and the calorimeter cluster. 
These different criteria can be used to obtain an acceptable signal to
background ratio. The selection of ``loose'' leptons excluding those that also
have a ``tighter'' identification is often used to define sideband samples to
extract background estimates from data. 
The triggering and reconstruction efficiencies are usually studied in
$Z\rightarrow \ell\ell$ events using the tag and probe method. 

Jets are reconstructed from all calorimeter towers. They usually show a
substantial contribution from the hadronic subsection and are usually broader
than signals from electromagnetic particles. At the Tevatron experiments
the ``improved legacy'' 
cone algorithms~\cite{Blazey:Baur:2000bi}
with radii of $0.4$ and $0.5$ are used by  CDF and D\O,
respectively. Quality cuts typically require that the energy of a jet is not
contained to more than $90\%$ in a single tower and deposits from electron and photon
candidates are removed. The jet energy reconstructed from the calorimeter
cells needs to be corrected for  a number of effects. These corrections
include imperfections of the calorimeter but also  energy
offsets due to contributions from the underlying event, multiple hadron
interactions and noise in the electronics. This correction, usually called the
Jet Energy Scale (JES), is obtained from precisely measured electromagnetic
objects by assuming momentum conservation in the transverse plane for
$\gamma+$jets events. 

Jets stemming from $b$-quarks can be identified due to the long lifetime of
about $1.5\ps$ of the $B$ hadrons in such jets. At the typical energies in top
quark events of $50$ to $100\GeV$ the mean decay length is of the order of $5\mm$.
This fact is exploited by computing the impact parameter for tracks or by
explicitly reconstructing a secondary vertex from the tracks that is displaced
from the primary vertex.
To identify $b$-jets CDF uses only tracks that fall within the cone radius of
the considered jet. A secondary vertex is reconstructed in two passes with
different track requirements~\cite{Acosta:2005zd}. 
The 2d decay length is computed as the distance of primary
to the secondary vertex. The significance, i.e. the decay length over its
uncertainty, is required to be larger than 3. If the direction from the
primary to the secondary vertex is opposite to the direction of the jet, the tag is
called a negative tag. These negative tags are useful to determine the purity
of mis-tag rates.
In {D\O} first track jets are built with a cone algorithm independent of the
calorimeter jets described above.  Secondary vertices are reconstructed with
tracks from a given track jet. Calorimeter jets are considered as identified $b$-jets,
if an identified track jet falls within a radius of $\Delta R<0.5$.
The most recent {D\O} tagging algorithm uses the
impact parameters of tracks matched to a given jet  and
information on the secondary vertex mass, the significance of displacement, and the number
of participating tracks  for any reconstructed secondary vertex
within the cone of the given jet. The information is combined in a  neural
network to obtain the output variable,
$\mathrm{NN}_B$, which tends towards one for
$b$-jets and towards zero for light jets~\cite{Scanlon:2006wc}.

The presence of a neutrino is inferred  from the 
momentum imbalance in the transverse plane, that occurs because 
neutrinos are invisible to the detectors.
If all objects of an event were measured the sum of the transverse momenta
should vanish. Thus the sum of the transverse momenta of all neutrinos
can be deduced from the missing transverse momentum needed to assure momentum conservation.
It is derived from the calorimetric measurements and their direction with respect to
the interaction region as the negative vector sum of the transverse energies 
and thus usually named  Missing Transverse Energy, $\met$.
The sum is taken over all calorimeter cells that remain after noise
suppression. In general corrections for identified objects with known energies
like electrons, muons or jets are applied.

Triggering and preselection of single top quark events and of
top quark pair events in the semileptonic and dileptonic channels are
based on reconstructed lepton, jet and $\met$ objects described above. Typically
the leptons and the missing transverse energy are required to have
$p_T>20\GeV$. Signal selection in addition requires the presence of
jets also with a typical momentum requirement  $p_T>20\GeV$. The number of
jets required of course depends on the channel under consideration. Some
analyses here improve the signal to background ratio by requiring one or
more identified $b$-jets. Others construct a likelihood for an event being top
quark like based on topological quantities to consciously avoid $b$-jet identification.
Additional cuts may be introduced to enhance the signal to background ratio or
to improve the data to Monte Carlo agreement. In the semileptonic selection 
D\O\ e.g. recently requires the leading jet to fulfil $p_T>40\GeV$ and avoids
events in which missing transverse momentum and the selected lepton are
aligned. 

Events of the all-hadronic channel have to be selected by requiring multijet
final states. Due to the overwhelming background from  multijet events
in these analyses usually stronger cuts are applied on the transverse momentum
of the jets.

%%% Local Variables: 
%%% mode: latex
%%% TeX-master: "EPJC_TopProperties"
%%% End: 

%
%\cleardoublepage
\section{Top Quark Mass Measurements}
\label{section:Masses}
% $Id: Masses.tex,v 1.73.2.31 2011-03-22 17:47:03 wicke Exp $
%\subsection{Introduction}
The mass is the only property of the top quark that is not fixed within the 
Standard Model of particle physics. The Yukawa coupling responsible for the 
coupling of the top quark to the Higgs boson and thus for the mass of the top quark is 
a free parameter of the Standard Model. 
This already illustrates the importance of measuring the top quark mass. 

Moreover Standard Model  predictions of electroweak precision observables depend on the value 
of the top quark mass via radiative corrections. By correlating the $W$ boson
mass with the top quark mass the mass range for the yet undiscovered Higgs
boson can be constrained. Measurements of the top quark mass are thus an
important preparation for discovering the Higgs boson and will serve as a
consistency check of the Standard Model  after its discovery.

In the following theoretical aspects of fermion mass measurements are
discussed, before the experimental methods used in the various decay 
channels are explained. Then issues of modelling
non-perturbative effects in $p\bar p$ collisions and their influence
on the existing measurements are discussed and
the combination of the various measurements to a final result is reviewed.
The conclusions contain a critical comparison of current results with
expectations and future prospects.

\subsection{Theoretical Aspects}
%    - Definition der Masse
%    - Renormierung von Massen
%    - Was ist der messbare Parameter
\newcommand{\msbar}{\overline{\mathrm{MS}}}
For free particles the physical mass is usually taken to correspond to  the pole of their
propagator, i.e. their value of the four-momentum squared,
$p^2=E^2-{\vec{p}}\,^{2}$. % , which occurs after a long distance of flight.
%
%Was ist mit massen definitionen für e.g. elektronen?
%
Because of confinement quarks cannot exist as free particles and
this definition becomes ambiguous. 

The definition of the pole mass is still
possible on an order by order basis in perturbation theory, but is considered
to be intrinsically ambiguous on the order of the confinement scale
${\cal O}(\Lambda_\mathrm{QCD})$~\cite{Laenen:2008jx}.
Mass definitions can also be obtained following other renormalisation schemes
such as the $\msbar$-mass in the  $\msbar$-scheme.
Other definitions have been suggested by \cite{Fleming:2007tv,Hoang:2008xm}.

The relation between the various mass definitions can usually be computed in
perturbation theory. For the top quark the numerical values of the different
definitions may differ
significantly. In NNLO for example the $\msbar$-mass of the top quark is about
$10\GeV$ smaller than the pole mass.

This large difference makes it important to understand which definition and
order of perturbation theory of the top quark mass is measured by the
experiments.
As we shall see below, all direct methods to determine the top quark mass use
Monte Carlo simulation to either extract the mass or 
calibrate the procedure. Thus the question really is, which definition of
the top quark mass is used as a parameter in these Monte Carlo generators.

\label{sec:mass-theo}
Unfortunately, it is not well understood which field theoretic mass 
definition the currently used
generators correspond to. Clearly, as the hard matrix elements of (most) 
generators is implemented in leading order, they correspondingly use the
leading order mass. It is usually argued that this
corresponds to the pole mass, because in the pole mass definition any shifts
of the position are to be absorbed in the mass definition. Monte Carlo
generators do not absorb corrections from the parton shower or the
hadronisation into the mass definition. And it is not clear in how far the
parton shower and the modelling of hadronisation alter the mass definition,
nor which approximation of QCD this mass parameter corresponds to.
Partial answers have been given in \cite{Hoang:2008xm},
but at this point a conceptual uncertainty remains that is considered to be
of the order of $1\GeV$.

Comparisons of the top quark mass from direct
measurements with electroweak  precision data to determine e.g. the Higgs boson
mass  %as shown in \fig{fig:ewww.mwmt}
currently assume the measured top quark  mass values corresponds to the pole
mass definition.  They thus have to be interpreted with care.

Experimentally precise measurements of the top quark mass are nevertheless useful. On the one
hand they are needed to set the mass parameter for simulating top quark events, which
will be important backgrounds for some LHC searches. On the other hand the
consistency between the various experimental methods and the top quark decay
channels gives confidence in these methods.
Finally, it seems feasible that  a theoretically more precise specification
of the top quark mass definition used in the Monte Carlo generators can be
derived in the future~\cite{DELPHI2003-061phys932}. Any bias from the interpretation as a
pole mass may then be corrected for. 

%%% Local Variables: 
%%% mode: latex
%%% TeX-master: "EPJC_TopProperties"
%%% End: 

\subsection{Lepton plus Jets Channel}
The lepton plus jets channel is considered to be the golden channel in top
quark mass measurements.
Due to a lepton and a neutrino in the final state it has a good signal to
background ratio.
In addition one of the top quark quarks is fully measured in the detector.

Several method have been applied by CDF and D\O\ to measure the top quark mass in this channel. 
The methods exploit the kinematics of the events to  different levels
and make different assumptions about details of the production mechanisms.

\subsubsection{Template Method}

The basic template method was already applied to lepton plus jets events in the
papers describing the first evidence and discovery of the top quark.

In this method a top quark mass is reconstructed for the each of the selected
events using the momenta measured for lepton and jets  and the transverse missing
energy. The distribution of the reconstructed masses is then compared
to template distributions from simulation. 
These templates are constructed from signal Monte Carlo
with varying top quark mass values and contain the expected amount of
background events. The method thus relies on good Monte Carlo modelling
of signal and background.

\subsubsubsection{CDF Run\,II Template Measurement}
\label{CDF-template-ljets-portion}

CDF has applied an improved template method in up to $5.6\ifb$ of data combining
the lepton plus jets channel and the dilepton
channel~\cite{Aaltonen:2008gj,CdfNote9578,CdfNote9679,CdfNote10273}. 
% Previous analysis l+jets only: Abulencia:2005aj
For clarity the two analysis parts will be described separately. 
The dilepton portion of this analysis can be found in
Section~\ref{CDF-template-dilepton-portion}. 

Lepton plus jets events are selected requiring a single high $p_T$ lepton,
large missing transverse momentum and at least four jets.  
Events are separated by the number of jets identified as $b$-jets based on the
transverse decay length of track inside the jet~\cite{\refcdfsecvtx}.
In case of only a single
identified jet only events with exactly four jets are considered. For events
with more than one identified $b$-jet more than four jets are allowed. 
%and the jet $p_T$ requirement on the fourth jet is loosened.
\begin{figure*}
  \centering
\includegraphics[width=0.42\textwidth,clip,trim=0mm 0mm 0mm 12.5mm]{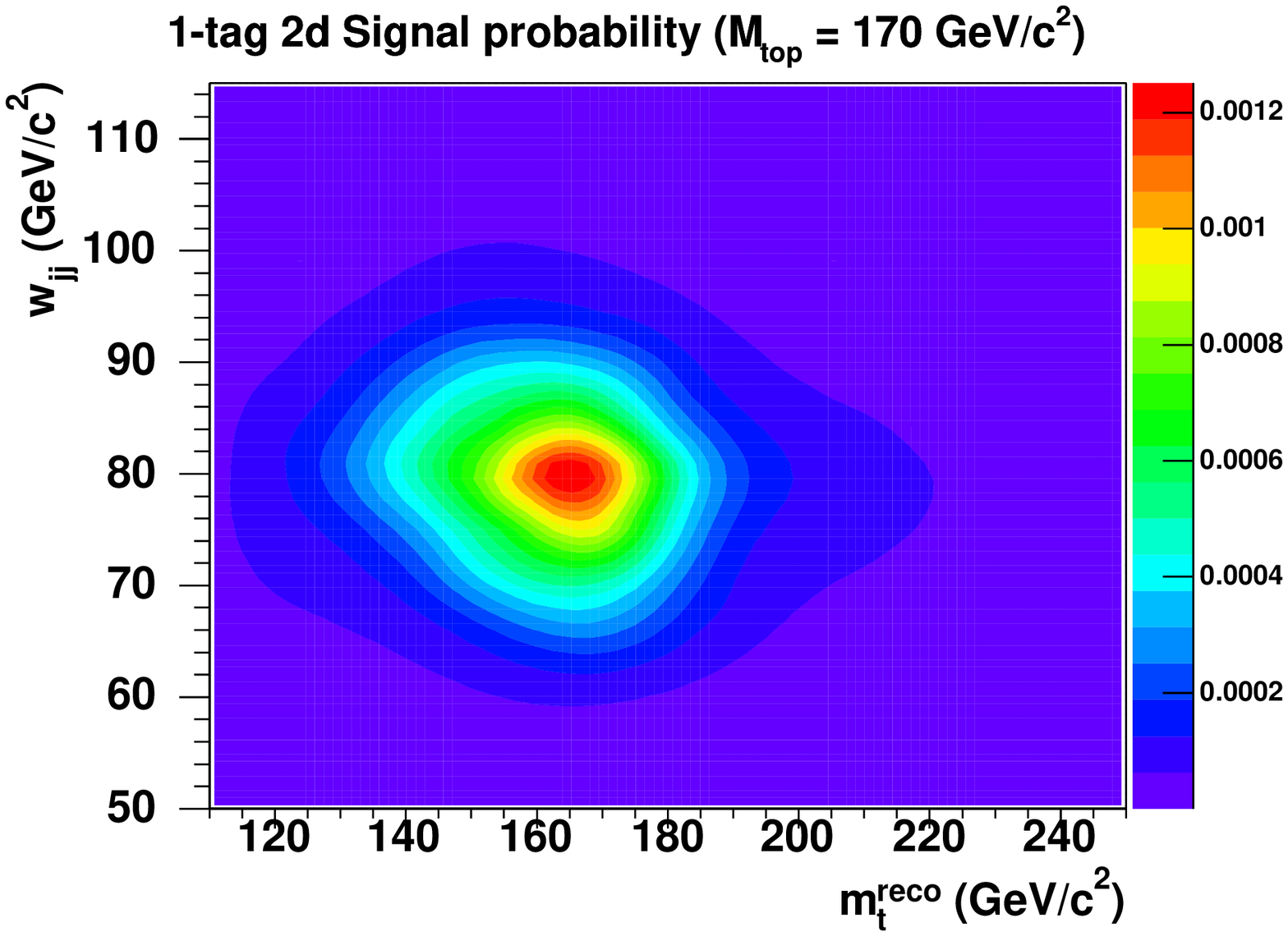}
\includegraphics[width=0.42\textwidth,clip,trim=0mm 0mm 0mm 12.5mm]{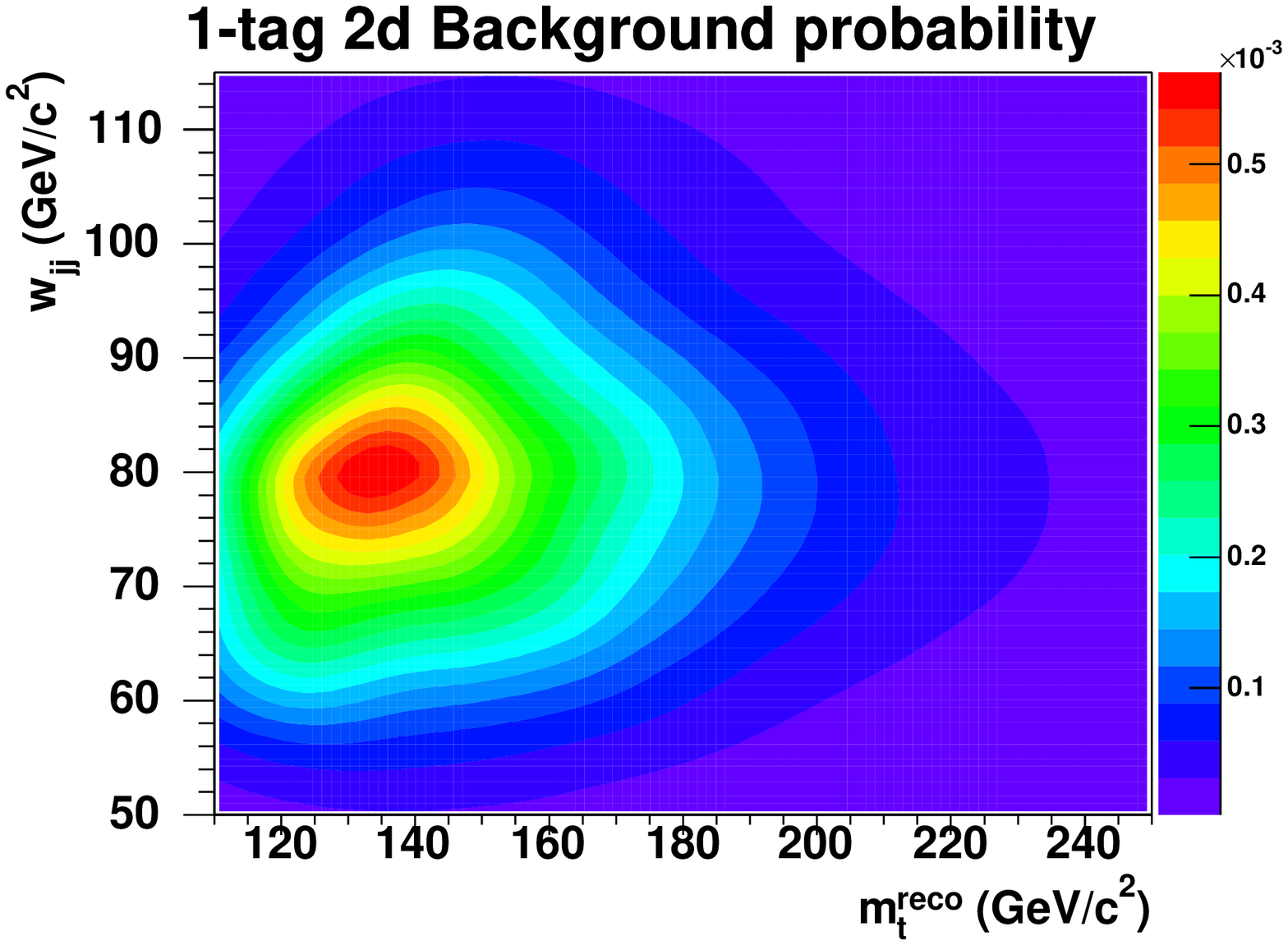}
  \caption{Probability densities for signal (left) and background (right) for
    nominal top quark  mass of $170\GeV$ and no jet energy scale shift as obtained
    for single $b$-tag events~\cite{CdfNote9578}.}
  \label{fig:cdf-template-densities}
\end{figure*}

In each event a top quark  mass $m_t^\mathrm{reco}$ is fitted using a constrained fit.
Besides the top quark  mass, the momenta of the top quark decay products (the quarks and
leptons) are fitted to the observed transverse jet and lepton momenta,
$p_T^{\mathrm{obs}}$, and the unclustered
energy, $\vec{U}_T^{\mathrm{obs}}$. 
The unclustered energy is the calorimetric energy in the transverse direction
not associated with any reconstructed object. 
Jet momenta are corrected to parton
level with CDFs common jet energy scale correction. 
The invariant masses of the $W$ boson decay products on both sides,
$M_{q\bar q}$ and $M_{\ell\nu}$, are constrained to be consistent with the nominal $W$ boson mass within the $W$
boson width.  The reconstructed top quark  mass $m_t^\mathrm{reco}$ is required to be
consistent within the top quark  width with the invariant mass of the top quark decay 
products on both sides, $M_{bqq}$ and $M_{b\ell\nu}$. The fit $\chi^2$ thus is
written as
\bea
\label{eq:cdf-constraint-chi2}
\chi^2&=&
\frac{(p_T^{\ell}-p_T^{\ell,\mathrm{obs}})^2}{\sigma_\ell^2}+
\sum_{i=1}^4\frac{(p_T^{i,q}-p_T^{i,\mathrm{jet}})^2}{\sigma_i^2}\nonumber\\
&&+\frac{(\vec{U}_T^{\mathrm{fit}}-\vec{U}_T^{\mathrm{obs}})^2}{\sigma_\ell^2}
+\frac{(M_{q\bar q}\!-\!M_W)^2\!\!}{\Gamma_W^2}
+\frac{(M_{\ell\nu}\!-\!M_W)^2\!\!}{\Gamma_W^2}\nonumber\\
&&+\frac{(M_{bqq}\!-\!m_t^\mathrm{reco})^2\!\!}{\Gamma_t^2}
+\frac{(M_{b\ell\nu}\!-\!m_t^\mathrm{reco})^2\!\!}{\Gamma_t^2}\quad\mbox{.}
\eea
The fitted unclustered energy, $\vec{U}_T^{\mathrm{fit}}$, is related to the
transverse momentum of the neutrino used in the computation of $M_{\ell\nu}$ and
$M_{b\ell\nu}$. This  $\chi^2$ is computed for all possible associations of
quarks to four jets, allowing $b$-jets only to be matched with $b$ quarks.
Only the top quark  mass value of the 
association with the best fit $\chi^2$ is kept. If this best
$\chi^2>9.0$  the event is rejected.

To constrain the jet energy scale simultaneously with the top quark  mass, the dijet
mass, $m_{jj}$, is measured from the non-$b$-tagged jets among the four
leading jets without applying the above kinematic fit. 
This is only unique for events with two identified jets. In
other events the two jets that yield the dijet mass closest to the $W$-boson
mass are chosen.

Thus for each event one top quark  mass value, $m_t^\mathrm{reco}$, and one dijet
mass, $m_{jj}$, is entering the following analysis.

Monte Carlo simulations are used to determine the expected behaviour of signal
events as well as the background contributions. The dominant $W+$jets
background is simulated with
\alpgen+\pythia~\cite{Mangano:2002ea,Sjostrand:2006za} 
with a normalisation derived
from data. The multijet background is modelled with samples containing
non-isolated leptons. Smaller background from single-top and diboson events are taken
from Monte Carlo with normalisation to theoretical cross-sections. All
backgrounds are assumed to have no dependence on the nominal top quark  mass, 
but are allowed to vary with the jet energy scale.
Signal samples for various nominal top quark  mass values, $m_t$, are generated using
\pythia.
All simulations are passed through the full CDF detector simulation and reconstruction.

These simulations are used to generate probability density functions,
$P^\mathrm{sig}$,
%(m_t^\mathrm{reco},m_{jj};m_t,\Delta_\mathrm{JES})$, 
to find a signal  event with
measured values $m_t^\mathrm{reco}$ and $m_{jj}$ given a nominal top quark  mass,
$m_t$ and jet energy scale shift, $\Delta_\mathrm{JES}$. 
Similarly a probability density for background events, $P^\mathrm{bkg}$, is
computed as function of the jet energy scale shift, only.
CDF uses a Kernel
Density Approach, where each simulated event contributes not only with its
reconstructed values $m_t^\mathrm{reco}$ and $m_{jj}$, but also in the
neighbourhood around these. The size of this neighbourhood is controlled by
smoothing parameters. The parameters are chosen dynamically: small near the maximum of a given
distribution where the statistics require less smoothing and larger in the tails. 
Examples of the resulting density functions are shown in \fig{fig:cdf-template-densities}.

With these probability densities the likelihood 
\linebreak
${\cal L}(m_t,\Delta_\mathrm{JES})$ is 
constructed. The number of signal and background events in the single $b$-tag
and the double $b$-tag sample are used as additional parameters with Gaussian
constraints on the background estimations obtained above. Also the jet energy
shift, $\Delta_\mathrm{JES}$, is constrained by a Gaussian term to be
consistent with zero within its nominal uncertainty, $\sigma_c$. 
\bea\label{eq:cdf-ljet-template-likelihood}
&&{\cal  L}(m_t,\Delta_\mathrm{JES})=\nonumber\\
&&\qquad\exp\left(-\frac{\Delta_\mathrm{JES}^2}{2\sigma_c}\right)\times{\cal
  L}_1(m_t,\Delta_\mathrm{JES})\times {\cal  L}_2(m_t,\Delta_\mathrm{JES})
\nonumber\\
&&{\cal  L}_i(m_t,\Delta_\mathrm{JES})=\nonumber\\
&&\qquad\exp\left(
-\frac{(b_i-b_i^\mathrm{e})^2}{2\sigma_{b_i}^2}\right)
\prod_\mathrm{events}
\frac{s_iP_i^\mathrm{sig}+b_iP_i^\mathrm{bkg}}{s_i+b_i}
\eea
Here $i=1,2$ indicates the subsamples with one or more than one 
identified $b$-jets,
respectively. $s_i$ and $b_i$ are the number of signal and background
events  and $b_i^\mathrm{e}$ the background expectations in the samples. 

The described likelihood formulae can only be evaluated at the discrete values
of $m_t$ and $\Delta_\mathrm{JES}$ for which simulations were run. To obtain
the likelihood for arbitrary values of $m_t$ and $\Delta_\mathrm{JES}$ a quadratic interpolation is used. 
The central result is obtained by maximising the likelihood and its
uncertainty is quoted as the (largest) mass shift corresponding to a likelihood 
change of $\Delta \log{\cal L}=0.5$.

Systematic uncertainties are evaluated by modifying several aspects of the
analysis described.  The dominating uncertainty in top quark  mass measurements is 
the jet energy scale. Through the simultaneous fit its contribution is 
part of the statistical uncertainty in this measurement. 
Residual effects remain through uncertainties in the $p_T$ and $\eta$
dependence. The uncertainty of modelling the top quark  pair signal events is evaluated
by comparing pseudo experiments generated with \pythia\ and \herwig. These two
systematic uncertainties contribute with about $0.7\GeV$ to the uncertainty and 
are the two largest single contributions to the systematic
uncertainty. Additional contributions in order of decreasing importance 
include the  uncertainties on modelling colour
reconnection~\cite{Sandhoff:2005jh,Skands:2007zg,Wicke:2008iz}, 
the background shape,  the parton
density functions.

\begin{figure}[t]
  \centering
\includegraphics[width=0.8\linewidth]{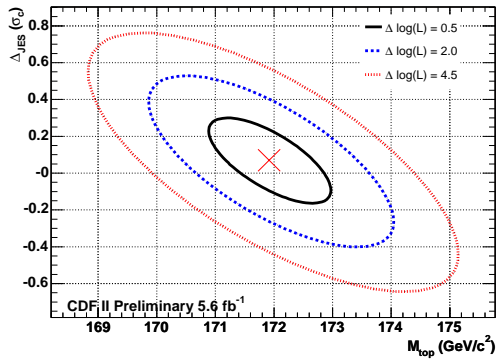}
  \caption{
    Likelihood contours for the combined lepton plus jet and
    dilepton fit of $5.6\ifb$ CDF data~\cite{CdfNote10273}.}
  \label{fig:cdf-template-contour}
\end{figure}
With $5.6\ifb$ of data using the lepton plus jets channel only CDF determines a
top quark  mass of~\cite{CdfNote10273} 
\beq
m_t=172.2\pm{}1.2_\mathrm{stat}\pm 0.9{}_\mathrm{syst}\GeV~\mbox{.} 
\eeq
For the jet energy scale shift the fit yields
$\Delta_\mathrm{JES}=(0.10\pm0.26)\sigma_c$, very consistent with the external
jet energy scale but with a significantly reduced uncertainty.
Combining the results with an analogous measurement in the dilepton channel
which uses a consistent jet energy scale shift
(c.f.~Section~\ref{CDF-template-dilepton-portion}) yields  
\beq
m_t=172.1\pm1.1_\mathrm{stat}\pm 0.9{}_\mathrm{syst}\GeV~\mbox{.}
\eeq
The likelihood contour of the two dimensional measurement of $m_t$ and
$\Delta_\mathrm{JES}$ is shown in \fig{fig:cdf-template-contour}.

\subsubsubsection{D\O\ Run\,II Template Measurements}
% Gibt es nicht auf dem Noteserver (2005 conf) \cite{Top Quark MassTemplate2005}

D\O\ has produced preliminary results for top quark  mass measurements using the
template method~\cite{d0note4574,TopMassTemplate2005}.
The method differs from the CDF method as 
the templates are not smoothened and in the use of a binned likelihood. 
As these results were not updated nor published in the last years they shall
not be described in more detail here.

\subsubsection{Ideogram Method}
\label{sec:ljets-ideogramm}

The ideogram method was transfered to top quark  mass measurements from a  method used
in the DELPHI $W$ boson mass measurement~\cite{Abreu:1997ic,Abreu:1999qk}. 

Also this method relies on an event by event reconstruction of top quark  mass
values. But in contrast to the template method the signal likelihood is
constructed event by event from the theoretically expected Breit-Wigner distribution smeared with
the experimental resolution of each individual event. Events with a
configuration that allows a more precise reconstruction thus contribute more
than events with a configuration that is difficult to reconstruct.

\subsubsubsection{D\O}
D\O\ has published an analysis of $425\ipb$ of data based on semileptonic top quark 
pair events~\cite{Abazov:2007rk}. The event selection requires an isolated
lepton, missing transverse energy and four or more jets. To identify $b$ quark
jets the decay length significance of a secondary vertex reconstruction is
used~\cite{Abazov:2005ex}. No cut is placed on the number of identified $b$-jets.

The top quark  mass in each event is reconstructed using a constrained fit that
determines the momenta of the top quark pair decay products ($\ell\nu b \bar b
q\bar q^\prime)$ from the measured momenta of the
charged lepton, the four leading jets and the missing 
transverse momentum and their uncertainties.
Constraints are placed on the invariant mass of the two light quarks and the two
leptons, respectively, which are required to be consistent with the $W$ boson
mass. The reconstructed top and anti-top quark masses are required to be equal.
The 12 possible assignments of jets to quarks that yield different constraints
are considered. In addition two possible solutions for the neutrino $z$
momentum are considered, which results in 24 top quark  mass values per event. 
As the common jet energy scale of D\O\ corresponds to the particle level,
i.e. what would be visible in an ideal detector, the fit uses jet-parton
mapping functions determined from \pythia\ simulation.
These mappings contain an overall scale factor, $f_\mathrm{JES}$, for the jet
energy scale that may modify the default jet energy scale.
% \begin{figure*}
%   \centering
% \includegraphics[width=0.8\textwidth]{figs/T07BF01.eps}  
%   \caption{Combined likelihood discriminant for $e+$jets (left) and $\mu+$jets
%     (right). The top quark  pair, $W$+jets and multijet background are normalised
%     according to a fit of their shapes to data~\cite{Abazov:2007rk}.
%           }
%   \label{fig:d0-ljets-ideogram-discriminant}
%\end{figure*}
\begin{figure*}
  \centering
\includegraphics[width=0.37\linewidth,clip,trim=0mm 0mm 95mm 0mm]{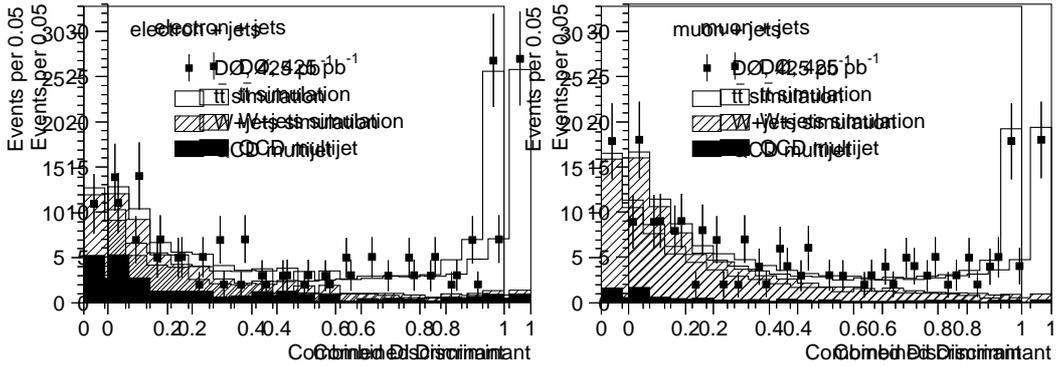}
%\\[2mm] 
~~~
\includegraphics[width=0.37\linewidth,clip,trim=85.5mm 0mm 10mm 0mm]{figs/T07BF01.eps}  
  \caption{Combined likelihood discriminant for $e+$jets (left) and $\mu+$jets
    (right). The top quark  pair, $W$+jets and multijet background are normalised
    according to a fit of their shapes to data~\cite{Abazov:2007rk}.
          }
  \label{fig:d0-ljets-ideogram-discriminant}
\end{figure*}

The described fit is repeated for various values of the jet energy scale 
factor, so that the resulting mass values and the fit $\chi^2$ are functions
of $f_\mathrm{JES}$.  Through the constraints to the nominal $W$ boson mass,
the fit $\chi^2$ is expected to be the smallest for the correct jet energy
scale (and top quark mass value). Events with no jet-parton combination reaching  
$\chi^2<10$ at the central jet energy scale are rejected at this point.

To determine the expected performance of the selection and the corresponding
background contamination, top quark  pair signal 
for various nominal top quark  mass values and $W+$jets background events are
generated using \alpgen+\pythia. The events are passed through the full detector
simulation and reconstruction. Multijet background is modelled with side band
data obtained from inverted lepton quality cuts. 
The sample composition is determined from a likelihood discriminant that
combines topological observables, with tracking based jet shape and $b$
tagging information, c.f.~\fig{fig:d0-ljets-ideogram-discriminant}. The
observables are selected to have low correlation to the top quark  mass.

An event by event likelihood  is now constructed to observe 
the top quark  masses reconstructed in the kinematic fit, the
discriminant and the number of identified  $b$-jets given the nominal values
of $m_t$ and $f_\mathrm{JES}$:
\bea
\label{eq:ljet-ideogram-likelihood}
{\cal L}_\mathrm{evt}(x;m_t,f_\mathrm{JES},f_\mathrm{top})
&=& f_\mathrm{top} ~ P_\mathrm{sig}(x;m_t,f_\mathrm{JES})\nonumber\\
&&+(1-f_\mathrm{top}) P_\mathrm{bkg}(x;f_\mathrm{JES})~\mbox{.}\qquad
\eea
Here $x$ represents the measured observables for the event under consideration
and $f_\mathrm{top}$ is the signal fraction of the corresponding sample.
The signal and background probabilities, $P_\mathrm{sig}$ and $P_\mathrm{bkg}$,
are proportional to the probabilities to find signal or background at the observed
discriminant value. % which is independent of the mass and the jet energy scale 
Because of the selection of observables in the discriminant, this factor
does not depend on the top quark  mass and the jet energy scale and is
factorised as:
\bea
P_\mathrm{sig}(x;m_t,f_\mathrm{JES}) &=&P_\mathrm{sig}(D)\,\,\,\tilde{P}_\mathrm{sig}(x;m_t,f_\mathrm{JES})\nonumber\\
P_\mathrm{bkg}(x;f_\mathrm{JES})     &=&P_\mathrm{bkg}(D)\,\tilde{P}_\mathrm{bkg}(x;f_\mathrm{JES})  ~\mbox{.}
\eea
The top quark  mass and jet energy scale 
dependent portion of the signal probability, $\tilde{P}_\mathrm{sig}$, is computed as the sum over all 24
jet-parton assignments and neutrino solutions. Their relative weight, $w_i$, is computed
from the $\chi^2$ probability of the kinematic fit and in presence of
identified $b$-jets includes the probability for the quarks associated to the
$b$-jets to produce a tagged jet. Because the result of the kinematic fit
depends on the jet energy scale factor the weights depend on $f_\mathrm{JES}$.
\bea
\label{eq:ljets-ideogram-weights}
\tilde{P}_\mathrm{sig}(x;m_t,f_\mathrm{JES})&=&\sum_{i=1}^{24}w_i(f_\mathrm{JES}) S(m_i,f_\mathrm{JES})\nonumber\\
\tilde{P}_\mathrm{bkg}(x;f_\mathrm{JES})&=&\sum_{i=1}^{24}w_i(f_\mathrm{JES}) B(m_i)
\eea
The background probabilities $B(m_i)$ are obtained from the simulated $W+$jets
events. The signal probabilities
\linebreak
$S(m_i,f_\mathrm{JES})$ are computed as the
convolution of a Breit-Wigner, ${\bf{BW}}$, that describes the theoretical distribution of
mass values and a Gaussian, $\bf{G}$, that represents the detector
resolution. 
As this ansatz is valid only for the correct jet-parton assignment and because
the wrong pairings do contain information about the top quark  mass, a second
contribution to describe wrong pairings is added:
\bea
\label{eq:ljet-ideogram-S}
S(m_i,f_\mathrm{JES})=f_c\!\! \int\limits_{m_\mathrm{min}}^{m_\mathrm{lax}}\!\!\!
{\bf{G}}(m_i,m',\sigma_i)  {\bf{BW}}(m',m_t)\,\,\mathrm{d}m'\nonumber\\
+(1-f_c) S_\mathrm{wrong}(m_i,m_t;n_\mathrm{tag})~\mbox{.}\quad
\eea
with $f_c$ being the fraction of events in which the weight is assigned to the
correct jet-parton pairing. For the width of the Gaussian the uncertainty, $\sigma_i$,
of the mass determination obtained in the constrained kinematic fit is
used. Thus events with a well-determined mass from the kinematic fit
contribute more than events with a less precise fit result.
The expected mass value distribution in wrong pairings, $S_\mathrm{wrong}$, 
also contains information about the true top quark  mass. 
It is determined from simulation as function of the number of identified $b$-jets.
Also the background function,  $B(m_i)$, is taken from simulation.

Due to the presence of wrong jet-parton assignments and background events the
described likelihood does not yield unbiased results for the jet energy scale
factor, $f_\mathrm{JES}$. The likelihood is thus corrected with an
$f_\mathrm{JES}$-dependent but mass-independent correction factor.
\begin{figure*}
  \centering
\includegraphics[width=0.86\textwidth]{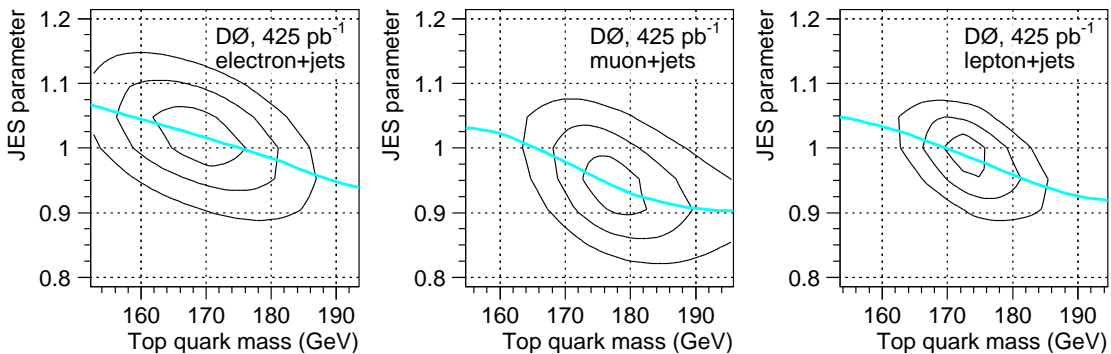}  
  \caption{Contours for the 2-dimensional likelihood of determined for $e+$jets
    (left), $\mu+$jets and the combined dataset in
    $425\ipb$~\cite{Abazov:2007rk}.
    The contours correspond to $\log$ likelihood
    differences of $0.5$, $2.0$ and~$4.5$.
          }
  \label{fig:d0-ljets-ideogram}
\end{figure*}

The likelihood for the complete sample of observed events is then simply the
product of the jet energy scale corrected likelihood,
${\cal L}_\mathrm{evt}^\mathrm{corr}$, for all individual events:
\beq
{\cal L}(m_t,f_\mathrm{JES},f_\mathrm{top})
=\prod_{\mathrm{events}}{\cal L}_\mathrm{evt}^\mathrm{corr}(x;m_t,f_\mathrm{JES},f_\mathrm{top})~\mbox{.}
\eeq
This likelihood is maximised simultaneously with respect to the top quark  mass,
$m_t$, the jet energy scale factor, $f_\mathrm{JES}$  and the signal fraction
$f_\mathrm{top}$.

Before this procedure is applied to data its performance  is
determined on large numbers of pseudo experiments with varying nominal
parameter values. The bias is determined as the mean difference between the nominal value and the
measured result. The correctness of the fit uncertainty is checked from the
distribution of pull values, i.e. the distribution of differences between the
measured values and their means normalised to the individual fit error. 
The bias and the width of the pull distributions for the fitted top quark  mass and
jet energy scale factor, $f_\mathrm{JES}$, are determined as function of
the nominal top quark  mass and jet energy scale factor simultaneously.
Linear corrections are applied to correct for the obtained biases and to
correct the uncertainty for  any deviation of the pull width from the ideal
value of one.

Systematic uncertainties for this measurement are determined from pseudo
experiments with events shifted according to the systematic variation under
study. %Individual systematic uncertainties are added quadratically. 
Due to the two dimensional fit the uncertainty of the overall jet energy scale is
contained in the uncertainty obtained from the likelihood fit. Residual
discrepancies between the data and Monte Carlo energy scales  still affect
the result and give the largest contributions to the systematic
uncertainties. D\O\ evaluates uncertainties due to $b$ quark fragmentation
modelling~\cite{Peters:2006zz} %,d0note5325 
and the calorimeter response to $b$-jets to $\pm 1.30$ and $\pm 1.15\GeV$,
respectively. The uncertainty from the $p_T$ of the jet energy scale yields
$\pm0.45\GeV$. Another large contribution of $\pm 0.73\GeV$ comes from the signal modelling which
is estimated by varying the fraction of high energy gluons produced in addition
to a top quark  pair. Further uncertainties include the influence of uncertainties on the trigger
efficiencies, the background modelling and the calibration.

The  ideogram method  with in-situ jet energy scale calibration
applied in $425\ipb$ of data~\cite{Abazov:2007rk} yields a top quark mass of 
\beq
m_t=173.7\pm4.4_\mathrm{stat+JES}{}^{+2.1}_{-2.0}{}_\mathrm{syst} \GeV~\mbox{,}
\eeq
the jet energy scale factor is determined to be $f_\mathrm{JES}=0.989\pm0.029_\mathrm{stat}$
consistent with the nominal value of $f_\mathrm{JES}=1.0$, which corresponds to the
calibration obtained in jet-photon events. The fit probability contour lines for the
individual lepton channels and the combined results are shown in~\fig{fig:d0-ljets-ideogram}.

\subsubsection{Matrix Element Method}
\label{sect:me-ljets} 
The Matrix Element method in the lepton plus jets channel 
was pioneered by D\O\ in Run\,I~\cite{Abazov:2004cs} and is
based on ideas described in \cite{Kondo:1993in,Dalitz:1998zn}. In this method
the likelihood is constructed according to the expected distribution from the
mass dependent top quark  pair production matrix element. With respect to the
previously described methods this
includes additional information from top quark  mass dependent kinematics into the measurement.

\subsubsubsection{D\O}
\label{d0-ljets-me-mass}
Results for the top quark  mass using the Matrix Element method have been regularly updated with D\O\ Run\,II
data~\cite{unknown:2006bd,Abazov:2008ds,d0note5750,d0note5877} with the latest result
including $3.6\ifb$ of data. The analyses select events from  semileptonic
top quark  pair decays by requiring a single isolated lepton, missing transverse
momentum and exactly four jets. In the recent analyses at least one of the jets needs to be
identified using D\O's neural network $b$ jet identification~\cite{Scanlon:2006wc}.
Vetoes are applied on additional leptons and
on events in which the lepton and the missing transverse momentum are close in
the azimuthal direction.

The top quark  mass and an overall jet energy scale factor are determined
simultaneously from an event by event probability that the
observed event may occur given the assumed values of the top quark  mass, $m_t$,  the jet
energy scale factor, $f_\mathrm{JES}$ and the signal fraction $f_\mathrm{top}$: 
\bea\label{eq:d0-ljet-me-master}
&&{P}_\mathrm{evt}(x;m_t,f_\mathrm{JES},f_\mathrm{top})
=\\
&&\quad A(x)\!\cdot\!\left( f_\mathrm{top} P_\mathrm{sig}(x;m_t,f_\mathrm{JES})
+(1\!-\!f_\mathrm{top}) P_\mathrm{bkg}(x;f_\mathrm{JES})\right)
\mbox{.}\nonumber
\eea
Here $x$ represents the measured momenta of the lepton, the jets and the
missing transverse momentum. 
$A(x)$ is the probability for the configuration $x$ to be accepted in the analysis.
The probabilities,
$P_\mathrm{sig}$ and $P_\mathrm{bkg}$, 
for signal and background events to be observed in the configuration $x$ 
given the set of parameters, $m_t$, $f_\mathrm{JES}$ and $f_\mathrm{top}$, are
computed from matrix elements of the dominating processes.

\begin{figure*}
  \centering
  \includegraphics[height=4.9cm]{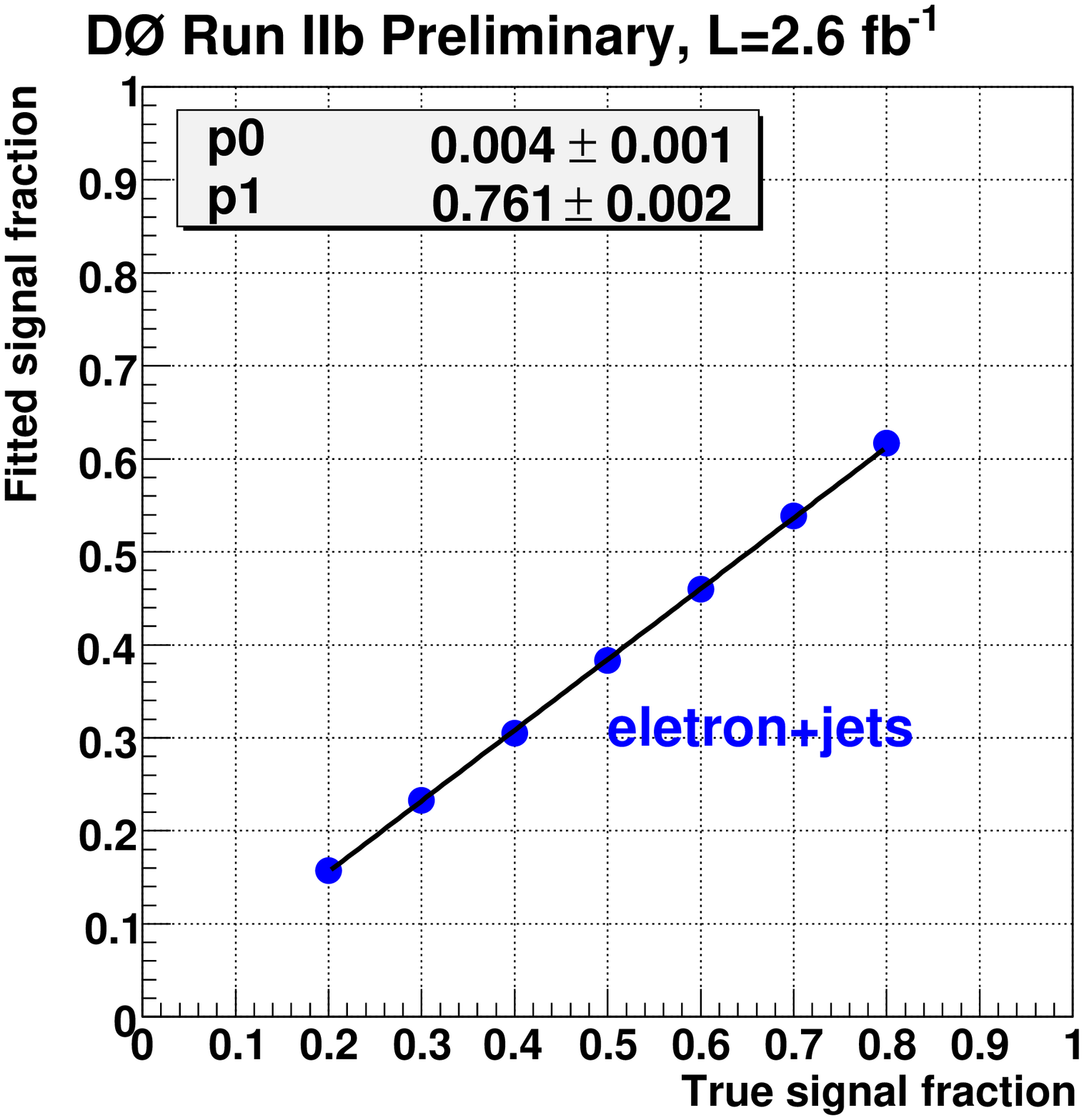}
  \includegraphics[height=4.9cm]{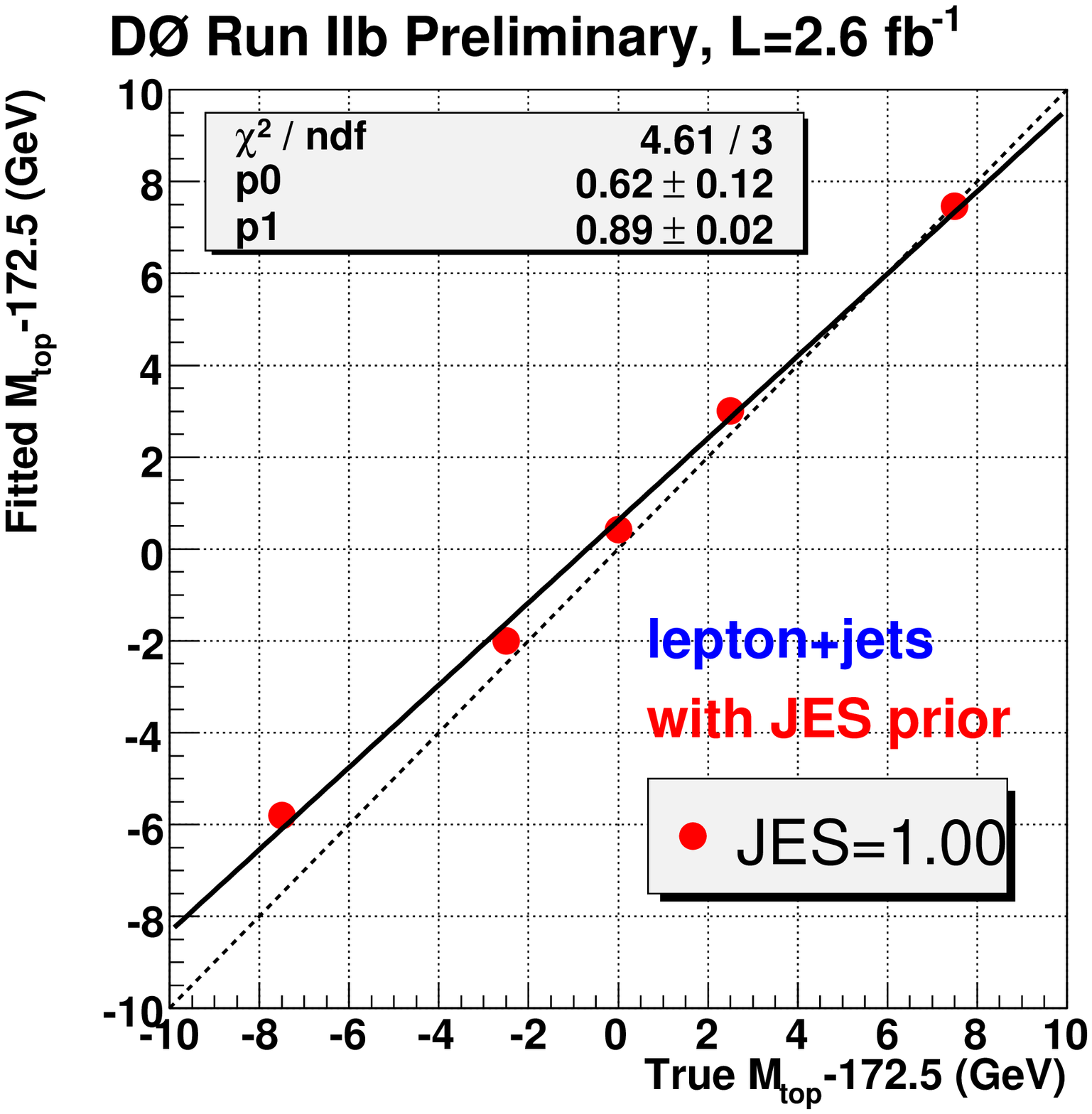}
  \includegraphics[height=4.9cm]{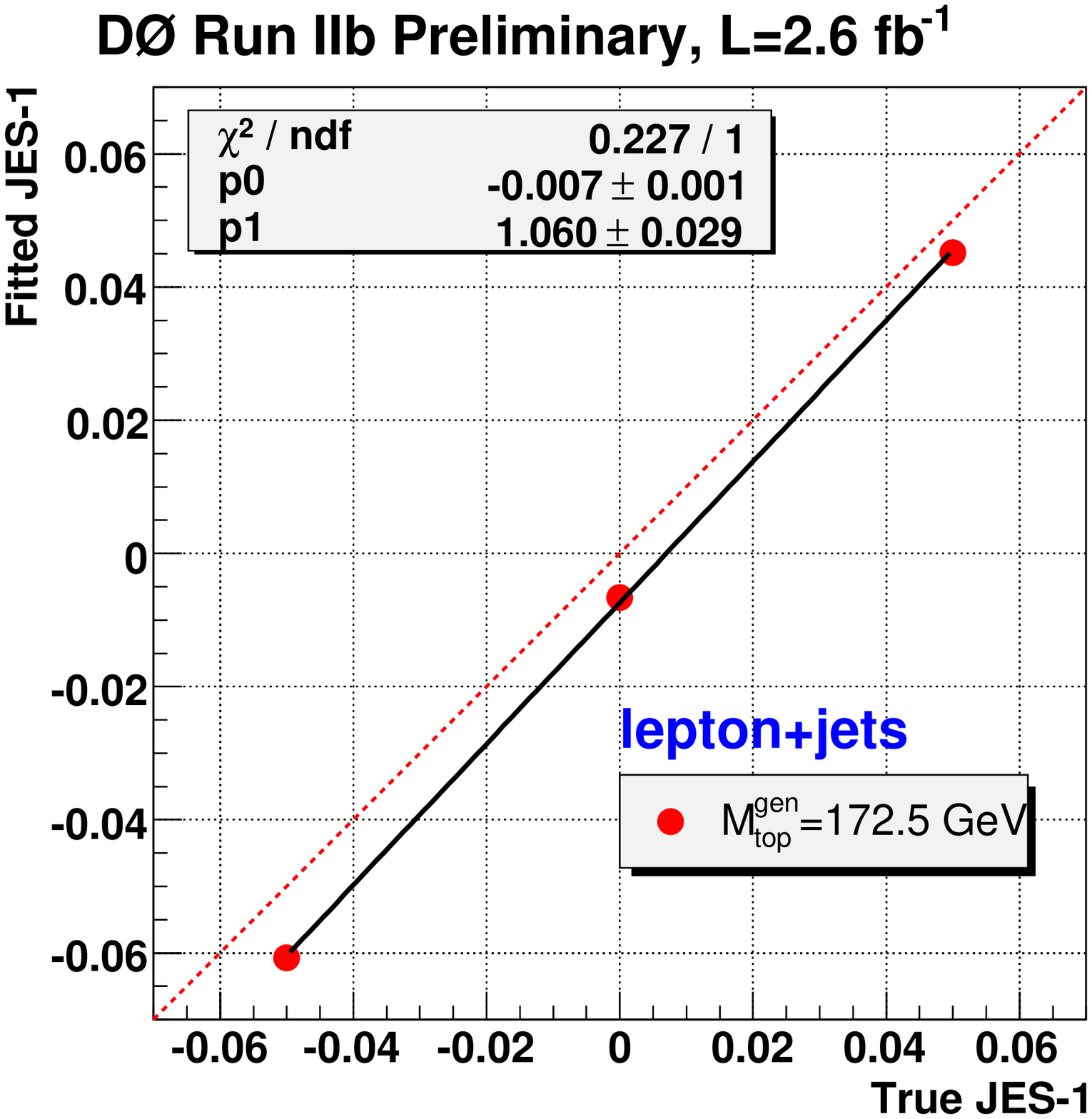}
  \caption{Calibration curves for the mass measurement with the Matrix Element method
 obtained for $2.6\ifb$ of Run\,IIb data by {D\O}. Left:
 Calibration of signal fraction determination in $e+$jets. Middle: top quark  mass
 calibration curve for nominal jet energy scale factor.
 Right: Calibration curve for
 jet energy scale
 factor~\cite{d0note5877}.}
  \label{fig:d0-me-mass-calib}
\end{figure*}

The signal probability is computed from the matrix element for top quark  pair
production and decay through quark anti-quark annihilation 
${\cal M}_{t\bar t}(y;m_t)$,
%={\cal M}_{q\bar q\rightarrow t\bar t\rightarrow \ell\nu b\bar b q q'}(y;m_t)$,
convoluted with the transfer function, $W(x,y;f_\mathrm{JES})$, that describes the
probability to observe a parton configuration, $y$, as the measured quantities, $x$.
\bea\label{eq:d0-me-Psig}
&& \hspace*{-0.8cm}P_\mathrm{sig}(x;m_t,f_\mathrm{JES})=
 \frac{1}{\sigma_\mathrm{obs}(m_t,f_\mathrm{JES})}\nonumber\\
&& \cdot\sum\limits_\mathrm{flavours}\int\limits_{}\!\mathrm{d}q_1\,\mathrm{d}q_2\,\mathrm{d}\Phi_6\,\,
 f(q_1) f(q_2)
\\&&\hspace*{0.17\linewidth}\cdot\quad
\frac{(2\pi)^4
|{\cal M}_{t\bar t}(y;m_t)|^2}{q_1 q_2 s}
 W(x,y;f_\mathrm{JES})\nonumber
\eea
The sum is over the possible flavours of the incoming quarks and the integral
over their momentum fractions, $q_1$ and $q_2$, as well as the 6-body phase space for the
outgoing particles, $\Phi_6$. $f(q_i)$ are the parton densities for
the incoming quarks and $s$ is the centre-of-mass energy squared. The
normalisation factor $\sigma_\mathrm{obs}(m_t,f_\mathrm{JES})$ is the
cross-section observable with the selection used, i.e. it includes effects of
efficiencies and geometric acceptance.

For the background probability a corresponding formula is used with the top quark 
pair matrix element replaced by the $W+$jets matrix element, which of course
is independent of $m_t$.  The contribution from  multijet events is considered to
have a similar shape and is not included separately.

Both, $P_\mathrm{sig}$ and $P_\mathrm{bkg}$ contain the same transfer
function, $W(x,y;f_\mathrm{JES})$. It is derived from full simulation for
individual jets and leptons. Only changes of  the size of the momenta but not
of the directions are considered.
Because it is not known which jet stems from which
parton, a weighted sum over all 24 possible assignments is made. The weight, $w_i$,
reflects the probability of the event's $b$-tags to be consistent
with the jet-parton assignment under consideration:
\beq\label{eq:d0-ljet-me-TF}
W(x,y;f_\mathrm{JES})\!=\!
W_\ell(x_\ell,y_\ell;f_\mathrm{JES})
\sum\limits_{i=1}^{24}\! w_i\!
\prod\limits_{j=1}^4\!
W_\mathrm{j}(x_j\!,y_{i,j};f_\mathrm{JES})\mbox{.}
\eeq
The values $x_\ell$ and $y_\ell$ are the measured and the assumed momenta of
the lepton, $x_j$ is the measured momentum of the $j$th jet and $y_{i,j}$ the
momentum of the matrix element parton associated to the  $j$th jet in the jet
parton association number $i$. $W_\ell$ and $W_\mathrm{j}$ are the transfer
functions for leptons and jets, respectively, which are zero when the
directions do not coincide.

The likelihood to observe the measured data is computed as the product of the
individual event probabilities, $P_\mathrm{evt}$:
\beq\label{eq:d0-ljet-final-like}
{\cal L}(m_t,f_\mathrm{JES},f_\mathrm{top})
=\prod_{\mathrm{events}}P_\mathrm{evt}(x;m_t,f_\mathrm{JES},f_\mathrm{top})~\mbox{.}
\eeq
For each assumed pair of the nominal top quark  mass and the jet energy scale factor, $m_t$
and $f_\mathrm{JES}$, the likelihood is maximised with respect to the top quark 
fraction, $f_\mathrm{top}$.
In~\cite{unknown:2006bd,Abazov:2008ds,d0note5750}  
the top quark  mass and jet energy scale factor are then determined by maximising the
two dimensional likelihood 
${\cal L}(m_t,f_\mathrm{JES},f^\mathrm{best}_\mathrm{top}((m_t,f_\mathrm{JES}))$.
In~\cite{d0note5877} the jet energy scale factor is constrained with a Gaussian
probability distribution to its nominal value and its uncertainty as obtained
from photon plus jet and dijet events.

Before the method is applied to data, its performance is calibrated in pseudo
experiments. Random events are drawn from a large pool of simulated  top quark  pair signal
and  $W+$jets background events with proper fluctuations of the signal and background
contribution such that the total number of events corresponds to the number of
events observed in data. The simulated events were generated with
\alpgen+\pythia\ and passed through the full D\O\ simulation and reconstruction.
This procedure is repeated 1000 times for several 
fixed nominal values of $m_t$, $f_\mathrm{JES}$ and $f_\mathrm{top}$. Thus for each of the
nominal values the signal fraction, the top quark  mass and the jet energy factor can be measured
with the described procedure 1000 times. The mean result for each set of
pseudo experiments with fixed nominal $m_t$, $f_\mathrm{JES}$ and
$f_\mathrm{top}$ are compared to the nominal values in the calibration curves
in~\fig{fig:d0-me-mass-calib}. The final result is corrected for any deviation
of these calibration curves from the ideal diagonal. Also the pull width
is computed and the statistical uncertainty corrected accordingly.

The leading source of systematic uncertainty contributing $\pm0.81\GeV$ (Run\,IIb)
stems from the modelling of differences in the detector response between light quark
and $b$ quark jets.  The next important contribution arises from uncertainties modelling
hadronisation and underlying event. It is estimated from the difference
between \pythia\ and \herwig\ and contributes with nearly $\pm0.6\GeV$. Also
the sample dependence of jet energy scale corrections in simulation contributes
with this size.
For the first time in \cite{d0note5877} this measurement includes an estimate of the uncertainty
due to colour reconnection
effects~\cite{Sandhoff:2005jh,Skands:2007zg,Wicke:2008iz}, 
which contributes $0.4\GeV$ to the uncertainty.
The squared sum of the individual contributions yields a total
uncertainty  of $\pm 1.4\GeV$ (Run\,IIb).
D\O\ has applied this analysis to their $3.6\ifb$ dataset separated by run
periods. The $1.0\ifb$  Run\,IIa result and the $2.6\ifb$  Run\,IIb result,
shown in \fig{fig:d0-me-mass-result}, are
combined with the BLUE method~\cite{\refblue} following the error categories
used by the Tevatron Electroweak Working Group~\cite{:1900yx} (see also
subsection \ref{sect:MassCombination} below) and yield
\beq
m_t=173.7\pm0.8_\mathrm{stat}\pm1.6_\mathrm{syst}\GeV~\mbox{,}
\eeq
where in this combined result 
the uncertainty due to the overall jet energy scale factor is
contained in the systematic uncertainty. This measurement is
currently D\O's most precise top quark  mass result. 
\begin{figure}[t]
  \centering
  \includegraphics[width=0.75\linewidth]{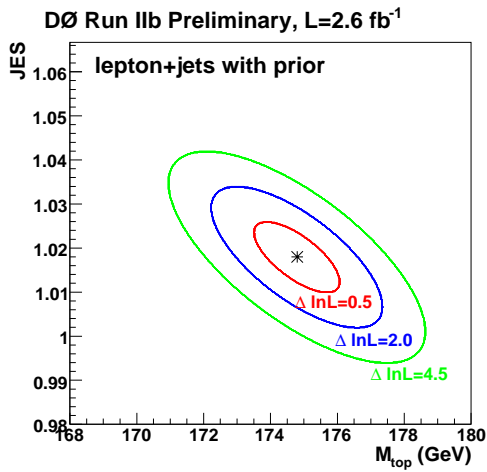}
  \caption{Result of 2-dimensional fit of top quark  mass and jet energy scale factor obtained
    with the Matrix Element method by D\O\ in $2.6\ifb$ of Run\,IIb
    data~\cite{d0note5877}. The ellipses correspond to $\log$ likelihood
    differences of $0.5$, $2.0$ and~$4.5$.
}
  \label{fig:d0-me-mass-result}
\end{figure}

\subsubsubsection{CDF}
The CDF collaboration is using the concept to measure the top quark  mass from a
likelihood based on the production matrix element in the lepton plus channel in several
variations~\cite{Abulencia:2007br,CdfNote9725,Abulencia:2005pe,CdfNote9135,CdfNote9427,CdfNote9692}. 

The recent analyses all base on an event selection that requires
a single isolated lepton, missing transverse momentum and exactly four jets,
at least one of which is required to be identified as $b$ jet.
Then the top quark  mass and an overall jet energy scale factor are determined
simultaneously from an event by event probability that the
observed event may occur given the true values of the top quark  mass, $m_t$,  the jet
energy scale factor, $f_\mathrm{JES}$ and the signal fraction
$f_\mathrm{top}$. The various CDF analyses differ in the construction of the
likelihood.

The \em CDF Matrix Element method (MEM)\em~\cite{Abulencia:2007br,CdfNote9725} follows closely the
procedure outlined in Section~\ref{d0-ljets-me-mass}. 
%As only events with
%identified $b$-jets are used, compared to D\O, no likelihood discriminant is used in the
%construction of the likelihood. 
The matrix element used to describe signal is
that of the $q\bar q \rightarrow t\bar t$ process with its decay. For background the
$W+4$jets matrix element is employed. Finally, the transfer functions of
\eq{eq:d0-ljet-me-TF} use only parton-jet assignments consistent with
$b$-tagging information and assumes the lepton measurement to be exact.
For the background probability, $P_\mathrm{bkg}$, the dependence on the jet
energy scale is taken from a parameterisation of the average likelihood response
rather than an explicit change of the transfer functions.

The \em Dynamical Likelihood method (DLM)\em~\cite{Abulencia:2005pe,CdfNote9135} 
\label{sect:dlm}
differs from the Matrix Element method in that it bases its likelihood only on
the signal contribution, $P_\mathrm{sig}(x;m_t,\Delta_\mathrm{JES})$. In this
signal term the squared matrix element is factorised into a production matrix
element, the \discretionary{(anti-)}{top}{(anti-)top} quark   propagators and their decay matrix elements: 
$|{\cal M}(a_1a_2 \rightarrow t\bar t\rightarrow \ell\nu b\bar b q q')|^2
=|{\cal M}_{a_1a_2 \rightarrow t\bar t}|^2\,{\cal P}_t{\cal P}_{\bar{t}}\,
  |{\cal D}_{t}|^2|{\cal D}_{\bar{t}}|^2$, 
which removes spin-correlations.
However, in contrast
to previously described measurements it includes gluon diagrams, i.e. $a_1 a_2$ can be $q\bar q$
or $gg$. For the treatment of background contributions the method fully relies
on results obtained in ensembles of pseudo-datasets similar to the calibration
step of the other methods. This correction is computed depending on the jet
energy scale correction and the background fraction.

The \em Matrix Element Method with Quasi-MC-Integration
(MTM)\em~\cite{Aaltonen:2008mx,CdfNote10191,Aaltonen:2010yz} 
%CdfNote9427,CdfNote9692
uses an again more complete matrix element to construct the signal
likelihood. The applied matrix element~\cite{Kleiss:1988xr} includes the
$q\bar q$ and the $gg$ production channels with full spin-correlations.
The method treats the background by subtracting the expected contribution of
the logarithm of the likelihood. 
Equation~(\ref{eq:d0-ljet-me-master}) is thus rewritten as
\bea
&&\log P_\mathrm{evt}(x;m_t,\Delta_\mathrm{JES})\nonumber\\
&& = \log P_\mathrm{sig}(x;m_t,\Delta_\mathrm{JES})-f_\mathrm{bkg}(q) \log P_\mathrm{bkg}(m_t,\Delta_\mathrm{JES})\qquad
\eea
with $\log P_\mathrm{bkg}$ being the average $\log P_\mathrm{sig}$ obtained in
background events. The expected background fraction, $f_\mathrm{bkg}$, is
computed from simulation and 
applied event by event as function of the output $q$ of a neural network
discriminant, c.f~\fig{fig:cdf-me-mtm3-discr}.
\begin{figure}[b]
  \centering
   \includegraphics[width=0.9\linewidth]{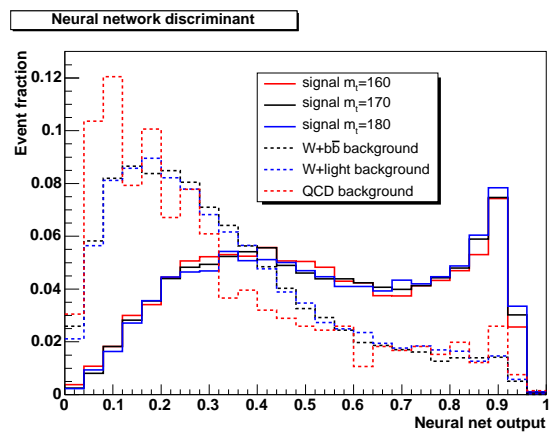}
  \caption{Expected performance of the neural network discriminant used in the
  MTM Matrix Element method to determine the event by event background fraction~\cite{CdfNote10191}.}
  \label{fig:cdf-me-mtm3-discr}
\end{figure}
Moreover the analysis removes events which are likely to be mismeasured,
e.g. due to extra jets, misidentification etc. For this events are required to have
% by requiring that their peak value of 
$\log P_\mathrm{evt}>10$ for at least some range of $m_t$.

\begin{figure*}
  \centering
  \includegraphics[height=4.1cm]{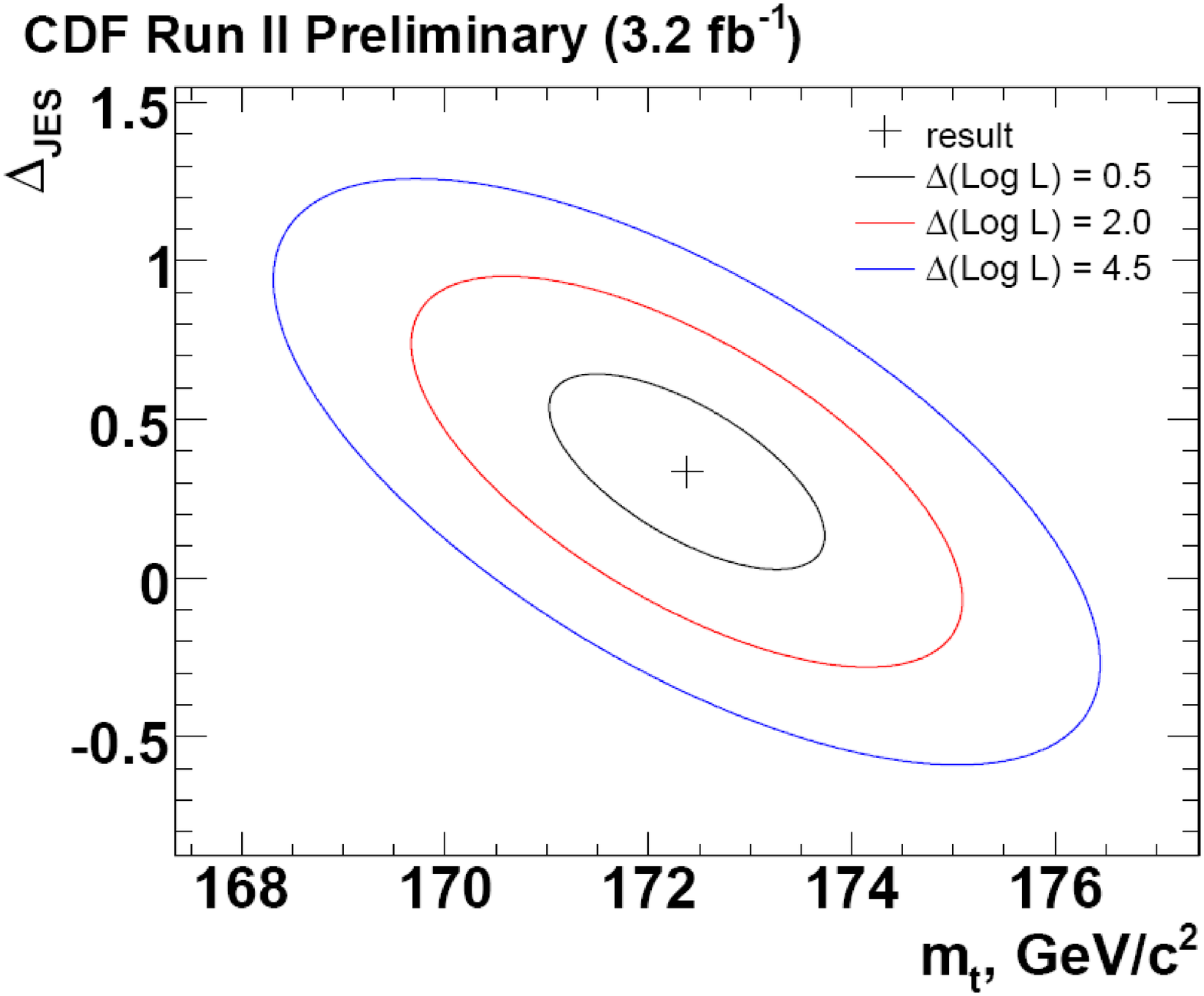}                 
  \includegraphics[height=4.1cm]{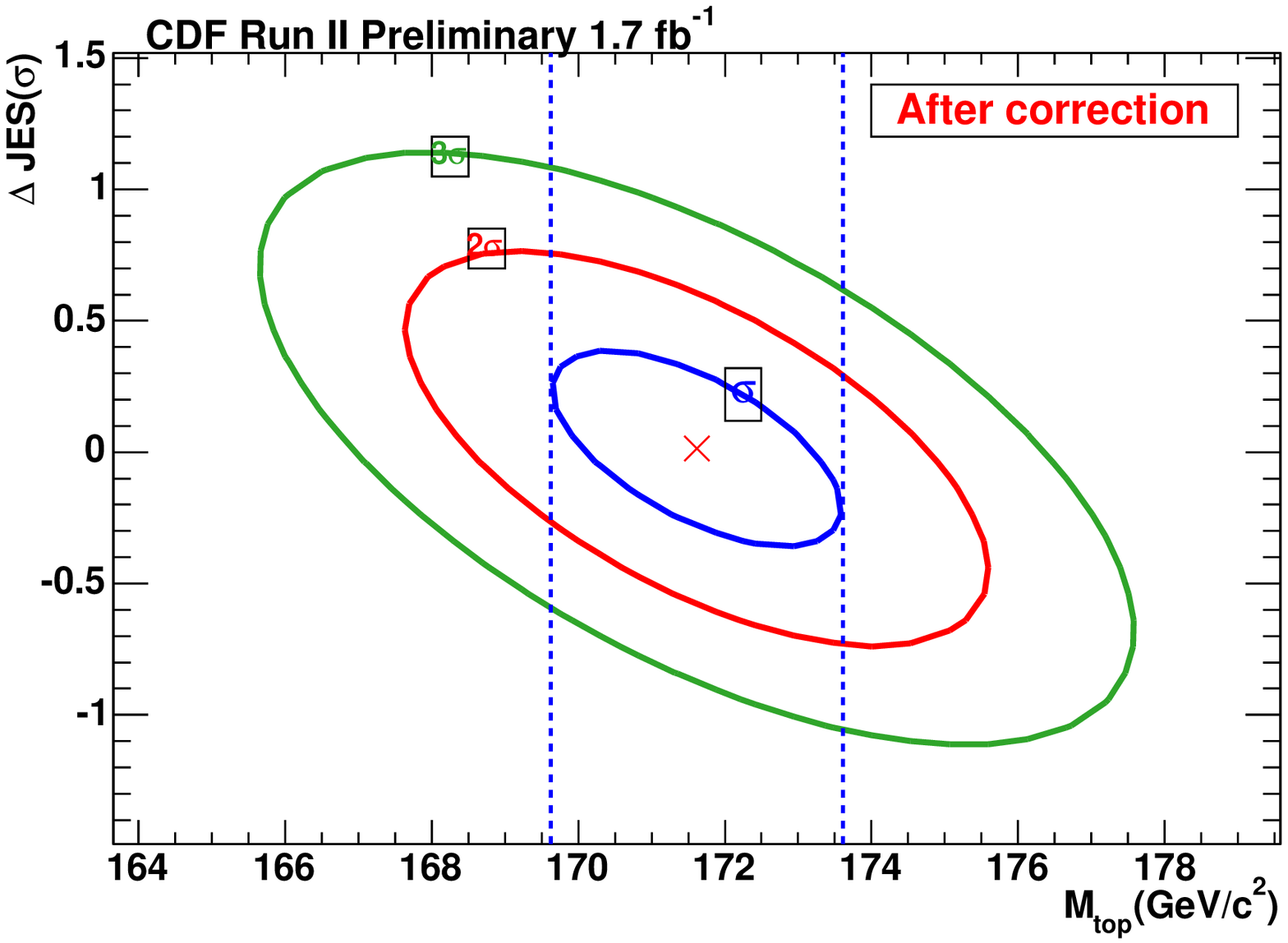}          
  \includegraphics[height=4.1cm]{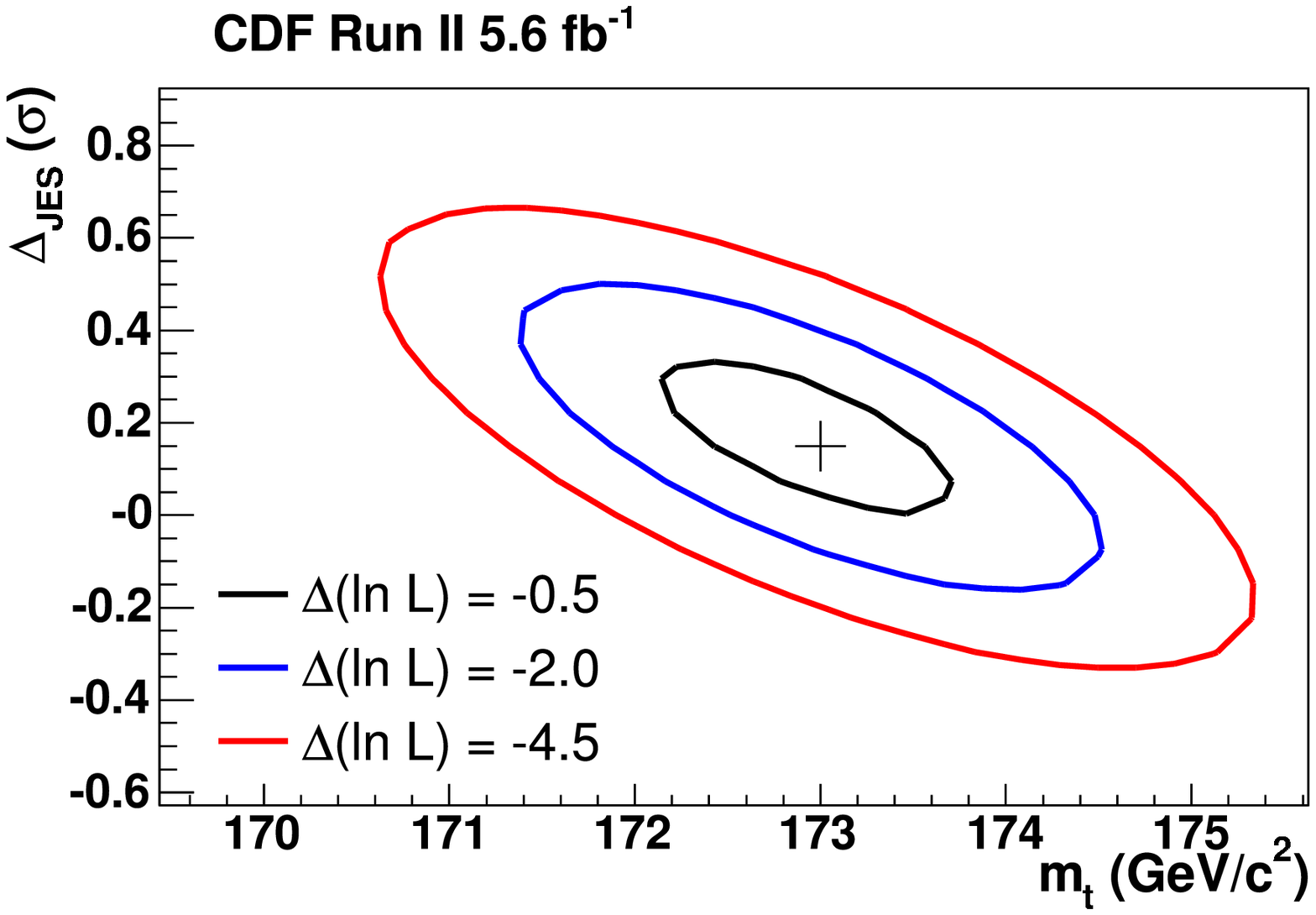}
%mtm3_p28_may10_data_contours_2dcalib.eps}
  \caption{Results of CDF top quark  mass measurements with in-situ jet energy
    calibration obtained in three variations of the Matrix Element method. 
Left: Matrix Element method \`a la D\O\ (MEM)~\cite{CdfNote9725}, Middle: Dynamical
Likelihood method (DLM)~\cite{CdfNote9135}, Right: Matrix
Element method~(MTM)~\cite{Aaltonen:2010yz}. The vertical axes show the results obtained
for the jet energy shift. In the left plot it is a scale factor; the middle
and right plot show it in units of the nominal jet energy uncertainty.
   }
  \label{fig:cdf-me-results}
\end{figure*}
In all methods the overall likelihood is computed from the various 
event likelihoods following \eq{eq:d0-ljet-final-like}. It is maximised to find the optimal top quark  mass
and jet energy scale parameters. For the MEM also the
background fraction is fitted.
The performance is then checked by applying the top quark  mass measurement on
ensembles of pseudo-experiments with various nominal values of the top quark  mass, $m_t$,
and the jet energy scale shifts, $\Delta_\mathrm{JES}$. The results are
corrected for any observed shifts and the uncertainties scaled to fit the
observed spread of results in the ensemble tests.  The observed shifts vary
between $1.4\GeV$ for the Dynamical Likelihood method~\cite{Abulencia:2005pe},
which relies on this calibration to describe the background effects,
and $0.09\GeV$ for the Matrix Element method~\cite{CdfNote9725}, which
has the most complete background term.

As all methods apply an in-situ jet energy calibration this important
systematic is already covered by the uncertainty of the fit
result. The residual effect of the jet energy scale is estimated by varying
the $p_T$ and $\eta$ dependence. The uncertainty of the  jet energy scale
of $b$-jets from various sources is considered separately. The uncertainty of
signal modelling is determined from comparison of \pythia\ and \herwig\ and by
varying the amount of initial and final state radiation produced in the parton
shower. Further uncertainties include an estimate of the background modelling
and treatment in each method. 
The DLM has the uncertainty due to initial and final state radiation as the most
important single contribution, while in the MEM and the MTM the residual jet energy
scale and the difference between generators used for signal simulation are the
two leading contributions. In MEM and MTM colour reconnection
effects~\cite{Sandhoff:2005jh,Skands:2007zg,Wicke:2008iz} are
estimated and yield significant contributions of $0.56\GeV$ and $0.37\GeV$, respectively.

The most recent results of the various Matrix Element like methods of CDF were
obtained at different integrated luminosities. All results yield in-situ jet energy
scale corrections consistent with the default calibration used in CDF. 
\beq
\begin{array}{llrcl}
\mbox{MEM:}
 &\mbox{$3.2\ifb$~\cite{CdfNote9725}:} & m_t & = & 172.4\pm1.4\pm1.3 \GeV\\
\mbox{DLM:}&\mbox{$1.7\ifb$~\cite{CdfNote9135}:}\quad & m_t & = & 171.6\pm2.0\pm1.3 \GeV\\
\mbox{MTM:} &\mbox{$5.6\ifb$~\cite{CdfNote10191,Aaltonen:2010yz}:} & m_t & = & 173.0\pm0.9\pm0.9 \GeV~\mbox{,}
\end{array}
\eeq
~\\
where the first uncertainty is the statistical one and includes the overall
jet energy scale uncertainty, the second uncertainty is the systematic one.
The two dimensional representation of these results which simultaneously
determine the jet energy scale are shown in~\fig{fig:cdf-me-results}.
These measurements of course use (partially) the same data, thus for the
combination only the most precise one is used, until their correlation is
determined. 
%\paragraph{Dyn. Likelihood}\cite{Abulencia:2005pe,CdfNote9135}
%\paragraph{Matrix Element}\cite{Abulencia:2007br}
%\paragraph{Multivariate/ME with QuasiMCIntegration/MTM}\cite{CdfNote9427}
%%% Template \cite{Aaltonen:2008gj}
% To be checked: \cite{CdfNote9245} %(still?) private

\subsubsection{Decay Length and Lepton Momentum Methods}

The top quark  mass measurements described so far, aimed to use the maximal available
information to yield the best statistical uncertainty for the given number of
events. However, all these measurements have significant 
uncertainties from the jet energy scale. Alternative observables that have
little or no dependence on the jet energy scale are therefore an interesting
complement.

In the lepton plus jets channel of top quark  pair
decays CDF has investigated two observables: 
the transverse decay length of $b$
tagged jets and the lepton transverse
momentum.
One analysis combines the mean values of both observables~\cite{Abulencia:2006rz,Aaltonen:2009hd}
while the other uses the mean and distribution of the lepton transverse
momentum alone~\cite{CdfNote9683}.

\subsubsubsection{Two observable mean value method}
For the method using both observables
the data selection requires one isolated lepton, missing transverse momentum
and at least three jets. For events with exactly three jets two of the jets need to be
identified as $b$ jets with CDFs secondary vertex
algorithm~\cite{\refcdfsecvtx}. For events with four or more jets only one
needs to have such a reconstructed secondary vertex.

The expected sample composition is determined from a combination of data and
simulation. Top quark  pair signal is simulated using \pythia\ with CTEQ5L parton
densities for various nominal top quark  mass values. Also single top quark  samples are
generated with various nominal top quark  mass values.
The dominating background $W+$jets is simulated with
\alpgen+\pythia, the multijet background  is modelled from a data sample with
modified lepton selection criterion. All simulated events are passed through
the full CDF detector simulation and reconstruction.

\begin{figure*}
  \centering
\includegraphics[width=0.325\textwidth,clip,trim=10mm 0mm 0mm 0mm]{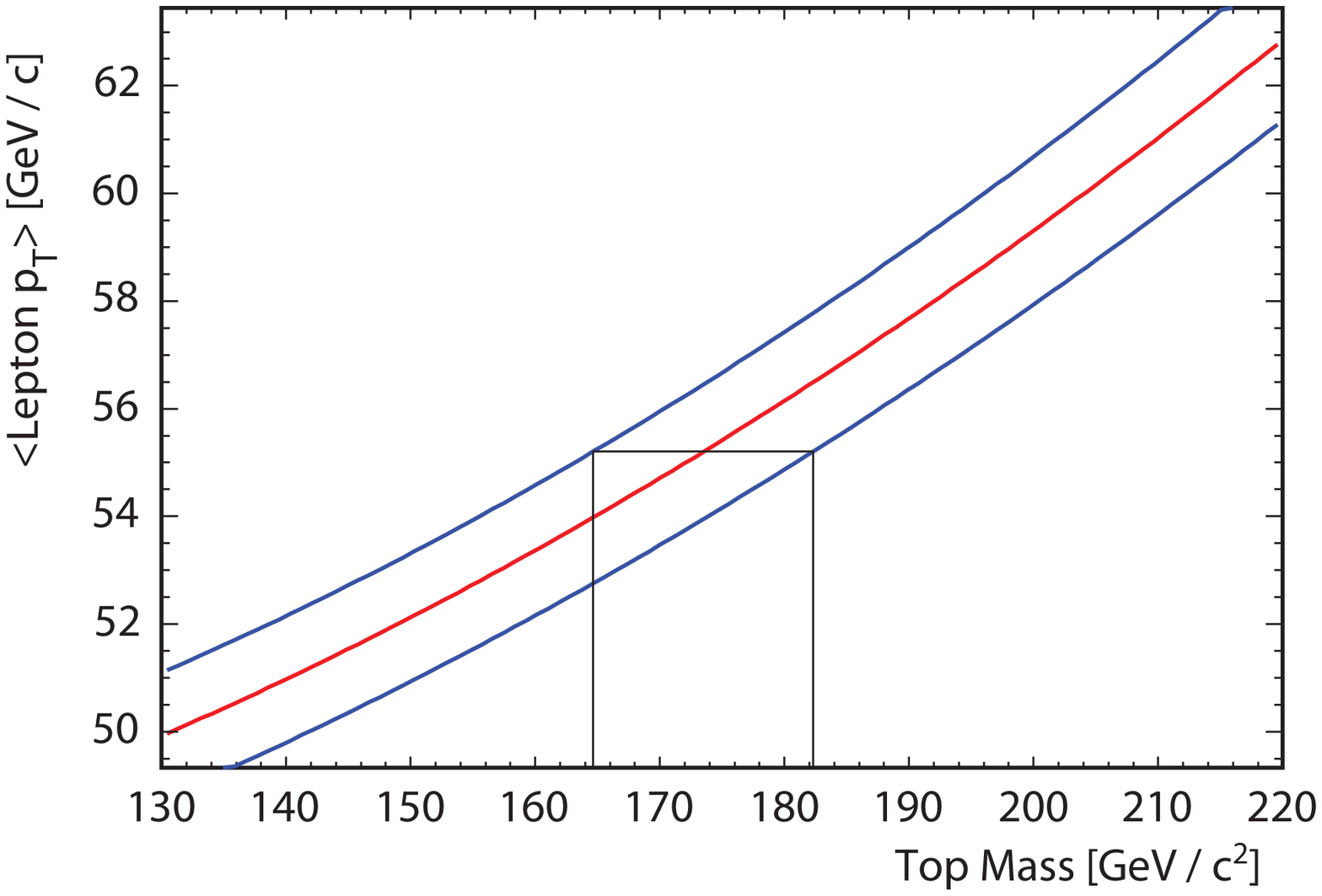}%
\includegraphics[width=0.325\textwidth,clip,trim=10mm 0mm 0mm 0mm]{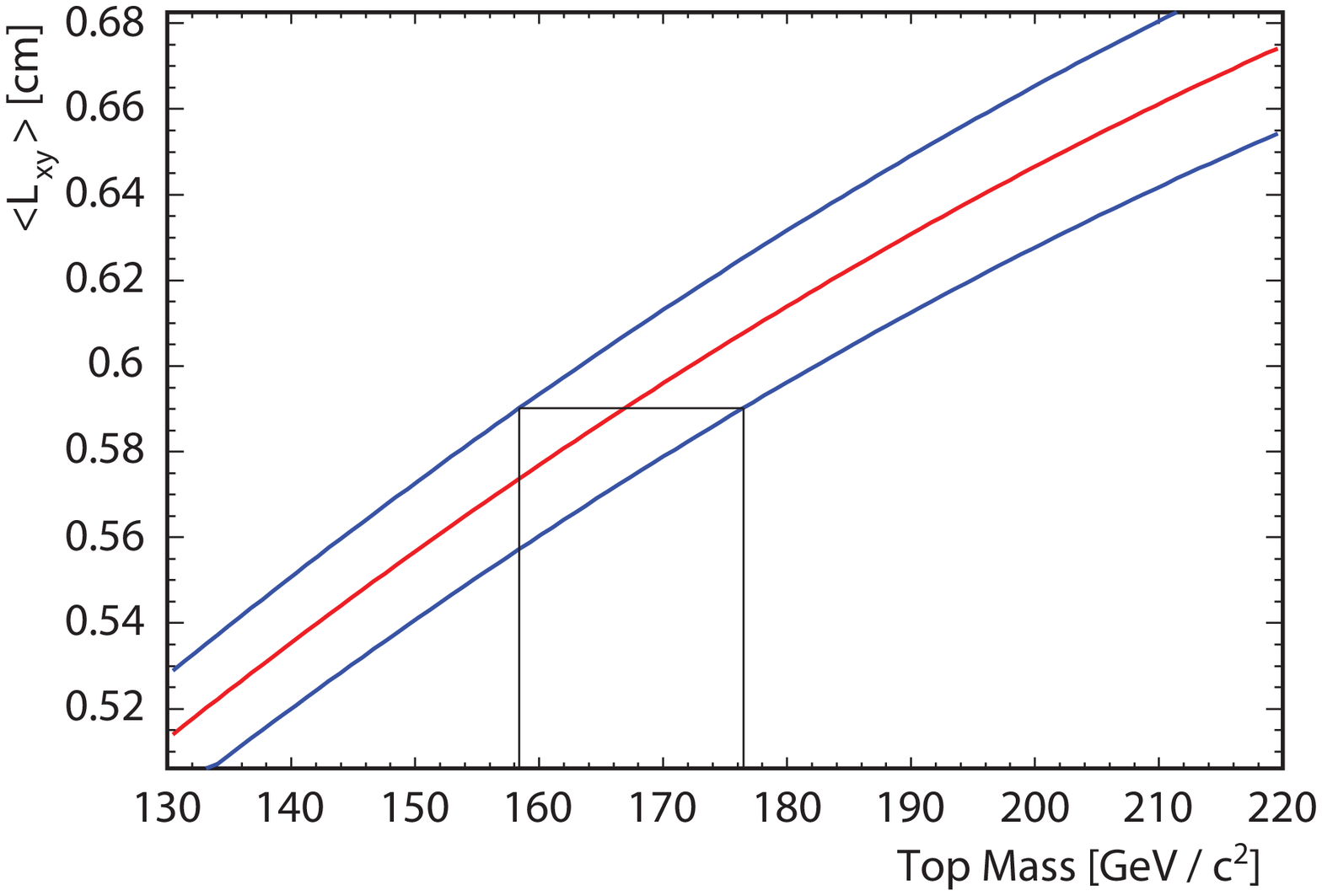}%
\includegraphics[width=0.325\textwidth,clip,trim=10mm 0mm 0mm 0mm]{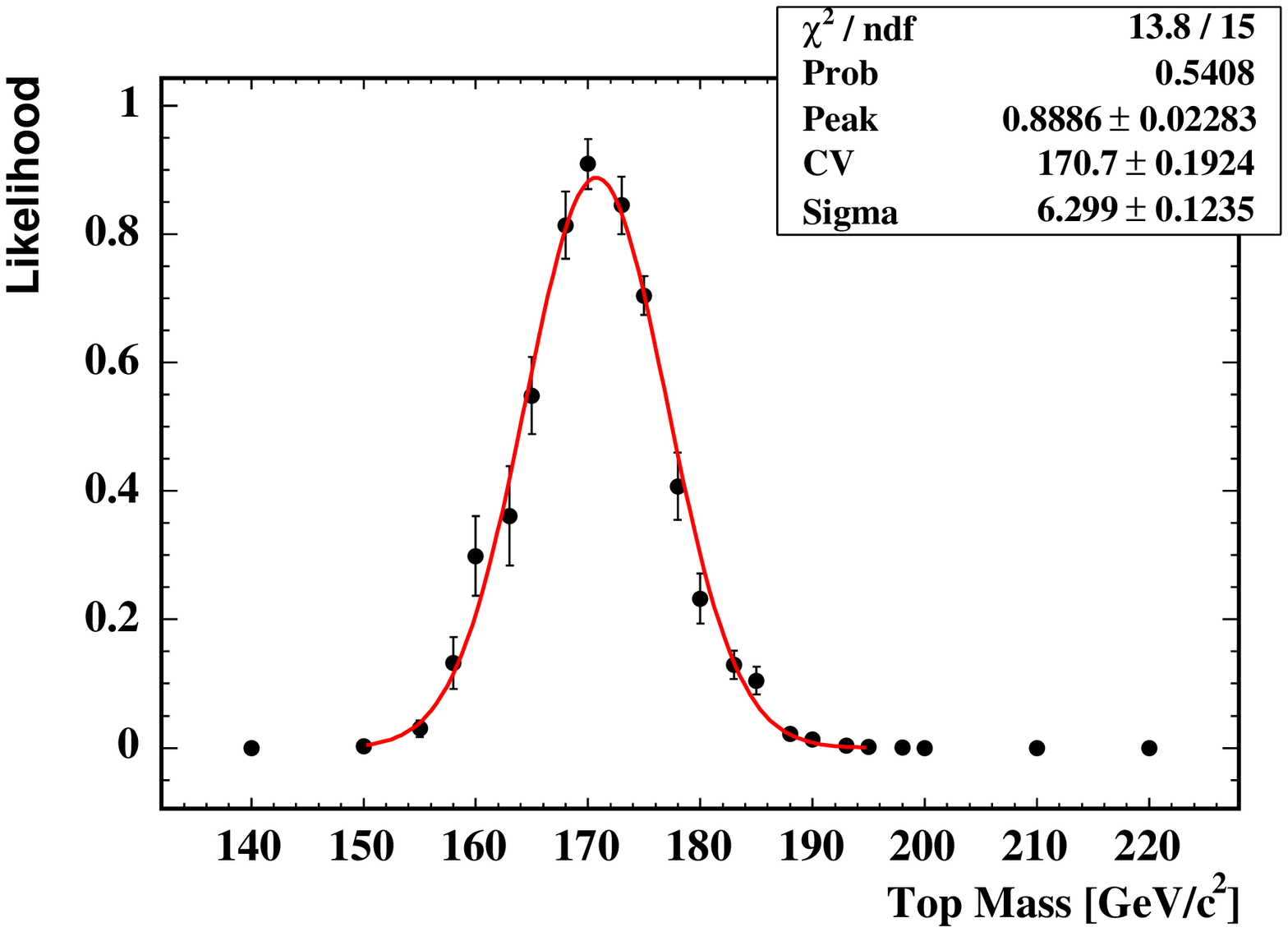}
  \caption{Left and middle: Calibration curves for the expected mean decay length, $\langle
L_{2D}\rangle$, and  mean lepton momentum, $\langle p^\ell_T\rangle$, as
function of the nominal top quark  mass (left and middle). Right: The likelihood for the
combined measurement~\cite{Aaltonen:2009hd}. }  
\label{cdf:mass-lxy}
\end{figure*}

For the top quark  mass determination the transverse decay length $L_{2D}$ of the $b$ tagged jets is measured with the vertex
algorithm used for the selection above.  The actual measurement is performed
using the means of the observed decay lengths, $\langle L_{2D}\rangle$, and the transverse
lepton momentum, $\langle p^\ell_T\rangle$. The expected contribution from
background is taken from the described background model. Because the 
observables chosen are sensitive to the details of the event kinematics the signal
simulation is corrected to yield parton distributions consistent with
CTEQ6M~\cite{Pumplin:2002vw}. In addition the total contribution from gluon
fusion is corrected as function of the top quark  mass. 
The performance of the simulation for determining the  decay length $L_{2D}$ of
the $b$ tagged jets is calibrated in a dijet control sample. 

With these corrections the expected means of decay lengths and transverse
lepton momentum and their expected statistical spread 
are determined as function of the nominal top quark  mass using
ensembles of pseudo-data. 
The  curves obtained are fitted by  quadratic
polynomials to obtain smooth curves. Figure~\ref{cdf:mass-lxy}~(left, middle) shows the
mean values $\langle L_{2D}\rangle$ and $\langle p^\ell_T\rangle$ and the
resulting top quark  mass ranges from the individual observables.
The combined result is obtained from the likelihood to find the observed mean
values at various nominal top quark  masses, c.f.~\fig{cdf:mass-lxy}~(right).

The dominant systematic uncertainties for the measurement from  $\langle
p^\ell_T\rangle$ arises from the uncertainty on the modelling of  initial and final state
radiation in the signal simulation, from the lepton momentum scale and the
shape of the background description.
In the measurement from $\langle
L_{2D}\rangle$ the uncertainty of the data to Monte Carlo correction for the
decay length dominates the uncertainties. 
The jet energy scale gives a non-negligible
contribution to the measurement from the average decay length
%because $b$-jets with lower transverse momentum
%yield a smaller mean decay length and 
through effects on the  event selection.

In $1.9\ifb$ of data CDF measures the top quark  mass from the mean decay length, $\langle
L_{2D}\rangle$, and the mean lepton momentum, $\langle p^\ell_T\rangle$ 
to be~\cite{Aaltonen:2009hd}
\beq
m_t=170.7\pm 6.3_\mathrm{stat}\pm 2.6_\mathrm{syst}\GeV~\mbox{,}
\eeq
where the jet energy scale uncertainty is included (as a small) contribution to the
systematic uncertainty. The larger statistical uncertainty (compared to the
Matrix Element methods) prevents a significant weight to the current
combination of top quark  masses~\cite{:1900yx}.

\subsubsubsection{Lepton momentum shape method}
\label{cdf-mt-leptonpt-ljets}
The analysis that concentrates on the transverse lepton momentum
alone~\cite{CdfNote9683} is based on events with a single isolated lepton,
missing transverse energy and at least four jets. At least one of the jets is
required to be identified as $b$-jet.

Top quark pair signal events are simulated for various assumed top quark mass
values. Backgrounds due to $W+$jets, $Z+$jets diboson and single top quark
production are described with simulation. Estimates for multijet background with fake
leptons and the normalisation for $W+$light flavour jets are derived from data.
These signal and background models are used to determine parametrised
templates for the distribution of the lepton $p_T$ spectrum as function of the
top quark mass. To extract the top quark mass these templates are compared to
the observed data using an unbinned maximum likelihood fit.
Leptonic $Z$ boson decays are used to calibrate the lepton~$p_T$. 

The uncertainty of the final result is dominated by the statistical
uncertainty. The systematic uncertainties are dominated by effects related to
signal and background modelling. The uncertainty on the fake lepton
description contributes  with $\pm1.8\GeV$; differences between generators for
the signal modelling contribute with $\pm1.8\GeV$ to the systematic uncertainty.

In $2.7\ifb$ of CDF data the top quark mass determined from the shape of the
lepton momentum distribution is~\cite{CdfNote9683}
\beq
m_t=176.9\pm8.0_\mathrm{stat}\pm2.7_\mathrm{syst}\GeV\quad\mbox{.}
\eeq
This result has been combined with a corresponding study of dilepton events~\cite{CdfNote9831},
c.f. section \ref{cdf-mt-leptonpt-dilepton}, to yield~\cite{CdfNote9881}
\beq
m_t=172.8\pm7.2_\mathrm{stat}\pm2.3_\mathrm{syst}\GeV\quad\mbox{.}
\eeq
The large statistical uncertainty of these methods prevents a 
significant contribution to the world average from such 
measurements at the Tevatron. Its independent and low systematics 
will make them more relevant at the LHC.

\subsection{Dilepton Channel}
Due to the small branching fraction the dilepton channel has much fewer
events, but the two charged leptons also yield a much cleaner signature. 
For a measurement of the top quark  mass the dilepton channel has the additional complication
that the kinematics are under-constrained due to the
two unmeasured neutrinos. Two of the six missing numbers can be recovered
from the transverse momentum balance, 
%i.e. the measurement of the missing transverse momentum, 
two more from requiring that the invariant mass of the charged lepton and its neutrino 
should be consistent with the $W$ boson mass in the top and the anti-top quark decay, a
fifth constraint can be obtained by forcing the top and the anti-top quark
masses to be equal. Thus for a full reconstruction of an event including the
top quark  mass one constraint is missing.

\subsubsection{Weighting Methods}
To completely recover the event kinematics additional assumptions can
be made on a statistical basis, i.e. by assuming the distribution of one or more kinematic
quantities. For a given event top quark  masses corresponding to certain values of these kinematic
quantities are weighted by the probability that these  kinematic values
occur.
%and are consistent with the observed kinematic. 
The top quark  mass reconstructed in the given event is then chosen 
such that it corresponds to the largest weight. 
%according to the weights. 
This basic idea leads to various so-called weighting methods
for measuring the top quark  mass in dilepton events that mainly differ in the
distribution that is assumed a priori. In the following the relevant methods
recently used by the two collaborations are described.

\subsubsubsection{CDF Neutrino Weighting}
\label{CDF-template-dilepton-portion}
One of these weighting methods is  the neutrino weighting method that
is used in CDFs combined lepton plus jets and
dilepton
analysis~\cite{Aaltonen:2008gj,CdfNote9578,CdfNote9679,CdfNote10273}. 
The lepton plus jets
part of this analysis is described in Section~\ref{CDF-template-ljets-portion}.

Dilepton events are selected by requiring two oppositely charged leptons,
missing transverse energy and exactly two energetic jets. In addition the
scalar sum of the transverse momenta, $H_T$, is required to be greater than
$200\GeV$. Events in which at least  one of the jets was identified as $b$ jet
are kept separately throughout the analysis. 

To reconstruct one top quark  mass for each selected event the distribution of
neutrino pseudo-rapidities is assumed a priori. The distribution is taken from
top quark  pair simulation and found to be Gaussian with an approximate width of one.
The weight is computed in two steps for each assumed top quark  mass. 

First, for fixed
values of the neutrino pseudo-rapidities, $\eta_{\nu}$ and $\eta_{\bar \nu}$, 
the event kinematics are reconstructed
using  constraints on the assumed top quark mass and the nominal $W$ boson
mass (ignoring the measurement of the transverse momentum). The weight of these
$\eta$ choices is constructed as the $\chi^2$ probability that the sum of the
reconstructed transverse neutrino momenta agree with the measurement of the
missing transverse momentum, $\mpt$:
\beq\label{cdf:vwt-weight}
w(m_t,\eta_{\nu},\eta_{\bar \nu})=\sum\limits_{i=1}^8 \exp - \frac{(\mpt-\vec{p}^{\,\nu}_T(i)-\vec{p}_T^{\,\bar{\nu}}(i) )^2}{2\sigma_T^2}~\mbox{.}
\eeq
Here, the sum adds the four possible sign choices that occur in solving
the above constraints and the two possibilities to assign the two measured
jets to the $b$ or $\bar b$ quark. 
$\vec{p}_T^{\,\nu}(i)$ and $\vec{p}_T^{\,\bar{\nu}}(i)$ are
the transverse neutrino momenta for the  choice $i$ and $\sigma_T$ is the
experimental resolution of the measurement of the missing transverse momentum.

The second step now folds the weights obtained for fixed pseudo-rapidities with
the a priori probability, $P(\eta_{\nu},\eta_{\bar \nu})$, that these pseudo-rapidities occur:
\beq
W(m_t)= \sum\limits_{\eta_{\nu},\eta_{\bar \nu}} P(\eta_{\nu},\eta_{\bar \nu}) w(m_t,\eta_{\nu},\eta_{\bar \nu})~\mbox{.}
\eeq
The top quark  mass that yields the highest $W(m_t)$ is the reconstructed
value, $m_t^{\mathrm{reco}}$, for the event under consideration. In addition
to this top quark  mass the variable $m_{T2}$ is used for further analysis. 
$m_\mathrm{T2}=\min\left[\max\left(m^{(1)}_T,m^{(2)}_T\right)\right]$, where
$m^{(i)}_T$ are the transverse mass of the top and the anti-top quark,
respectively. The minimisation is performed over all possible neutrino
momenta consistent with the missing transverse energy. Previous versions of the
analysis instead used the scalar sum of the jets and lepton transverse momenta and the missing
transverse momentum, $H_T$.
\begin{figure}
  \centering
  \includegraphics[height=5cm]{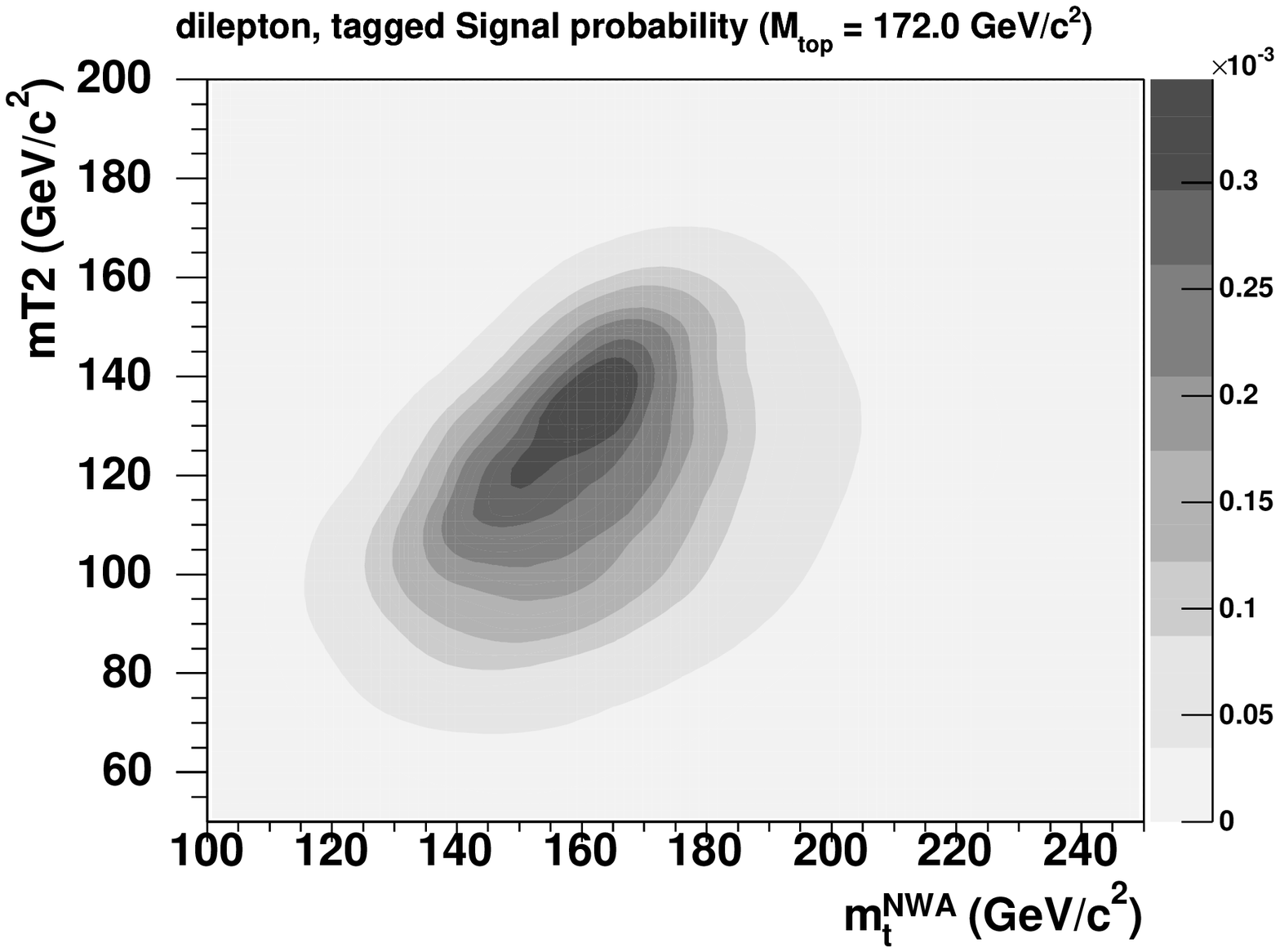}\\
  \includegraphics[height=5cm]{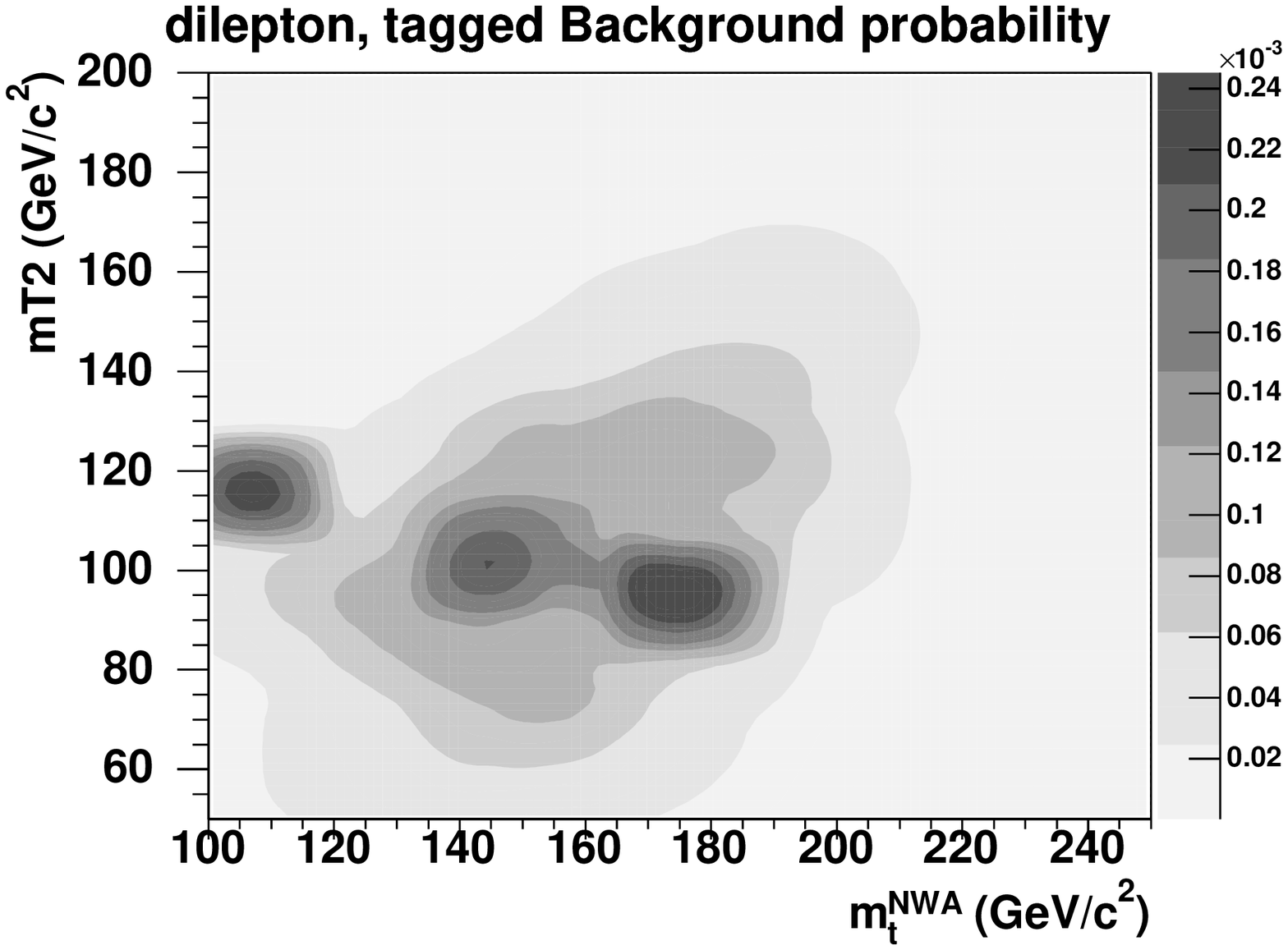}
  \caption{Probability density in the  $m_t^{\mathrm{reco}}$-$m_\mathrm{T2}$ 
    plane for a dilepton signal at $m_t=172\GeV$ (top) and background
    (bottom) at nominal jet energy scale~\cite{CdfNote9679}.}
  \label{fig:cdf-mt-nwa}
\end{figure}

Simulation is now used to determine the expected distribution of
$m_t^{\mathrm{reco}}$ and $m_\mathrm{T2}$. The backgrounds in the dilepton sample stem 
from fake events with a jet misidentified as lepton, from Dell-Yan and from
diboson production. The dominating fake background is modelled from data. Drell-Yan events
are simulated with \alpgen+\pythia, diboson events with \pythia.
The simulated events are passed through the CDF detector simulation and
reconstruction.

These simulations are used to generate probability density functions for the
two observables, $m_t^{\mathrm{reco}}$ and  $m_\mathrm{T2}$, 
as function of the nominal top quark  mass, $m_t$, and the jet energy scale
shift, $\Delta_\mathrm{JES}$, see~\fig{fig:cdf-mt-nwa}. From the probability densities
a likelihood is constructed that is maximised to find the final result. 
The construction of the probability densities and the likelihood and the
determination of the resulting measured top quark  mass follows the lepton plus jets
analysis described in Section~\ref{CDF-template-ljets-portion}
The dependence on the jet energy shift is only used for simultaneously fitting
the lepton plus jets and dilepton events. 
Systematic uncertainties for the pure dilepton measurement are dominated by
jet energy scale uncertainties, followed by the uncertainty of modelling
colour reconnection effects.

In $5.6\ifb$ of data CDF determines the top quark  mass~\cite{CdfNote10273} in
dilepton events using the neutrino weighting method and $m_\mathrm{T2}$ to be:
\beq
m_t=170.3\pm2.0_\mathrm{stat}\pm3.1_\mathrm{syst}\GeV~\mbox{.}
\eeq
A combined fit including lepton plus jet events and in-situ jet energy scale
determination the dataset yields
\beq
m_t=172.1\pm 1.1_\mathrm{stat}\pm 0.9_\mathrm{syst}\GeV~\mbox{.}
\eeq
With the combined fit the dilepton result profits from the constraint of
the jet energy scale. The overall precision is still dominated by the lepton
plus jets result.

\subsubsubsection%[CDF Neutrino $\phi$ Weighting Methods]%
{\boldmath CDF Neutrino $\phi$ Weighting Methods}

The neutrino $\phi$ weighting method uses the distribution of neutrino azimuthal directions,
$\phi$, as an a priori distribution. CDF has analysed up to $2.9\ifb$ of dilepton
events to measure the top quark  mass~\cite{Aaltonen:2009tza} with this method. 
%Aaltonen:2008gj,CdfNote9456 

Events are selected by requiring one isolated well-iden\-tified lepton, one
oppositely charged isolated track, missing transverse momentum and at least
two jets. Vetoes are applied on events where the missing transverse momentum is
close to a jet, on cosmic events, conversions and $Z$ boson events.

For each value pair $(\phi_\nu,\phi_{\bar{\nu}})$ on a grid of azimuthal
directions the top quark  mass is reconstructed using a constrained fit that
determines the top quark  mass, $m_t$, the lepton and (anti-)$b$ quark momenta 
using the measured lepton, track and jet momenta, the missing transverse momentum
and the constraints on the $W$ boson mass and the equality of the top and
anti-top quark mass. The $\chi^2$ is defined similarly to
\eq{eq:cdf-constraint-chi2}, but uses Breit-Wigner distributions rather than
Gaussians for the mass constraints. For this fit two possibilities exist to 
assign the two measured jets to the $b$ and anti-$b$ quarks. In addition 
the quadratic nature of the $W$ boson mass constraints gives a fourfold
ambiguity. Of the corresponding  8 top quark  mass results at
each $(\phi_\nu,\phi_{\bar{\nu}})$ pair  only the one with the lowest
$\chi^2$ is kept. To arrive at a single mass value for the event under
consideration the masses obtained at the various  azimuthal
directions are averaged weighted with their $\chi^2$ probability.
Only events with a weight of at least $30\%$ of the maximum are considered in
the average.
% The analyses in~\cite{Aaltonen:2008gj,CdfNote9456} evaluate

The top quark  mass is finally measured by comparing the distribution of mass values
reconstructed for each event in data to templates in a likelihood fit. 
The templates are derived from simulation. For the dilepton signal \herwig~\cite{Corcella:2000bw} is
used at many nominal top quark  mass values. The background is simulated with
\pythia\ for the Drell-Yan background and
\alpgen+\herwig\ for the fakes from $W+$jets production. The diboson background
was simulated with \pythia\ and \alpgen+\herwig~\cite{Aaltonen:2008gj}. Signal
and background templates were parametrised to obtain smooth templates. The
signal templates yield a smooth dependence on the nominal top quark  mass.

The likelihood for $N$ events is built as the product over the probabilities
that each event agrees with the sum of the signal and background templates,
$P_\mathrm{sig}$ and $P_\mathrm{bkg}$, weighted according to the
signal and background contributions, $s$ and $b$. The likelihood contains a
Poissonian term on the total number of expected events and a Gaussian
constraint on the background contribution, $b$, to be consistent with the a
priori expectation, $b^\mathrm{e}$:
\bea\label{eq:cdf-vwt-likelhood}
{\cal  L}(m_t) &=&
e^{
-\frac{(b-b^\mathrm{e})^2}{2\sigma_{b}^2}
}
\left(
\frac{e^{-(s+b)}(s+b)^N}{N!}\right)\nonumber\\
&&\prod_\mathrm{i=1}^N
\frac{sP_\mathrm{sig}(m_i;m_t)+bP_\mathrm{bkg}(m_i)}{s+b}~\mbox{.}
\eea
In addition a term, ${\cal L}_\mathrm{param}$, that describes the parametrisation uncertainties of the
template curves is included into the likelihood. Minimising the total likelihood yields
the top quark  mass results, $m_t$, and estimates for the signal and background
contributions,  $s$ and $b$. The statistical uncertainty is obtained by
finding the top quark  masses  corresponding to a $\log$ likelihood decrease of $0.5$.
Due to the inclusion of  ${\cal L}_\mathrm{param}$ in the likelihood this includes the
uncertainties from the template parametrisation.

The systematic uncertainties are evaluated by measuring the top quark  mass in
ensembles of pseudo data generated from simulation with modifications that
reflect  a one sigma change of the systematic under consideration.
The systematics are dominated by the jet
energy scale that has to be taken from the external calibration of the CDF
calorimeter. It contributes $\pm 2.9\GeV$. The next to leading contributions from the
background composition and the $b$ jet energy scale contribute less than $20\%$
of this uncertainty. 

In $2.9\ifb$ of CDF data with the described lepton plus track selection the method
yields~\cite{Aaltonen:2009tza}
\beq
m_t=165.5^{\,+3.4}_{\,-3.3~\mathrm{stat}}\pm3.1_\mathrm{syst}\GeV~\mbox{.}
\eeq
The method shows similar statistical precision as the neutrino weighting method
and similar sensitivity to the jet energy scale systematics.

\subsubsubsection{D\O\ Neutrino Weighting Method}
\label{NWTD0}
\begin{figure*}
  \centering
\includegraphics[width=0.4\textwidth]{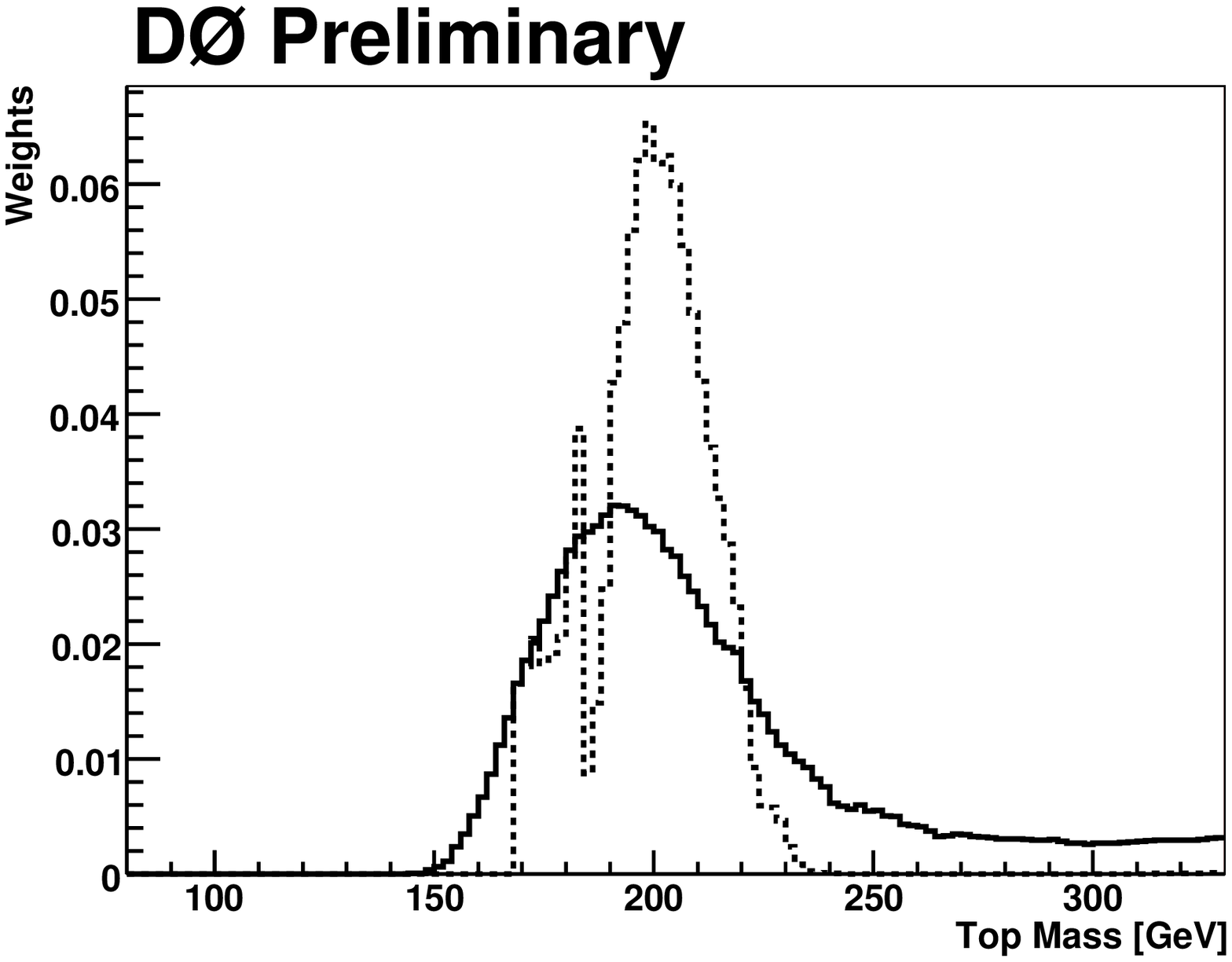} 
% {
% \tiny\unitlength=1mm
% \begin{picture}(0,0)
% \put(11,8){$\sigma_w$}
% \put(11,5){[\GeV]}
% \put(61,4.5){$\mu_w$}
% \put(57.65,1.5){[\GeV]}
% \end{picture}
% }%
%\includegraphics[width=0.455\textwidth,clip,trim=100mm 0mm 0mm 0mm]{figs/T73F07.eps} 
\includegraphics[width=0.48\textwidth,clip,trim=0mm 0mm -8mm 12mm]{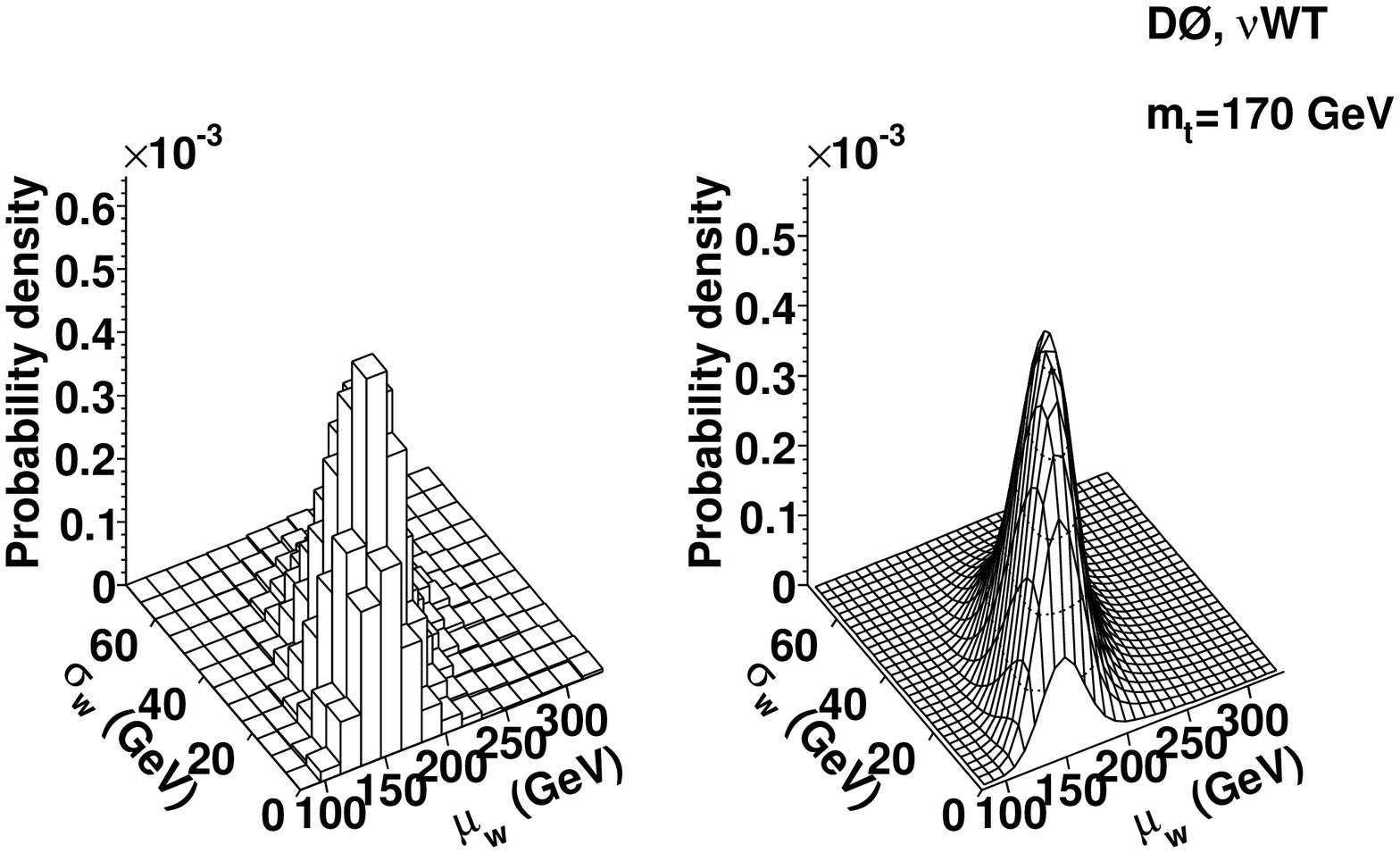}

  \caption{Normalised weight distribution for an example simulated event
    with $m_t=175\GeV$ before (dashed) and after (full line) smearing lepton
    and jet momenta~\cite{d0note5746} (left).
    Example of the template histograms (middle) and  functions (right) of $\mu_w$ vs. $\sigma_w$ 
    evaluated at $m_t=170\GeV$ for the $e\mu$ channel~\cite{Abazov:2009eq}.}
  \label{fig:d0-vwt-weight}
  \label{fig:d0-vwt-template}
\end{figure*}
D\O\ has applied the Neutrino Weighting method in up to $5.3\ifb$ 
of data~\cite{Abazov:2006bg,d0note5746,Abazov:2009eq}. Events are selected by requiring two
isolated, oppositely charged, identified leptons or an isolated, identified lepton and an isolated
track of opposite charge. In addition the events are required to have two
energetic jets and missing transverse energy. For lepton plus track events at
least one of the jets needs to be identified as $b$-jet.
Vetoes on the lepton pair (or lepton plus track) invariant
mass and the scalar sum of transverse momenta %and on sphericity 
are applied to 
reject $Z+$jets and other backgrounds depending on the identified lepton types. 

The event kinematics is reconstructed by assuming the neutrino rapidity
distribution which according to simulations is expected to be Gaussian. Scanning through the range
of possible top quark  masses an event weight is computed as function of the top quark  mass. 

First, for given values of the neutrino rapidities and the top quark mass a constrained
fit is performed to determine the momenta of the top quark decay products,
$b\bar b \ell^+\ell^-\nu\bar\nu$. As constraints the $W$ boson mass and the
assumed top quark mass are used, ignoring the measured values of the missing
transverse momentum. An individual weight, $w(m_t,\eta_\nu,\eta_{\bar\nu})$, 
is computed from a $\chi^2$ term that compares
the  sum of neutrino momenta in the transverse
plane  with the measured missing transverse momentum, c.f.~\eq{cdf:vwt-weight}.
Then the a priori Gaussian distribution of the 
neutrino rapidities are folded into a total weight, $W_i(m_t)$,
by adding the $w(m_t,\eta_\nu,\eta_{\bar\nu})$ at 10 values of the neutrino
rapidity with appropriate unequal distance.

To take detector resolution for jet and lepton energies into account the
determination of $W_i(m_t)$ is repeated for a number of jet and lepton momenta
fluctuated according to their experimental resolution. The weight averaged
over these fluctuations,
$W(m_t)=\langle W_i(m_t)\rangle$, shows a much smoother distribution and yields
fit results for a wider range of top quark  masses, see~\fig{fig:d0-vwt-weight}~(left).

To determine the top quark  mass the mean of the weight distribution and its
variance are computed for each individual event:
\beq
\mu_w = \int m_t W(m_t)\,\mathrm{d}m_t~\mbox{,}\quad
\sigma_w^2 = \int m_t^2 W(m_t)\,\mathrm{d}m_t - \mu_w^2~\mbox{.}
\eeq
Compared to using $\mu_w$ alone, including $\sigma_w$ to the following
extraction of the top quark mass yields an $16\%$ improvement in the
statistical uncertainty. Previous analyses of D\O\ used even more detailed
information about $W(m_t)$  and thus the statistical information was exploited
slightly better but at the cost of higher
complexity~\cite{Abazov:2006bg,Abazov:2009eq}. 

The distribution of $\mu_w$ and $\sigma_w$ values in data is now compared to
templates derived from simulation. 
Top quark  pair signal events were generated with \alpgen+\pythia\ for 
 various nominal top
quark masses. Background contributions from $Z/\gamma$+jets  are simulated
using \alpgen+ \pythia,  diboson events with \pythia. All simulated events are
passed through the full D\O\ detector simulation and reconstruction.
For $Z/\gamma$+jets events with $Z/\gamma\rightarrow e^+ e^- $ or $\mu^+
\mu^-$ the amount of fake missing transverse momenta and of fake isolated
leptons or tracks is derived from control samples in data and used in the
normalisation of the samples.
Template histograms are obtained from the above signal and background estimates.
The observed data distributions are compared directly to 
the template histograms or are compared to 
parameterised smooth functions fitted to the templates, see \fig{fig:d0-vwt-template}~(right) for an example.
This yields two methods, a histogram based and a function based method.

The final top quark  mass result is extracted by maximising a likelihood as
function of the number of signal and background events as well as the top
quark  mass. The 
likelihood describes the agreement of the data with the templates, a Gaussian
constraint on the expected amount of background and a Poisson term for the
total number of observed events as in \eq{eq:cdf-vwt-likelhood}. 

The performance and precision of the method is tested on a large number of
pseudo data with known nominal top quark  mass composed from the signal and background
models. Calibration curves of the nominal vs. the measured top quark  mass 
are obtained by comparing the average of the measured top quark  masses on many such
pseudo data with their nominal top quark  mass.  
The observed offsets of around $1\GeV$ and deviations of the pull distribution
from the normal distribution are corrected for.

As for the other dilepton measurements the dominating systematic uncertainty
stems from the energy scale uncertainties for
the two $b$ jets in the event. In this analysis
it contributes about $\pm1.5\GeV$. Sub-leading uncertainties arise from limited
template statistics and  from  the difference between modelling the top quark
pairs with pure \pythia\ and using \alpgen+\pythia. 

The histogram based method and the method using parametrised templates show a
correlation of only $85\%$. They are thus averaged using the BLUE method.
Using dilepton events selected in $1\ifb$ of data D\O\ determines the top
quark mass  to be~\cite{Abazov:2009eq}
\beq
m_t=176.2\pm4.8_\mathrm{stat}\pm2.1_\mathrm{syst}\GeV~\mbox{.}
\eeq
An updated result including up to $5.3\ifb$ of data and combining with the  Matrix Weighting
method  is described at the end of the next section.
Despite the large systematic uncertainty from the jet energy scale the
measurement is statistically limited at this luminosity.

\subsubsubsection{D\O\ Matrix Weighting Method}
\label{D0MWT}
Another method to measure the top quark mass in dilepton events 
is the Matrix Weighting method. This has been applied by D\O\ in Run\,II to
$5.3\ifb$ of data~\cite{Abazov:2009eq}. The event selection requires two
identified oppositely charged leptons, large missing transverse momentum and
at least two jets. As above, vetoes on the lepton pair invariant
mass and the scalar sum of transverse momenta %and on sphericity 
are applied to 
reject $Z+$jets and other backgrounds depending on the identified lepton types.

To reconstruct the top quark  mass in each event this method assumes the distribution
of lepton energies in the top quark  rest frame~\cite{Dalitz:1991wa} and the parton
density functions. 
First %In this method
the kinematics is reconstructed as function of the top quark  mass. At each assumed
top quark mass a kinematic fit is applied to reconstruct the top quark decay
products from the measured quantities. The two leading jets are identified
with the $b$ quarks, the missing transverse momentum is required to be
consistent with the sum of neutrino momenta in the transverse plane. In
addition kinematic constraints from the $W$ boson mass and the assumed top quark
mass are applied. With the reconstructed kinematics a weight as function of
the top quark mass is computed:
\beq
w_{j}(m_t)=f(x_1)f(x_2)P(E^*_{\ell^+}|m_t)P(E^*_{\ell^-}|m_t)~\mbox{.}
\eeq
Here the quark parton densities, $f(x)$, at the quark and anti-quark momentum
fractions, $x_1$ and $x_2$, are explicitly included in the weight. For the
neutrino weighting described above they are included implicitly 
in the expected $\eta$ distribution. 
%Here $f(x)$ are the quark parton densities for in proton. $x_1$ and $x_2$ are
%the quark and anti-quark momentum fractions in the proton and anti-proton,
%respectively. 
$P(E^*_{\ell^\pm}|m_t)$ are the probabilities that (given the hypothetical top
quark mass, $m_t$)  
the reconstructed energy of the lepton $\ell^\pm$  in the rest frame 
of the corresponding top quark is $E^*_{\ell^\pm}$.
The distribution of these energies is taken from the top quark  pair production and
decay matrix element~\cite{Dalitz:1991wa}. 
% \begin{figure*}
%   \centering
%   \includegraphics[width=0.48\textwidth]{figs/T55F05b.eps}
%   \includegraphics[width=0.48\textwidth]{figs/T55F05a.eps}
%   \caption{Left: Distribution of peak mass values in $1\ifb$ of D\O\ data compared to simulation
%     with $m_t=175\GeV$. Right: Likelihood curve of comparing the peak mass
%     values to templates~\cite{d0note5463}.}
%   \label{fig:d0-mwt}
% \end{figure*}

In each event there are up to four possibilities to solve the constraint from
the $W$ boson mass and two possibilities to assign the two jets to the two $b$
quarks, thus up to eight values of $w_{i}(m_t)$ may exist. The total event
weight is computed as the sum of these. 
\beq
W_i(m_t)=\sum\limits_j w_{j}(m_t)
\eeq
As for the neutrino weighting method resolution effects are included by
recomputing the $W_i$ with the measured quantities fluctuated according to
their resolutions and averaging the results:
$W(m_t)=\langle W_i(m_t)\rangle $.
In each event the top quark  mass which maximises this smeared weight, $W(m_t)$, is
used as the estimator of the top quark mass for this event. 
%The distribution
%of peak mass values expected and observed in data is shown in~\fig{fig:d0-mwt}~(left).

The distribution of the top quark  mass estimators observed in data is compared to
templates obtained from simulation for a range of nominal top quark  mass values.
The simulations use \pythia\ to generate top quark  pair signal and diboson
background events. Backgrounds from $Z+$jets are generated with
\alpgen+\pythia. The generated events are passed through full D\O\ detector
simulation and reconstruction. For the comparison and to extract the final 
top quark mass a binned likelihood is used.
%\beq
%{\cal L}(m_t)=\prod\limit_{i=1}^{n_\mathrm{bins}} p_i^{n_i}
%\eeq

The method is calibrated by using ensembles of pseudo data constructed from
the simulated events for various nominal top quark  masses. The obtained bias in the
determination of the central result and its  statistical error 
are corrected for.
Ensemble tests are also used to determine the effect of systematic
uncertainties. 
As for the other dilepton mass measurements the systematic uncertainty is
dominated by uncertainties related to  the jet energy scale.  
They account for $\pm1.2\GeV$. 

In $1.0\ifb$ D\O\ measures the top quark mass using the Matrix Weighting
method in dilepton events to be~\cite{d0note5463,Abazov:2009eq}.%, c.f.~\fig{fig:d0-mwt}~(right)
\beq
m_t=173.2\pm4.9_\mathrm{stat}\pm2.0_\mathrm{syst}\GeV~\mbox{.}
\eeq
With respect to the D\O\ neutrino weighting this Matrix Weighting method yields
a slightly worse statistical precision, but has a smaller dependence on the
jet energy scale. D\O\ has averaged the two weighting methods described taking
correlations into account using the BLUE method~\cite{Abazov:2009eq}. 
The combined result updated to $5.3\ifb$ yields~\cite{d0note6104}:
\beq
m_t=173.3\pm2.4_\mathrm{stat}\pm2.1_\mathrm{syst}\GeV~\mbox{.}
\eeq
As different events contribute to the two methods to a different amount, this allows a
significant improvement of the statistical uncertainty over the individual results.

\subsubsection{Matrix Element Methods}
An alternative to the weighting methods of measuring the top quark mass in
dilepton events is the Matrix Element method described above already for the lepton
plus jets channel, see Section~\ref{sect:me-ljets}. 
Instead of assuming individual distributions to add effective
constraints, the full knowledge from the matrix element is used to check the
agreement of the measured events with different top quark mass assumptions.

\subsubsubsection{CDF Matrix Element Method}
CDF has applied the Matrix Element method to the dilepton channel in several
analyses~\cite{Abulencia:2005uq,Abulencia:2006mi,CdfNote8951}. 
The analyses require two oppositely charged leptons, missing transverse energy
and two or more energetic jets. 

The probability to observe a given event depends on the top quark mass and
is computed from folding the matrix element with parton densities and detector
resolutions along Eqs.~(\ref{eq:d0-ljet-me-master})
and (\ref{eq:d0-me-Psig}).  In a dilepton sample of course the integration
needs to include the unmeasured momenta of \em two \em neutrinos.  The
transfer functions that connect the measured quantities with the parton
momenta correspondingly contains terms for only two jets.
%No transfer function is applied for lepton momenta.
The momenta of the charged leptons are kept unsmeared.

In the $1\ifb$ analysis~\cite{Abulencia:2006mi} an additional contribution to the
transfer function was used to connect  unclustered energy and
energy from additional sub-leading jets with
the transverse momentum of the top quark  pair system. This more accurate treatment
was found to yield better statistical power, but also larger systematics.
In the recent update to $1.8\ifb$ \cite{CdfNote8951} such a term is no longer
used, instead events with more than two jets are rejected.

The matrix element used for top quark  pair events corresponds to the
quark annihilation process. Background contributions are computed using the matrix elements for
$Z/\gamma+$jets, $WW+$jets and $W+3$jets. The contributions are added
according to their expected fractions in the selected dataset, depending
on the number of identified $b$-jets in the event.

The probability for the full sample, $P(m_t)$, is the product of the event by event
probabilities. As the jet energy scale can not be constrained in dilepton
events the external nominal CDF energy scale is used.
The mean,  $m_t^\mathrm{raw}\!=\!\int\!m_tP(m_t)\,\mathrm{d}m_t$, is
used as the raw measured mass.
The result is calibrated using ensembles of pseudo data with varying nominal
top quark masses. The pseudo data are modelled using \herwig\
for signal events, 
\alpgen+\pythia\ for $Z/\gamma+$jets and pure \pythia\  for
diboson samples. Misidentified leptons are modelled from data.
The observed bias and pull width observed by applying the above procedure on
the ensembles are used to correct the raw mass measurement and its statistical 
uncertainty.

The systematic uncertainties are again dominated by the jet energy scale
uncertainty. In the $1.0\ifb$ analysis~\cite{Abulencia:2006mi} the additional
inclusion of additional jets is found to increase 
the dependence on the jet energy scale by $15\%$. In the $1.8\ifb$
analysis~\cite{CdfNote8951} the jet energy scale accounts for a $\pm 2.6\GeV$
mass uncertainty. The next leading uncertainties in this analysis stem from
simulation statistics for the background description, which of course can be
addressed, and the uncertainty in the calibration. Differences of signal
simulation with \pythia\ vs. \herwig\ give only a marginally smaller contribution.

In $1.8\ifb$ of dilepton events CDF measured the top quark mass with the
Matrix Element method  as~\cite{CdfNote8951}
\beq
m_t=170.4\pm3.1_\mathrm{stat}\pm3.0_\mathrm{syst}\GeV~\mbox{.}
\eeq
With the increased statistics the method of~\cite{Abulencia:2006mi}
that yields a statistical improvement at the cost of increased systematics is
no longer applied. 

\subsubsubsection{D\O\ Matrix Element Method}

D\O\ has applied the Matrix Element method to measure the top quark mass in
dilepton events to $3.6\ifb$~\cite{d0note5897}. The event selection requires
one isolated electron and one isolated muon of opposite charge and at least two jets.

The probability to observe a given event is computed as function of the 
top quark mass by folding the matrix element with parton densities and detector
resolutions along Eqs.~(\ref{eq:d0-ljet-me-master})
and (\ref{eq:d0-me-Psig}). The integration includes the additional unknowns
due to the second unseen neutrino and the unknown transverse  momentum of the 
top quark  pair system.
The weight function, $W(x,y)$, contains terms to
describe the detector resolution for two leptons and two jets. The two
possible jet to (anti-)$b$-quark assignments are summed.
For the signal description D\O\ uses the matrix element of quark annihilation
to top quark  pairs and the subsequent decay. The backgrounds are all described with
the matrix element for $Z+$jets process with $Z\rightarrow
\tau^+\tau^-\!$.\  Special transfer functions were used to
relate the measured electron and muon momenta from a tau decay 
to the tau momentum. 

The likelihood for the observed data sample as function of the assumed top
quark mass, ${\cal L}(m_t)$, 
is the product of the event probabilities, c.f.~\eq{eq:d0-ljet-final-like}.
As dilepton events cannot constrain the jet energy scale, the externally
determined nominal jet energy scale is used.
The measured top quark  mass is the one that maximises the sample likelihood ${\cal L}(m_t)$.
Its statistical uncertainty is taken from the masses that yield an ${\cal
  L}(m_t)$ decreased by half a unit from the maximum. 

The performance of the method is evaluated using ensembles of pseudo data that
are created from simulation. Top quark  pair signal and $Z+$jets background events
were generated with \alpgen+\pythia, diboson events are generated with pure
\pythia. The samples were normalised to theoretical cross-sections. All
simulated events are passed through the full D\O\ detector simulation and
reconstruction. 
The bias and widths of the pull distribution  observed in the ensemble tests 
are used to correct the above measured numbers. 

Systematics are dominated by uncertainties related to the jet energy scale of
the two $b$ jets in the event. They account to about $\pm2\GeV$ of the total
systematic uncertainty. 

D\O\ has applied the Matrix Element method to determine the top quark mass to the
Run\,IIa ($1.1\ifb$) and Run\,IIb ($2.5\ifb$) run periods 
separately and combined the individual results
using the BLUE method~\cite{\refblue} with the error categories
used by the Tevatron Electroweak Working Group~\cite{:1900yx} (see also
subsection \ref{sect:MassCombination} below):
\beq
m_t= 174.8\pm3.3_\mathrm{stat}\pm2.6_\mathrm{syst}\GeV~\mbox{.}
\eeq
Combinations with the other D\O\ results in the dilepton are available
in~\cite{d0note5897} also.

\subsubsection{Lepton Momentum Method}
\label{cdf-mt-leptonpt-dilepton}
The so far described methods of determining the top quark mass in the dilepton
channel, suffer from uncertainties of the jet energy scale. This uncertainty
dominates here the uncertainties much more than in the lepton plus jets
channel, where the in-situ calibration allows to constrain the jet energy scale. 
Measurements with lepton based quantities thus yield complementary results.

CDF has performed an analysis of $2.8\ifb$ of data in the dilepton channel
using the transverse momentum of the leptons to extract the top quark
mass~\cite{CdfNote9831}. 
Events are selected by requiring two isolated leptons of opposite charge and large missing transverse momentum.
For the signal sample at least two jets are required, one of which needs to be
identified as $b$ jet.

The background in this selection is dominated by fake signals from multijet
events, which is estimated from events with same charge leptons.
Contributions from diboson and $Z/\gamma$ events are simulated with \pythia\ passed through
full detector simulation and reconstruction. Events of $W+b\bar b+$jets are
generated with \alpgen+\pythia.

To extract the top quark mass from the selected
events the distribution of the lepton transverse
momenta is employed. The expected distribution of the leptons' transverse
momenta is parametrised as function of the top quark mass. A binned likelihood
is used to compare the observed data to parametrised templates.
The observed distribution and the best fit templates are shown in
\fig{fig:CDF-dilept-mt-lpt}. 
\begin{figure}
  \centering
  \includegraphics[width=0.9\linewidth]{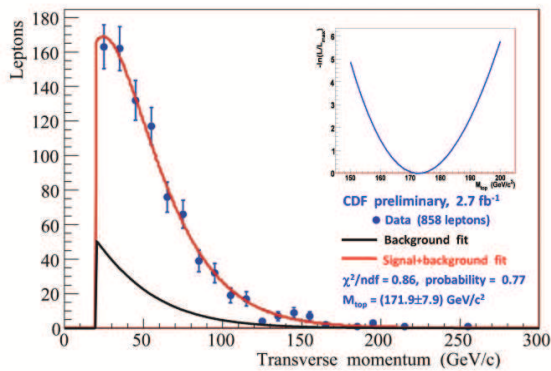}
  \caption{Distribution and fit of lepton transverse momenta observed in $2.7\ifb$ of
    CDF data using dileptonic events. For fitted contribution of the background is
    shown in black. The inset shows the corresponding likelihood
    as function of the top quark mass~\cite{CdfNote9831}.}
  \label{fig:CDF-dilept-mt-lpt}
\end{figure}
Systematic uncertainties of this measurement are dominated by effects related to signal
simulation. The choice of the generator and the uncertainty in modelling initial and
final state radiation contribute with $1.5\GeV$ and $1.3\GeV$, respectively.
The momentum scale contributes with only $0.7\GeV$.

In this preliminary result based on $2.7\ifb$ of data the top quark mass 
determined from  the shape of the lepton $p_T$ distribution is~\cite{CdfNote9831}
\beq
m_t=154.6\pm13.3_\mathrm{stat}\pm2.3_\mathrm{syst}\GeV~\mbox{.}
\eeq
The combination with the corresponding study of semileptonic events~\cite{CdfNote9683},
c.f. section \ref{cdf-mt-leptonpt-ljets}, yields~\cite{CdfNote9881}
\beq
m_t=172.8\pm7.2_\mathrm{stat}\pm2.3_\mathrm{syst}\GeV\quad\mbox{.}
\eeq
The statistical power of these methods is not
sufficient to yield significant contributions to the average from
measurements at the Tevatron. Due to its independent and low systematics 
it will become more relevant at the LHC, though.

\subsection{All Hadronic Channel}

The full hadronic channel is the most difficult of all top quark  pair decay modes. 
The channel has the advantage of the largest branching fraction and fully
reconstructed events, but the huge
background from multijet production makes it difficult to separate signal from
background.

D\O\ has published a measurement of the top quark  mass using the all hadronic decay
channel in Run\,I using the template method~\cite{Abazov:2004ng}, but not yet
published a result  with Run\,II. 

\subsubsection{Template Method}

CDF has applied a template method with in-situ jet energy calibration to
$2.9\ifb$ of data~\cite{Aaltonen:2010pe}. The event selection requires between
six and eight jets and vetoes on large missing transverse momentum and on
identified leptons in the event.
To further suppress the huge multijet
background without top quarks a neural network is employed which uses 13
input variables. These input variables include event shape observables like
sphericity, the minimal and maximal two and three jet invariant masses 
and observables based on the
transverse jet energies and the jet angles.
Only events above a threshold value of the neural
network output are accepted. The threshold has been optimised for 
the most precise top quark mass result and depends on the number of $b$ tagged
jets in the events. Finally, at least one of the six leading jets 
needs to be identified as $b$ jets using CDFs secondary vertex tagger. 

The background remaining after the selection is estimated from data. Jet by
jet tag rate functions are measured in events with exactly four jets and
parameterised as  function of the jet transverse energy, the track
multiplicity and number of vertices in the event.  
These jet tag rates are applied to the selected data with six to eight jets, 
but before requiring $b$ tagged jets. 
Corrections for the fact that heavy flavour quarks are produced in pairs and
for the signal content in the pre-tag samples are applied.
This background model was verified on data with the neural network cut inverted.

The top quark  mass is reconstructed in each selected event by performing a
constrained kinematic fit to the six leading jets. The quark energies are
allowed to vary within the experimental resolution  of the  measured jet
momenta. The two pairs of light quark momenta  are constrained to the $W$ boson
mass within the $W$ boson width. The invariant mass of the 
two triples of quarks from the top and
anti-top quark  decays are constrained to agree with the top quark  mass
within the nominal top quark  width, which is assumed as $\Gamma_t=1.5\GeV$.
The top quark  mass and the quark energies are free parameters of the fit.
The corresponding fit $\chi^2$ can thus be written as follows:
\bea
\label{eq:cdf-alljets-constraint-chi2}
\chi^2&=&
\frac{(M^{(1)}_{q\bar q}\!-\!M_W)^2\!\!}{\Gamma_W^2}
+\frac{(M^{(2)}_{q\bar q}\!-\!M_W)^2\!\!}{\Gamma_W^2}
\nonumber\\
&&
+\frac{(M^{(1)}_{bq\bar q}\!-\!m_t^\mathrm{reco})^2\!\!}{\Gamma_t^2}
+\frac{(M^{(2)}_{\bar{b}q\bar q}\!-\!m_t^\mathrm{reco})^2\!\!}{\Gamma_t^2}
\nonumber\\
&&
+\sum_{i=1}^6\frac{(p_T^{i,q}-p_T^{i,\mathrm{jet}})^2}{\sigma_i^2}
\,\mbox{.}
\eea
Of the possible jet to parton assignments required for the above fit, only
those that assign $b$ tagged jets to $b$ quarks are used. With six jets this
still allows thirty permutations for events with one identified $b$ jet and
six permutations in
presence of two tagged jets.  The top quark  mass obtained  
from the jet-parton assignment that yields the lowest $\chi^2$ is used as the
reconstructed mass for the event under consideration.
To obtain a handle on the jet energy scale, also the $W$ mass is reconstructed
in each event. This is done by repeating the above fit with also the $W$ boson
mass as a free parameter. Again the jet-parton assignment with the lowest
$\chi^2$ is chosen for further analysis.

\begin{figure}
  \centering
\includegraphics[width=0.4\textwidth,clip,trim=0mm 0mm 0mm 15mm]{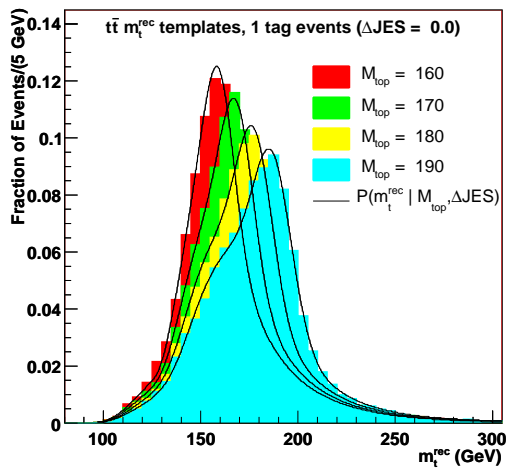}
  \caption{Signal templates for various nominal top quark masses 
%    (top) and jet energy scale shifts (bottom) 
    obtained for the sub-sample of events with
    exactly one $b$ tagged jet. The lines correspond to the fitted functions
    used in the likelihood computation~\cite{Aaltonen:2010pe}.}
  \label{fig:cdf-mtop-alljets-template}
\end{figure}

The distribution of these mass values reconstructed in each of the data events
is compared to templates for various nominal top quark  masses. These templates are
constructed by applying the above fitting procedure to events from top quark  pair
signal simulation for various nominal top quark  masses 
and to the described background background estimate. For the
background estimate  the reconstructed top quark and $W$ boson masses are
entering the distributions with weights computed from the tagging
probabilities and the corrections described above.
Signal templates are created not only for various top quark  mass values but also for
a range of jet energy scale shifts, $\Delta_\mathrm{JES}$. To obtain smooth
signal templates the distribution obtained at discrete values of the nominal
top quark mass and jet energy scale shift are fitted by functional
forms. See \fig{fig:cdf-mtop-alljets-template} for examples of such templates
and their parametrisation.

Now the results obtained in data are compared to these templates
with a likelihood that is maximised  with
respect to the top quark  mass, $m_t$, the jet energy scale shift, $\Delta_\mathrm{JES}$, and the number of signal
and background events. The likelihood consists of a term describing the
agreement with the $W$ boson mass templates %, ${\cal L}_{\mathrm{JES}}$, 
and of terms for the subsamples containing one or more than one $b$ tagged jets, 
${\cal  L}_{1}$ and ${\cal  L}_{2}$: 
\beq
{\cal L}(m_t,\Delta_\mathrm{JES})=
%{\cal L}_{\mathrm{JES}}(m_t,\Delta_\mathrm{JES})
\exp\left(-\frac{\Delta_\mathrm{JES}^2}{2\sigma_c}\right)
{\cal L}_1(m_t,\Delta_\mathrm{JES})
{\cal L}_2(m_t,\Delta_\mathrm{JES})\mbox{.}
\eeq
The individual terms are constructed very similar to the CDF 
template method lepton plus jets, see \eq{eq:cdf-ljet-template-likelihood}.

\begin{figure}[b]
  \centering
\includegraphics[width=0.8\linewidth,clip,trim=0mm 0mm 0mm 9mm]{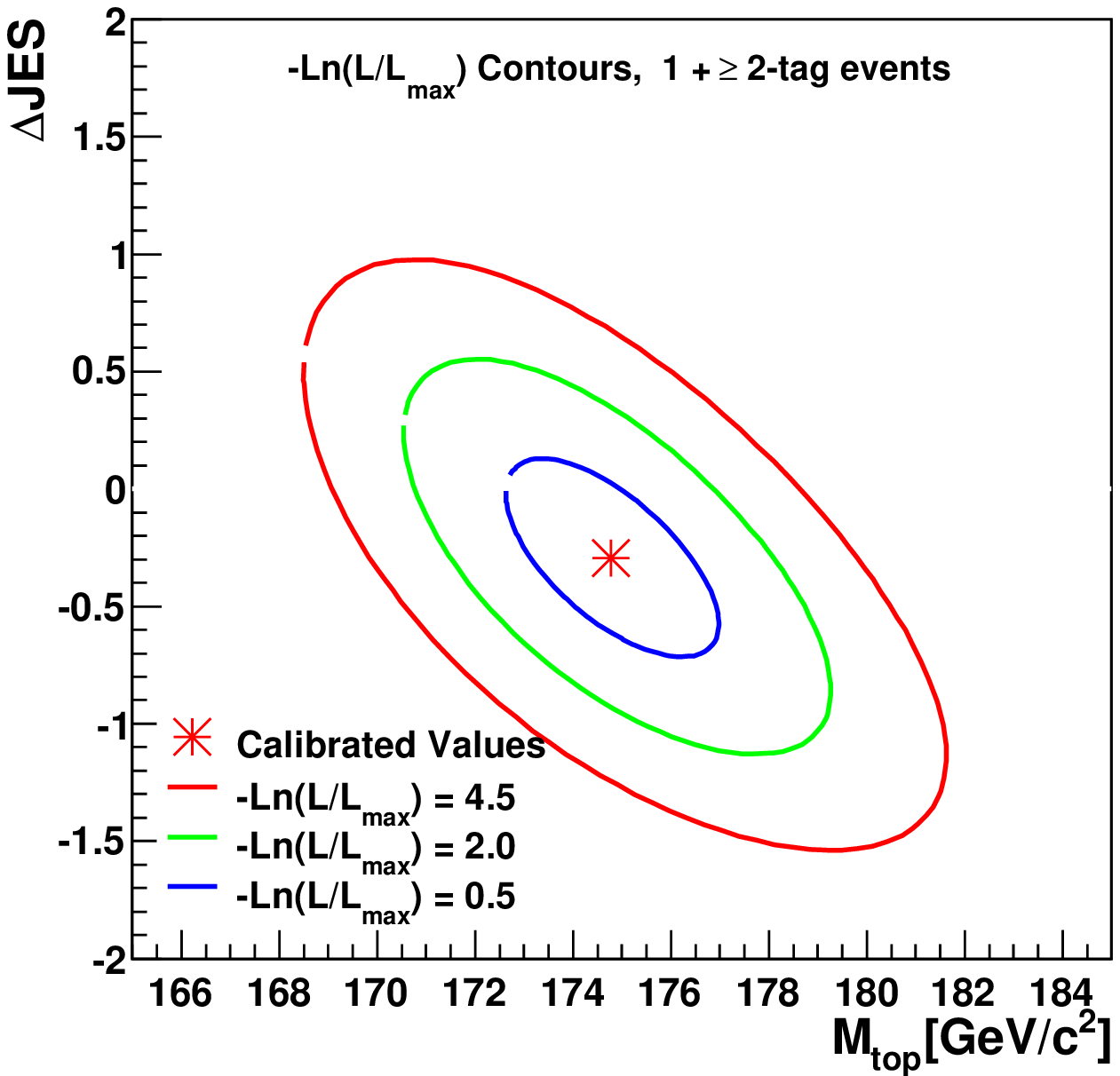}
  \caption{Top quark mass and jet energy scale shift observed with  the template method by CDF in
    $2.9\ifb$ using the all hadronic decay channel. The curves correspond to 
    points of equal likelihood distance from the optimum~\cite{Aaltonen:2010pe}.}
  \label{fig:cdf-mtop-alljets-result}
\end{figure}
To estimate the performance of the method, the procedure is applied to
ensembles of pseudo-data for various nominal values of the top quark mass and
of the jet energy scale shift. 
The pseudo-data are constructed from the simulated signal and
the data based background templates.
Then the nominal top quark mass for an ensemble of pseudo-data is compared to
the average of the results obtained for each of the pseudo-data in that
ensemble. Similarly, a calibration curve for the jet energy scale shift is
constructed. These calibration curves show excellent linearity and only very
minor overall offsets. In addition the uncertainty estimate is verified with the
width of the pull distribution and corrected accordingly.

Systematic uncertainties are estimated as the average effect
determined on ensembles of pseudo-data with systematic effects included.
The largest single uncertainty is estimated to stem from residual (mass and
jet energy dependent) biases from the template parametrisation, not covered by
the calibration procedure (${}^{+0.8}_{-0.4}\GeV$). Residual effects of the
jet energy scale follow with $\pm0.5\GeV$. This analysis also evaluates the
effect of changing the underlying event model from a default~\cite{tunea} 
to a new model including colour reconnections~\cite{Sandhoff:2005jh,Skands:2007zg,Wicke:2008iz} 
and finds an uncertainty
of $0.4\GeV$. 

In $2.9\ifb$ of data CDF determines the top quark mass with in-situ jet energy
calibration from events in the all hadronic channel to be~\cite{Aaltonen:2010pe}
\beq
174.8\pm2.4_{\mathrm{stat+JES}}{}^{+1.2}_{-1.0~\mathrm{syst}}\GeV~\mbox{.}
\eeq
The measured two dimensional likelihood is shown in
\fig{fig:cdf-mtop-alljets-result}. 
The jet energy scale shift determined is consistent with CDFs nominal value,
but has a smaller total uncertainty.

\subsubsection{Ideogram Method}

Also the ideogram method described for the case of lepton plus jet
events in Section~\ref{sec:ljets-ideogramm} is well suited to be applied in
the all hadronic channel. 

CDF applied this method of measuring the top quark mass in up to $1.9\ifb$ of
data~\cite{Aaltonen:2006xc,CdfNote9265}. The first steps of the analysis are
very similar to the template based analysis for the all-hadronic channel.
The selection of events requires exactly six jets. A neural network  is used
to further suppress background from multijet production with no top quarks.
For the signal sample at least two of the jets are required to be identified
as $b$ jets using CDFs secondary vertex algorithm.

The background contribution after this selection is computed by applying jet by jet tag
rate functions on the data before the $b$ tag requirement. The required tag rate
functions are derived on events with four jets, where the signal contamination
is negligible.

Then a constrained kinematic fit is performed to simultaneously determine 
the top quark mass and the $W$ boson mass %, $m_t$ and $M_W$,
for the event under consideration. 
Transfer corrections are applied to correct differences between the jet energy
inside the chosen cone radius of 0.4 and originating quark. These transfer
corrections optionally implement a jet energy scale shift.
The constraints of the fit are that the quark momenta  yield identical top quark
masses and identical $W$ boson masses in the two decay chains of the event, 
c.f.~\eq{eq:cdf-alljets-constraint-chi2}. 
The result for all possible jet
to parton assignments which assign $b$ tagged jets to (anti-)$b$ quarks
are kept for the further analysis.

With these numbers in the ideogram method an event by event likelihood is
computed that describes the probability to observe the reconstructed top quark
and $W$ boson masses for the various jet to parton assignments given the true
top quark mass and jet energy scale shift. This likelihood consists of 
probabilities, $P_\mathrm{sig}$ and $P_\mathrm{bkg}$,
corresponding to the signal and the background hypothesis,
c.f.~\eq{eq:ljet-ideogram-likelihood}.
The signal and background probabilities obtained for each of the jet to parton assignment are
weighted by their  $\chi^2$ probability, $w_i$, which depends on the jet
energy scale applied. In addition CDF adds a term to the signal probabilities
to describe events with no correct jet to parton assignment of the selected
six jets:
\bea
\label{eq:allj-ideogram-weights}
{P}_\mathrm{sig}(x;m_t,\Delta_\mathrm{JES})&=&f_\mathrm{nm}\sum_{i=1}^{N}w_i(\Delta_\mathrm{JES})S(m_{t,i},M_{W,i},\Delta_\mathrm{JES})\nonumber\\
&& + (1-f_\mathrm{nm}) S_\mathrm{nm}(m_{t,i},M_{W,i},\Delta_\mathrm{JES})
\nonumber\\[0.7em]
{P}_\mathrm{bkg}(x;\Delta_\mathrm{JES})&=&\sum_{i=1}^{N}w_i(\Delta_\mathrm{JES}) B(m_{t,i},M_{W,i})~\mbox{.}
\eea
$N$ is the number of possible jet parton assignments, i.e. 6 for doubly tagged
events and 18 for events with three $b$ tagged jets. $\Delta_\mathrm{JES}$ the
jet energy scale shift in units of the nominal jet energy scale uncertainty,
$m_{t,i}$ and $M_{W,i}$ are the fit results for the jet parton assignment
number $i$ and $1\!-\!f_\mathrm{nm}$ is the fraction of events with no correct jet
parton assignment. % and is determined to be 41\%.

The signal probability, $S$, is written as the convolution of the theoretically
expected Breit-Wigner, ${\bf BW}$, and the experimental smearing represented
by a  Gaussian, ${\bf G}$. Here two Breit-Wigner functions are
needed, one for the top quark and one for the $W$ boson, and the
Gaussian is two dimensional, constructed including correlations between the
top quark and the $W$ boson mass extracted from the kinematic fit:
\bea\label{eq:allj-ideogram-S}
&&S(m_{t,i},M_{W,i},\Delta_\mathrm{JES})=\nonumber\\
&& \int\!\!\!\!\int\!
{\bf{G}}(m_{t,i},M_{W,i},m',M',\sigma_i)\qquad \nonumber\\
&&\qquad\qquad{\bf{BW}}(m',m_t){\bf{BW}}(M',M_W)\,\,\mathrm{d}m'\mathrm{d}M'~\mbox{.}\quad
%\nonumber\\
% ANDERS ALS {D\O} &&+(1-f_c) S_\mathrm{nm}(m_i,m_t;n_\mathrm{tag})
\eea
The background probability, $B$, and the probability for the signal events with
no matching, $S_\mathrm{nm}$, are derived from simulation. 
The final likelihood is then written as the product of the event likelihoods
times a prior likelihood for the jet energy scale shift to describe the
external calibration
\beq
{\cal  L}(m_t,\Delta_\mathrm{JES})=
\exp\left(-\frac{\Delta_\mathrm{JES}^2}{2\sigma_c}\right)
\prod\limits_\mathrm{events}{\cal L}_\mathrm{evt}(x;m_t,\Delta_\mathrm{JES})
\eeq
and maximised with respect to the top quark mass, $m_t$, and the jet energy
scale shift, $\Delta_\mathrm{JES}$.

The performance of this method is evaluated in ensemble tests. Ensembles of
pseudo data are generated for a grid of nominal top quark masses and jet
energy scale shifts using the \pythia\ generator and full CDF detector
simulation and reconstruction. 
The mean response of the analysis on these ensembles is
used to correct the top quark mass and jet energy scale shift obtained  in
data. 
\begin{figure}[t]
  \centering
  \includegraphics[width=0.9\linewidth]{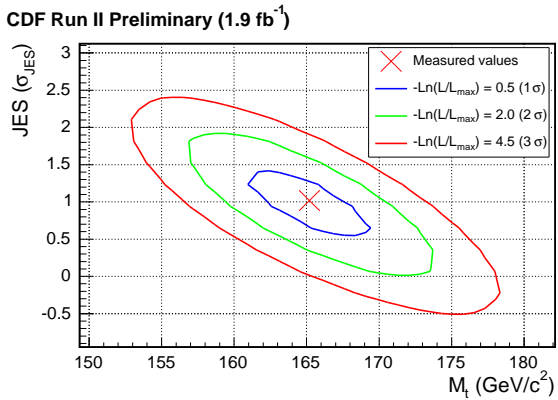}
  \caption{Top quark mass and jet energy scale shift observed with the
    ideogram method by CDF in
    $1.9\ifb$ using the all hadronic decay channel. The curves correspond to 
    points of equal likelihood distance from the optimum~\cite{CdfNote9265}. }
  \label{fig:mt-allj-ideogramm}
\end{figure}

Systematic uncertainties are evaluated on ensembles of pseudo data with the
corresponding systematic variation. A variation of initial and final state
parameters of the simulation yields the dominating contribution of
$\pm1.2\GeV$, followed by the difference obtained from comparing \pythia\ to
\herwig ($\pm 0.8\GeV$).  The residual jet energy scale contributes with
$\pm 0.7\GeV$.

In $1.9\ifb$ of data CDF measures the top quark mass from all hadronic events
to be~\cite{CdfNote9265}
\beq
m_t=165\pm4.4_\mathrm{stat+JES}\pm1.9_\mathrm{syst} \GeV~\mbox{.}
\eeq
Figure~\ref{fig:mt-allj-ideogramm} shows the result with points of equal
likelihood distance from the optimum in the top quark mass vs. jet energy
scale shift plane.

\subsection{Top Quark  Mass from Cross-Section}
As discussed in Section~\ref{sec:mass-theo} the definition of quark masses is inherently ambiguous. 
The quoted mass values are often considered to be given in the pole mass
definition. However, this definition usually yields significant higher order
corrections. For Monte Carlo simulations the order of the corrections included
in the determination cannot be easily computed due to the presence of parton
shower cut offs and modelling of hadronisation.

Deriving the top quark mass from the top quark cross-section measurements
avoids using the simulation for calibration and allows the determination of
the top quark mass using a well defined mass definition in an understandable approximation.

Both  predicted and the  measured cross-section depend on the true top
quark mass. In the theoretical prediction this dependence stems from the
change in phase space with varying top quark mass. 
For the experimental results this dependence is introduced
through the selection efficiencies which vary with the amount of energy available
for the decay products.

\begin{figure}[b]
  \centering
\includegraphics[width=0.9\linewidth]{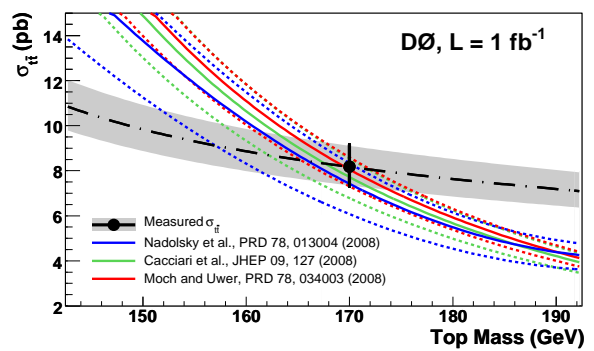}
  \caption{Comparison of the theoretical top quark pair 
  cross-section to the experimental results ($\ell+$jets, dilepton and
  $\tau+$lepton) with their top quark mass
    dependence. The lines show parametrised dependence and the error
    bands~\cite{D0Note5742conf}.} 
  \label{fig:mt-from-xsec}
\end{figure}
D\O\ has compared their cross-section measurement for up-to $1\ifb$ of data 
to theoretical predictions as function of the top quark pole mass for various
theoretical approximations~\cite{Abazov:2008gc,Abazov:2009si,Abazov:2009ae}.  

For the extraction of the top quark mass the 
dependence of the theoretical predictions and the experimental cross-section
results are parametrised with a polynomial. Both the theoretical and the experimental
cross-section uncertainties are assumed to be Gaussian to build a likelihood
that then allows to find the top quark pole mass that yields the best
agreement and its uncertainty. This assumption is justified also for the
theoretical uncertainty as it is dominated by PDF uncertainties. 

The measured cross-section is compared to several theoretical predictions in 
\fig{fig:mt-from-xsec}: A pure NLO prediction~\cite{Beenakker:1988bq},
approximate next-to-next-to leading order~\cite{Moch:2008ai,Kidonakis:2008mu} and a 
next-to-leading order 
with resummation of leading and next-to-leading soft
logarithms~\cite{Cacciari:2008zb}. 
All are evaluated with the CTEQ6.6 parton density
distribution~\cite{Nadolsky:2008zw}.

In this method statistical and systematic uncertainties are already combined. 
Using the combined cross-section in $\ell+$jets, dilepton and $\tau+$lepton, 
D\O\ derives in $1\ifb$~\cite{Abazov:2009ae}:
\beq
\begin{array}[b]{lrcl}
\mbox{NLO}~\mbox{\cite{Beenakker:1988bq,Nadolsky:2008zw}}\qquad  & m_t& =& 165.5\pm 6.0\GeV\\
\mbox{NNLO}_\mathrm{approx}~\mbox{\cite{Moch:2008ai}}\qquad  & m_t& =& 169.1\pm 5.6\GeV\\
\mbox{NNLO}_\mathrm{approx}~\mbox{\cite{Kidonakis:2008mu}}\qquad  & m_t& =& 168.2\pm 5.6\GeV\\
\mbox{NLO+NLL}~~\mbox{\cite{Cacciari:2008zb}}               & m_t& =& 167.5\pm 5.7\GeV~\mbox{.}\\
\end{array}
\eeq
The results on the top quark pole mass agree well with the world average of direct measurement, but
have a much larger uncertainty. %, see \fig{fig:mt-from-xsec}~(right). 
As these numbers refer to the same data, the differences between the results
of  about $\pm 2\GeV$ has to be attributed to theoretical differences and may be
an indication of the sensitivity of the pole-mass definition to higher order corrections.

Recently in a first determination of the $\msbar$-mass of the top quark has
been presented~\cite{Langenfeld:2009wd}. Based on the data of the same {D\O}
cross-section result~\cite{Abazov:2009ae} a running top quark mass of
$\bar{m}_t=160.0\pm 3.3\GeV$ is extracted using the approximate \mbox{NNLO}
prediction~\cite{Moch:2008ai}. To compare this number to the pole
mass result quoted above it has to be converted to the pole mass. This
conversions strongly depends on the order used. In leading order no change
occurs, while using the NNLO formula yields ${m}_t=168.2\pm 3.6\GeV$.
(For simplicity the slightly asymmetric uncertainties were symmetrised by the
author.)
It should be noted that the $\msbar$-mass determined
from different orders of the $\msbar$ cross-section calculation change by
less than $1\GeV$ between leading order and next-to-next-to-leading order,
while the corresponding pole mass of the same order changes by nearly $10\GeV$.

\subsection{Modelling of Non-Perturbative Effects}
% CR & Top Mass

Modelling non-perturbative effects in simulations is a notoriously difficult
task. Beside the effect of hadronisation, which is believed to be well
understood since LEP times, in hadron-hadron collisions multiple parton
interactions yield a different kind of non-perturbative effects. 
The uncertainties on the top quark mass due to their modelling uncertainties
have only recently been considered.

In the context of hadron collisions at the Tevatron several efforts
to tune the default model present in \pythia\ to data~\cite{tunea} resulted in
several parameter tunes, known as Tune A, Tune DW, Tune D6, etc. These tunes
required large values of parameters describing the colour correlations between
the multiple parton interactions and the hard process. This is often
interpreted as an implicit hint for the need of colour reconnections, i.e. a
modification of the colour flow obtained from the simulation of the hard and
semi-hard process and parton showers.

Newer versions of \pythia\ implement an extended
\linebreak 
model to describe the
underlying event including variants of the \pythia\ model with explicit
colour-reconnection~\cite{Sandhoff:2005jh,Skands:2007zg}. These are tuned  
and their influence on the top quark  mass measurements is
studied with a toy top quark mass analysis~\cite{Skands:2007zg,Wicke:2008iz}.
Compared to the variants of the default \pythia-model the new model and its
variants yields shifts in the extracted
top quark mass of up to $1\GeV$. The greater portion, namely $0.7\GeV$,
of this shift can be assigned to
differences between the different parton showers used in the old and the new
underlying event models. Less than $0.5\GeV$ is attributed to non-perturbative
effects. 
It should be noted here that additional alternative models of colour reconnection suitable 
for hadron collisions exists~\cite{Rathsman:1998tp,Webber:1997iw}, but were not yet
studied in the context of the top quark mass.

While the smallness of the non-perturbative portion of the derived shift on
the top quark mass 
confirms the prejudice that non-perturbative effects are
expected at the order of (a few times) the confinement scale,
$\Lambda_\mathrm{QCD}$, the sizable shift due to parton shower details is very
worrying. Only very recent top quark mass measurements include an uncertainty
on the colour reconnection effects (see the previous subsections) and confirm
the estimated uncertainty due to modelling of non-perturbative effects. The greater shift due to the parton shower
differences is still not included.

It has recently been realised that the two types of models differ mainly in
their $b$-quark jet energy scale and that the shape of $b$-jets does not agree
with the models using the $p_T$-ordered parton shower~\cite{lina:2010,Acosta:2005ix}.
The simultaneous determination of the light
quark jet energy scale and the $b$-quark jet energy scale as suggested by 
\cite{Fiedler:2007tf,Fiedler:2010sg}
may thus help to resolve the issue.

\subsection{Combination of Top Quark Mass Results}
\label{sect:MassCombination}
%\cite{Azzi:2004rc,unknown:2005cg,unknown:2006qt,:2005cc}

The Tevatron experiments regularly combine their Run\,I and Run\,II top quark mass results to
take advantage of the increase in statistical power. The most recent
combination was performed in July 2010~\cite{:1900yx}.

For each measurement that enters the combination a detailed break down of
errors is performed. Uncertainties that are believed to have correlations of
one measurement with any other measurement of the same or the other
collaborations are separated so that these correlations can be taken into
account in the combination. 
%The jet energy scale uncertainties, e.g. are 
%separated into contributions from the in-situ calibration, 
%contributions which stem from the external jet energy scale determination,
%uncertainties due to modelling of $b$ quark jets,
At this point only measurements from independent dataset or selections are
used in the combination. The evaluation of partial correlations, as they would
appear when using results form multiple methods on 
the same dataset and channel, are thereby avoided.
The various uncertainty contributions are considered 
to be either fully correlated or to be uncorrelated. 

The average is then computed with the best linear unbiased estimator
(BLUE)~\cite{\refblue}
and yields~\cite{:1900yx}:
\beq
m_t= 173.3\pm1.1\GeV~\mbox{.}
\eeq
Uncertainties on this estimation due to the approximations of the procedure
including a cross-check with reduced correlation coefficients was estimated to
be much smaller that $0.1\GeV$.

\subsection{Mass difference between Top and Anti-top Quark}

According to the CPT-theorem local quantum field theories require the mass of any particle to be equal to
that of its anti-particle. A measurement of the mass difference
between top and anti-top-quarks is a unique test of the validity of the
CPT-theorem in the quark sector, as other quark masses are more difficult to 
assess due to hadronisation.  
Both Tevatron experiments have extended their studies of the top quark mass to
determine this mass difference.

\subsubsection{Matrix Element Method}
D\O\ extends the Matrix Element method using $1\ifb$ of data~\cite{Abazov:2009xq}. 
The analysis selects events with one isolated lepton, transverse missing energy and exactly four
jets. At least one of the jets is required to be identified as $b$-jet.

The Matrix Element method used to measure the top quark mass as described
in Sect.~\ref{d0-ljets-me-mass} is extended by explicitly keeping a separate
dependence on the top quark mass, $m_t$, and the anti-top quark mass, $m_{\bar t}$
in Eqs.~(\ref{eq:d0-ljet-me-master}) to (\ref{eq:d0-ljet-final-like}). 
For this the leading order matrix element is
rewritten to explicitly depend on $m_t$ and $m_{\bar t}$. 
At the same time the dependence on the overall jet energy scale factor,
$f_\mathrm{JES}$, is dropped. 
This yields a likelihood ${\cal L}(m_t,m_{\bar t},f_\mathrm{top})$ corresponding to \eq{eq:d0-ljet-final-like}.
The likelihood to
observe a mass difference of  $\Delta= m_t- m_{\bar t}$ is obtained by
integrating over all possible average mass values $m_\mathrm{avg}=(m_t+ m_{\bar t})/2$.
\beq
{\cal L}(\Delta)=\int\!\mathrm{d}m_\mathrm{avg}\,{\cal L}(m_t,m_{\bar t},f^\mathrm{best}_\mathrm{top}(m_t,m_{\bar t}))
\eeq
Here $f^\mathrm{best}_\mathrm{top}(m_t,m_{\bar t})$ is the fraction of top
quark events  fitted 
for the considered values of $m_t$ and $m_{\bar t}$.
The measured mass difference $\Delta$ is the one that maximises this likelihood.

As for the mass measurements the method is calibrated using
pseudo-experiments. The pseudo-data in these tests are constructed
from $t\bar t$ signal events and $W+$jets background. For the signal 
various values of $m_t$ and $m_{\bar t}$ were generated with a modified version of
\pythia\ that allows to set \mbox{$m_t\ne m_{\bar t}$}. It was
confirmed that the mass difference measured in the various pseudo
datasets is very close to the generated value. The small deviation in the
reconstructed value and the derived uncertainty are corrected for in the final result.

Many uncertainties that are important for the mass measurement  cancel in the
determination of the top anti-top quark mass difference. This includes the
uncertainties due to the jet energy scale and the uncertainty due
to a difference in the detector response between light and $b$ quark jets.
Instead the dominating systematic uncertainty in the measurement of the mass
difference is that of modeling additional
jets in events of top quark pairs. This signal modeling uncertainty is
estimated from data and simulation. 
\begin{figure}[b]
  \centering
\includegraphics[width=0.48\linewidth]{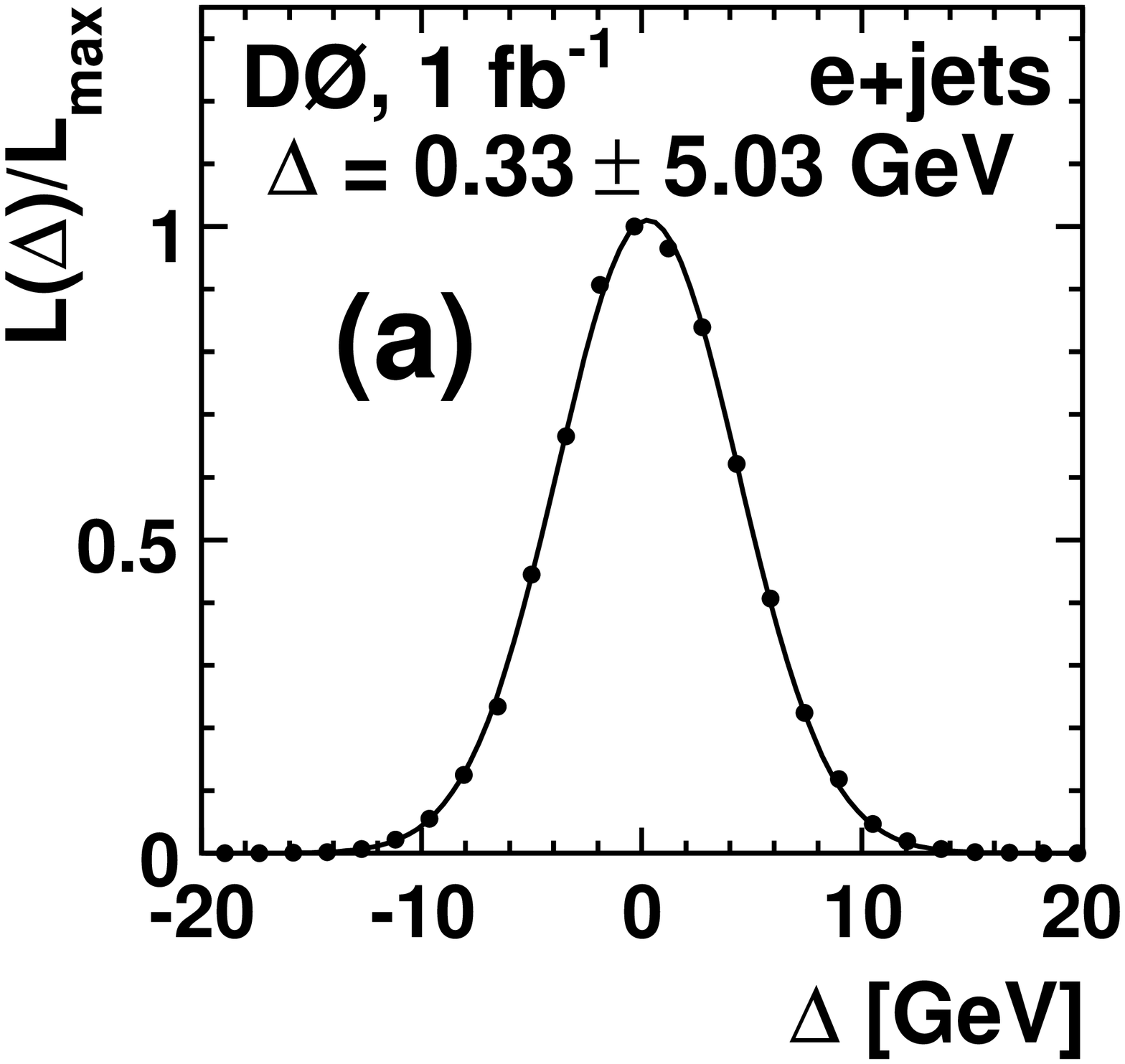}  
\includegraphics[width=0.48\linewidth]{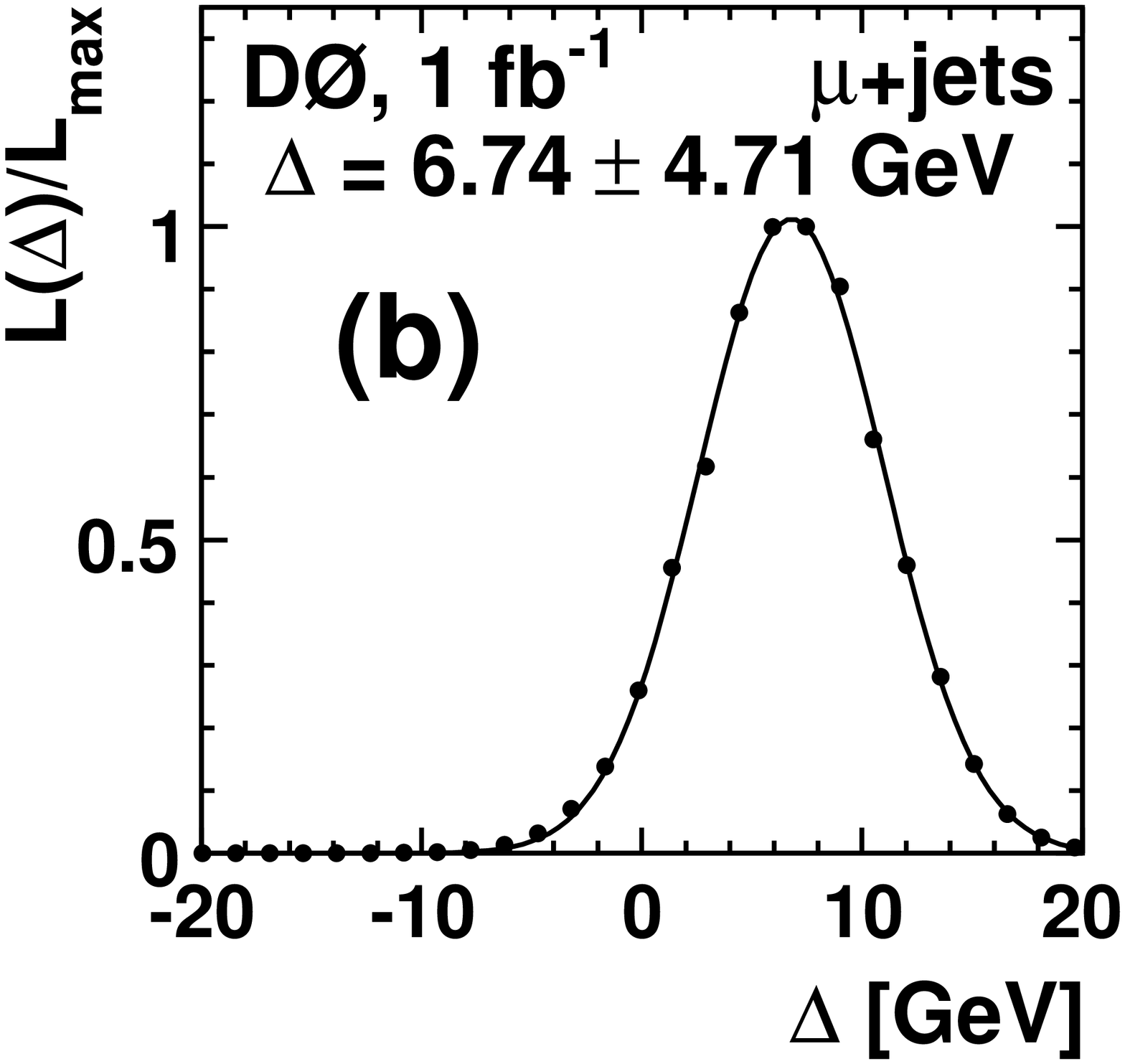}
  \caption{Normalised likelihood distribution as function of the top to
    anti-top quark mass difference, $\Delta$, as measured by D\O\ in the
    electron and muon plus jets channels, separately~\cite{Abazov:2009xq}.}
  \label{fig:d0massdiff}
\end{figure}
Using simulation the amount of events with more than four jets is varied to
agree with data. In addition, pseudo experiments are performed in which 
appropriate data events with more than four jet are added to the standard
signal simulation with various amounts.
%not used  elsewhere in the measurement.
%In addition real events with more than four jets are not used in the
%measurement are used to in pseudo datasets to study their effect on the
%result. 
In combination D\O\ finds an uncertainty of $0.85\GeV$ due to signal modelling.
Because of the necessity to distinguish top from anti-top quarks also the
uncertainty in the lepton charge determination and the uncertainty in
modelling the differences of the calorimeter response for $b$ and $\bar b$
jets is evaluated. The latter gives a sizable contribution of 0.4\GeV\ in the D\O\ result,
but its determination is limited by the statistics of the corresponding
simulation.
Due to the negligence of multijet background in the likelihood computation and
its calibration the multijet contamination yields another contribution of $0.4\GeV$.
The total systematic uncertainty is $1.2\GeV$. 

D\O\ applies the described method using $1\ifb$ of data in the $e+$jets
and the $\mu+$jets channel separately. The corresponding likelihood functions
are shown in \fig{fig:d0massdiff}.
The two channels are then combined by a
weighted average which yields
\beq
\Delta=3.8\pm3.7\GeV
\eeq
in good agreement with the CPT-theorem expectation of zero~\cite{Abazov:2009xq}.
As a cross-check the likelihood 
${\cal L}(m_t,m_{\bar  t},f^\mathrm{best}_\mathrm{top}(m_t,m_{\bar t}))$ 
is integrated over all possible mass differences and used to determine the
standard top quark mass. From this the relative difference is obtained to be
\beq
\quad\Delta/m_t = (2.2\pm 2.2)\%\quad\mbox{.}
\eeq

\subsubsection{Template Method} % (CDF)
For the measurement of the top quark mass difference~\cite{CdfNote10173}
CDF has expanded the template method with full reconstruction of the top quark
decay products described in Sect.~\ref{CDF-template-ljets-portion}.
The event selection requires one isolated energetic lepton, missing transverse
energy and at least four energetic jets in the final state. Events are categorised by
the charge sign of the lepton and the number of identified $b$-jets. Only for events with more than one
identified $b$-jets more than four high $p_T$ jets are allowed.
 
In each event the mass difference is reconstructed with a fit constrained by
the top quark decay kinematics. The $\chi^2$ for this fit is described by
\eq{eq:cdf-constraint-chi2} when replacing $m_t^\mathrm{reco}$ with
$172.5\GeV\pm \Delta/2$.  If the top quark decays hadronically and the
anti-top quark leptonically, the positive sign applies for term including
$M_{bqq}$ and the negative sign for the term including $M_{b\ell\nu}$. 
If the top quark decays leptonically instead, the signs are exchanged so that
$\Delta$ represents the reconstructed mass difference in all
cases: $\Delta=m_t-m_{\bar{t}}$.

The $\chi^2$ is minimised for all possible associations of jets to the four
quarks requiring that identified $b$-jets are only associated to $b$-quarks.
For further analysis the values of $\Delta^{(1)}$ and $\Delta^{(2)}$ corresponding to the association
with the smallest and the second smallest $\chi^2$ are used. Events where the lowest $\chi^2$ is
larger than 3.0 (9.0) are rejected for events without (with) $b$-tagged jets.

Simulations are used to compute the expected signal contributions as function
of the two reconstructed mass difference values for various nominal mass
differences using \madgraph+\pythia. 
The contribution of multijet background is modelled using data with loosened
lepton identification criteria. 
The distribution of mass differences reconstructed in
$W+$jets background is modeled with \alpgen+\pythia. Its normalisation is
derived from data. Further smaller backgrounds are taken from simulation
normalised to the NLO cross-sections.

As for the mass determination (Sect.~\ref{CDF-template-ljets-portion}) 
the signal simulations are used to derive a probability density,
$P^\mathrm{sig}$, to reconstruct a pair $\Delta^{(1)},\Delta^{(2)}$  for a signal event 
with a given nominal mass difference. These two dimensional functions depend
on the lepton charge, %see \fig{fig:cdf-massdiff-2dprob}, 
which is the
reason to separate samples of different lepton charge.
Similarly a probability density for
background events, $P^\mathrm{bkg}$, is computed. 
Then a likelihood for the six
subsamples, $i=1...6$, is computed as
\bea\label{eq:cdf-ljet-mdiff-template-likelihood}
%\nonumber\\
{\cal  L}_i&=&
\exp\left(
-\frac{(b_i-b_i^\mathrm{e})^2}{2\sigma_{b_i}^2}\right)
\prod_\mathrm{events}
\frac{s_iP_i^\mathrm{sig}+b_iP_i^\mathrm{bkg}}{s_i+b_i}
\eea
with $s_i$ and $b_i$ being the number of signal and background events in the
corresponding sample;  $b_i^\mathrm{e}$ and $\sigma_{b_i}$ are the corresponding background
expectations and its error. 

The final likelihood, i.e. the product of all ${\cal L}_i$, is available only
at discrete values of the nominal mass difference. For the final maximisation
of the likelihood it is interpolated with polynomial smoothing. 
CDF found their method to be bias free in pseudo experiments. A small
correction to the derived statistical uncertainty is applied. 

Systematic uncertainties are dominated by the signal modeling uncertainty of
$0.7\GeV$, which  is
derived by comparing \madgraph\ to pure \pythia\ and by comparing the
results of using the \herwig\ to the \pythia\ parton shower. The next to
dominating uncertainties stem from the description of multi hadron
interactions and differences in the detector response for $b$ and $\bar b$
quarks.
The former is derived by verifying the agreement of the simulation with data
as function of the number of primary vertices.
The latter is determined by comparing the $p_T$ balance in di-$b$-jet events.
These effects yield systematic uncertainties of $0.4\GeV$ and $0.3\GeV$,
respectively.

Using $5.6\ifb$ of data CDF finds the top quark to anti-top quark mass
difference to be~\cite{CdfNote10173}
\beq
\Delta = -3.3\pm 1.4_\mathrm{stat}\pm1.0_\mathrm{syst}
\eeq
which agrees with the CPT-theorem at a little less than two standard
deviations.
The analysis assumes explicitly an average (anti-)top quark mass of $172.5\GeV$.
%%% Local Variables: 
%%% mode: latex
%%% TeX-master: "EPJC_TopProperties"
%%% End: 

\subsection{Conclusions and Outlook to LHC}
In the Standard Model of particle physics 
the top quark mass is an a priori unknown parameter. The Tevatron experiments
have employed many different techniques to determine its value.  Emphasis has
been put on the reduction of the experimental uncertainties. The Ideogram and
the Matrix Element methods aim at the maximal utilisation of the
available experimental information. The simultaneous 'in situ' determination of the jet
energy scale along with the top quark mass addresses the dominant systematic
uncertainty. Also the next dominant uncertainty, the $b$-jet energy scale, can
in principle be determined in situ, however, more statistics is needed to
achieve an improvement over the current uncertainty. 

These results constrain the Standard Model prediction for processes that involve top
quarks either as real particles or in virtual loops. Besides the real
production of top quarks, electroweak precision data are sensitive to the value
of the top quark mass. 
Even before the discovery of the top quark the electroweak precision data
constrained the top quark mass~\cite{lepewwg1994}. Now such indirect values
are compared with the direct measurements at the Tevatron to verify the
consistency of the Standard Model~\cite{:1900yx,:2005ema}.
Figure~\ref{fig:ew-mwmt} shows the direct measurement of the $W$ boson and
the top quark mass compared to the indirect top quark mass from electroweak precision
measurements. It indicates the Standard Model expectation as function of the Higgs
mass. Also the Standard Model agreement with the global electroweak results is
usually computed. Figure~\ref{fig:ew-blueband} shows the $\chi^2$ 
fit as function of the Higgs boson mass which also determines the
indirect constraints on the allowed Higgs boson mass. 

\begin{figure}[b]
  \centering
\vspace*{-3mm}\includegraphics[width=0.93\linewidth,clip,trim=0mm 14mm 0mm 18mm]{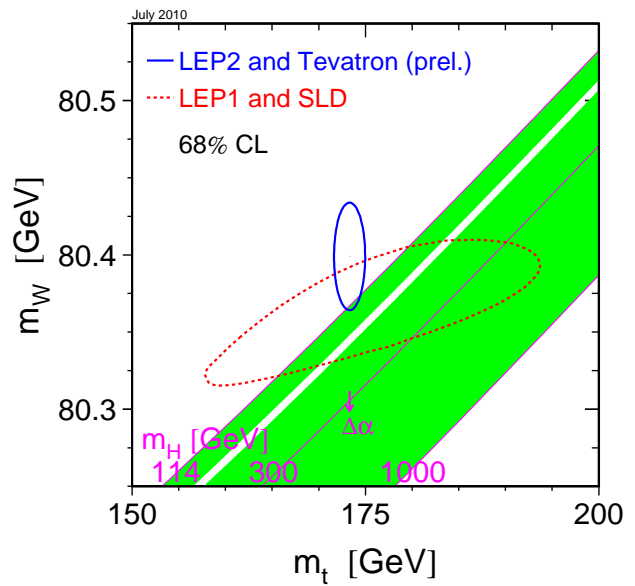}
  \caption{Mass of the $W$ boson vs. the top quark mass. The smaller
    (blue) contour indicates the one 68\% C.L. of the direct measurements, the
    larger (red) contour the indirect results. The expectation in the Standard
    Model forms diagonal lines that depend on the Higgs boson
    mass~\cite{:1900yx}.
  }
  \label{fig:ew-mwmt}
%\vspace*{3.75mm}
\end{figure}
\begin{figure}
  \centering
  \includegraphics[width=0.93\linewidth,clip,trim=0mm 12mm 0mm 17mm]{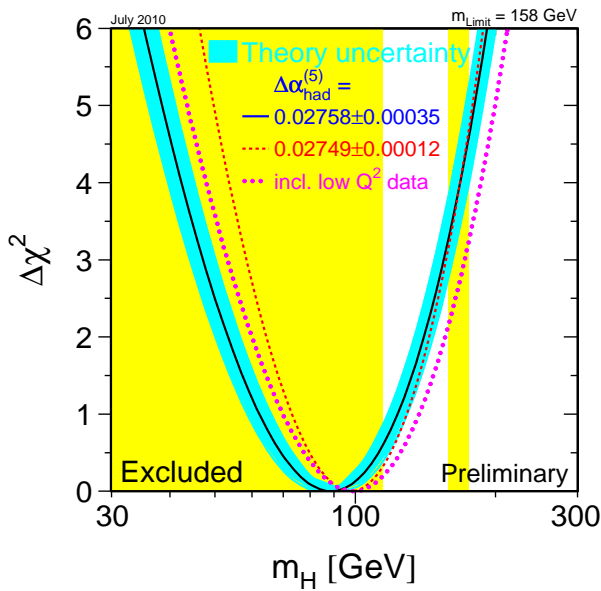}
  \caption{Quality of the agreement of precision measurements with the
    Standard Model prediction as function of the assumed Higgs boson mass
    after a global fit of the parameters to data. 
    The yellow region indicates the direct exclusion limit. The central result
    is shown as a back line with blue band. Alternative fits are shown in
    other colours~\cite{:1900yx}.}
  \label{fig:ew-blueband}
\end{figure}

% Diskussion der theoretischen Fragen
% im Zusammenhang mit Aublick auf erwartete Genauigkeit
% am Tevatron und beim LHC

Implicitly, such comparisons assume that the experimentally measured top quark mass
values are given in the pole mass scheme. This assumption, however, is not
confirmed to the level of the experimental precision of the combined result
from direct top quark mass measurements. 
Therefore conclusions based on such comparisons need to be drawn with great care. 
Hopefully, in the future it will be possible to evaluate the exact top quark mass definition
used in the Monte Carlos and  used in the calibration of top quark
mass measurements. Any bias in the current usage of the value can then be
corrected for and the
quoted experimental uncertainties can be used in these comparisons unchanged.
Until the mass definition of the simulations used for calibration can be determined,
it will be an important task to develop top quark
mass measurements that are based on well defined mass schemes.

The LHC has started data taking in 2010 at an initial centre-of-mass energy of
$7\TeV$. First results show a tremendous quality of the data taken by the
experiments and top quark pair production was already seen in about 
$3\ipb$~\cite{Aad:2010ey,Khachatryan:2010ez}. With the planned LHC luminosity
of $1\ifb$  the LHC experiments will become competitive to the Tevatron
experiments within 2011.  Systematic uncertainties on the top quark mass using the
methods developed essentially at the Tevatron were expected to be below
$2\GeV$ even before the initial data
taking~\cite{ATLAS_TDR2,Ball:2007zza,Aad:2009wy}. 
With the performance shown in the first year of running a result comparable to
the current Tevatron result can be expected. The statistical uncertainty in
these results will still be noticeable. 

A truly negligible statistical uncertainty can only be expected after a
successful LHC data taking at the design centre-of-mass energy of
$14\TeV$ or close. This energy can be expected only after a long year shutdown. 
Only then the correspondingly enhanced production
cross-section will open the door to methods applicable only at the LHC.
The top quark mass determination involving leptonic $J/\psi$ decays from the $b$-quark
jet~\cite{Kharchilava:1999yj,Chierici:2006dh} is expected to be less dependent
on the jet energy scale allowing to cross-check the in situ methods.
In production with very high transverse top quark momentum 
the top decay products are clearly separated from the antitop quark decay products. 
These events are theoretically better
understood and allow a mass measurement in a well defined mass
definition~\cite{Hoang:2008xm,Hoang:2008zz}.  
% The restriction to top quarks with very high transverse
% momentum is theoretically better understood and allows a more precise
% definition of the meaning of the top quark mass
At a design LHC this might resolve the puzzle of interpreting
the current experimental top quark mass.

%%% Local Variables:
%%% mode: latex
%%% TeX-master: "EPJC_TopProperties"
%%% End:

%\cleardoublepage
\section{Interaction Properties}%{ of the Top Quark}
\label{section:interaction}
The Standard Model fixes the properties of top
quark for all three interactions considered in the Standard Model. 
To establish that the top quark discovered at the Tevatron is in fact the
Standard Model top quark it is important to verify the expected properties
experimentally and to set limits on 
possible deviations.  This subsection will consider interaction properties, i.e. 
measurements of top quark properties and possible deviations from the Standard
Model that do not assume 
explicit presence of non-Standard Model  particles. 
First measurements of top quark properties regarding the weak interaction
shall be described. Then the verification of the electrical charge is
summarised followed by properties of the top quark production through the strong force. 
Measurements that involve non-Standard Model particles are covered in Section~\ref{sec:newPhysics}.

\subsection[$W$ Boson Helicity]{\boldmath $W$ Boson Helicity}

One of the first properties of the electroweak interaction with top quarks
that was measured is that of the $W$ boson helicity states in top quark decays.
Within the Standard Model top quarks decay into a $W$ boson and a $b$ quark.
To check the expected $V-A$ structure of this weak decay the
$W$ boson helicity is investigated. 
Only left-handed particles are expected to  couple to the $W$ boson and thus
 the $W$ boson
can be either left handed ($-$) or longitudinal ($0$). For the known $b$ and
$t$ quark masses the
fractions should be $f_-=0.3$ and $f_0=0.7$, respectively. The right
handed ($+$) contribution is expected to be negligible.

Depending on the $W$ boson helicity ($-,0,+$) the charged lepton in the $W$ boson decay prefers
to align with the $b$ quark direction, stay orthogonal or escape in the
opposite direction.
%opposite to the $b$ quark.
Several observables are sensitive to the helicity: the  transverse
momentum of the lepton, %$p_T^\mathrm{lept}$, %
the lepton-$b$-quark invariant mass, $M_{lb}$,
and the angle between the lepton and the $b$ quark directions, $\cos\theta^*$. For best
sensitivity at Tevatron energies $\cos\theta^*$ is measured in the $W$ boson rest-frame.

All of these observables have been used to measure the $W$ boson helicity fractions
at the Tevatron. The most recent and thus most precise results use $\cos\theta^*$
as the observable.

\subsubsection*{CDF}
\paragraph[Analysis based on $M^2_{lb}$]{\boldmath Analysis based on $M^2_{lb}$}
%\cite{Abulencia:2006iy}%Mlb 
%an older version includes lepton pt.

\begin{figure*}
  \centering
\includegraphics[height=5cm]{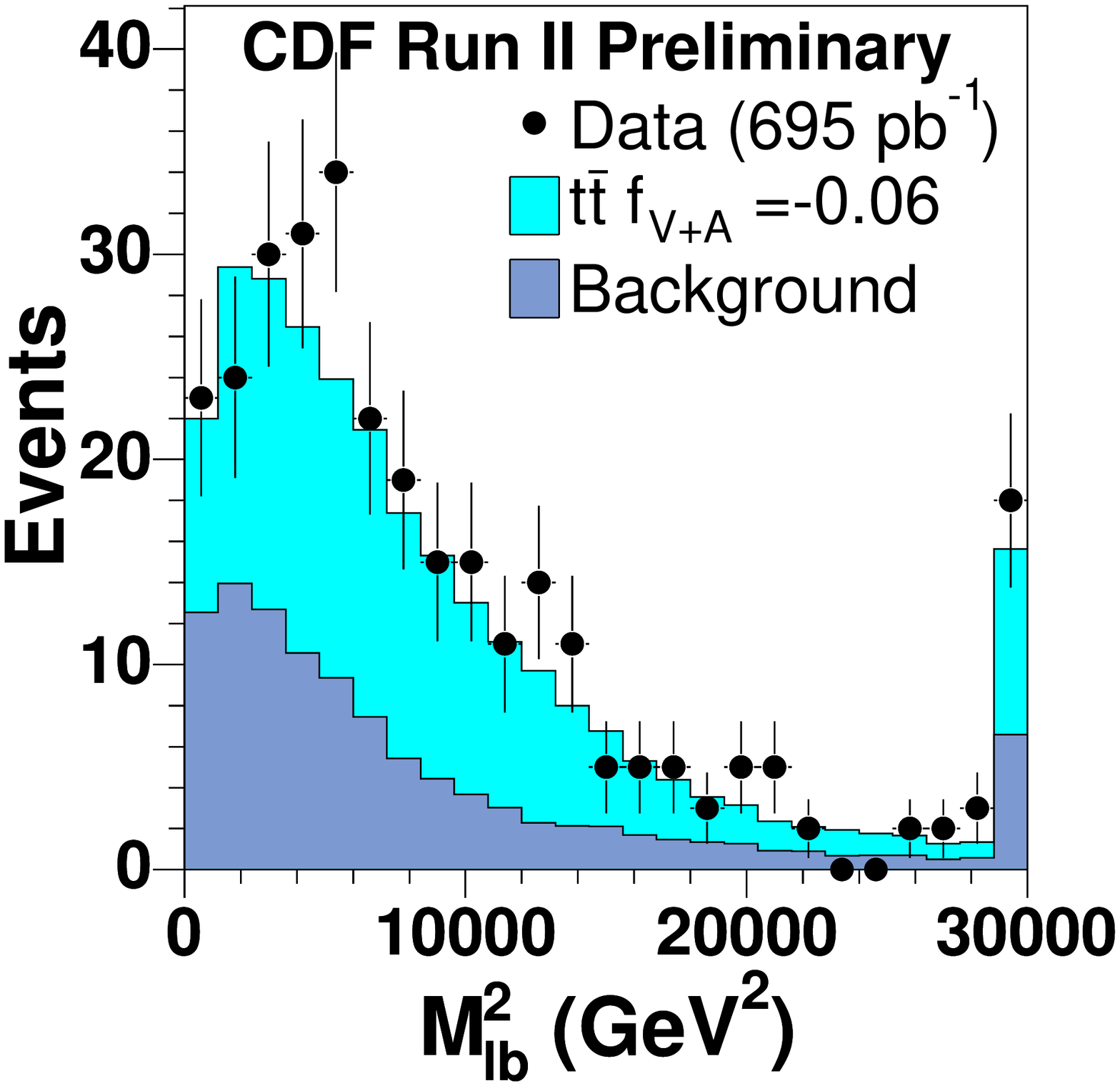}  
\includegraphics[height=5cm]{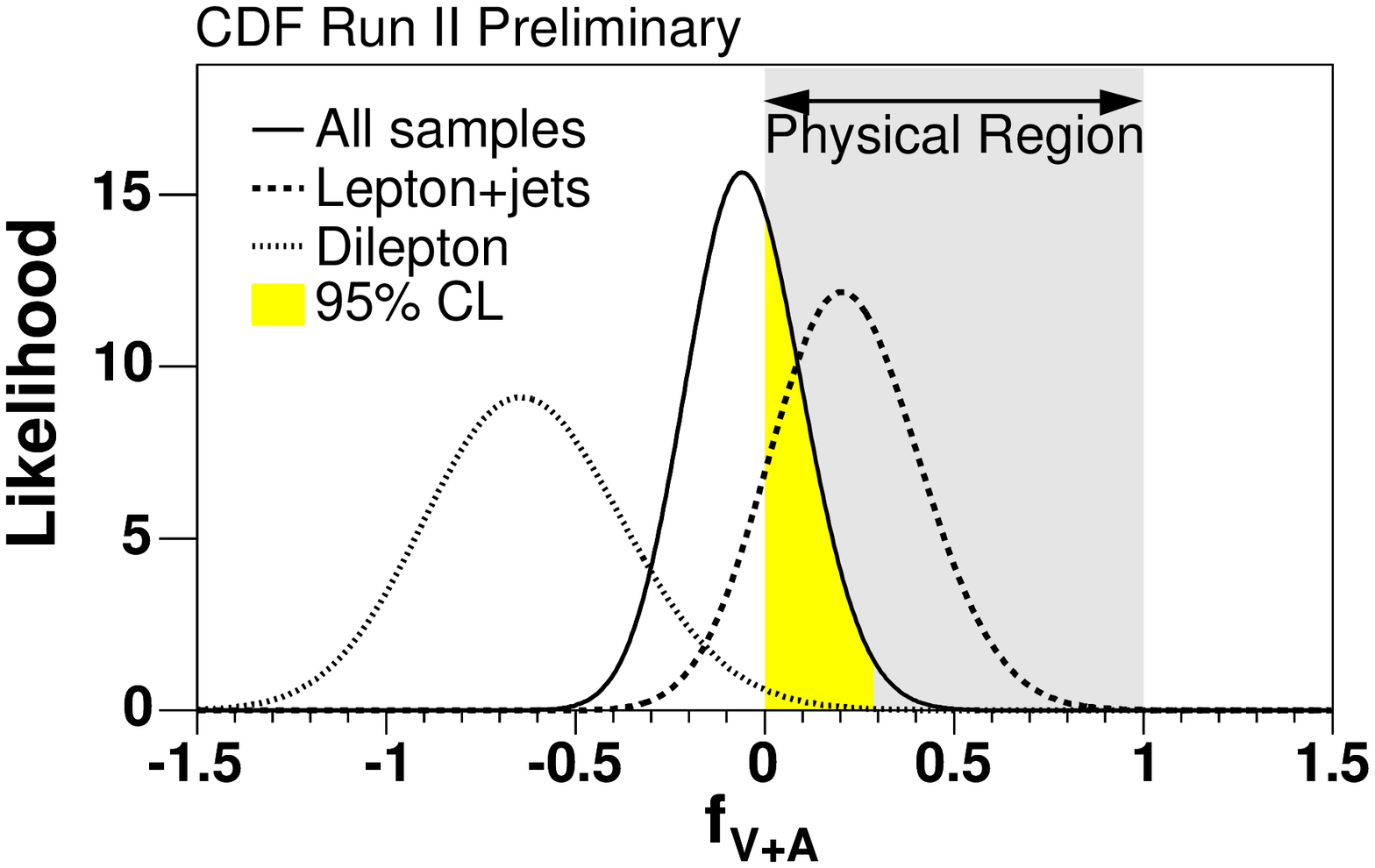}  
  \caption{Distribution of the squared  lepton-$b$-quark invariant mass 
    measured by CDF compared to best fit expectation~(left). Likelihood
    distributions as function of the $V\!+\!A$ fraction $f_{V+A}$ for the
    individual and the combined channels. The shaded yellow region indicates
    the allowed 95\% confidence range~(right)~\cite{Abulencia:2006iy}.
}
  \label{fig:cdf:mlb-1}
  \label{fig:cdf:mlb-2}
\end{figure*}
The lepton-$b$-quark invariant mass, $M_{lb}$, was used in an analysis of
approximately $700\ipb$~\cite{Abulencia:2006iy}. Events with a lepton plus jets signature
containing an isolated lepton,
missing transverse energy and at least 3 jets with identified $b$-jets were
studied separately for one and two identified jets. In addition events with a
dilepton signature containing two identified leptons with opposite electrical
charge, missing transverse energy and at least 2 jets are investigated.

For each selected event the squared invariant lepton-$b$-quark mass, $M^2_{lb}$, is computed. 
In lepton plus jets events with a single identified $b$-jet the
computation is unambiguous, however, the
identified $b$-jet and the lepton are from the same top quark in only half of
the cases, see~\fig{fig:cdf:mlb-2}~(left). 
For events with two identified $b$-jets a two dimensional
distribution of $M^2_{lb}$ is constructed. One dimension being $M^2_{lb}$
computed with the higher energetic $b$-jet, the other with the lower energetic
$b$-jet. 
In the dilepton events the two dimensional histogram is filled twice,
i.e. using each of the leptons.

The contribution of background events in the lepton plus jets samples is  modelled by \alpgen\ $W+b\bar b$
events and by multijet events obtained from a control data sample.
In the singly tagged sample multijet events contribute 15\%, in the doubly tagged
sample they are neglected. For the dilepton sample the background is described
by about 50\% $Z+$jets from \alpgen, 30\% W+jets with one jet misidentified as
lepton using a single lepton sample and applying the misidentification rate.
The final 20\% of background are from diboson samples, $WW$ and $WZ$. 

The signal expectation is simulated with \alpgen\ and \pythia\ assuming a top
mass of $m_t=175\GeV$ for both $V\!-\!A$ and $V\!+\!A$ coupling to the $W$ boson. 
A binned log likelihood fit is used to extract the fraction of $V\!+\!A$ coupling
in the top decay, $f_{+}$. The likelihood uses Poisson  probabilities to
describe the expected number of events in each bin of the $M^2_{lb}$
distributions. The parameters of the Poisson distributions are taken from the
described simulation and are smeared with nuisance parameters for top pair
cross-section and the total background contribution in each of three samples.
These nuisance parameters are constrained to their nominal values with 
Gaussian probability distribution.

The procedure was verified using a large number of pseudo experiments with
different nominal contributions from $V\!+\!A$ decays. The fit was found to be
stable and unbiased.
Systematic uncertainties are dominated by the uncertainty on the jet energy
scale, followed by uncertainties of the background shape and normalisation as
well as the limited Monte Carlo statistics. 

The likelihood distributions obtained with this procedure are shown in \fig{fig:cdf:mlb-1}~(right).
The left-handed $W$ boson fraction in the top quark decay is measured to be $f_+=-0.02\pm
0.07$ in agreement with the Standard Model expectation of a negligible
contribution.  The upper limit is computed using Bayesian statistics with a flat
prior for $f_+$ between $0$ and $1$ and yields $f_+<0.09$ at 95\% C.L. 
Results for individual samples differ by at most $1.8$ standard deviations.

\paragraph{Analyses based on \boldmath$\cos\theta^*$}
The angle, $\cos\theta^*$,  measured in the $W$ boson rest-frame
between the lepton and the $b$ quark direction 
yields calculable distributions of this angle for each of the possible $W$
boson helicities ($-,0,+$):
\beq
d_0(\theta^*)=\frac{3}{4}\left(1-\cos^2\theta^*\right)\mbox{,}\qquad
d_\pm(\theta^*)=\frac{3}{8}\left(1\pm\cos\theta^*\right)\mbox{.}
\eeq
Measuring  $\cos\theta^*$ thus allows to reconstruct the contribution of each
of these helicities in top quark decays.

CDF has used this observable in several  analyses to
measure the $W$ helicity fractions in top
decays~\cite{Aaltonen:2008ei,CdfNote10333}. %CdfNote9215,CdfNote9431,
In~\cite{Aaltonen:2008ei} events with an isolated lepton, missing transverse
energy and at least four jets including at least one identified $b$-jet  of $1.9\ifb$ are investigated.
Two different methods  of reconstructing the full top pair
kinematics are used to then compute the $W$ boson rest-frame and $\cos\theta^*$ for
each individual event.
\begin{figure*}
  \centering
\includegraphics[width=0.48\textwidth]{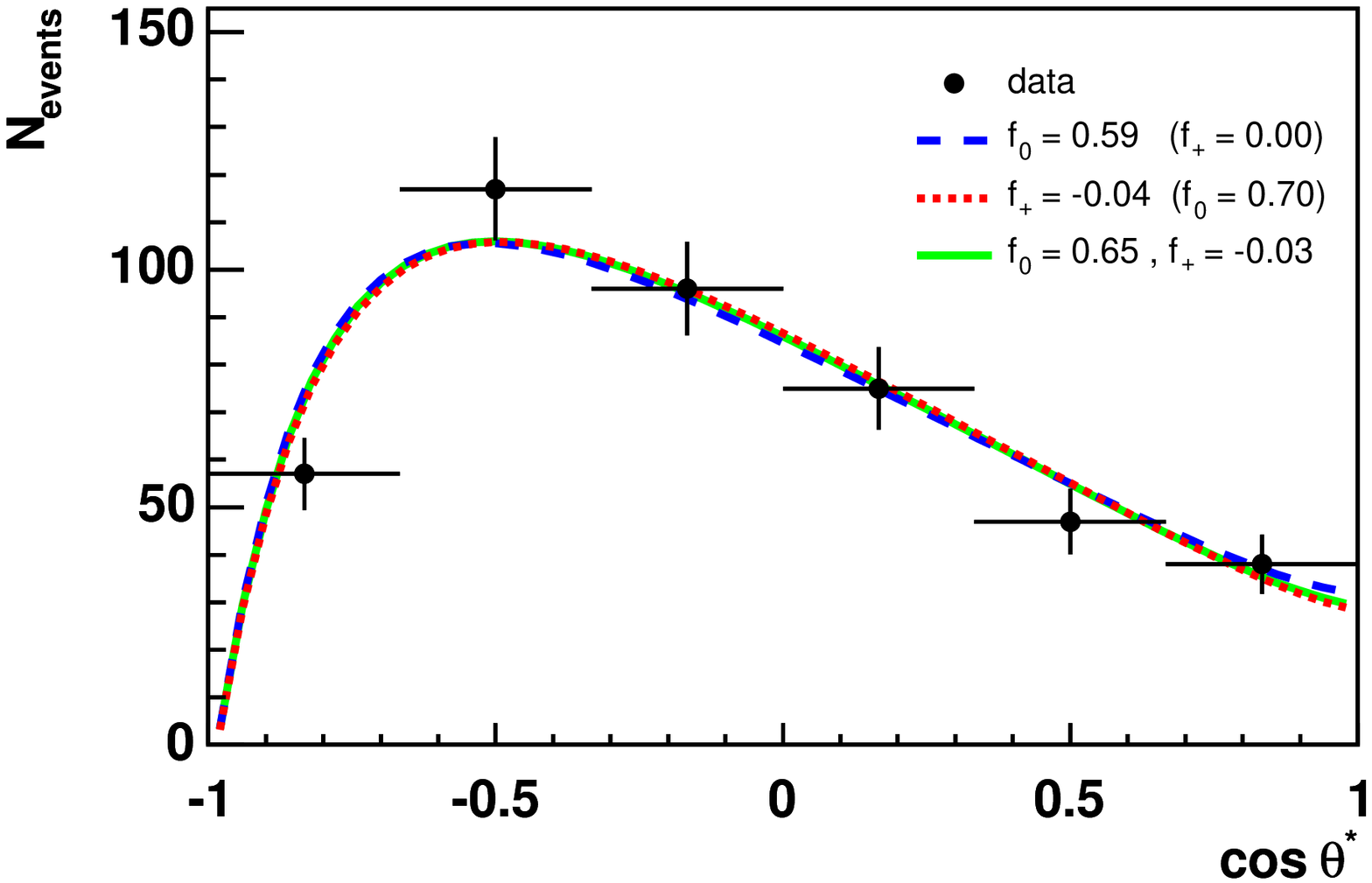}
\includegraphics[width=0.48\textwidth]{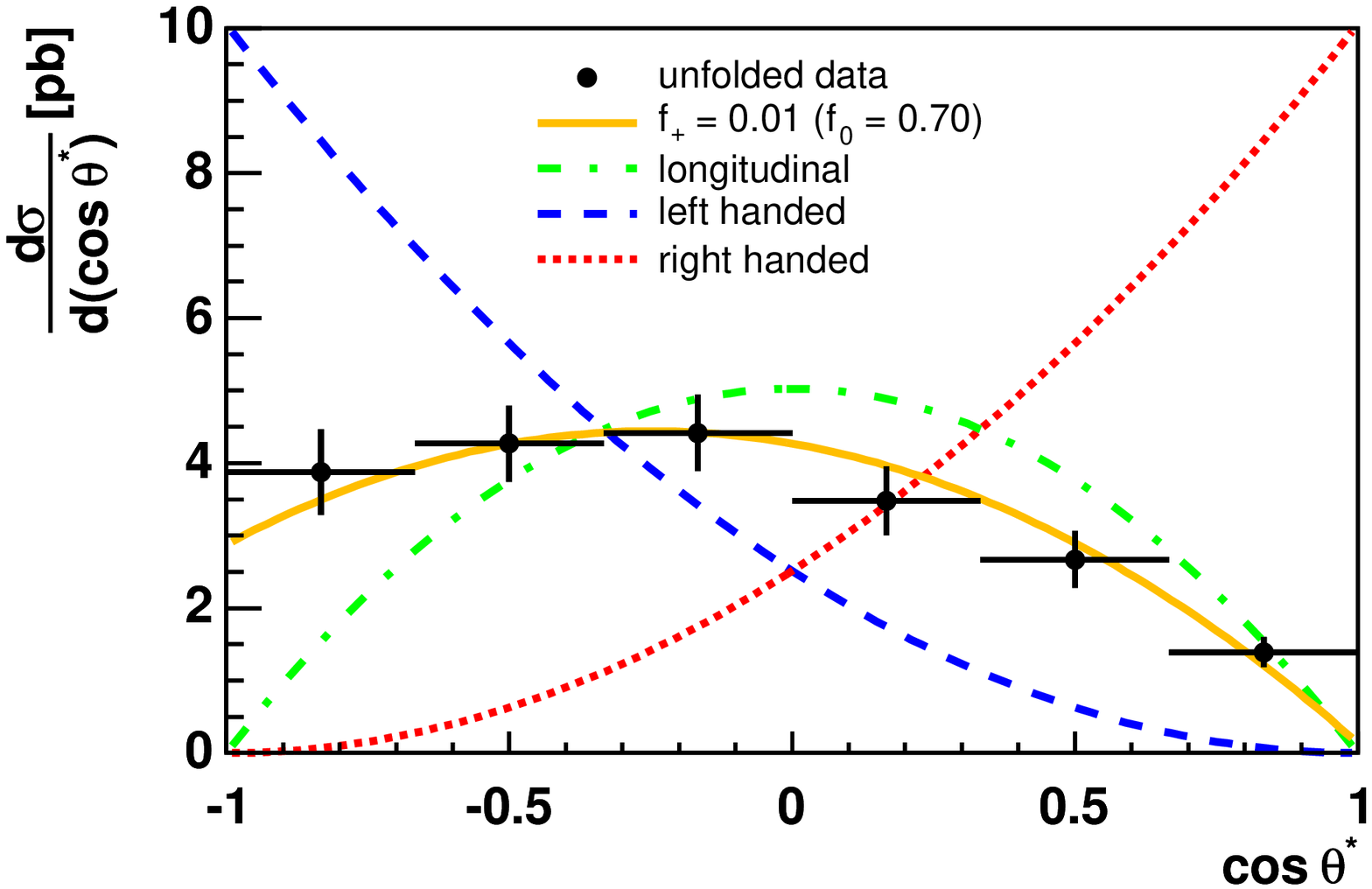}  
  \caption{Distribution of $\cos\theta^*$ as measured by CDF in $1.9\ifb$ compared to
    various predictions~\cite{Aaltonen:2008ei}. The left plot shows the data measured for the
    ``template'' method compared to the template prediction with the best fit
    parameters. The right plot shows the de-convoluted data of the
    ``convolution'' method compared to theoretical curves for individual
    helicity types and the best fit combination.}
  \label{fig:cdf-whel-costheta}
\end{figure*}

One method recovers the unmeasured neutrino momentum
from the missing transverse energy and from solving the quadratic equation
following from the $W\rightarrow \ell\nu$ decay kinematics when using the
nominal $W$ boson mass. This part of the analysis (for reasons that will become clear
below) is called the ``convolution'' method.
The other method (which will be called ``template'' method) 
uses a constrained kinematic fit to determine the lepton and
parton momenta, where these momenta are allowed to float within the
experimental uncertainties of the measured quantities. Constraints are built
from the $W$ boson mass and the equality of the top and anti-top quark masses
constructed from the fitted lepton and parton momenta.
Both methods require an association of the measured jets to the partons of the
top quark pair topology and use the quality of the constrained fit to select this
assignment.

The reconstructed $\cos\theta^*$ distribution is used in two different
likelihood fits to determine the helicity fractions $f_0$ and $f_+$ either
individually, fixing the other one to the Standard Model value, or
simultaneously. Both methods require simulation to construct their log
likelihood function.
\pythia\ and \herwig\  are used to simulate Standard Model top quark pair
production. Modified versions of \herwig\ and
\madevent+\pythia~\cite{Alwall:2007st} are used to
generate samples with varied helicity fractions. The $W+$jets background is
simulated with \alpgen+\herwig\ and normalised to the amount of data before
$b$-tagging and after removing all other background and $t\bar t$ signal events.
Multijet background is taken from a control sample. Additional minor backgrounds
from diboson, $Z+$jets and single top quark production are taken from simulation 
normalised according to their theoretical cross-sections.
%\cite{Abulencia:2006in}

In the  ``template'' method the background expectations are combined with
signal templates for the three different helicity states. The helicity
fractions are taken from an unbinned likelihood fit with proper correction for
acceptance effects. For this method an additional cut on the scalar sum of all
transverse energy was required, $H_T>250\GeV$.

The ``convolution'' method uses the signal simulation to derive acceptance
functions which are then convoluted with the theoretically predicted number of
events in each bin of $\cos\theta^*$. The helicity fractions are then taken from  a
binned likelihood fit after subtracting the background estimation from data.

Systematic uncertainties for both methods were determined from pseudo datasets
with templates modified according to the systematic effect under
consideration. The jet energy scale uncertainty is among the dominating
systematic effects in both methods. Only for the $f_0$ in the two dimensional fits
the initial/final state radiation  uncertainties are more important. In general
the ``convolution'' technique yields slightly larger systematic uncertainties.
Data are compared to the estimated results for the optimal fit parameters in
\fig{fig:cdf-whel-costheta}. For the ``convolution'' method (right) the data shown are
de-convoluted and compared to the pure theory prediction.

The results of the methods are combined using the BLUE
method~\cite{\refblue} with the statistical correlation between the two methods 
determined in pseudo experiments and the systematic uncertainties considered
completely correlated. The combined results yields an
improvement of about 10\% compared to the individual methods. 
The model independent 2d fit yields
\bea
f_0 & =& \Plus 0.66\pm 0.16 \mathrm{(stat)}\pm 0.05  \mathrm{(syst)} \nonumber\\
f_+ & =& -0.03\pm 0.06 \mathrm{(stat)}\pm 0.03  \mathrm{(syst)}
\eea
with a correlation of $-82\%$ between $f_0$ and $f_+$.
Upper limits on the positive helicity fraction, $f_+$ are not set.
\bigskip

In \cite{CdfNote10333} CDF investigates the dilepton channel in $4.8\ifb$ to
determine the $W$ boson helicity fractions from $\cos\theta^*$.
The selection requires two isolated leptons and at least  two jets one of
which is required to be identified as $b$-jet. 

To reconstruct $\cos\theta^*$ the neutrino momenta are deduced by assuming the
top quark pair decay kinematics with an intermediate $W$ boson. Requiring
that the momenta of the corresponding decay products are compatible with the
$W$ boson mass  and a top quark mass of $175\GeV$ and using the missing
transverse energy allows to determine the
neutrino momenta up to an 8-fold ambiguity. To account for the experimental
resolutions this determination is repeated with appropriately smeared input
momenta.
Of the eight (smeared) solutions the one that yields the smallest invariant $t\bar
t$ mass is used. %to reconstruct $\cos\theta^*$. 
The distribution of measured reconstructed $\cos\theta^*$ values is compared
to templates with a binned likelihood function. Templates for the signal are
obtained for various values of $f_0$ and $f_+$ using a customised \herwig\
generator. In addition templates for background processes of diboson,
Drell-Yan and fake leptons are added to obtain
the complete expectation.

Significant systematic uncertainties come from the jet energy
scale and from the uncertainty of the modelling of initial and final state
radiation. This special study is also suffering from limited template
statistics.

In dilepton events of $4.8\ifb$ CDF determines the helicity fractions 
to be~\cite{CdfNote10333} 
\bea
f_0 & =& \Plus 0.78\pm 0.20 \mathrm{(stat)}\pm 0.06  \mathrm{(syst)} \nonumber\\
f_+ & =& -0.11\pm 0.10 \mathrm{(stat)}\pm 0.04  \mathrm{(syst)}\quad\mbox{.}
\eea
Additional improvements are expected from  relaxing the requirement of
identified $b$-jets.

\paragraph{Analysis using the Matrix Element Method}
In addition to the two analyses with explicit reconstruction of the top quark
kinematics described above, CDF has performed an analysis that uses the Matrix Element technique
to measure the longitudinal $W$ boson helicity fraction, $f_0$, in $1.9\ifb$ of data~\cite{CdfNote9144,CdfNote10004}.
The analysis selects events with one isolated lepton, large missing
transverse energy and at least four energetic jets including at least one jet
identified as  $b$-jet. 
%In addition a cut on the scalar sum of all transverse
%energy, $H_T>200\GeV$, is applied.

For each selected event a likelihood, $L(f_0,f_+|\{j\})$, to observe the measured
quantities, $\{\vec j\}$ (the lepton and jet momenta and the missing
transverse energy), is computed as function of the  $W$ helicity
fractions. This likelihood consists of a signal term which depends on $f_0$
and $f_+$, and a background term which is independent of these:
\bea
&&L_i(f_0,f_+|\{\vec j\})=
\nonumber\\
&&\quad 
f_\mathrm{top} P_{t\bar t}(\{\vec j\};f_0,f_+)+ \left(1-f_\mathrm{top}\right) P_{W+\mathrm{jets}}(\{\vec j\})
\quad
\eea
The signal and background probabilities, $P_{t\bar t}$ and
$P_{W+\mathrm{jets}}$, are computed by integrating the differential parton level
signal and background cross-sections according to the leading order
cross-sections for $q\bar q\rightarrow t\bar t$ and $W+4$jet production,
respectively. These integrations account for the experimental resolutions.
The product of likelihoods for all selected events is evaluated for discrete
values of $f_0$ and $f_+$. At each point the signal fraction, $f_\mathrm{top}$, is chosen to
minimise the total likelihood.
The result for the longitudinal helicity fraction and its statistical
uncertainty are taken from the minimum of the  log likelihood and its change by $0.5$ units.

The method is validated and calibrated using simulation for top pair signal
and various background processes. Samples of various nominal $f_0,f_+$ values were
created by reweighting the top pair events according to the expected
$\cos\theta^*$ distribution. Applying the above method to various pseudo
datasets with various nominal $f_0,f_+$ values yields a calibration curve.
The observed slope of less than one is explained by the incomplete description of
signal and background each with only a single leading order matrix element.
The final results are corrected with this calibration curve.

The largest systematic uncertainty for this method stems from the uncertainty
in simulation. It is estimated by checking the difference between \pythia\ and
\herwig. In addition, uncertainties due to initial and final state radiation, different
PDF sets, the jet energy scale uncertainty and various other experimental effects are considered.

Using the described Matrix Element method for $2.7\ifb$ and assuming
$m_t=175\GeV$  CDF finds 
\bea
f_0 & =& \Plus 0.88\pm 0.11 \mathrm{(stat)}\pm 0.06  \mathrm{(syst)} \nonumber\\
f_+ & =& -0.15\pm 0.07 \mathrm{(stat)}\pm 0.06  \mathrm{(syst)}
\eea
with a correlation of $-59\%$. A variation of the assumed top quark mass
shifts the results for $f_0$  by $0.017$ and for $f_+$ by $-0.010$  per $\pm 1\GeV$
of shift in the top quark mass from the central value~\cite{CdfNote10004}. 

\subsubsection*{{D\O}}
The {D\O} collaboration has investigated a total of  $5.3\ifb$ to determine the $W$ boson
helicity fractions based on the reconstruction of the decay angle
$\cos\theta^*$ in lepton plus jets and in dilepton
 events~\cite{Abazov:2006hb,Abazov:2010jn}. 
In the lepton plus jets events the hadronic decay is utilised to measure
$\left|\cos\theta^*\right|$ which helps to measure the longitudinal fraction,
$f_0$. 

Events are selected by requiring an isolated lepton, missing transverse energy
and at least 4 jets. No second lepton is allowed in the event. Dilepton events
are selected with two isolated charged leptons with opposite charge, large
missing transverse energy and at least two jets. Some additional cuts are
applied to suppress $Z\rightarrow \ell\ell$ events in the $ee$ and
$\mu\mu$ channels and to assure a minimal transverse energy in the $e\mu$
channel. All channels use a multivariate likelihood discriminant based on
kinematic observables and the neural network $b$-jet identification to improve the
purity of the top quark pair signal.
\begin{figure}
  \centering
  \includegraphics[width=0.8\linewidth,clip,trim=0mm 0mm 17mm 0mm]{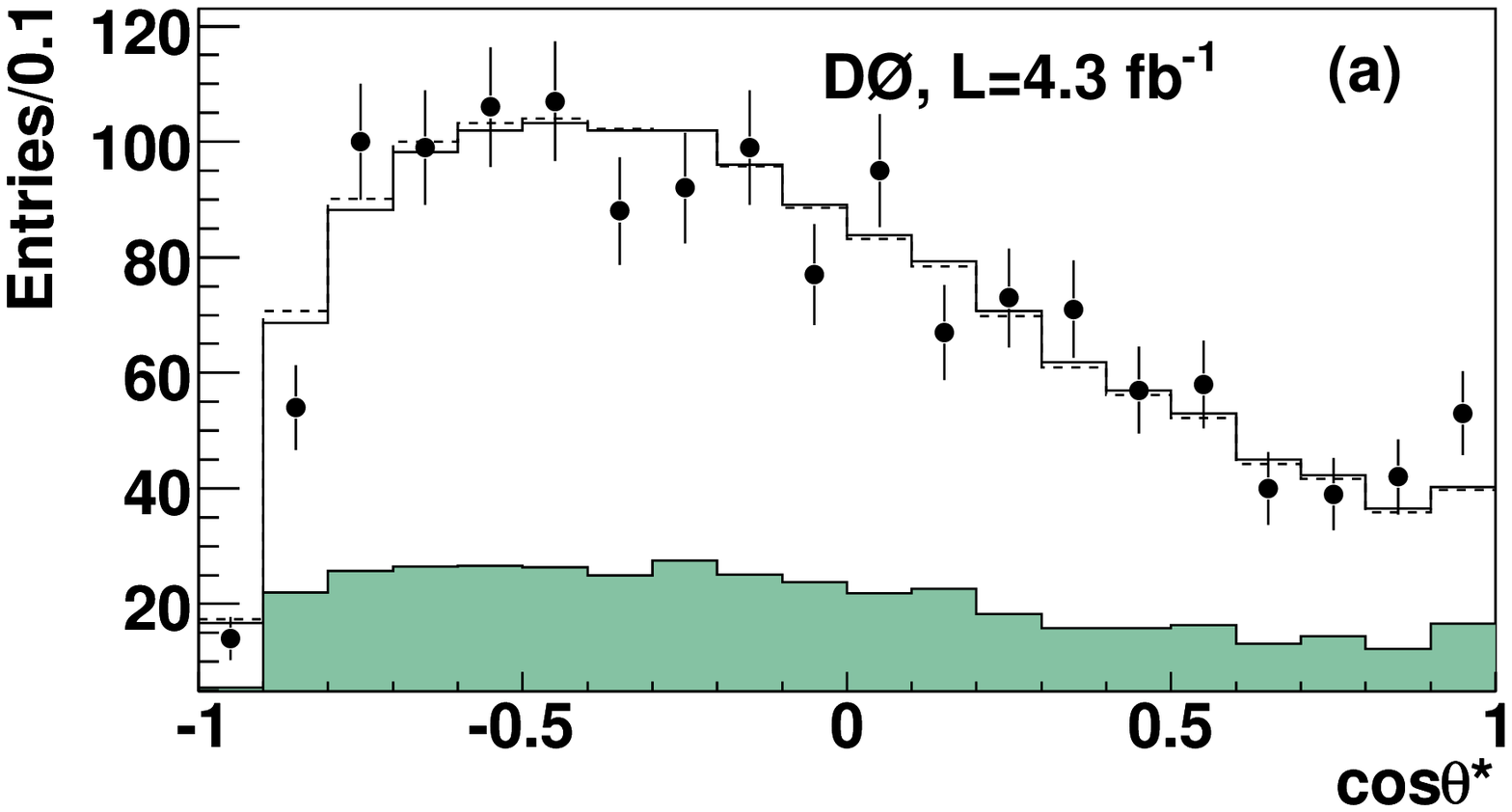}
  \includegraphics[width=0.8\linewidth,clip,trim=0mm 0mm 17mm 0mm]{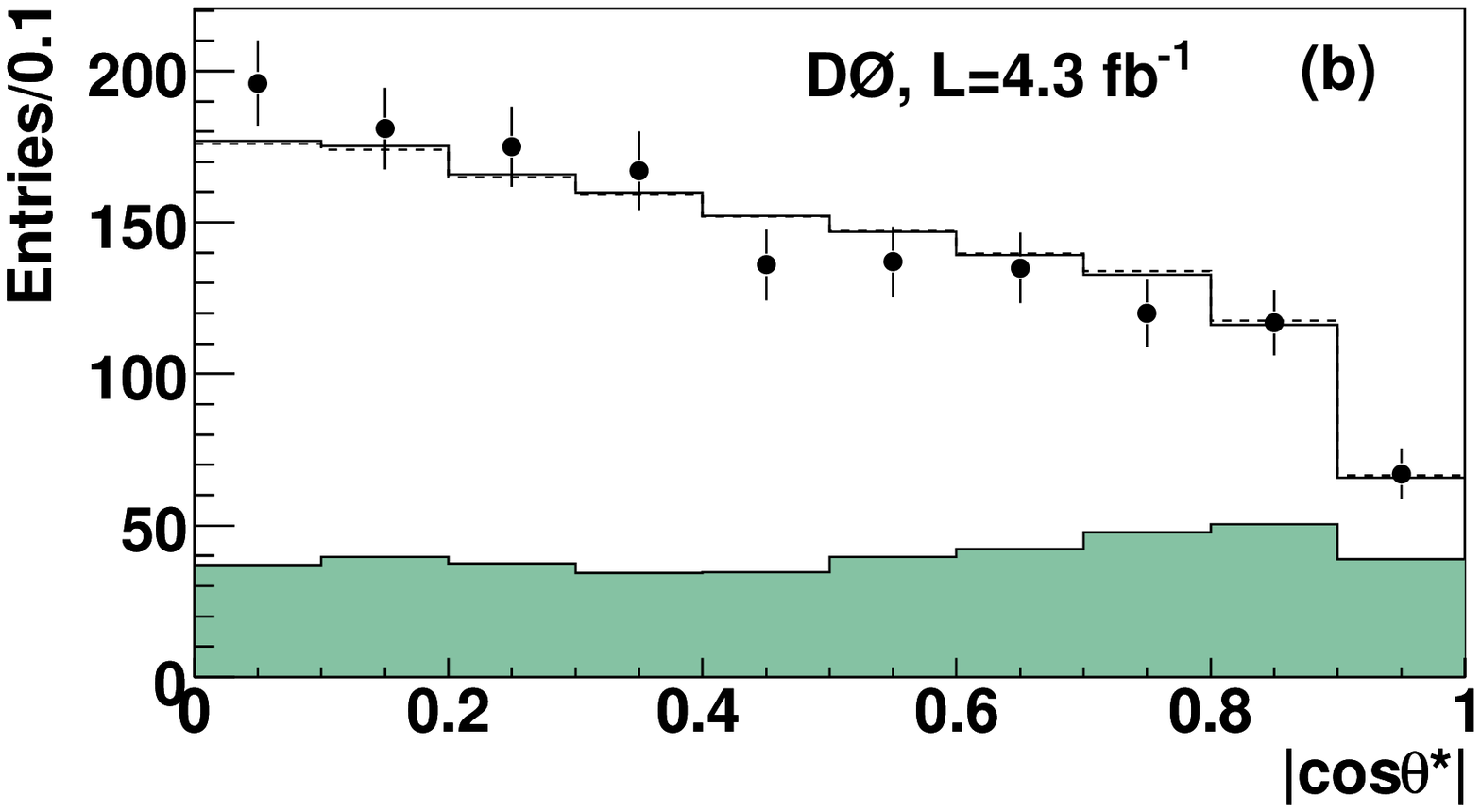}
  \includegraphics[width=0.8\linewidth,clip,trim=0mm 0mm 17mm 0mm]{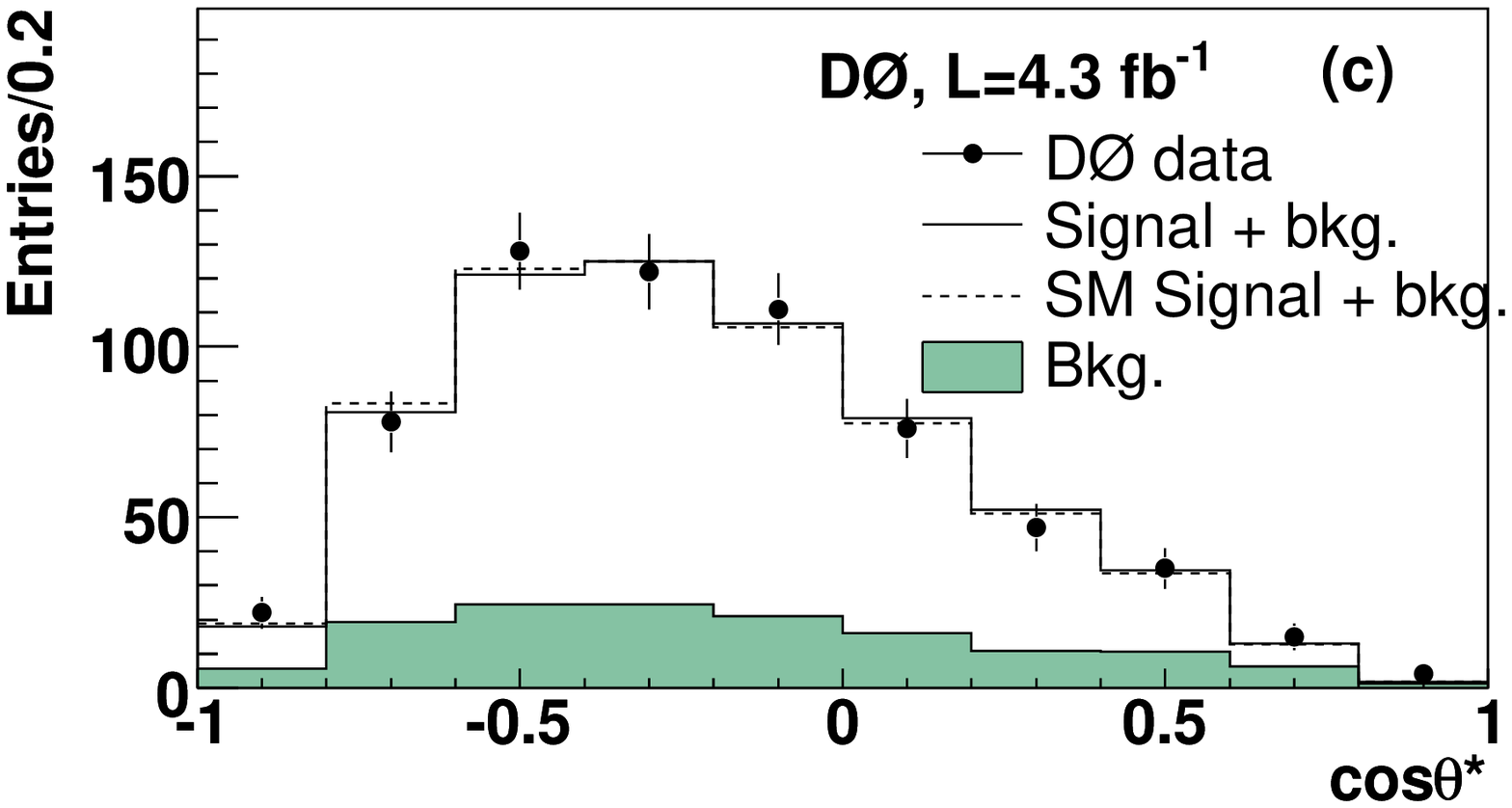}%
  \caption{$W$ decay angle distributions as measured in $4.3\ifb$ of D\O\
  Run\,IIb data. The top plot shows the distribution obtained from the leptonic decay
  in the lepton+jet events, the middle plot the one obtained from the hadronic
  decay. The bottom plot shows the distribution obtained from dilepton events.
  In addition to the data shown with error bars all plots contain 
  the best fit prediction shown as full line and the Standard
  Model as a dashed line histogram. The shaded area represents the background
  contribution~\cite{Abazov:2010jn}.}
  \label{fig:d0:whel-costheta}
\end{figure}

In lepton plus jets events the decay kinematics of the  top quark pairs  is reconstructed using
a constrained fit that determines the momenta of top quark and the $W$ boson
decay products from the measured jets and lepton momenta as well as from the missing transverse
energy. Only the four leading jets in $p_T$ are used.
The fit requires the momenta of $W$ boson decay products  to be consistent
with the nominal $W$ boson mass and the momenta of the top quark decay
products to yield a top quark mass of $172.5\GeV$. Of the 12 possible jet-parton
assignments the one with the highest probability of being the correct one is
chosen. This probability for each association is computed using the fit $\chi^2$ as well as the
output value of the neural network $b$-tagger for the four jets 
and its consistency with the light or heavy quark assignment under consideration.
From the chosen solution $\cos\theta^*$ is computed from the leptonic side and
a second measurement of  absolute value from the hadronic side,
see~\fig{fig:d0:whel-costheta}~(top and middle).

The kinematics of dilepton events can be solved assuming the top quark mass with a
fourfold ambiguity. In addition the two possible assignments of the two leading
jets in $p_T$ to the $b$ quarks are considered. For each of these solutions
the decay angle $\cos\theta^*$ is computed. To explore the full phase space
consistent with the measurements, the measured jet and charged lepton momenta
are fluctuated according to the detector resolution and $\cos\theta^*$ is
computed for each fluctuation. The average  over all solutions
and all fluctuations is computed for each jet to find two $\cos\theta^*$
values per event. The resulting distribution for  Run\,IIb data is shown in  \fig{fig:d0:whel-costheta}~(bottom).

The expected distribution for signal top pair events is simulated using
\alpgen+\pythia\ with various  $V\!-\!A$ to $V\!+\!A$ ratios. These samples are then
reweighted to form samples corresponding to the three $W$ boson helicity states. 
Important backgrounds in the selected event sample are $W$+jets and multijet events in the lepton
plus jets channel, $WW+$jets and $Z+$jets in the dilepton samples. Backgrounds
of a single weak boson with jets are simulated with \alpgen+\pythia, diboson
samples are generated with \pythia. The multijet
contribution is estimated from data for each bin of the $\cos\theta^*$ distribution.

From the data and the estimated signal and background contributions a binned likelihood,
${\cal L}(f_0,f_+)$,  is computed for the observed data to be consistent with the sum of 
the backgrounds and the estimates for the three $W$ boson helicity states.
The normalisation of the background is kept as a nuisance parameter with a Gaussian
constraint to its nominal value. The measured helicity fractions, $f_0$
and $f_+$, are those that minimise this likelihood.

Systematic uncertainties are evaluated in ensemble tests. Pseudo datasets are
drawn from models with systematic variations and compared to the standard
templates to find the resulting shift in the obtained helicity fractions. 
Dominating uncertainties stem from the modeling of $t\bar t$, which are
determined by exchanging the standard \alpgen+\pythia\ with pure \pythia,
\herwig\ and \mcatnlo.
Further significant contributions to the systematic uncertainties 
stem from background modeling, from the limited template statistics, from jet
energy scale, jet energy calibration and jet reconstruction.
An uncertainty due to the top quark mass uncertainty of $1.4\GeV$ is also included.

Combining the datasets of Run\,IIa and Run\,IIb, which are analysed separately,
D\O\ finds
\bea
f_0&=&0.669\pm0.078\mathrm{(stat)}\pm0.065\mathrm{(syst)}\nonumber\\
f_+&=&0.023\pm0.041\mathrm{(stat)}\pm0.034\mathrm{(syst)}\quad\mbox.
\eea
The correlation between the two numbers is about $-0.8$. The result shows a good
consistency between the Run\,IIa and Run\,IIb datasets as well as between the
dilepton and lepton plus jet channel. As for the other good agreement with the
Standard Model expectation.

%%% Local Variables: 
%%% mode: latex
%%% TeX-master: "EPJC_TopProperties"
%%% End: 

\subsection[The CKM Element $V_{tb}$]{\boldmath The CKM Element  $V_{tb}$ }
Another aspect of the weak coupling is that of flavour changing charged
currents.  The Standard Model explains these through the CKM
matrix that needs to be determined from experiment. The elements
$V_{td}$ and $V_{ts}$ of this matrix have been determined from
experiment assuming the Standard Model and 
allow to infer
$0.9990<\left|V_{tb}\right|<0.09992$~\cite{\refpdg,Alwall:2006bx} using
unitarity of a $3\times 3$ CKM matrix.
Physics beyond the Standard Model may invalidate these assumptions and leave
$\left|V_{tb}\right|$ unconstrained~\cite{Alwall:2006bx}. The value of $\left|V_{tb}\right|$
directly influences the single top quark  production cross-section and the ratio
between top quark decays to $b$-quarks ($t\rightarrow bW$) and to light-quarks
($t\rightarrow qW$ with $q=d,s$). Both effects have been studied by the
Tevatron experiments to constrain the CKM elements related to the top quark.

\subsubsection{Single Top Quark }
The single top quark  cross-section from Standard Model sources is proportional
to~$\left|V_{tb}\right|^2$.  
Both experiments have used this relation to convert
their cross-section measurements to determinations of the CKM
element~\cite{Abazov:2009ii,Aaltonen:2009jj,Group:2009qk}. %{Aaltonen:2008sy,Abazov:2008kt}. 
In addition to the uncertainty of the single top quark  cross-section measurement,
theoretical uncertainties on the Standard Model cross-section need to be taken
into account. 

The combination of the CDF and D\O\ measurements yields a cross-section for single
top quark production of $2.76^{+0.58}_{-0.47}\pb$~\cite{Group:2009qk} assuming
a top quark mass of $170\GeV$ in modelling the signal efficiencies.
To extract the CKM matrix element it is assumed that  $\left|V_{tb}\right|$
is much larger than  $\left|V_{td}\right|$ and  $\left|V_{ts}\right|$, so that
the top quark decay is dominated by the decay to $Wb$ and no significant
production through $d$ and $s$ quarks in the initial state takes place. 
No assumption about the unitarity of the CKM matrix is made.

With theses assumptions the analyses performed to determine the cross-section can remain
unchanged for the determination of $\left|V_{tb}\right|$. Attributing the full
deviations of the experimental result from the SM prediction to the value of $\left|V_{tb}\right|$ the
combined CDF and D0 single top quark production yields~\cite{Group:2009qk}:
\beq
\left|V_{tb}\right|=0.88\pm 0.07 \quad\mbox{or}\quad \left|V_{tb}\right|>0.77\quad\mbox{at $95\%$C.L.}
\eeq

% From the combined single top quark  cross-section of three analyses using $3.2\ifb$ CDF measures
% $\left|V_{tb}\right|=0.91\pm{0.11}(\mathrm{stat+syst})\pm
% 0.07(\mathrm{theory})$~\cite{Aaltonen:2009jj}. Restricting the allowed values
% to the physical range between zero and one yields a lower limit of
% $\left|V_{tb}\right|>0.71$ and 95\%~C.L.

% D\O\ analysed $2.3\ifb$.  Allowing for an anomalously
% enhanced coupling one finds
% $\left|V_{tb}f_1^L\right|=1.07\pm0.12$~\cite{Abazov:2009ii}, where $f_1^L$ is the
% enhancement factor of the $Wtb$ coupling. 
% For the Standard Model value of $f_1^L=1$ and restricting
% $\left|V_{tb}\right|$ its physical range, a lower limit of $\left|V_{tb}\right|>0.78$ is obtained.

% Because the conversions of single top quark  production cross-section to
% $\left|V_{tb}f_1^L\right|$ leave
% the single top quark  analyses unchanged they assume that only the $bW\rightarrow t$
% contributes to single top quark  production, that the top quark decay is dominated
% by the decay to $Wb$ and that the kinematics of the production and decay is as expected by
% the Standard Model. An anomalous coupling strength of the $V-A$ coupling at
% the $tbW$ vertex could be allowed, but for extraction of $\left|V_{tb}\right|^2$
% it is left at its Standard Model value.

%The particle data group has concluded a slightly stronger limit after
%averaging the single top quark  cross-section results of the two experiments: 
%$\left|V_{tb}\right|>0.74$~\cite{\refpdg}.

\subsubsection{Top Quark  Pairs}
\label{sect:Rb}
In top quark  pair decays the number of identified $b$ jets is used to measure the
branching fraction for $t\rightarrow bW$. This fraction can be expressed in  terms
of the CKM matrix elements
\beq
R_b=\frac{\left|V_{tb}\right|^2}{\left|V_{td}\right|^2+\left|V_{ts}\right|^2+\left|V_{tb}\right|^2}
\quad\mbox{,}
\eeq
assuming that the top quark decay is restricted to Standard Model quarks.
\subsubsubsection{CDF}
CDF has investigated $160\ipb$ of data using both events with lepton plus jets
and dilepton events~\cite{Acosta:2005hr}. 
For the lepton plus jets sample events are selected by requiring an isolated
lepton, missing transverse momentum and four jets. Dilepton events consist of
two charged leptons, missing transverse momentum and two jets. Both samples are
classified according to the number of jets that are identified as $b$ jets.

The background in the lepton plus jets sample is dominated by $W+$jets
events and multijet events with fake electrons. 
The multijet background is estimated from data using a range of control samples. 
The $W+$jets background is simulated using the
\alpgen+\herwig\ generator.  In the subsamples with one or
more than one identified $b$-jet, it is normalised using data before $b$-tagging. The
fraction of heavy flavour in these samples is scaled according to
the Monte Carlo to data ratio in a control sample of inclusive jet events. 
For the description of $W+$jets events without identified $b$ jets, a neural
network based on kinematic observables is used, which enriches $W+$jets
background at low values and top quark  pair signal at high values. 
The neural network was trained on \alpgen+\herwig\ samples.
The distribution of neural
network output values measured in data is compared to the simulation of
$W+$jets and $t\bar t$ to fit the signal and background contribution. The
shape of the multijet background is included at the rate determined above.
%Lepton+jets xsec: \cite{Acosta:2005am}

For the dilepton sample the main backgrounds stem from Drell-Yan, diboson and
from $W+$jets events with fake leptons. The Drell-Yan background for $ee$ and
$\mu\mu$ is simulated using \pythia\ normalised to the number of $Z$ bosons in a mass
window around $M_Z$. Other electroweak backgrounds are fully taken from simulation.
The $W+$jets backgrounds are taken from the \alpgen+\herwig\ simulation
applying lepton fake rates, that are determined in a complementary jet sample. 
A tag rate probability for generic QCD jets is used to find the contribution
of fake lepton events to the various $b$-tag subsamples.
%Dilepton xsec: \cite{Acosta:2004uw}

The distribution of the number of $b$-tags for top quark  pair events depends on the
branching fraction, $R_b$. It is determined from events generated with
\pythia\ and passed through full CDF detector simulation  (as all
simulations above).

Finally, a Poisson likelihood for the observed data to agree with the expectation is
constructed as function of  $R_b$. Gaussian functions with nuisance parameters 
are used to take systematic uncertainties including correlations between the
samples and the $b$-tag bins into account. 
The dominant uncertainty comes from the background estimate in the $0$-tag samples
and the $b$ quark identification efficiency.

In the analysed $160\ifb$ CDF finds
$R_b=1.12\pm0.2(\mathrm{stat})_{-0.13}^{+0.14}(\mathrm{syst})$~\cite{Acosta:2005hr}.
The Feldman-Cousins approach~\cite{Feldman:1997qc,\refpdg} is used to compute
a lower limit of 
\beq
R_b>0.61\quad \mbox{or}\quad \left|V_{tb}\right|>0.78 \quad\mbox{at $95\%$ C.L.}
\eeq
where the conversion to the limit on the CKM element
 is done assuming three generations and unitary of
the CKM matrix, only.

\subsubsubsection{D\O}
D\O\ has measured the top quark  branching fraction, $R_b$, in conjunction with the top quark 
pair cross-section using $0.9\ifb$~\cite{Abazov:2008yn}. Events are selected 
for the lepton plus jets channel requiring an isolated lepton,
missing transverse momentum and at least three jets. In data $b$ jets are identified
using a neural network tagger.  
\begin{figure*}
  \centering
\includegraphics[width=0.3\linewidth]{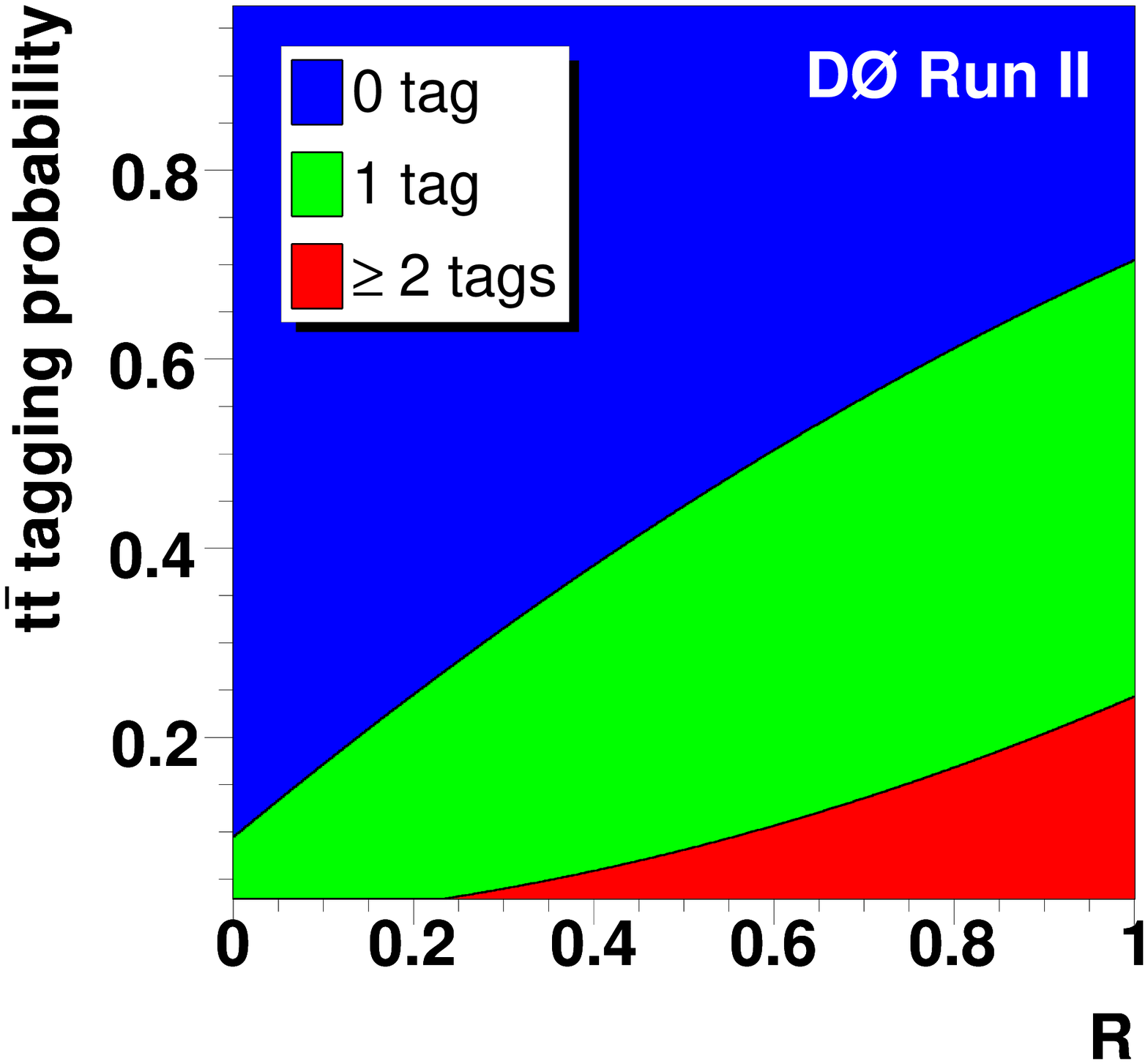}%
\includegraphics[width=0.3\linewidth]{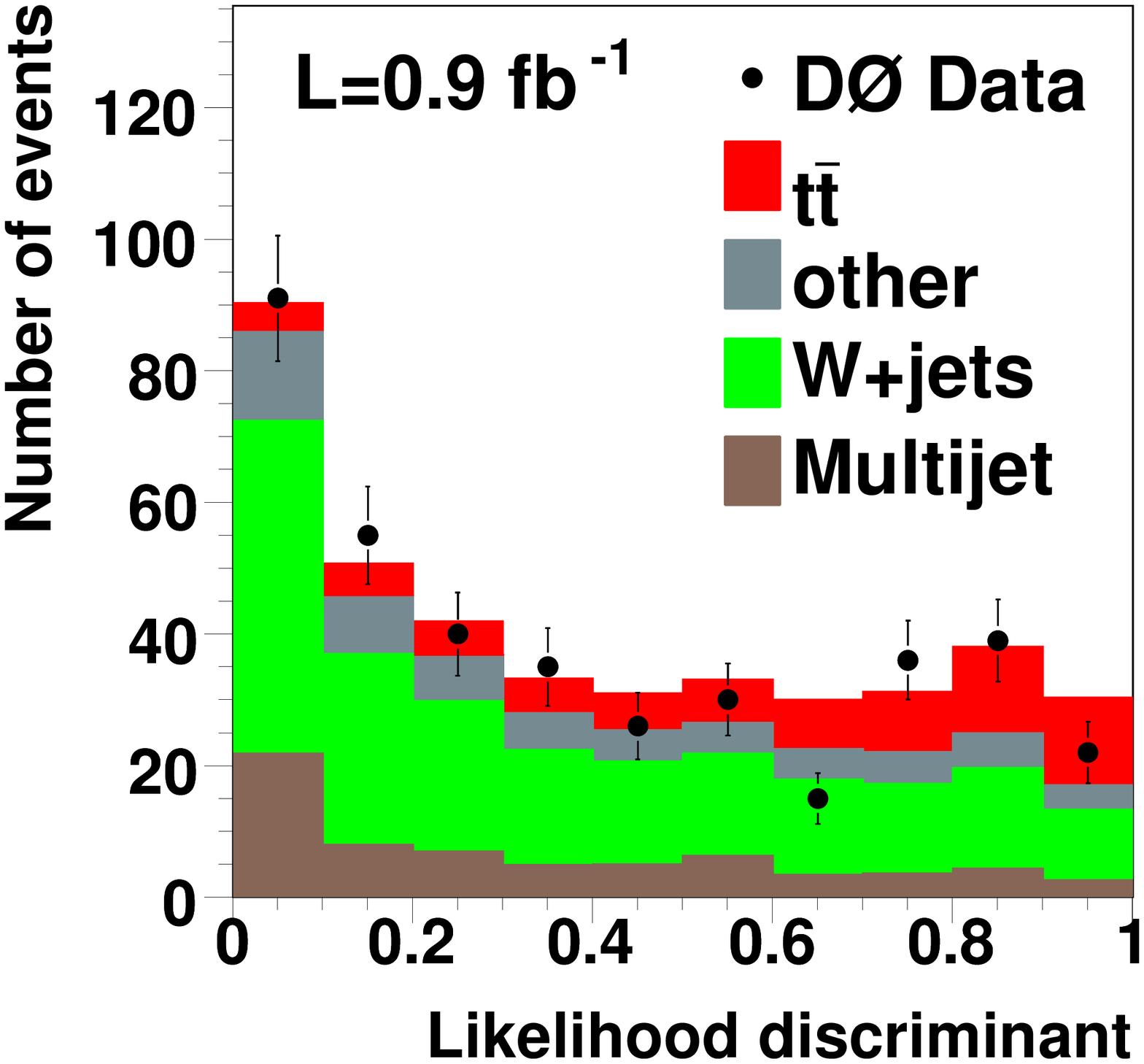}%
\includegraphics[width=0.3\linewidth]{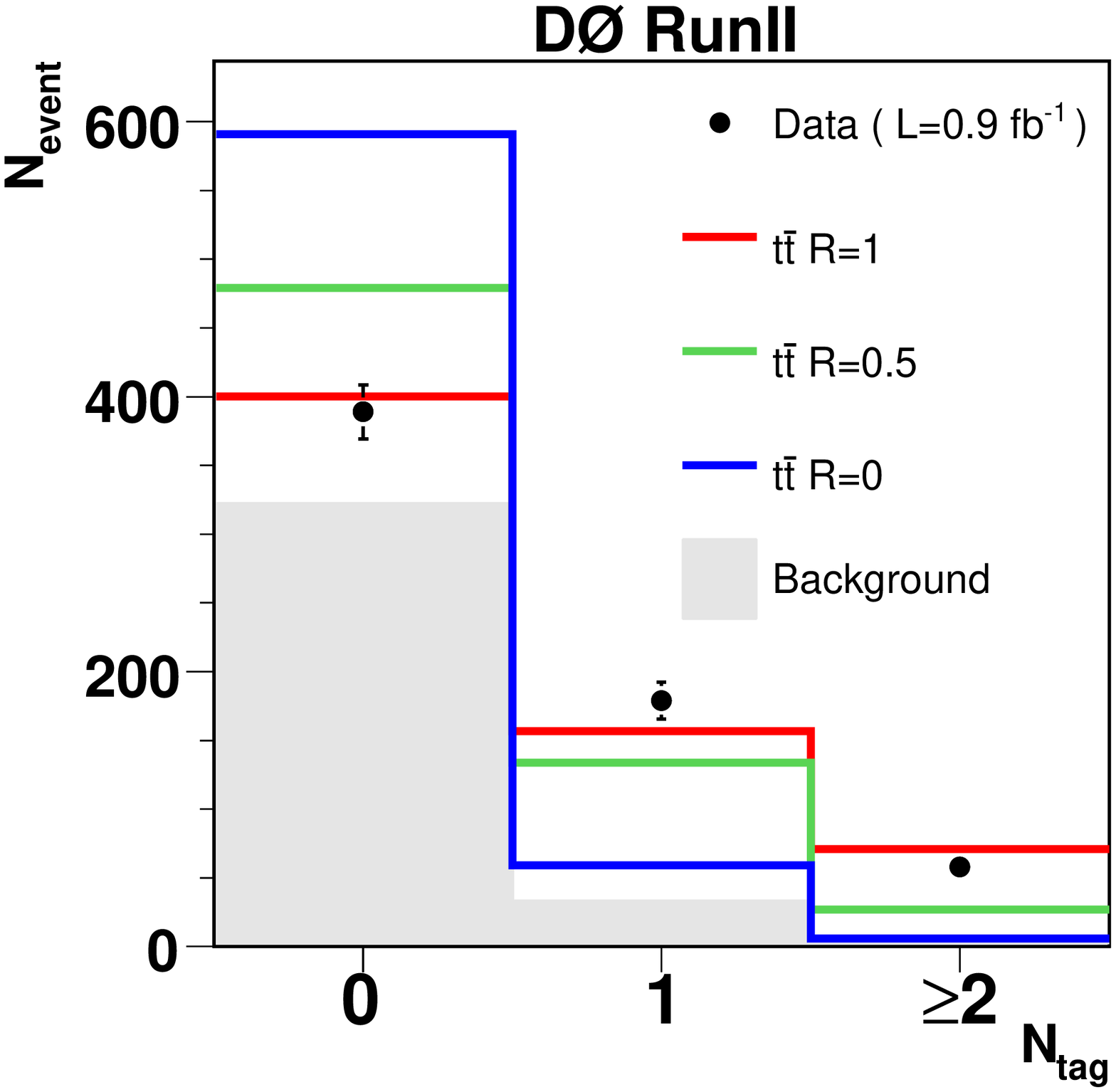}%
  \caption{\label{fig:d0:rb}%
Left: Probability to have zero, one or more identified $b$ jets in top quark  pair
events as function of the top quark  branching fraction, $R_b$. Middle: Data compared
to contribution from various backgrounds as function of the topological
likelihood discriminant. Right: Observed number of events as
function of the number of identified $b$ jets compared to expectations for
various values of $R_b$~\cite{Abazov:2008yn}.
}
\end{figure*}

Top quark  pair signal is simulated with \pythia\ including samples in which one or
both top quarks decay to a light quark and a $W$ boson. 
The dominating $W+$jets background is simulated using \alpgen+\pythia.
Its heavy flavour content of the $W+$jets background was corrected
according to a measurement in a control sample.
The fake lepton background from multijet events is fully estimated from data. 
Additional smaller backgrounds from  diboson, single top and $Z+$jets are
simulated using \pythia, \singletop\ and  \alpgen+\pythia, respectively, 
and normalised to their NLO cross-sections.
All simulations are passed through the D\O\ detector simulation and reconstruction.
In the simulation tag rate functions, determined on control samples in data, are
used to describe the probability for a given jet to be identified as $b$ jet.
Figure~\ref{fig:d0:rb} (left) illustrates the probability to have  zero, one
or more identified $b$ jets in top quark  pair events as function of the top quark 
branching fraction obtained from simulation. 

The event samples are separated by lepton type, number of jets ($3$ or $\ge 4$) and
number of identified $b$ jets ($0$, $1$ or $\ge2$). The $0$-$b$-tag sample
with four or more jets is further split in bins of a topological likelihood
discriminant to obtain additional separation between $W+$jets background 
and top quark  pair signal events, c.f.~\fig{fig:d0:rb} (middle).

To simultaneously determine the top quark  pair production cross-section,
$\sigma_{t\bar t}$,  and the branching fraction, $R_b$, 
a binned likelihood is constructed. Poisson distributions according to 
the expected event count as function of $\sigma_{t\bar t}$ and $R_b$  
is used for each sample and discriminant bin. 
The normalisation for $W+$jets is fixed globally by subtracting all other backgrounds
as well as top quark  pair estimates from data.
Systematic uncertainties are
included using nuisance parameters with Gaussian constraints.

For the determination of $R_b$ the measurement is dominated by the statistical
uncertainty. Systematic uncertainties are dominated by the uncertainty of the
$b$ tagging efficiency. In contrast to the CDF measurement, due to the global
determination of the $W+$jets normalisation, the uncertainty on the size of
this background is no significant source of uncertainty.

In $0.9\ifb$ of lepton plus jets data D\O\ obtains
$R_b = 0.97^{+0.09}_{-0.08}\mbox{(total)}$~\cite{Abazov:2008yn} consistent
with the expectation of the Standard Model. In \fig{fig:d0:rb} (right) the
observed number of events is compared to expectations for various values of $R_b$.
Limits on $R_b$  obtained using
the Feldman-Cousins procedure yield 
\beq
R_b>0.79 \quad\mbox{at $95\%$~C.L.}
\eeq
This limit is 
converted to a limit on the ratio of $\left|V_{tb}\right|^2$ to the
off-diagonal elements:
\beq
\frac{\left|V_{tb}\right|^2}{\left|V_{td}\right|^2+\left|V_{ts}\right|^2}>3.8
\quad\mbox{at $95\%$~C.L.}
\eeq 
The only assumption entering this limit is that top quarks
cannot decay to quarks other than the known Standard Model quarks. Thus it is
valid  even in presence of an additional generation of quarks as long as the
$b'$ quark is heavy enough.

%%% Local Variables: 
%%% mode: latex
%%% TeX-master: "EPJC_TopProperties"
%%% End: 

\subsection{Flavour Changing Neutral Currents}
Flavour changing neutral currents (FCNC) do not appear in the SM
at tree level and are suppressed in quantum loops~\cite{Glashow:1970gm,DiazCruz:1989ub,Eilam:1990zc,Mele:1998ag}. 
However, anomalous couplings could lead to enhancements of FCNC in
the top quark sector and their observation would be a clear sign of new 
physics~\cite{Fritzsch:1989qd,AguilarSaavedra:2004wm}.

The Tevatron experiments have looked for FCNC both in top quark
decays~\cite{Aaltonen:2008aaa} and in the production
of  (single) top quarks~\cite{Aaltonen:2008qr,Abazov:2007ev,Abazov:2010qk}. 
Limits on the single top production through anomalous couplings were also 
set with LEP and HERA 
data~\cite{Heister:2002xv,Abdallah:2003wf,Achard:2002vv,Abbiendi:2001wk,Chekanov:2003yt,Aktas:2003yd,H1:FCNC2007,Aaron:2009vv}.

\subsubsection[Top Quark Decay through $Z$ Bosons]{\boldmath Top Quark Decay through $Z$
  Bosons}

In an investigation of data with a total luminosity  of $1.9\ifb$ CDF looks
for top quark pairs that show a flavour changing neutral current decay through a
$Z$ boson~\cite{Aaltonen:2008aaa}.  The analysis aims to identify events in which the $Z$ boson decays
leptonically and the second top quark decays through a $W$ boson into hadrons.  
The event selection thus looks for a pair of leptons and at least four jets.
The leptons need to be of the same flavour and
have opposite charge. Their invariant mass  is required to be
within $15\GeV$ of the $Z$ boson mass. The cuts on the total transverse mass and 
the transverse energy of the leading and sub-leading jets were optimised in
simulation.  Events failing these cuts are used as control sample, events
passing these cuts are split into events without any identified $b$ jet and
events with at least one  identified $b$ jet.

The dominant background with this selection stems from Standard Model $Z+$jets
production, which is simulated using \alpgen. %+\herwig (*** oder Pythia???***).
Further but much smaller background contributions stem from Standard Model top quark
pair production, and diboson production. 
The signal of top pairs with FCNC decay was simulated using \pythia. The
events are reweighted to yield helicities of 65\% longitudinal and $35\%$ left-handed $Z$ bosons.

To separate signal from background the mass of the $W$ boson is reconstructed
from two jets, the top quark mass is reconstructed by adding a third jet and
a second top quark mass is reconstructed from the $Z$ boson with the fourth jet. 
A $\chi^2$ variable is built from the differences of the reconstructed masses
to the nominal $W$ boson and top quark masses, respectively. The $\chi^2$ of
the jet parton assignment that yields the lowest $\chi^2$ is used to build a
distribution of $\chi^2$ values.

The estimated shapes of the various backgrounds and the signal events are used
as templates that are fitted to the distribution measured in data. The main
parameters of the fit are the branching fraction, ${\cal B}(t\rightarrow Zq)$,
and the normalisation of the
dominating $Z+$jets background (in the control sample). Further  parameters 
describe the difference of the background normalisation between the signal and
the control samples (with a Gaussian constraint), the $b$ quark identification
fraction and the jet energy scale
shift. The latter is considered to cover all shape changing affects.  
\begin{figure}[t]
  \centering
\includegraphics[width=0.8\linewidth]{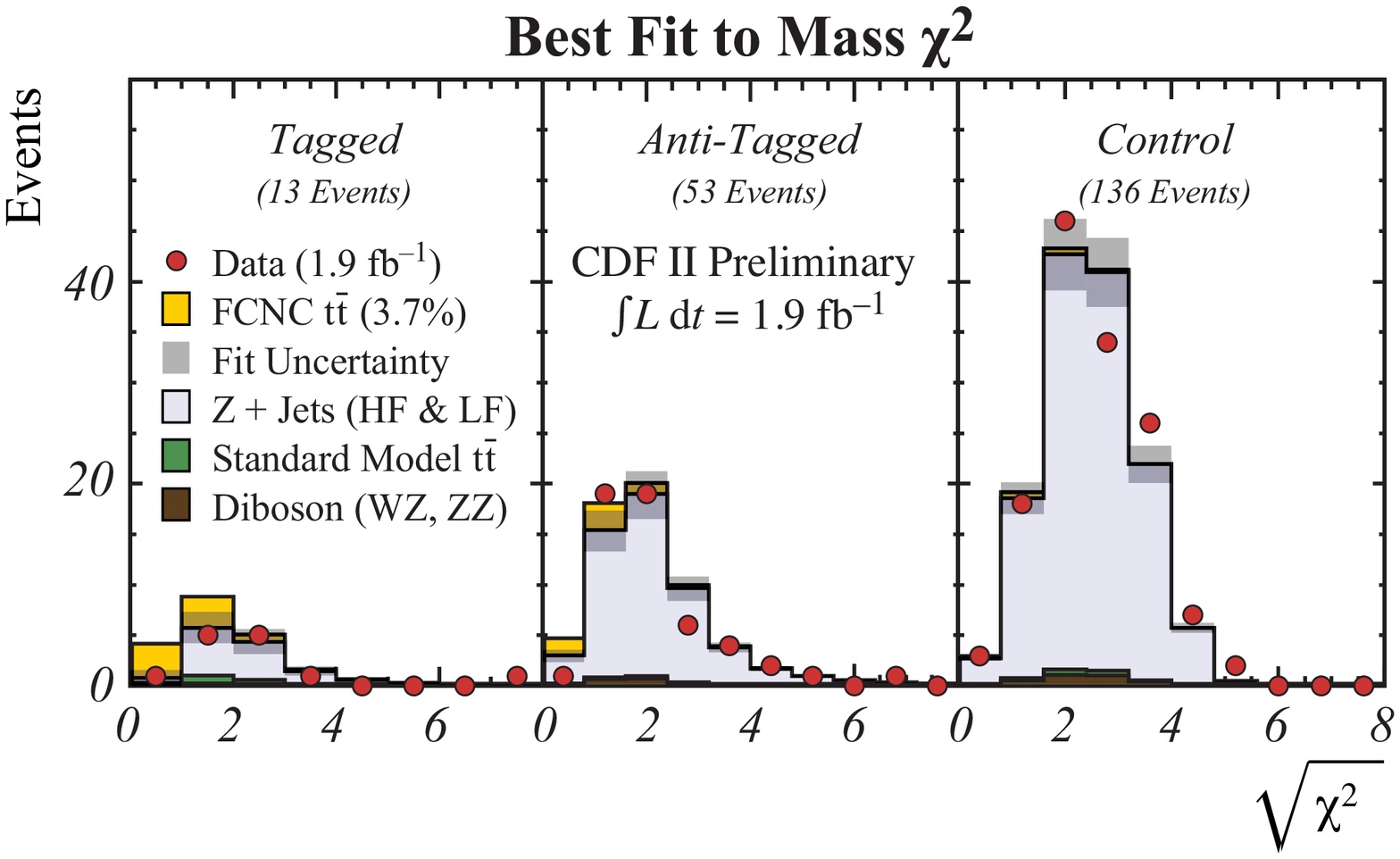}  
\includegraphics[width=0.8\linewidth]{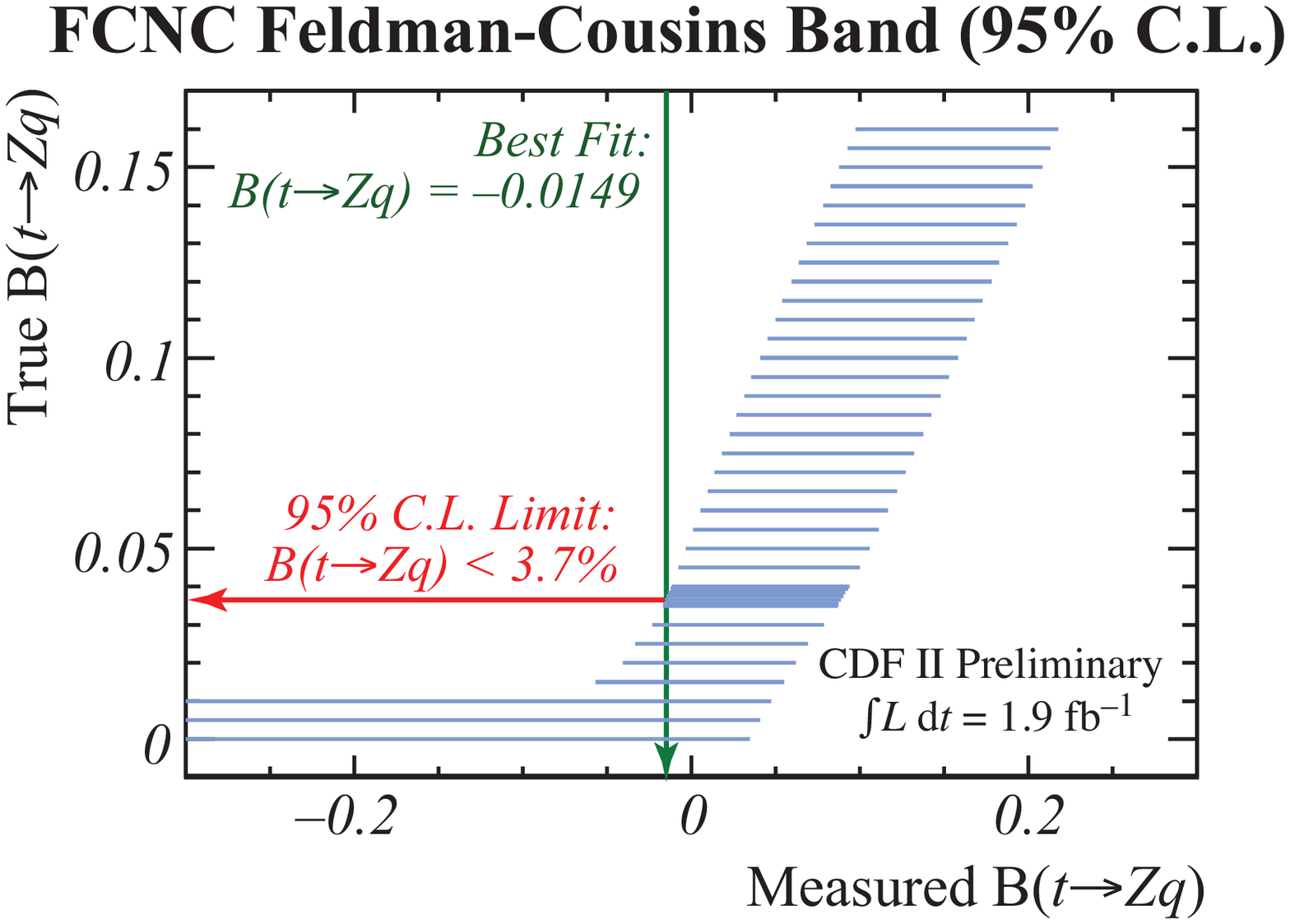}
  \caption{Top: Mass $\chi^2$ distribution observed in the two signal and one
    control sample compared the the best fit expectation plus a signal of
    $3.7\%$ branching to $Z$ boson.  Bottom:  Feldman-cousins band of $95\%$
    C.L. with the measured branching fraction ${\cal B}(t\rightarrow Zq)$~\cite{Aaltonen:2008aaa}.}
  \label{fig:cdf-fcnc-toppairs}
\end{figure}

The distribution of observed $\chi^2$ values is shown with the best fit of the
signal and background templates in \fig{fig:cdf-fcnc-toppairs}.
Data agree well with the Standard Model templates and thus limits  on the
branching fraction $t\rightarrow Zq$ are set. For this the
Feldman-Cousins method is applied and yields  \mbox{${\cal B}(t\rightarrow
Zq)<3.7\%$} at $95\%$~C.L.~\cite{Aaltonen:2008aaa}
The Run\,I result in addition sets a limit on  flavour changing neutral currents in the
photon plus jet mode of \mbox{${\cal B}(t\rightarrow
\gamma u)+{\cal B}(t\rightarrow
\gamma c)<3.2\%$}~\cite{Abe:1997fz}.

\subsubsection{Anomalous Single Top Quark Production}

While the above study of top quark decays addresses a flavour changing neutral
current through the $Z$ boson, investigations of the production of single top quark
events can be used to restrict anomalous gluon couplings.

\subsubsubsection{CDF}

In an analysis of $2.2\ifb$ CDF looks for the production of single top quarks
without additional jets, $u(c)+g\rightarrow t$. To select such events with a
leptonic top quark decay, one isolated lepton, transverse missing energy and
exactly one hadronic jet are required. The jet must be identified
as $b$ jet. Additional cuts are used to reduce the backgrounds without a $W$ boson as in
the single top quark analyses~\cite{Aaltonen:2008qr}.

To describe the expected background from Standard Model processes diboson and
top quark pair events are simulated with \pythia\ and normalised to the NLO
cross-sections. Single top quark events are simulated using \madgraph+\pythia~\cite{Stelzer:1994ta,Maltoni:2002qb}.
Finally, processes of weak vector bosons are simulated with
\alpgen+\pythia. In these samples the heavy flavour contribution is enhanced
according to the findings in a control sample. The total normalisation of the
$W+$jets samples is taken from sideband data. 
The signal of FCNC production of single top quark is simulated using
\toprex+\pythia~\cite{Slabospitsky:2002ag}.

Due to the large background from $W+1$jet data, a neural network is employed to
differentiate between FCNC and Standard Model production. 
Fourteen observables, which each allow a significance of more than $3\sigma$ in
discriminating signal and background, were chosen as inputs to the neural
network. They utilise kinematical properties of the measured quantities and the
reconstructed $W$ boson as well as the output of a special flavour separation
neural network. The neural network is trained on samples with equal amount of signal
and background.
It is then applied to the individual signal and background samples to obtain 
templates for all simulated physics processes considered, see~\fig{fig:cdf-fcnc-singletop}~(left).

To determine the possible contribution of FCNC single top quark production
the background templates are added according to their expected contribution and a binned maximum
likelihood fit is used to measure the contribution due to FCNC production.
Systematic uncertainties are parametrised in the likelihood function with Gaussian constraints.
They are dominated by uncertainties on the cross-sections of the background
samples normalised to NLO and the selection efficiency for signal events.

\begin{figure*}
  \centering
\includegraphics[height=5cm]{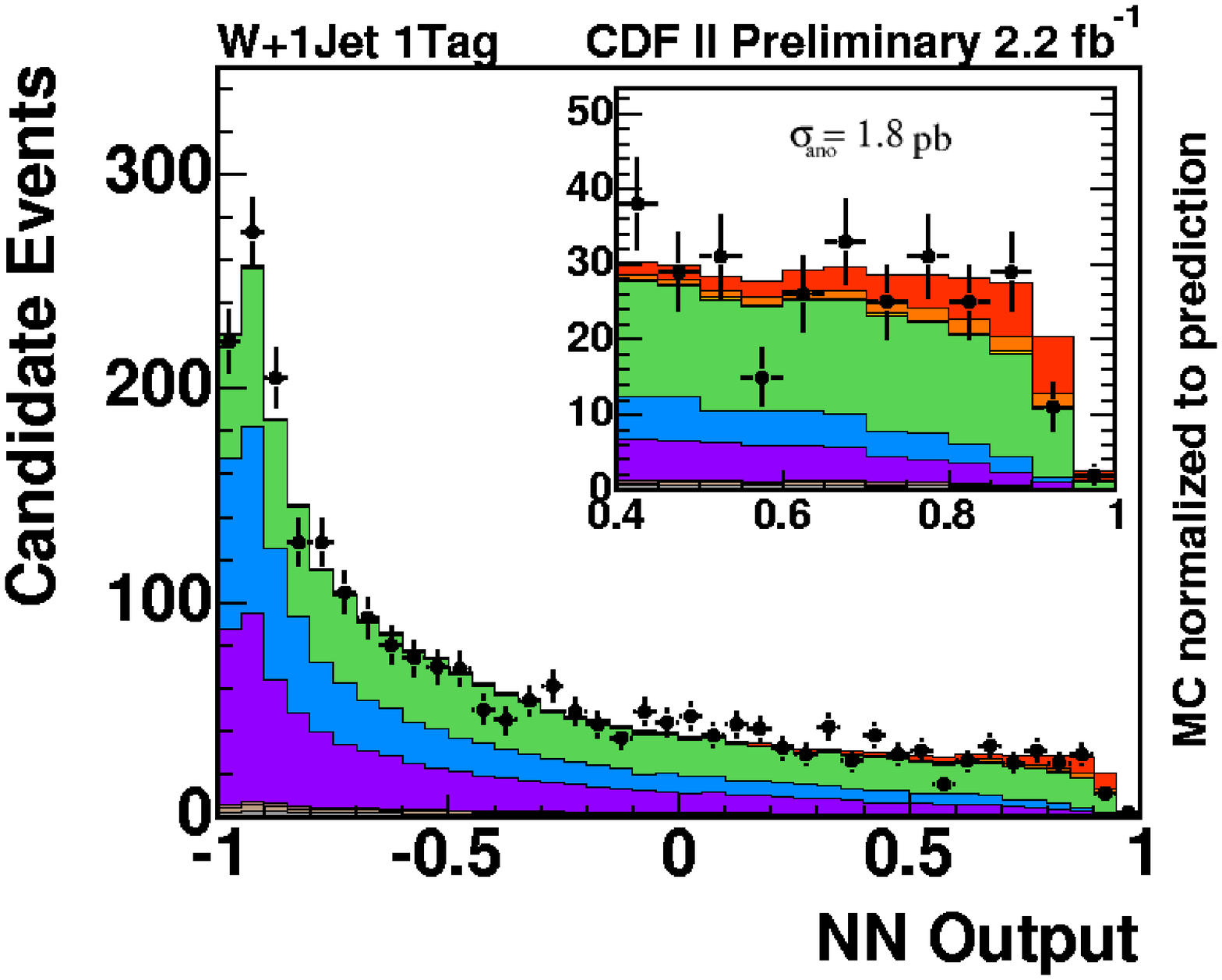}
\includegraphics[height=5cm]{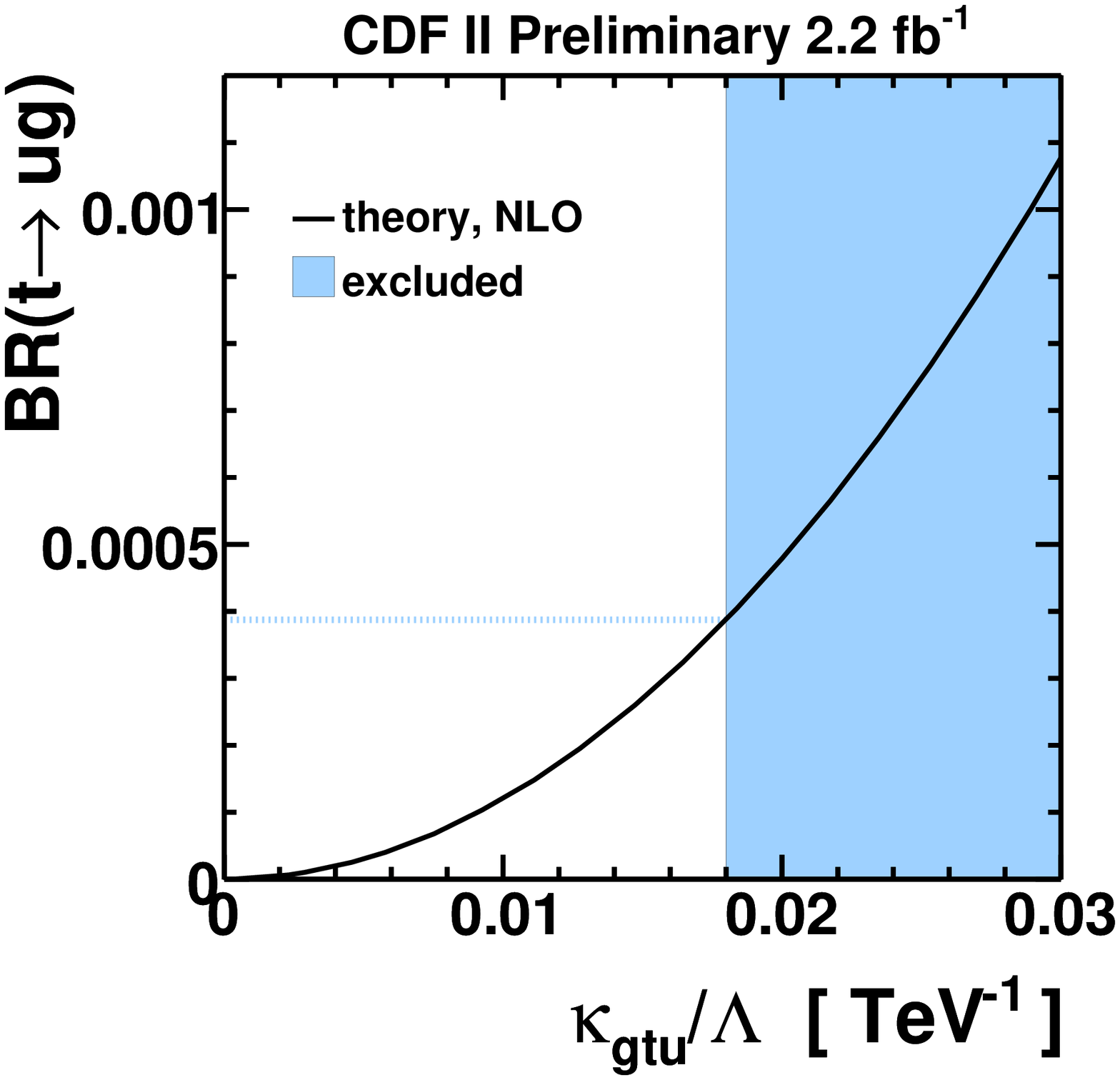}
\includegraphics[height=5cm]{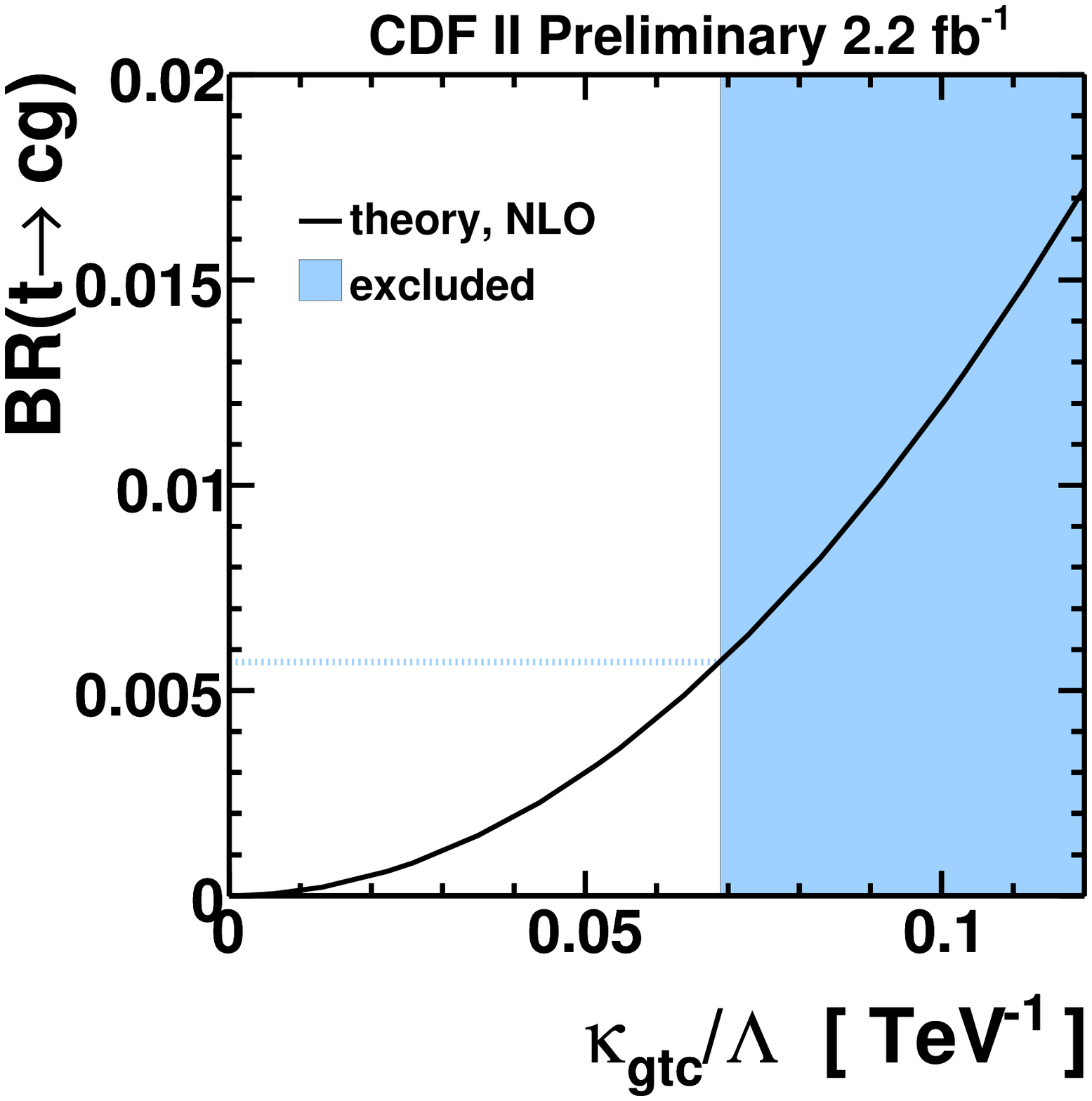}
  \caption{Neural network output compared to $2.2\ifb$ of CDF data (left). Upper
    limits on the anomalous branching fractions derived from the limits on the
    anomalous couplings to $u$-quarks (middle) and $c$-quarks (right)~\cite{Aaltonen:2008qr,CdfNote9251}.}
  \label{fig:cdf-fcnc-singletop}
\end{figure*}

CDF finds no significant contribution of FCNC single top quark production in
$2.2\ifb$ of data. The limit on the allowed production cross-section
$\sigma_t^{\mathrm{FCNC}}$ is set using Bayesian statistics with a flat prior
for positive cross-sections and yields $\sigma_t^{\mathrm{FCNC}}<1.8\pb$ at
$95\%$~C.L.~\cite{CdfNote9251}.  
This cross-section limit is converted to limits on FCNC top quark-gluon coupling
constants following~\cite{Liu:2005dp,Yang:2006gs}. Assuming that only one of
the couplings differs from the Standard Model expectation CDF finds
$\kappa_{gtu}/\Lambda<0.018\TeV^{-1}$ or
$\kappa_{gtc}/\Lambda<0.069\TeV^{-1}$.
Expressed in terms of the top quark branching fraction through this processes these limits
correspond to ${\cal B}(t\rightarrow u+g)<3.9\cdot 10^{-4}$ and 
${\cal B}(t\rightarrow c+g)<5.7\cdot 10^{-3}$ as shown in \fig{fig:cdf-fcnc-singletop}~(middle
and right).
These small limits justify the approximation of pure Standard Model decays
made in simulating signal samples above. 

\subsubsubsection{D\O}

D\O\ has set limits on the FCNC anomalous couplings of the
top quark in up to $2.3\ifb$ of data~\cite{Abazov:2007ev,Abazov:2010qk}. 
The analyses investigate the singly production of a top quark in
association with at least one additional jet. The event selection requires
an isolated charged lepton,  missing transverse momentum and at least two
jets. Exactly one of the jets has to be identified as $b$ jet. With this the
selection follows closely the selection used  for the measurements of the single
top quark production~\cite{Abazov:2006uq,Abazov:2008kt,Abazov:2009pa,Abazov:2009ii}.

In the most recent analysis the single top quark samples for the SM and the FCNC signal
processes are simulated with the {\singletop} generator~\cite{Boos:2006af,Boos:2004kh}.
%Previous was \comphep+\pythia~\cite{Pukhov:1999gg,Boos:2004kh,Belyaev:2000wn}  
Background contributions from top quark pair, $W+$jets and  $Z+$jets production are
simulated using \alpgen+\pythia\ and diboson production is simulated by
\pythia. All samples are passed through \geant\ to simulate the D\O\ detector
and then reconstructed with the standard event reconstruction.
The SM samples of single top quarks, top quark pairs, $Z+$jets and dibosons
are normalised to the NLO (or better) cross-sections.
Events from $W+$jets production are normalised to  data before
$b$-tagging accounting for  other simulated backgrounds and multijet
background. The multijet background is described using data in which the
lepton candidates fail one of the lepton identification cuts.

To separate FCNC from Standard Model prediction the method of a Bayesian neural network (BNN) is
employed. A large number of  variables is  used as input to the neural
network. These variables describe object and
event kinematics, top quark reconstruction, jet width and angular
correlations. For training the BNN the two FCNC processes including $tgu$ and
$tgc$ couplings, respectively, are combined into a single training sample. 
Separate BNNs are trained for each lepton flavour and jet multiplicity.
The distribution of BNN outputs observed in data agrees well with the pure SM
expectation, thus limits on  the allowed anomalous gluon couplings are computed.

\begin{figure}[b]
  \centering
\includegraphics[width=0.7\linewidth]{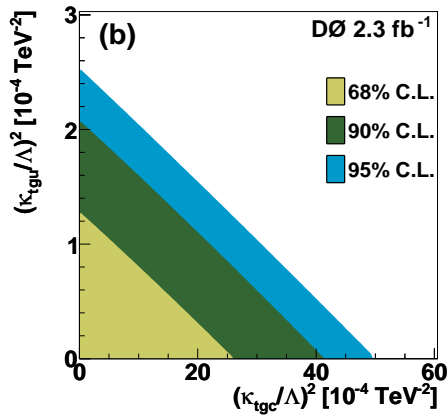}
  \caption{
    Exclusion contour of the quadratic FCNC couplings obtained by D\O\
    in a two dimensional approach using $2.3\ifb$~\cite{Abazov:2010qk}.}
  \label{fig:d0-fcnc-singletop}
\end{figure}

These limits are computed using a Bayesian approach. A likelihood for the
distribution of neural network outputs observed in data to occur 
is computed from the events expected in the Standard
Model and the FCNC production of single top quark as function of the anomalous
gluon couplings $\kappa_{gtu}/\Lambda$ and  $\kappa_{gtc}/\Lambda$. 
The likelihood for each bin is based on a Poisson distribution. 
Systematics are taken into account by smearing the Poisson parameters with a
corresponding Gaussian distribution.
The dominant uncertainties stem from shape 
changing effects like those from the jet energy scale and the modelling of $b$
quark identification.
In addition  normalisation uncertainties for the background simulations and
the overall luminosity uncertainty give a significant contribution.

The likelihood is folded with a prior flat in the FCNC cross-sections and
exclusion contours are computed as contours of equal probability that contain
$95\%$ of the volume.  The two dimensional limits on the squared couplings
observed by D\O\ in $2.3\ifb$~\cite{Abazov:2010qk}
are shown in \fig{fig:d0-fcnc-singletop}. One dimensional limits are obtained
by integrating over one of the two anomalous couplings and yield
$\kappa_{gtu}/\Lambda<0.013\TeV^{-1}$ and
$\kappa_{gtc}/\Lambda<0.057\TeV^{-1}$, corresponding to branching fractions of
${\cal B}(t\rightarrow u+g)<2.0\cdot 10^{-4}$ and 
${\cal B}(t\rightarrow c+g)<3.9\cdot 10^{-3}$. 
These D\O\ results are the currently most stringent branching fraction limits.

%%% Local Variables: 
%%% mode: latex
%%% TeX-master: "EPJC_TopProperties"
%%% End: 

\subsection{Top Quark Charge}
% $Id: Charge.tex,v 1.15.2.10 2011-03-25 09:19:15 wicke Exp $
The top quark's electrical properties should be fixed by its charge.
However, in reconstructing top quarks the charges of the selected objects
are usually not checked. Thus an exotic charge
value of $\left|q_t\right|=4e/3$ is not excluded by standard analyses.
Furthermore the amount of photon radiation off the top quark may a priori
differ from the expection based on the top quark charge.
The Tevatron experiments have searched for deviations from the SM expectation
in both aspects of the top quark charge.

\subsubsection{Exotic Top Quark Charge}
To distinguish between the Standard Model and the exotic top quark  charge it is
necessary to reconstruct the charges of the top quark decay products,
the $W$ boson and the $b$ quark. The $W$ boson charge can be taken from the charge
of the reconstructed lepton, but finding the charge of the $b$ quark
is more difficult.

\subsubsubsection{D\O}

%\subsubsection{Data Selection and Background Description}
{D\O} has performed an analysis of $\ell+$jets events with at least two
$b$-tagged jets in $370\ipb$ using a jet charge technique to determine the
charge of the $b$ jets~\cite{Abazov:2006vd}.
Semileptonic events are selected following the cross-section analysis by requiring exactly one isolated
lepton, transverse missing energy and four or more jets. At least two of the
jets must be identified as $b$ jets using a secondary vertex tagging
algorithm.

%\subsubsection{Jet Charge}
The charge of a jet can be defined as the sum of the charges of all
tracks inside the cone of that jet. In this analysis the sum has been
weighted with the component of the track momenta transverse to the jet
momentum, $p_\perp$:
\beq
 Q_\mathrm{jet}:=\frac{\sum{q_i\cdot p^{\kappa}_{\perp,i}}}{\sum{p^{\kappa}_{\perp,i}}}
\quad\mbox{with}\quad \kappa=0.6
\eeq
where the sums run over all tracks, $i$, within the jet under consideration
and $q_i$ is the charge sign of the track $i$.
Because particles may easily escape the jet cone such a jet charge 
fluctuates strongly from event to event, so only statistical
statements can be made.
It is crucial to determine the expected distribution of
$Q_\mathrm{jet}$ in the case of $b$ or $\bar b$ quark and, because a
significant fraction of charm quarks gets flagged by the secondary
vertex tagger, also for the $c$ and $\bar c$ quarks. These expected
distributions, c.f.~\fig{fig:d0-charge}~(left), are derived from dijet
data using a tag and probe method. 

\begin{figure}
  \centering
  \includegraphics[width=0.6\linewidth]{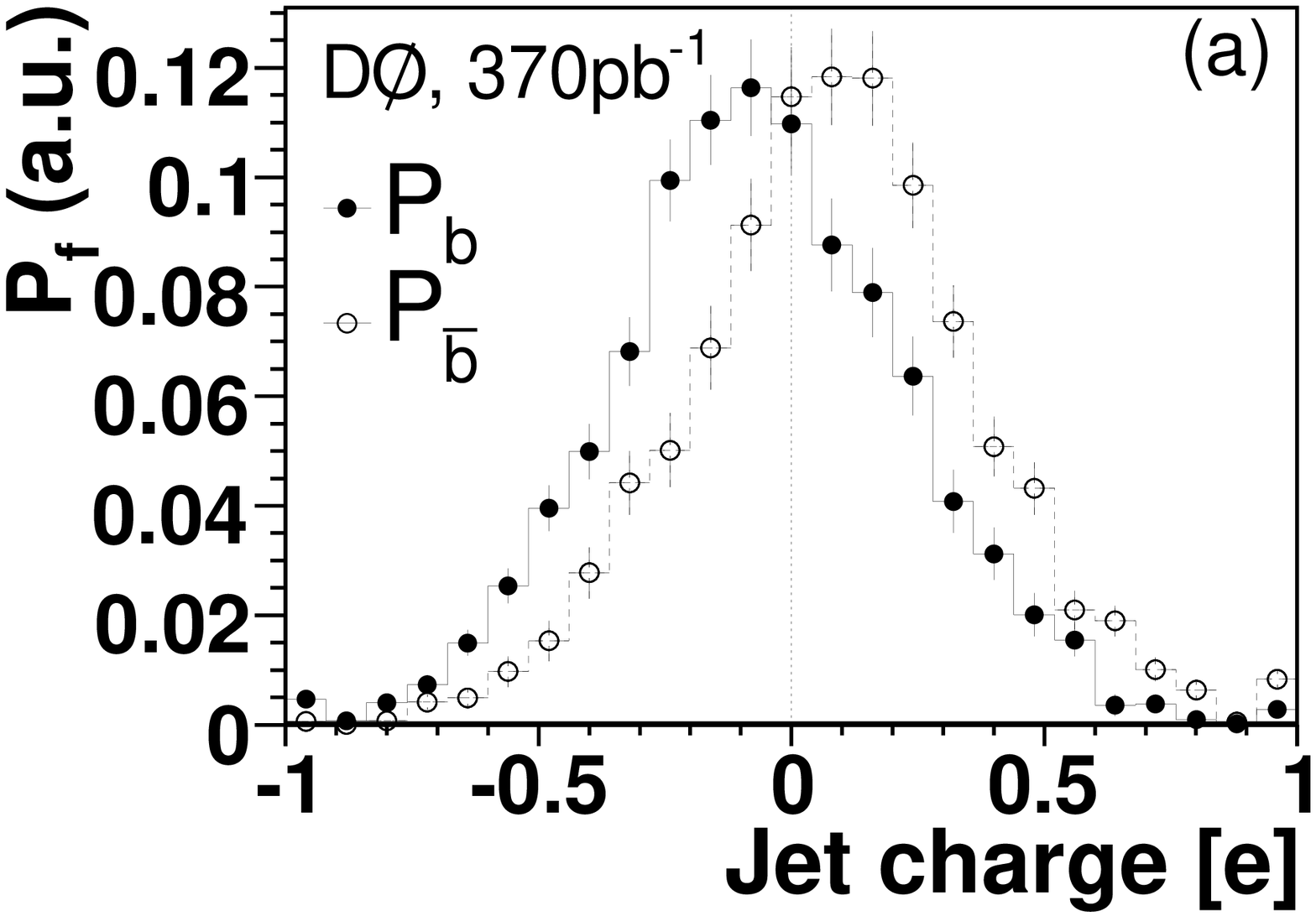}
  \includegraphics[width=0.6\linewidth]{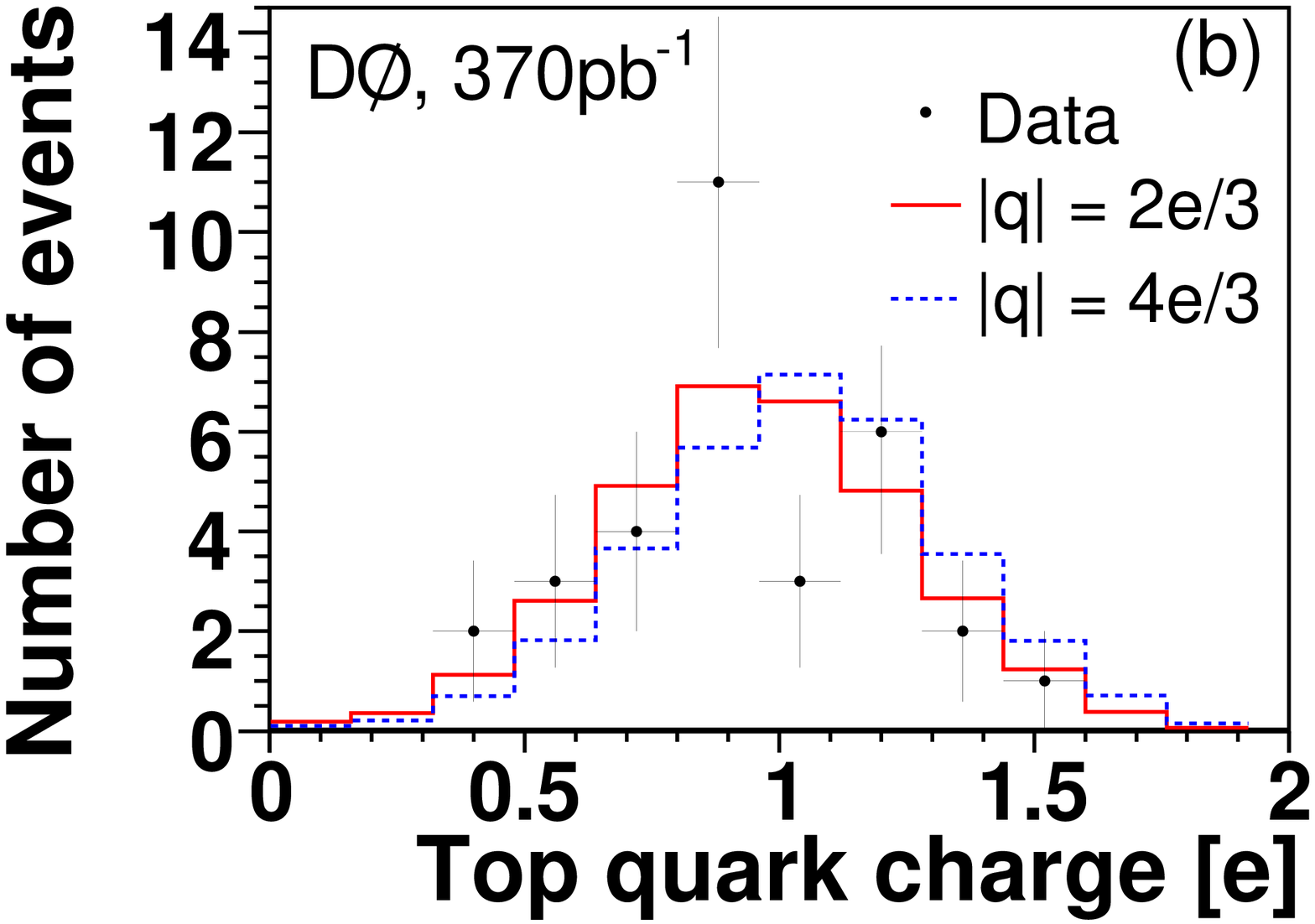}
  \caption{Expected jet charge distribution for $b$ and anti-$b$
    quarks (left). Measured absolute top quark  charge compared to Standard Model and exotic
    models (right)~\cite{Abazov:2006vd}.}
  \label{fig:d0-charge}
\end{figure}
%\subsubsection{Top Quark  Charge Results}
To determine the top quark  charge an
assignment of $b$-jets to the leptonic or hadronic event side is
necessary. This analysis uses the quality of a fit to the $t\bar t$
hypothesis, which uses the $W$ boson and top quark masses as constraints, to
select the best possible assignment.
The jet charge for the $b$ jet on the leptonic (hadronic) side,
$q_{b_l}$ ($q_{b_h}$) is then combined with the charge of the measured
lepton $q_l$ to define two top quark  charge values per event:
%\bea
%~~\quad
%  Q_\mathrm{lep}=\left|q_l+q_{b_\mathrm{l}}\right|&\qquad&
%  Q_\mathrm{had}=\left|-q_l+q_{b_\mathrm{h}}\right|\quad\mbox{.}
%\eea
$Q_\mathrm{lep}=\left|q_l+q_{b_\mathrm{l}}\right|$ and $Q_\mathrm{had}=\left|-q_l+q_{b_\mathrm{h}}\right|$.
The distribution of the measured top quark  charges is compared to 
templates simulated for the Standard Model and the exotic case, where the exotic case has
been obtained by inverting the jet charge, see \fig{fig:d0-charge}~(right). 
The top quark  pair events are simulated using
\alpgen+\pythia. 
The dominating $W+$jets background is simulated using \alpgen+\pythia\ with a
normalisation to data. Multijet templates are derived from data alone. 
All simulated events are passed through full D\O\ detector simulation. 

An unbinned likelihood ratio accounting also for remaining background
yields a $p$-value for the exotic case of $7.8\%$ and a Bayes factor
of $B_f=4.3$ favouring the Standard Model charge scenario~\cite{Abazov:2006vd,d0-charge-clarification}.

\subsubsubsection{CDF, Jet Charge}

The corresponding analysis by CDF  investigates $1.5\ifb$ with events from the semileptonic
and the dileptonic decay channel~\cite{CdfNote8967}. The former are selected requiring an
isolated charged lepton, missing transverse momentum and at least four
jets. Two of the jets are required to be identified $b$ jets using CDFs
secondary vertex algorithm. Dilepton events are selected by asking two
oppositely charged leptons, missing transverse momentum and at least two jets,
one of which needs be identified as $b$ jet.

Compared to the D\O\ analysis the jet charge is computed slightly
differently. Instead of the transverse momentum the scalar product of the jet
and the track momentum is used to weigh the measured charges:
\beq
 Q_\mathrm{jet}:=\frac{\sum{q_i\cdot (\vec{p_i}\cdot\vec{p}_{\mathrm{jet}})^{\kappa}}}{\sum{(\vec{p_i}\cdot\vec{p}_{\mathrm{jet}})}}
\quad\mbox{with}\quad \kappa=0.5.
\eeq
Depending on the sign of $Q_\mathrm{jet}$ the identified $b$ jet is considered
to stem from the $b$ or the
$\bar b$ quark. The purity of this assignment is calibrated on dijet events with
two identified $b$ jets. One 
of the jets is required to contain a muon, that serves as tag in the tag and
probe method. The resulting purity is corrected for effects due to muons from
secondary decays, for $B$ meson mixing and for light or $c$-quark jets
misidentified as $b$ jets. 
% The first two are described by MC, the latter is taken from template fits to
% data.

\begin{figure}[b]
  \centering
  \includegraphics[width=0.88\linewidth]{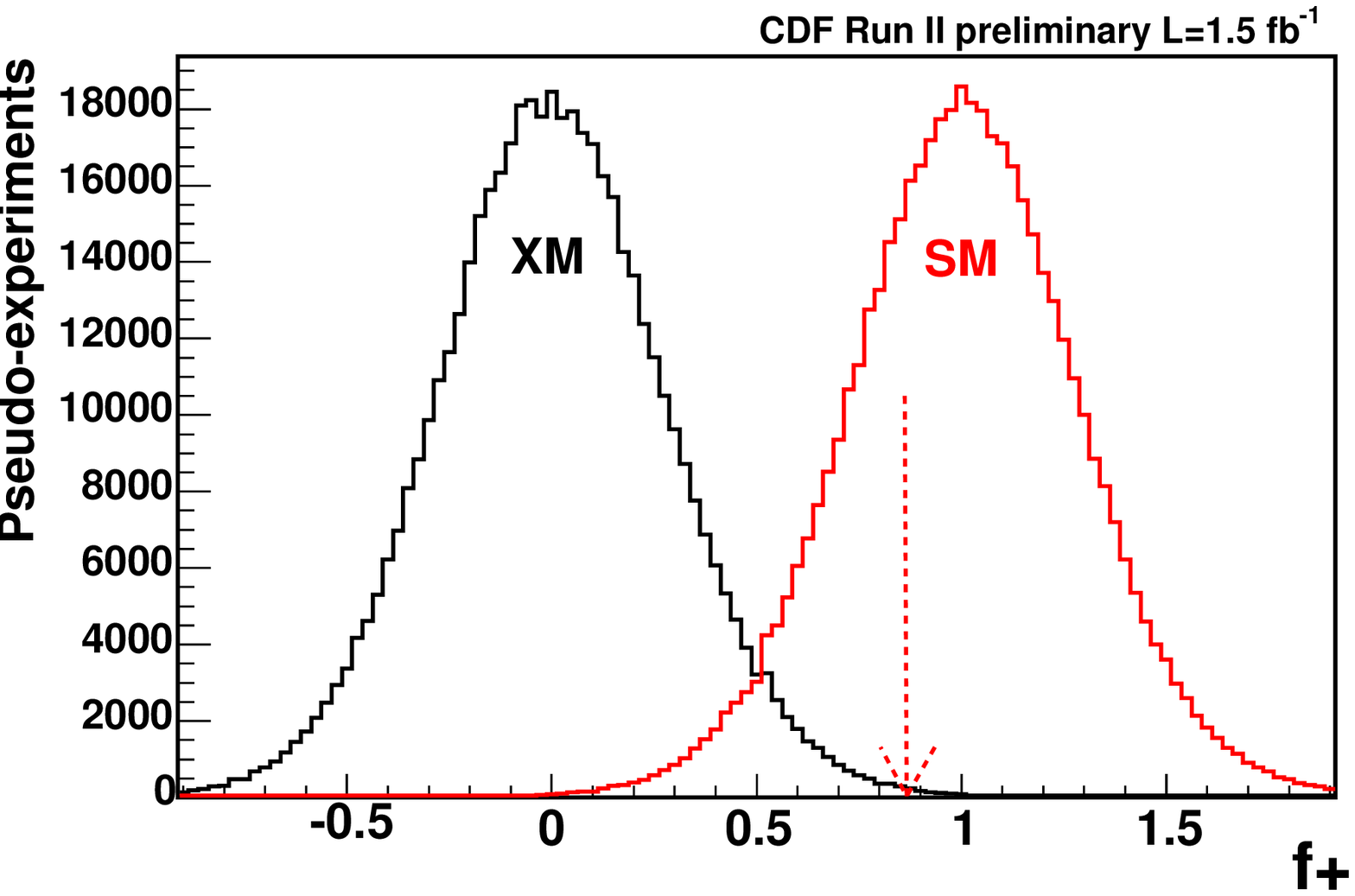}
  \includegraphics[width=0.88\linewidth]{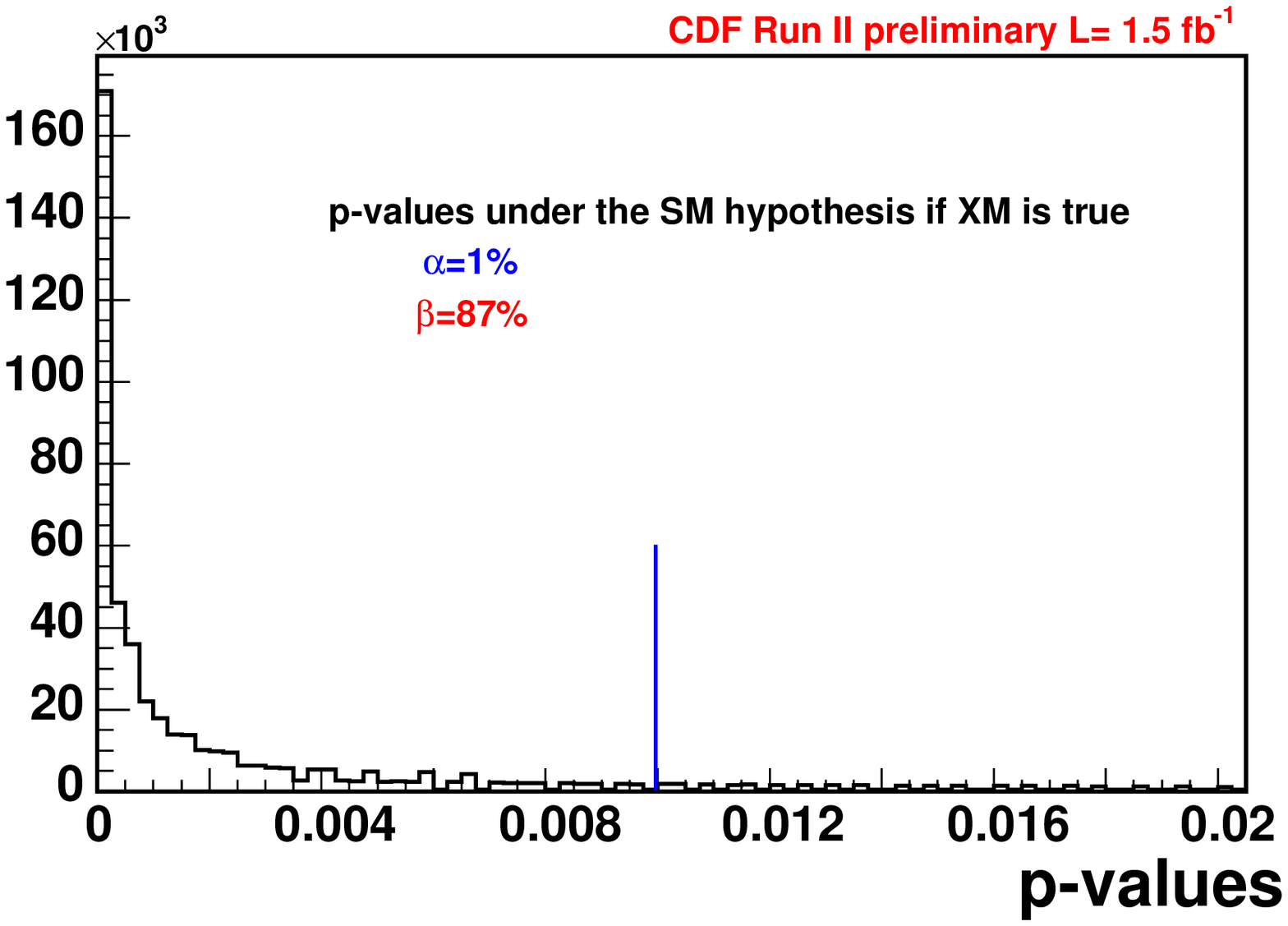}
  \caption{Top: Distribution of the obtained fraction of Standard Model like
    pairs, $f_+$, in ensembles with the exotic and the Standard Model top quark
    charge. 
    Indicated is the value measured in data: $f_+=0.87$. % which corresponds to a  $p$-value of $0.31$.
    Bottom: Distribution of $p$-values for the Standard Model hypothesis to be true
    obtained in ensembles of the exotic model. Indicated
    is the a priori chosen limiting probability of incorrectly rejecting
    the Standard Model if Standard Model is true, $1\%$,  and the corresponding probability of rejecting the 
    Standard Model if the exotic model is true, $87\%$~\cite{CdfNote8967}.}
  \label{fig:cdf-charge}
\end{figure}
To compute the charge of the top and anti-top quark the jets need to be
associated to the leptons. In the semileptonic channel a kinematic fit with
constraints on the top quark mass and the $W$-boson mass is used. The
jet-parton association with the lowest $\chi^2$ of this fit is kept. 
In the dilepton channel the invariant mass of each pair of one lepton and one
jet, $M_{lb}^2$, is computed. The combination which does not produce the
largest value of $M_{lb}^2$ is used.
In both channels cuts on $\chi^2$ and  $M_{lb}^2$, respectively, are used to
enhance the purity of correct assignments.

Each event can now be classified as Standard Model like or as exotic model
like. To obtain a statistical interpretation a likelihood is computed 
as function of the fraction of Standard Model like signal pairs,
$f_+$. Nuisance parameters that represent the number and purity of signal and
background events are optimised for each value of $f_+$. 
The systematic uncertainties considered include effects from the choice of
parton density function, 
the uncertainties in the simulation of initial and final state radiation, the
jet energy scale and the choice of the generator.
All systematic uncertainties are included in the statistical treatment through
their effect on the nuisance parameters. 

Background predictions are obtained as for other CDF lepton plus jets and dilepton
analyses based on a mixture of simulation and data. For this analysis each
background is checked for a correlation between the charge of the signal lepton
and the jet charge value of the corresponding $b$ jet. Such a correlation
could occur for the semileptonic channel from the
$b\bar b$ background when the lepton from the $b$ decay passes the lepton
selection criteria and from single top quark  events. In both cases the correlation
is found to be small and consistent with zero  within uncertainties.

In $1.5\ifb$ of data CDF finds the most likely value of the fraction of Standard Model like signal events
to be $f_+=0.87$. This corresponds to a  $p$-value of $31\%$~\cite{CdfNote8967},
see also~\fig{fig:cdf-charge}. 
Because this is larger than the  a priori chosen limiting probability of $1\%$ to falsely reject
the Standard Model hypothesis, CDF claims to confirm the Standard Model
hypotheses. The confidence limit corresponding to this $1\%$ choice is computed
as $87\%$. CDF computes the Bayes factor to be $2\log(B_f)=12$, which shows that this
analysis with $1.5\ifb$ yields a much stronger exclusion of the exotic
hypothesis than the D\O\ analysis of $370\ipb$ described above.  

\subsubsubsection{CDF, Soft Lepton Tag}
With more data an alternative method to determine the charge of the $b$ quarks
is to concentrate on the leptonic decays of the $b$ quarks. CDF applies this
method on $2.7\ifb$~\cite{CdfNote9939}. 

Events are selected requiring an
isolated energetic lepton, missing transverse energy 
and at least four jets. Vetoes are applied against
additional isolated energetic leptons, conversion electrons, 
cosmic muons, and $Z$ bosons. In the four jets at least one soft lepton and one
secondary vertex tag must be reconstructed. The soft lepton tag is optimised
to select leptonic $b$ quark decays, $b\rightarrow \ell\nu X$, 
but to suppress cascade decays, $b\rightarrow c\rightarrow\ell\nu X$.
With this the charge of the $b$ quark can be deduced from the measured
soft lepton charge.

To resolve the ambiguities of assigning the measured jets and energetic
leptons to the (anti-)top quark decay products a kinematic fitter is applied. 
For the computation of the top quark charge the solution with the best fit
quality is used. Only solutions which assign the tagged jets to $b$ quarks are
considered. Events with a bad fit quality are rejected, if  $\chi^2<9$ or
$\chi^2<27$  for events with one or two tagged jet, respectively.

To estimated the expected performance of the analysis, signal events are
generated with \pythia\ using {\sc EvtGen}~\cite{Lange:2001uf} for the decays. 
Background from $W+$jet events are generated using \alpgen+\pythia\ correcting
the heavy flavour contribution by a factor derived in $W+1$jet
events. These samples are normalised to the data before requiring any $b$ tags. 
The contribution of diboson single top, $Z+$jets and
Drell-Yan events is considered using simulation normalised to the theoretical
or experimentally measured cross-sections.  Contributions from multijet
events with fake leptons are estimated from data.
The total background estimation is $B=2.4\pm0.8$ events. 
The purity of the charge
determination is determined from these simulations and then calibrated in data
using pure $b\bar b$ events.

In $2.7\ifb$ of data CDF finds a total of 45 events, $N_\mathrm{SM}=29$ of
which are reconstructed as 
SM like and $N_\mathrm{XM}=16$ with the exotic charge
value~\cite{CdfNote9939}. The statistical significance of this result is
obtained by considering a large number of pseudo
experiments. Figure~\ref{fig:cdf-charge-slt} shows the expected distribution in
terms of the asymmetry
\beq
A=\frac{1}{D_s}\frac{N_\mathrm{SM}-N_\mathrm{XM}+ B D_b}{N_\mathrm{SM}+N_\mathrm{XM}- B }
\eeq
where $D_s$ and $D_b$ are the dilution factors obtained in the calibration
step for signal and background, respectively.
\begin{figure}[b]
  \centering
  \includegraphics[width=0.85\linewidth]{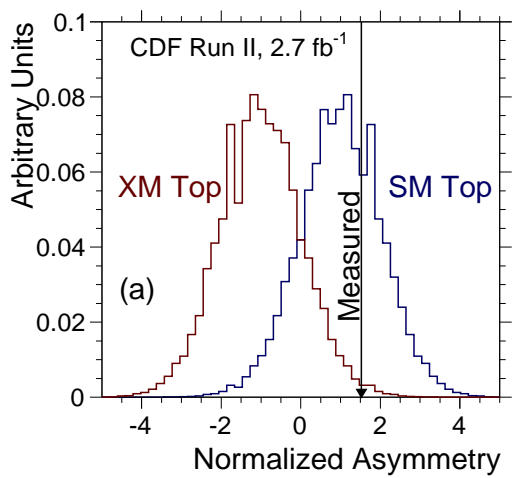}
  \caption{Distribution of the asymmetry observed in a large number of pseudo
    experiments for the SM and the exotic charge case. The measured result is
    indicated by a vertical line~\cite{CdfNote9939}.}
  \label{fig:cdf-charge-slt}
\end{figure}
The SM scenario corresponds to positive values, the exotic model to negative values.
In $69\%$ of the pseudo experiments assuming the SM the asymmetry turned out
to be smaller than the measured value. Only in $0.9\%$ of the pseudo
experiments assuming a fully exotic top quark showed an asymmetry larger
than the measured value.
From these studies the exotic charge model can be excluded at $95\%$C.L.

\subsubsection{Photon Radiation of Top Quarks}
The top quark charge is also expected to define the amount of photon radiation
off the top quark. The measurement of this process is thus a complementary
way of verifying the electrical charge  of the top quark. 
The CDF collaboration has searched for top quark pairs with an additional
photon in $1.9\ifb$ as part of a more general search~\cite{Aaltonen:2009iw}.

For the search of the $t\bar t\gamma$ final state, events are selected that
contain an isolated energetic photon, one isolated 
energetic lepton, large missing transverse
energy and at least three jets. One of the jets is required to be identified
as $b$-jet.

With this selection the dominant backgrounds stem from jets misidentified as
photons and from light-jets misidentified as $b$-jets. 
The former is estimated from data by measuring the photon fake rate in jet
events as function of $\met$. This fake rate is then applied on events passing
the selection without the photon requirement. 
Also the amount of fake $b$-jets is determined from data and then applied to
the events without the  requirement of a $b$-jet identification.
Additional backgrounds including $W\gamma+$jet and diboson events are
estimated using \madgraph+\pythia\ normalised to NLO cross-sections.
\madgraph+\pythia\ passed through the full detector simulation and the
reconstruction is also used to determine the signal selection efficiencies.

In $1.9\ifb$ of data CDF finds a total of 16 events which pass the described
selection. The expectation including SM $t\bar t\gamma$ events is $11.2\pm2.2$
events. Attributing the full difference between the expectation and the data to
the $t\bar t\gamma$ process, CDF measures the cross-section for radiation
off top quark pair events of
\beq
\sigma_{t\bar t\gamma}=0.15\pm 0.08\pb
\eeq
with the quoted uncertainty being dominated by the statistical
uncertainty.
The result is in agreement with the SM expectation of 
$0.080\pm0.011\pb$~\cite{Aaltonen:2009iw}. 

%%% Local Variables: 
%%% mode: latex
%%% TeX-master: "EPJC_TopProperties"
%%% End: 

\subsection{Spin Correlations}
% $Id: SpinCorr.tex,v 1.1.2.12 2011-03-22 17:47:06 wicke Exp $
\label{Sect:SpinCorr}

At the level of the hard interaction the spins between the top and the antitop
quark are correlated in top quark pair production. Because top quarks decay
before they hadronise these correlations are conserved in the weak decay and
thus ``good'' quantum-mechanic observables
\cite{Bigi:1986jk,Barger:1988jj,Stelzer:1995gc,Bernreuther:2001rq,Bernreuther:2010ny}.
The degree of correlation depends on the production and decay processes. Its measurement
thus probes the details of the production mechanism (when assuming a SM weak
decay).  The amount of correlation also depends on the reference axes used to
define the top and anti-top quark spin states. At the Tevatron the
spin-correlations expected in the SM are largest in the so-called off-diagonal
and beam bases. 

Both Tevatron experiments have measured the spin correlation of top quark
pairs in different channels and using different spin bases.

\subsubsection{Dilepton Channel}
In the dilepton channel both experiments use the normalised double differential
cross-section with respect to the angle of flight directions of the two leptons
as their observable. 
The angle of flight,  $\cos\theta_\pm$, is measured with respect to the
spin quantisation axis chosen in the analysis. 
This double differential cross-section depends on the spin correlation
coefficient $\kappa$:
\beq
\frac{1}{\sigma}
\frac{\mathrm{d}^2\sigma}{\mathrm{d}\cos\theta_+\mathrm{d}\cos\theta_-}
=
\frac{1}{4}\left(1+\kappa\cos\theta_+\cos\theta_- \right)
\label{eq:spincorr:diffxsec}
\eeq
In the off-diagonal
and beam bases the SM predicts a correlation coefficient $\kappa$  of about
$0.8$.

\subsubsubsection{CDF}
The CDF analysis of the dilepton channel is based on
$2.8\ifb$~\cite{CdfNote9824}. The event selection requires
two leptons of opposite charge, large missing transverse energy
and at least two energetic jets. A veto is applied 
on Drell-Yan events and on special multijet fake configurations.

The analysis selects the off-diagonal basis for the definition of the 
flight directions. In this basis the spin quantisation axis is defined in the $t\bar t$
centre-of-mass frame. %The flight directions are measured with
It is defined as the direction which deviates from the direction of flight of
the \discretionary{(anti-)}{top}{(anti-)top}  
quark by the angle $\xi$ in clockwise direction. The value of $\xi$
is computed as function of the top quark velocity, $\beta$, and the angle
between the top quark and proton flight direction, $\theta^*$: 
%(in the $t\bar t$ centre-of-mass frame). 
\beq
\tan \xi = \sqrt{1-\beta^2}\,\tan\theta^*
\eeq
The directions of flight with respect to the direction defined
by $\xi$ are measured in the rest frames of the top and the anti-top quark.
The analysis uses the angles obtained for
the leptons ($\cos\theta_+,\cos\theta_-$) and the (anti-)$b$-quarks
($\cos\theta_b,\cos\theta_{\bar b}$).

To reconstruct these angles in an individual event 
a full kinematic reconstruction of the top quark
pair and its decay to $l^+l^-\nu\bar\nu b\bar b$ 
is necessary. In the dilepton channel 
with its two neutrinos this requires the six constraints.
The reconstructed $W$ boson and the top quark masses 
need to be consistent with their nominal values (two constraints each).
In addition the sum of the  transverse momenta of the reconstructed
neutrinos are required to agree with the measured missing transverse energy,
$\vec\met$. 
Due to the quadratic nature of the corresponding equations these constraints
yield up to four solutions for the unmeasured neutrino momenta, $\vec p_\nu$
and $\vec p_{\bar\nu}$.

To further improve the reconstruction a kinematic fit is applied that varies
the reconstructed (anti-)$b$  quark energies within the experimental resolution of
the $b$ jet measurements and the reconstructed sum of transverse neutrino
momenta within the resolution of $\vec\met$. Besides these resolutions the
kinematic fit includes the probability densities for the distribution of the
$p_z$, the $p_T$ and the invariant mass of the top quark pair in its
likelihood. Of the initially up to four solutions for the neutrino momenta, 
$\vec p_\nu$ and  $\vec p_{\bar\nu}$, the one which gives the best likelihood
is  used to compute the flight directions in each event.

Simulations are used to determine the expected outcome of this measurement as
function of the spin correlation. For the $t\bar t$ 
signal simulation \pythia\ is used. Because the \pythia\ samples do not
contain spin correlations between the generated top quarks, the generated
events are weighted proportional to  $1+\kappa\cos\theta_+\cos\theta_-$
according to the generated values of $\cos\theta_+$ and
$\cos\theta_-$. Background of dibosons and  Drell-Yan are simulated using
\pythia\ and \alpgen+\pythia, respectively.
Background due to fake leptons is simulated
from data with a single energetic lepton and jets. 
In these events  one jet is artificially interpreted
as a fake lepton. Both signal and  background templates are smoothened by a
polynomial function, where for the signal template the dependence on $\kappa$
is kept. 

The measured two-dimensional distributions of the reconstructed
($\cos\theta_+,\cos\theta_-$) and ($\cos\theta_b,\cos\theta_{\bar b}$) values
are now compared to the templates to determine the spin correlation coefficient
$\kappa$ using a likelihood fit. The measured value is corrected slightly
according to a calibration performed on ensembles of pseudo experiments. 

In $2.8\ifb$ of dilepton top quark pair events CDF measures the spin correlation
coefficient in the off-diagonal basis to be
\beq
\kappa=0.32^{+0.55}_{-0.78}\quad.
\eeq
The total uncertainty is by far dominated by the statistical error. 
The leading systematics stems from the uncertainty on  
relative contribution of signal and background events. 
At the available statistics this results poses only a very minor constraint on the
real spin correlations.

\subsubsubsection{D\O}
The measurement of the spin correlation by D\O\ uses dilepton events of
up to $4.2\ifb$ of data~\cite{D0Note5950}. Events are selected requiring two
oppositely charged leptons, large missing transverse energy and at least two
jets. Drell-Yan events are vetoed near the $Z$ boson resonance. 

In this study the flight directions in \eq{eq:spincorr:diffxsec} are computed
using the beam basis. The spin quantisation axes for the beam basis are the
directions of the proton and anti-proton in the $t\bar t$ rest frame.
The flight directions with respect to the quantisation axes are measured in
the top and anti-top quark rest frame, respectively.
D\O\ uses the direction of flight angles for the two measured leptons, $\cos\theta_+$ and $\cos\theta_-$.

For the full reconstruction of the top quark pair kinematics D\O\ applies the
Neutrino Weighting Method described in the context of top quark mass
measurement in Sect.~\ref{NWTD0}. For the spin correlation 
measurement the weight, $W$, is
computed for a fixed top quark mass of $175\GeV$, but scanning possible values of 
the flight angles to find its dependence on the product $\cos\theta_+
\cos\theta_-$,  i.e. for each event $W=W(\cos\theta_+ \cos\theta_-)$.
For further analysis the mean, $\mu$, of the weight function in each event is used: 
\beq
\mu = \int x W(x)\,\mathrm{d}x
%\quad\mbox{with}\quad x=\cos\theta_+\cos\theta_-
~~.
\eeq

The distribution of $\mu$ values found in the selected events is now 
compared to templates obtained from simulation. 
Signal templates for a range of $\kappa$ values  are obtained by reweighting
\pythia. Templates for the $Z/\gamma$ events is generated with
\alpgen+\pythia, while for diboson processes \pythia\ is used alone. 
Both background types are normalised to the theoretical cross-sections.
To enhance the bad simulation of the $Z$ boson $p_T$ distribution, a
reweighting of these events has been applied.
The measured spin correlation parameter $\kappa$ is obtained by a binned
likelihood fit of the templates to the measured data.

D\O\ has studied the performance of this method in ensembles of pseudo
experiments using various nominal values of $\kappa$. This calibration is used
in the Feldman-Cousins procedure~\cite{Feldman:1997qc,\refpdg} 
to obtain the final results. 
Using dilepton events D\O\ determines the spin correlation coefficient in the
beam axis 
\beq
\kappa=-0.17^{+0.64}_{-0.53}\quad.
\eeq
The uncertainty is dominated by statistics. The leading systematic uncertainty
is found to come from signal modelling. This includes a dependence on the assumed
top quark mass as well as differences observed when replacing \pythia\ by
\alpgen+\pythia\ or \mcatnlo. The observed spin correlation is in slight
tension with the SM expectation of about $0.8$.

\subsubsection{Lepton plus Jets Channel}
CDF has also determined  top quark spin correlations in the semileptonic decay
channel~\cite{CdfNote10048,CdfNote10211}. 
In $5.3\ifb$ %a total of 725
events  with one energetic lepton, large missing
transverse momentum and at least 4 jet are selected; at least one of the jets
is required to be identified as $b$-quark jet.

This analysis uses the the beam and the helicity bases. 
The spin quantisation axes are defined in the $t\bar t$ rest frame.
For the beam basis the beam direction is used, for the helicity basis
the \discretionary{(anti-)}{top}{(anti-)top}  quark directions are taken.
Then the angles between the flight directions of the top quark decay products, $b\bar b \ell\nu
qq'$, and the quantisation axis 
are reconstructed in the \discretionary{(anti-)}{top}{(anti-)top} quark rest frame. The analysis
considers the lepton direction, $\cos\theta_\ell$, the bottom
quark from the hadronic top quark decay, $\cos\theta_b$, and the down-like
quark, $\cos\theta_d$. 

For the determination of these angles the full event kinematics of the top
quark pair decay is reconstructed using a kinematic fitter which constrains
the top quark mass to $172.5\GeV$.
Only events with a reasonable fit quality are considered in the analysis. 
In the following analysis the distributions of the $\cos$ of the
reconstructed angles are compared to templates. For the reconstructed value of
$\cos\theta_d$ the jet which is closest to the $b$-jet in the $W$ rest frame
is used. This is found to  be the down-type quark in about $60\%$ of the cases.

For the analysis of the angles in the helicity basis, samples of the four
possible helicity combinations of the  top and anti-top quark are generated
with a modified version of Herwig. From these four combinations, equal
helicity and opposite helicity samples are obtained assuming parity and $CP$ conservation.
In the analysis of the angles with respect to the beam basis the signal samples
are obtained by reweighting \pythia\ events according to the generated decay
angles. 
In addition templates for the expected background are produced, including
multijet, $W+$jets and diboson events.

To determine the spin correlation the two dimensional distribution of
$\cos\theta_\ell\cos\theta_d$ vs. $\cos\theta_\ell\cos\theta_b$ is fitted with
the described templates using a binned likelihood fit. 
The contribution of the same spin and the opposite
spin templates are allowed to float freely, the background contribution is
allowed to float but constrained to the expectation within its errors. 
The determined fractional contribution of the opposite spin contribution,
$f_o$, is converted to the spin correlation coefficient using $\kappa=2f_o-1$.

In $5.3\ifb$ CDF determines the spin correlation in the   helicity and in the
beam bases to be

\bea
\kappa_\mathrm{helicity}&=&0.48\pm0.48_\mathrm{stat}\pm0.22_\mathrm{syst}\nonumber\\
\kappa_\mathrm{beam}    &=&0.72\pm0.64_\mathrm{stat}\pm0.26_\mathrm{syst}\quad.
\eea

The measurement is clearly dominated by the limited statistical
uncertainty. Of the systematic uncertainties the uncertainty on the signal
modelling is by far dominating. 

%%% Local Variables: 
%%% mode: latex
%%% TeX-master: "EPJC_TopProperties"
%%% End: 

\subsection{Charge Forward Backward Asymmetry}
% $Id: Asymmetry.tex,v 1.19.2.12 2011-03-22 17:46:36 wicke Exp $

At the Tevatron the initial state of proton anti-proton is not an eigenstate
under charge conjugation. Thus in principle 
also the final state may change under this operation. 
In QCD, however, such a charge asymmetry appears only at next-to-leading
order and arises mainly from interference between contributions symmetric and
anti-symmetric under the exchange of top and anti-top
quarks~\cite{Halzen:1987xd,Nason:1989zy,Beenakker:1990maa,Kuhn:1998kw,Bowen:2005ap}. 

Experimentally, CDF and {D\O} investigated forward backward asymmetries~\cite{d0:2007qb,Aaltonen:2008hc,D0Note6062conf,Aaltonen:2011kc}
\beq 
A_{\mathrm{FB}}=\frac{N_{\mathrm F}-N_{\mathrm B}}{N_{\mathrm F}+N_{\mathrm
  B}}
\eeq
where $N_{\mathrm F}$ and $N_{\mathrm B}$ are the number of events observed in the
forward and backward direction, respectively.
The forward and backward directions are either defined in the
laboratory frame, i.e. according to the sign of the rapidity of the top quark,
$y_t$, or can be defined in the frame where the top quark pair system rests along
the beam axis, i.e. according to the sign of the rapidity difference between
top and anti-top quark, $\Delta y=y_t-y_{\bar t}$. The two different
definitions of forward and backward yield two different asymmetries that are
labelled $A_{\mathrm{FB}}^{p\bar p}$ and $A_{\mathrm{FB}}^{t\bar t}$ according to their rest frame of definition.
In the Standard Model at NLO asymmetries are expected to be  
$0.05$ and $0.08$, respectively~\cite{Antunano:2007da}, but at NNLO 
significant corrections are predicted for the contributions from $t\bar t+X$~\cite{Dittmaier:2007wz}.

The smallness of the asymmetries expected within the Standard Model make them a
sensitive probe for new physics.

\begin{figure*}
  \centering
  \includegraphics[width=0.25\linewidth]{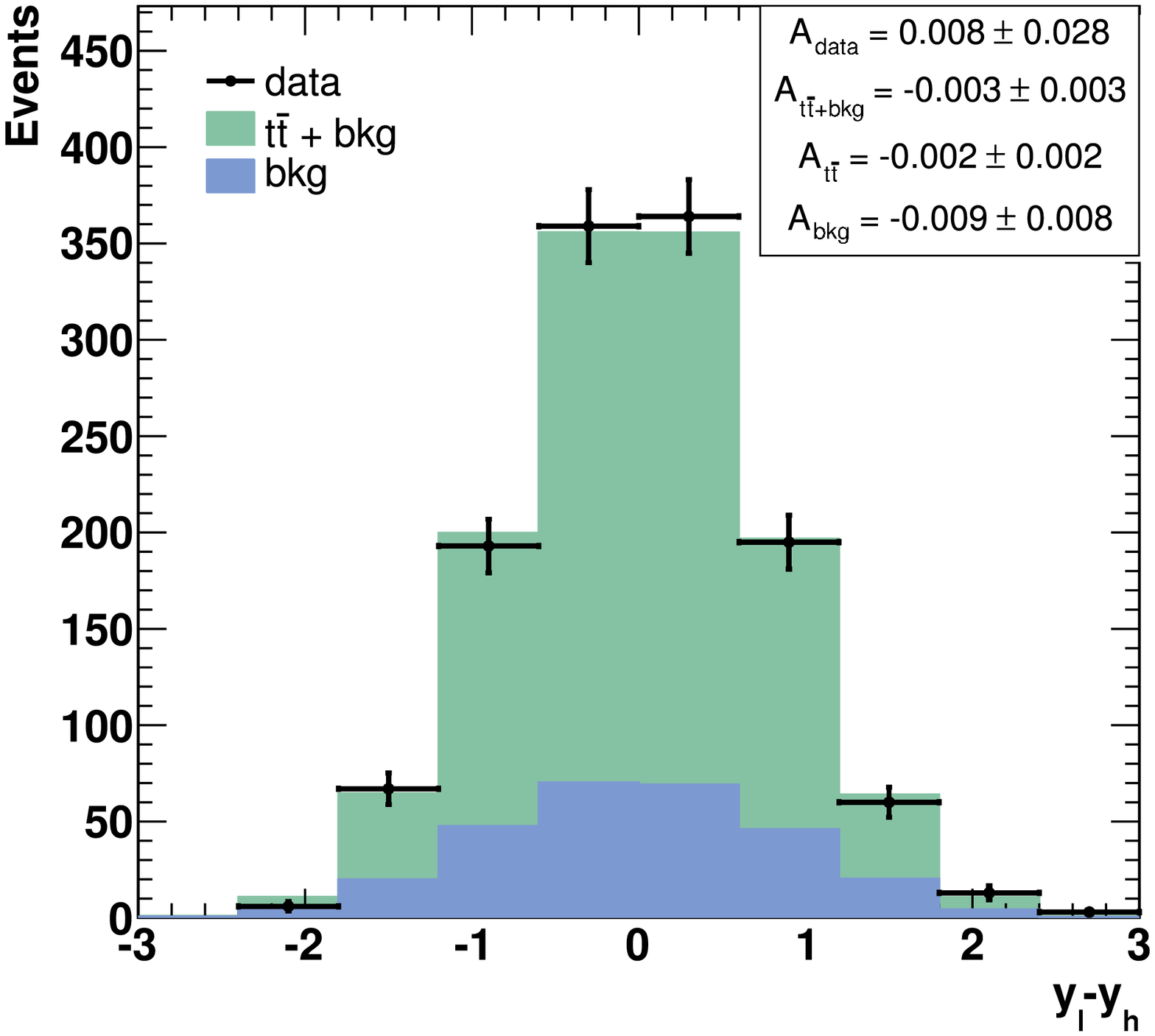}\qquad\quad
  \includegraphics[width=0.25\linewidth]{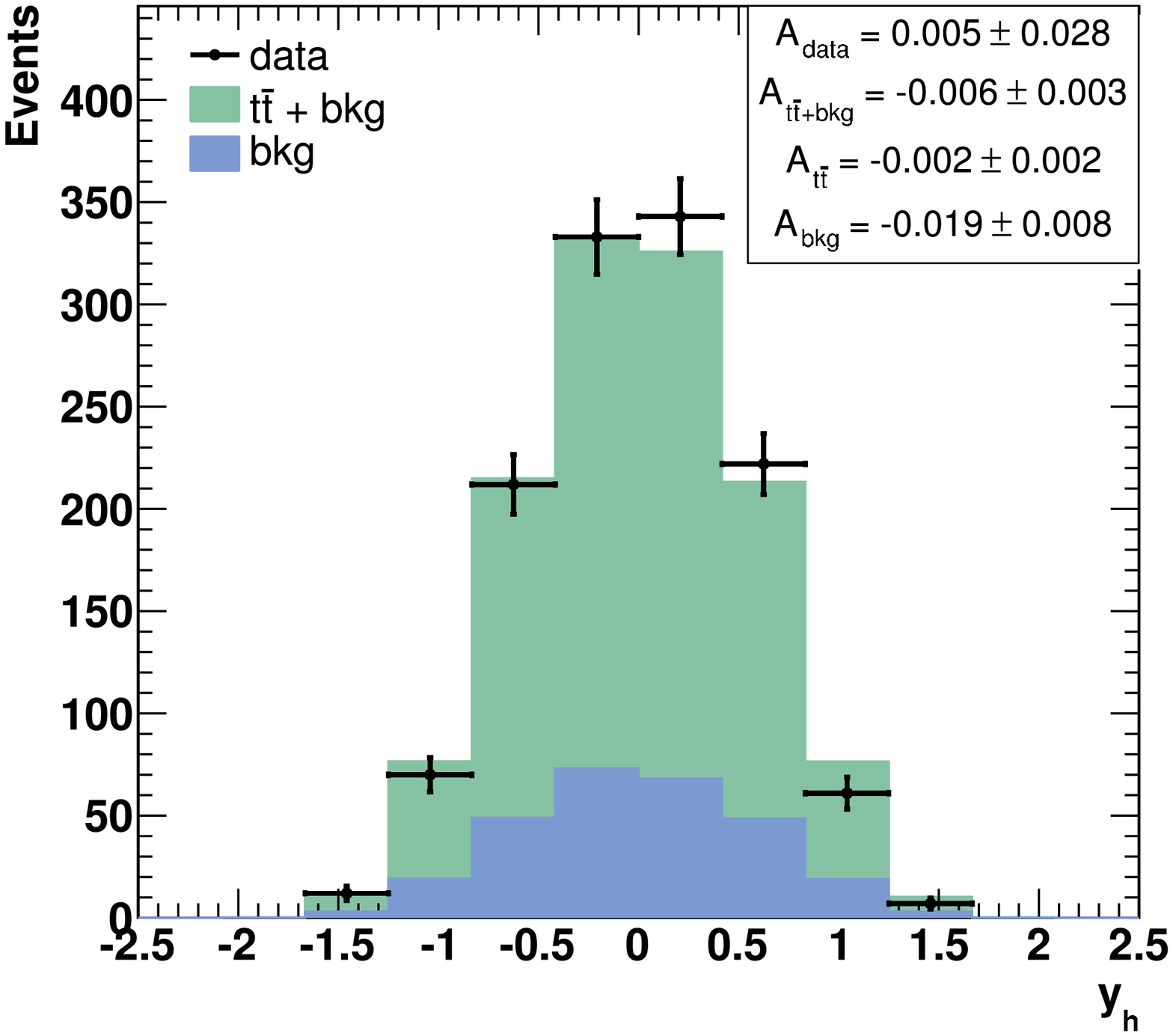}\qquad%\qquad
  \includegraphics[width=0.343\linewidth]{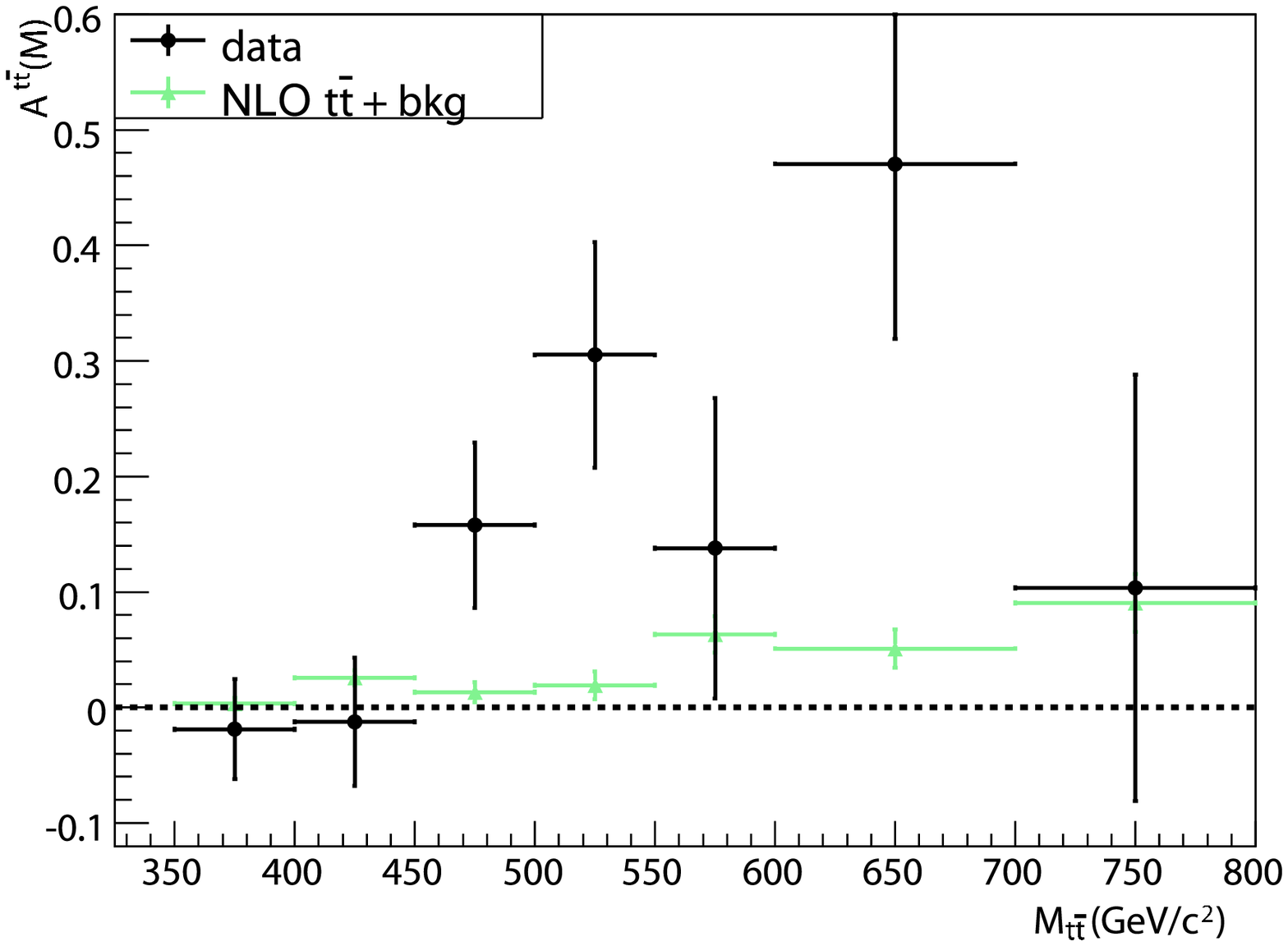}%\qquad
  \caption{Distribution of the top quark rapidity (left) and rapidity
    difference (middle) as measured by CDF in $5.3\ifb$ compared to the
    Standard Model prediction. Right: The functional dependence of the rapidity
    difference of the top quark pair invariant mass, $M_{t\bar t}$~\cite{Aaltonen:2011kc}. }
  \label{fig:cdf-afb}
\end{figure*}
\subsubsection*{CDF}
The CDF collaboration has investigated up to $5.3\ifb$ of data and measures both
charge asymmetries defined above from top quark pairs with semileptonic
decay~\cite{Aaltonen:2008hc,Aaltonen:2011kc}. 
The event selection requires an isolated lepton, missing transverse energy and
at least four hadronic jets, one of which must be identified as
$b$ jet. 

The top and anti-top quark kinematics are reconstructed from the 
jet momenta, the lepton momentum and the missing transverse momentum  using
mass constraints from the $W$ boson and the top quark. 
The reconstructed values of these masses are constrained by the nominal values of $M_W=80.4\GeV$
and $m_t=172.5\GeV$ and $b$ tagged jets are assigned to $b$ quarks only~\cite{Aaltonen:2011kc}. 
Of the possible jet parton assignments the one with the best fit
probability is taken.
The rapidity of the hadronically decayed
\discretionary{(anti-)}{top}{(anti-)top}  
%(anti-)top 
quark, $y_h$, is multiplied by minus the charge, $Q_\ell$, of the
lepton to obtain the top quark rapidity: $y_t=-Q_\ell\, y_h$.
For the $t\bar t$ frame asymmetry, $A_{\mathrm{FB}}^{t\bar t}$
the rapidity difference is computed 
$\Delta y=y_t-y_{\bar t}$ as $\Delta y= Q_\ell\,(y_\ell-y_h) $, with 
$Q_\ell$ and $y_h$ as above and $y_\ell$ being the rapidity of the
leptonically decayed \discretionary{(anti-)}{top}{(anti-)top}   quark.

The NLO Standard Model expectation of top quark pair production 
is done with \mcfm~\cite{Campbell:1999ah} and the next-to-leading order
generator \mcatnlo~\cite{Frixione:2002ik} which contain a small asymmetry.
Leading order signal simulation without asymmetry from \pythia\ is used to
check for any detector or selection asymmetry.
The dominating background events of $W+$jets are simulated with \alpgen+\pythia,
diboson backgrounds and single top quark events are simulated with \pythia\ and
\madevent, respectively. The normalisation of the $W+$jets background 
and contributions from misreconstructed multijet events
are estimated from data. 

The uncorrected rapidity and rapidity difference distributions measured in
data are compared to the expectations in \fig{fig:cdf-afb}. These
distributions differ from the true particle level shape due to acceptance and
reconstruction effects. After background subtraction CDF
derives the particle level distributions 
inverting the acceptance efficiencies and migration probability matrices as
derived from \pythia\ simulation with zero asymmetry using a reduced number of
only four bins. 
%By doing so the unfolding assumes acceptance efficiencies
%generated with \pythia\ to be valid. 
The final asymmetries are computed from these unfolded distributions.

The dominating systematic uncertainties are the background normalisation
and shape. For $A_{\mathrm{FB}}^{p\bar p}$ the
amount of initial and final state radiation contributes significantly,
while for $A_{\mathrm{FB}}^{t\bar t}$ the jet energy scale is the next leading uncertainty.
Further uncertainties from the parton
distribution functions, due to colour reconnection and the MC generator are considered.

The final asymmetries measured in the update to $5.3\ifb$ of CDF data 
are~\cite{Aaltonen:2011kc}:
\bea
A_{\mathrm{FB}}^{p\bar p} &=& 0.150 \pm 0.050_{\mathrm{stat}} \pm 0.024_{\mathrm{syst}} \nonumber\\
A_{\mathrm{FB}}^{t\bar t} &=& 0.158 \pm 0.072_{\mathrm{stat}} \pm 0.017_{\mathrm{syst}}\quad\mbox{.}
\eea
These values are  somewhat larger than the $0.038$ and $0.058$
expected in the Standard Model at NLO, respectively, but agree within two standard deviations. 

CDF also investigated the dependence of the asymmetry on several other
topological and kinematic properties. 
Considering the two ranges of $\Delta y$ yields:
\bea
A_{\mathrm{FB}}^{t\bar t}(|\Delta y|<1.0)&=& 0.03 \pm 0.12 \quad \mbox{(SM: $0.39$)}\nonumber\\
A_{\mathrm{FB}}^{t\bar t}(|\Delta y|\ge 1.0)&=& 0.61 \pm 0.26\quad \mbox{(SM: $0.123$)}\mbox{.}
\eea
Clearly, the deviation from the expectation is driven by the
effect at large rapidity differences.

The functional dependence of the asymmetry on the invariant mass of the top
quark pairs is shown in \fig{fig:cdf-afb}~(right). Separating this result in
two bins yields%
\bea
A_{\mathrm{FB}}^{t\bar t}(M_{t\bar t}\!<\!450\GeV)&=& -0.12 \pm 0.15  \quad \mbox{(SM: $0.04$)}\nonumber\\
A_{\mathrm{FB}}^{t\bar t}(M_{t\bar t}\!\ge\! 450\GeV)&=& +0.48 \pm 0.11 \quad \mbox{(SM: $0.09$)}\mbox{.}\quad
\eea
The events with high invariant mass show a  deviation  $3.4$ standard
deviations from the NLO prediction obtained with~\mcfm.

CDF completes their study by verifying that the asymmetries are consistent
with CP conservation by separately considering events with positively and
negatively charged leptons. The slightly enhanced asymmetry observed in the
inclusive measurement thus seems to stem from effects at large rapidity
difference and high invariant mass of the top quark pair system.

\subsubsection*{D\O}
The D\O\ collaboration has investigated up to $4.3\ifb$ of data to measure
$A_{\mathrm{FB}}^{t\bar t}$ in semileptonic top quark pair events~\cite{d0:2007qb,D0Note6062conf}. The event
selection requires exactly one isolated lepton, missing transverse momentum and at least
four jets, the hardest of which must have $p_T>40\GeV$. At least one of the
jets is required to be identified as $b$ jet with D\O's neural network tagger.

The top quark pair kinematics is reconstructed by fitting the momenta of the
top quark decay products to the measured jet and lepton momenta  and the
missing transverse energy with constraints on the reconstructed $W$ boson
and top quark mass to $80.4\GeV$ and $172.5\GeV$, respectively.
Only the $b$ jets and the remaining three leading jets are used. The possible
jet parton assignments are reduced by assigning identified $b$ jets only to
$b$ quarks. In the final analysis only the assignment with the best fit
probability is used. The rapidity difference with the correct sign is
determined from the rapidities reconstructed of the leptonic and the hadronic
side ($y_\ell$ and $y_h$) and the lepton charge, $Q_\ell$:\, $\Delta y= Q_\ell\,(y_\ell-y_h)$.

To estimate the dominant background of $W+$jets
production a set of observables well described by the simulation is used to
construct a likelihood discriminant that does not depend on $\Delta y$. 
The expected shape of top quark pair signal and the $W+$jets background 
in the distribution of the discriminant and on
the asymmetry is determined from \mcatnlo\ and \alpgen+\pythia\ simulation,
respectively,  passed through D\O\ detector simulation and reconstruction.
The effect of multijets events  on the asymmetry and the discriminant is
determined from data that fail the lepton identification.
Other backgrounds were checked to have negligible effects.

The final reconstructed  asymmetry, $A_{\mathrm{FB}}^{t\bar t}$, in signal events
is determined by maximising the combined likelihood of the observed
discriminant distribution and the distribution of the sign of $\Delta y$ as function of the signal
and background contributions and of the signal asymmetry. 
The dominant systematic uncertainties on the asymmetry are the jet energy
calibration and the
asymmetry reconstructed in $W+$jets events.
% and the modelling of multiple proton interactions. 
All of them are much smaller than the statistical uncertainty.

In $4.3\ifb$ of data D\O\ finds a final observed asymmetry~\cite{D0Note6062conf} of 
\beq
A_{\mathrm{FB}}^{t\bar t~\mathrm{obs}}=0.08 \pm 0.04_\mathrm{stat} \pm 0.01_\mathrm{syst}\quad\mbox{.}
\eeq
To keep the result model independent and in contrast to the CDF results this
number is not corrected for acceptance and resolution effects.
Instead it needs to be compared to a theory prediction for the phase space
region accepted in this analysis which is corrected for dilution effects.  
For NLO QCD and the cuts used in this analysis D\O\ evaluates  $A_{\mathrm{FB}}^{t\bar t}=0.01^{+0.02}_{-0.01}$.  
Thus as for CDF this  result corresponds to an asymmetry that is slightly higher than
expected in NLO QCD, but not by more than two standard deviations.
\begin{figure}[t]
  \centering
  \includegraphics[width=0.9\linewidth]{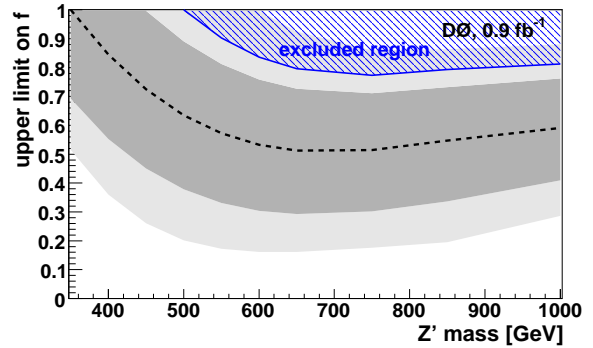}
  \caption{Limits on a possible fraction, $f$, of resonant top quark pair production
    through a $Z'$ boson obtained from the measurement of the forward backward
  asymmetry in D\O~\cite{d0:2007qb}.}
  \label{fig:d0-afb}
\end{figure}

In addition to the QCD expectation in the published analysis~\cite{d0:2007qb} 
D\O\  provides a parameterised procedure 
to compute the asymmetry expected  for an arbitrary model of new
physics. As an example the measurement's sensitivity to top quark pair production
via a heavy neutral boson, $Z'$, with couplings proportional to that of the
Standard Model $Z$ boson is studied. \pythia\ is used to obtain a prediction of
this kind of top quark pair production and due to the parity violating decay yields
large observable asymmetries of $13$ to $35\%$ depending on the assumed $Z'$
boson mass. Limits on the possible fraction of heavy $Z'$ production are
determined as function of the $Z'$ boson mass using the Feldman-Cousins approach.
These limits are shown in \fig{fig:d0-afb} and can be applied to wide $Z'$
resonance by averaging the appropriate mass range.

%%% Local Variables: 
%%% mode: latex
%%% TeX-master: "EPJC_TopProperties"
%%% End: 

\subsection{Differential Cross-Section}
%Mtt, Jiri
Measurements of the differential cross-sections of top quark pair production can be
used to verify the production mechanism assumed in the Standard Model. 
Due to the required unfolding these measurements are especially cumbersome.
The CDF collaboration has measured the differential cross-section with respect
to the invariant top quark pair mass, 
$
\frac{\mathrm{d}\sigma_{t\bar t}}{\mathrm{d}M_{t\bar t}}(M_{t\bar t})
$, using $2.7\ifb$ of data~\cite{Aaltonen:2009iz}. 
The event selection requests a lepton with high
transverse momentum, large missing transverse momentum and at least four
jets. At least one of the jets needs to be identified as $b$-jet.

The invariant mass of the top quark pairs is reconstructed from the four-momenta of 
the four leading jets in $p_T$, the four momentum of the lepton and the missing
transverse energy. The $z$-component of the neutrino is not
reconstructed  but used as if it was zero~\cite{cdf-dsigmadmtt-url}.

The dominating background in this selection stems from $W+$jets production. Its
kinematics simulated with \alpgen+\pythia\ correcting heavy flavour
contribution for differences between data and Monte Carlo. 
The required normalisation is measured in data before applying the $b$-jet 
requirement~\cite{Acosta:2005am}.  
Multijet background is extrapolated from data with low missing transverse
momentum. The smaller backgrounds of diboson, $Z+$jets and single top quark is
fully taken from simulation using \pythia, \alpgen+\pythia\ and \madgraph, respectively.
All simulated events are passed through the CDF detector simulation and
reconstruction.

To obtain the differential cross-section from the background subtracted 
distribution of observed $M_{t\bar t}$ values, acceptance effects and smearing
effects from the reconstruction need to be corrected for.  The required acceptance correction is computed
from signal simulation with \pythia. Factors to correct for
differences between data and Monte Carlo observed in control samples are
applied for the lepton identification and $b$-jet identification rates.
The distortions of the reconstructed distribution are unfolded using using the
singular value decomposition~\cite{Hocker:1995kb} of the response matrix that
is obtained from simulations.

Relevant systematic uncertainties arise from the background normalisation, the
acceptance, parton density distributions, the used Monte Carlo generator and
jet energy scale. The relative contributions of the uncertainties strongly
depend on  $M_{t\bar t}$. To reduce the uncertainty on the jet energy scale
an in-situ calibration of the jet energy scale is performed. This uses the
invariant dijet mass reconstructed from the two non-$b$-tagged jets most
consistent with $M_W$.

The differential cross-section obtained in $2.7\ifb$ of data using the
semileptonic decay mode is shown in \fig{fig:cdf-diffxsec-dist}~\cite{Aaltonen:2009iz}. The consistency
with the Standard Model expectation is computed using Anderson-Darling
statistics~\cite{AndersonDarling:1952}. The observed $p$-value is 0.28,
showing good agreement with the Standard Model.
\begin{figure}
  \centering
  \includegraphics[width=0.45\textwidth]{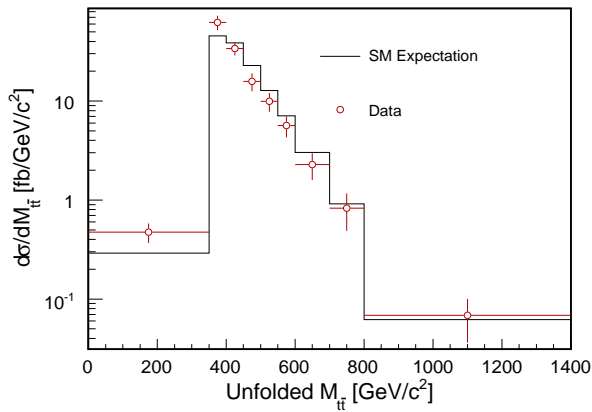}
  \caption{Differential top quark pair production cross-section measured by CDF in
    $2.7\ifb$ of data using the semileptonic decay mode. Indicated are the
    total uncertainties for each bin, excluding the overall luminosity
    uncertainty of $6\%$~\cite{Aaltonen:2009iz}.}
  \label{fig:cdf-diffxsec-dist}
\end{figure}
\begin{figure}
  \centering
  \includegraphics[width=0.45\textwidth]{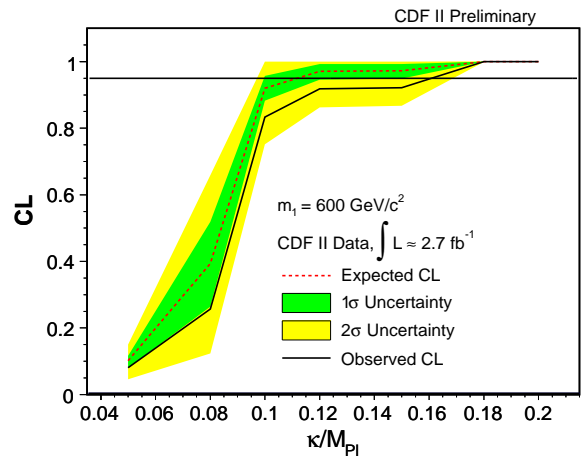}
  \caption{Expected and observed limit on
    $\kappa/M_{\mathrm{Pl}}$ in a Randall-Sundrum model obtained from ${\mathrm{d}\sigma_{t\bar t}}/{\mathrm{d}M_{t\bar t}}$~\cite{Aaltonen:2009iz,cdf-dsigmadmtt-url}.}
  \label{fig:cdf-diffxsec-rslimit}
\end{figure}

Finding no evidence for physics beyond the Standard Model limits on gravitons
in a Randall-Sundrum model~\cite{Randall:1999ee} decaying to top quarks are
set using the CL$_s$ method~\cite{Junk:1999kv,Read-in-James:2000et}. 
Signal is modelled with \madgraph+\pythia\
assuming a first resonance with a mass of $600\GeV$. 
The Anderson-Darling statistics is used as test statistics in the CL$_s$
method. 
For the ratio of the warping parameter over the Planck mass CDF finds
$\kappa/M_{\mathrm{Pl}}<0.16$ at $95\%$~C.L, see \fig{fig:cdf-diffxsec-rslimit}.

The invariant top quark pair mass was used in further analyses by both CDF and D\O\
to search for new physics. These results are described in Section~\ref{sect:BSM:Mtt}.

%%% Local Variables: 
%%% mode: latex
%%% TeX-master: "EPJC_TopProperties"
%%% End: 

\subsection{Gluon Production vs. Quark Production}
Top pair production at the Tevatron with $\sqrt{s}=1.96\TeV$ takes place
either through quark anti-quark annihilation or through gluon fusion. The
former is expected to dominate with the gluon fusion contributing about
$15\%$. Due to the large uncertainties of the large-$x$ gluon density in the
proton  the exact size of the gluon contribution is rather uncertain~\cite{Kidonakis:2003qe,Cacciari:2003fi,Moch:2008ai}. 

Two properties of the two production processes allow to separate them and to
measure their relative contributions. 
Close to threshold the spin states of the gluon fusion are $J=0$,~$J_z=0$,
while the $q\bar q$ annihilation yields  $J=1$,~$J_z=\pm
1$~\cite{Arens:1992fg}. This yields 
angular correlations between  the charged leptons in the dilepton channel.
Alternatively one can exploit the difference in the amount of gluon radiation
from quarks and gluons: The gluon fusion processes are expected to contain
more particles from initial state radiation.
CDF has used both features to measure the gluon fraction of top quark pair
production. 

\subsubsection{Angular Correlation Methods}
\subsubsubsection{Dilepton Channel}
% The two processes of production have different spin state. 
% The gluon fusion leads to $J=0$,~$J_z=0$, while the $q\bar q$ annihilation
% yields  $J=1$,~$J_z=\pm 1$ with the $z$ axis along the direction of initial
% state parton direction, i.e. approximately the beam directions.
% The difference in spin states is observable 
% in azimuthal correlations of the charged leptons in the top pair dilepton
% final state~\cite{Arens:1992fg}.

CDF has investigated $2.0\ifb$  of data with an event signature of top quark pair
dilepton events~\cite{CdfNote9432}. 
The selection requires two oppositely charged leptons, at
least one of which must be isolated, and at least two jets. The scalar sum of
the lepton and jets transverse energies must exceed $200\GeV$. Additional cuts
are placed to reject cosmic particles, leptons from photon conversion 
and $Z$ boson events.

The azimuthal angle between the two leptons is measured in
each event. Then a template method is used to measure the fractional
contribution of the different production mechanisms. 
The expected behaviour of signal events is simulated using \herwig\ with the
top quark mass set to $m_t=175\GeV$ and the CTEQ5L parton distribution
function. \pythia\ and \mcatnlo\  are used in systematic studies.
Backgrounds are dominated by diboson production and $Z$ boson events with
tauonic decay. These are simulated by \pythia. 
In addition the background from events with only one
true lepton and a jet misidentified as lepton are described using data.
All simulated events are passed
through the full CDF detector simulation and reconstruction. 
 
The angular distributions obtained for events 
produced by $q\bar q$ annihilation  and $gg$  fusion and the sum of
backgrounds are separately fitted with smooth functions, that then serve as
signal and background templates. 

\begin{figure}[t]
  \centering
   \includegraphics[width=0.7\linewidth]{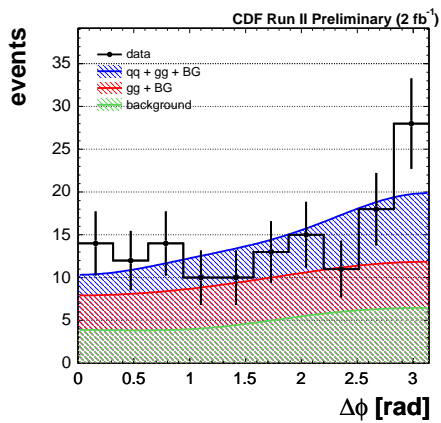}
  \caption{%Left: 
    Distribution of azimuthal angle between the two leptons, $\Delta
    \phi$, observed in $2.0\ifb$ of CDF data, compared to the best fit
    template curves%
%    Right: Feldman-Cousins band of measured vs. true gluon
%    fusion fraction. The value measured in data is indicated a vertical 
%    line.
~\cite{CdfNote9432}. 
   }
  \label{fig:cdf-fgg}
\end{figure}
The measured fraction of top quark pairs produced through gluon fusion is obtained from an
unbinned likelihood fit of these templates to the observed data, 
c.f.~\fig{fig:cdf-fgg}~(left).
Systematic uncertainties include  uncertainties on the template shapes, the
acceptance differences between $q\bar q$ annihilation and gluon fusion,
the used matrix element, initial and final state radiation and PDF
uncertainties. The uncertainties are determined as function of the nominal
gluon production fraction and several of them may contribute up to $10\%$. 
All systematic uncertainties are included in the determination of the
Feldman-Cousins band, see~\fig{fig:cdf-fgg}~(right), which is used to obtain
the final result with errors.

In the investigated $2.0\ifb$ of dilepton events CDF obtains a gluon fusion 
fraction of $0.53\pm0.37$~\cite{CdfNote9432}. The total uncertainty is dominated by
statistical uncertainties and is not yet able to restrict the theoretical
uncertainties on the gluon fusion production.

\subsubsubsection{Lepton plus Jets Channel} %:2007kq,
Angular correlations are also used by CDF in an analysis of $0.96\ifb$ with
lepton plus jets 
events~\cite{Abulencia:2008su}. %Phys. Rev. D 79, 031101(R) (2009) 
The events are required to contain one energetic lepton, large missing
transverse energy and at least four jets. One of the jets is required to be
identified as $b$ jet through the presence of a secondary vertex.

In each event the decay chain of the top quark pair decay is reconstructed
from the four leading jets 
using a kinematic fit with constraints on the $W$ boson and the top quark mass,
c.f. Section~\ref{CDF-template-ljets-portion}.
Only jet parton assignments which associate the tagged $b$ jet(s) with a
$b$ quark are considered. The one with the best fit quality is used for
further analysis.

From the reconstructed top quark pair decay, eight observables are used to
feed a neural network that is trained to distinguish between $q\bar q$ and $gg$
production. The observables are the (cosine of the) angle between the top
quark momentum and the beam direction of the incoming proton, the top quark
velocity and the (cosines of the) 
six angles of the top quark decay product as defined in the
off-diagonal spin basis, c.f. Section~\ref{Sect:SpinCorr}.
The first two observables are contributing about one third of the total
sensitivity, each. The six decay angles yield the remaining third.

The neural network is trained separately for events with one and with more
than one identified $b$ jet.
Simulation of the two signal production
processes is done with \herwig, the dominant
background of $W+$jets is generated with \alpgen+\herwig.
The generated events are passed through the full detector simulation and
reconstruction chain of CDF. 
To obtain the  measured $gg$ production fraction, $f_{gg}$,
templates for the neural network output are constructed from
the simulation as function of $f_{gg}$ and a
likelihood to observe the measured data as function of $f_{gg}$ is maximised.
The Feldman-Cousins approach~\cite{Feldman:1997qc,\refpdg} 
is applied to restrict the final result to the
physically allowed range.

To determine the systematic uncertainties various pseudo experiments with
systematically varied signal and/or background templates are studied. 
The deviation of $f_{gg}$ obtained in these samples from the standard template
result is considered as systematic uncertainty. 
The dominant uncertainty is found to stem from background shape and
composition as well as from the differences between leading and
next-to-leading order simulation of the signal.

In $0.96\ifb$ of 
lepton plus jets 
events CDF is able to set limits on the $gg$ production fraction
of $f_{gg}<0.61$ at $95\%$C.L.~\cite{Abulencia:2008su}.
This method yields independent information, but is not as sensitive as the
following method.

\subsubsection{Soft Track Method}
Another method that CDF applies to measure the fraction of top quark pair
production through gluon fusion relies on differences that occur because gluons
have a higher probability to radiate than quarks~\cite{:2007kq}. The
analysis is based on top quark pairs with semileptonic decays in $0.96\ifb$ of CDF data.

As sensitive observable the average number
 of soft tracks per event, $\left<N_\mathrm{trk}\right>$, with
$0.9\GeV<p_T<2.9\GeV$ in the central detector region $|\eta|\le 1.1$ is used.
In simulation this is shown to have linear relation to the average number of
gluons, $\left<N_g\right>$, 
in the hard process. This relation is calibrated on two samples
with different gluon content: $W+0$jets events for low gluon content and dijet
for high gluon content. $W+1$jet events are used as cross-check.

The dijet event sample for calibration is selected requiring a leading jet with
transverse momentum between $80$ and $100\GeV$ and a second recoiling jet with 
$|\Delta\phi|\ge 2.53$. Vetoes are applied on lepton 
candidates and missing transverse energy. The $W+$jets sample is selected
requiring an isolated lepton, large missing transverse energy. For the signal
top quark pair sample in addition at least four jets are required. 
At least one of these jets  must be identified as $b$-jet. In the $W+$jets
and top  quark pair samples vetoes on additional lepton candidates and on leptons
consistent with photon conversion or cosmic rays are applied. 

After calibration of the relation between $\left<N_\mathrm{trk}\right>$ and
$\left<N_g\right>$, the fraction of events with a high gluon content, $f_g$, is
determined using a binned likelihood fit. The fit result is corrected according to the
expected background contribution. 
%This background expectation is computed from 
%average number of events with high gluon content in background,
%$f_g^\mathrm{bkg}$. 
The fraction of events with high gluon content in the background,
$f_g^\mathrm{bkg}$, is extrapolated from the  $W+$jets
sample with up to three jets to the four or 
more jet sample. The expected amount of background in the selected signal
sample is determined following the neural network based method
in~\cite{Acosta:2005am}. 
The obtained  high gluon content in top quark pair events, $f_g^{t\bar t}$, in a
last step is corrected for the differences in acceptance between the
gluon fusion and the $q\bar q$ annihilation processes.

The systematic uncertainties of this measurement are dominated by
uncertainties of the calibration procedure and were determined by varying the
corresponding parameters in the analysis. 

In the dataset of $0.96\ifb$ CDF determines  a gluon fusion 
fraction of top quark pair production in semileptonic events of
$%\frac{\sigma(gg\rightarrow t\bar t)}{\sigma(p\bar p\rightarrow t\bar t)}=
0.07 \pm 0.14_\mathrm{stat} \pm 0.07_\mathrm{syst}$~\cite{:2007kq}. This number corresponds
to an upper limit of $0.33$ at $95\%$~C.L, well in agreement with the Standard
Model expectations. Also this measurement is statistically limited.
A combination of this result with the  result of the 
Angular Correlation Method described in the
previous section yields an about $10\%$ improvement on the upper limit~\cite{Abulencia:2008su}.

%%% Local Variables: 
%%% mode: latex
%%% TeX-master: "EPJC_TopProperties"
%%% End: 

\subsection{Top Quark Width and Lifetime}
The top quark width and its lifetime are related by Heisenberg uncertainty
principle. In the Standard Model the top quark  width is expected to be $1.34\GeV$
corresponding to a very short lifetime of about $5\cdot 10^{-25}\,\mathrm{s}$. 
Experimentally, these predictions have been
challenged for deviations in very different analyses. CDF constrains the
lifetime from the distribution of reconstructed top quark  mass
values and from the distribution of lepton track
impact parameters. D\O\ combines the measured $t\rightarrow Wb$ branching
fraction and the single top $t$-channel cross-section to measure the top quark width.

\subsubsection{Top Quark Mass Distribution}
The limit on the top quark width was obtained by CDF  from the the 
distribution of reconstructed top quark mass
values from top quark  pairs decaying to lepton plus jets in
up to $4.3\ifb$  of data%\cite{CdfNote8953}
~\cite{Aaltonen:2008ir,CdfNote10035}. After selecting events with one lepton,
large missing transverse momentum and at least four jets. One of the jets is
required to be identified as $b$-jet. 

In these events the top quark  mass is reconstructed using a kinematic fit that
determines the four-momenta of the top quark  decay productions from the measured jet
and lepton momenta and the transverse missing energy. The fit uses constraints
that force the $W$ boson decay products to build the $W$ boson mass within the
width of the $W$ boson and the reconstructed top and anti-top quark masses to be
equal within the  top quark width. In the ambiguous
association of jets to partons identified $b$-jets  are only associated to
$b$-quarks. Of the remaining associations and the two solutions for the
neutrino $z$-momentum, the one with the best $\chi^2$ is used.
It was checked that the use of
the constraint of the equality of the top quark masses 
width does not destroy the sensitivity to the true width.

To find the measured value of the top quark width the distribution of top quark
masses reconstructed with the best association in each event is compared to
parametrised templates with varying nominal width. Templates for top quark  pair signal events 
were generated using \pythia\ with $m_t=172.5\GeV$~\cite{CdfNote10035}.
Background contributions of $W+$jets
% \em and \em multijet contributions  % only ~\cite{Aaltonen:2008ir}.
are modelled %with $W+$jets events from 
\alpgen+\pythia.
Multijet contributions from data with non-isolated leptons. %only update ~\cite{CdfNote10035} 
Single top quark  and diboson events are simulated with \madgraph. The template
distributions for discrete values of the nominal top quark width are
parametrised to obtain smooth template functions that can now be interpreted
as probability densities. The measured top quark width %, $\Gamma_t^\mathrm{fit}$, 
is determined in an unbinned likelihood fit. 

Recently, in addition to the reconstructed top quark mass, the invariant mass of the
jets assigned to the hadronic $W$ decay is considered as an observable. This
allows a simultaneous fit of the top quark width and the jet energy scale~\cite{CdfNote10035}.
%DW: But then again it isn't really used!!

The Feldman-Cousins approach~\cite{Feldman:1997qc,\refpdg}  is used
to determine the final result excluding the unphysical values of negative
widths that may occur in the fit. The  jet resolution followed by colour
reconnection effects  yield the biggest
single contribution to the systematic uncertainties.   are
propagated to the final Feldman-Cousins band by convoluting their effects with
the fitted width function. 

Including all systematics this CDF analysis of $4.3\ifb$ yields an upper limit
of the top quark width $\Gamma_t<7.5\GeV$ at $95\%$ C.L. which corresponds to 
$\tau_t>8.7\cdot10^{-26}\,\mbox{s}$. At $68\%$C.L. the top
quark width is determined as $0.4\GeV<\Gamma_t<4.4\GeV$~\cite{CdfNote10035}.

\subsubsection{Lepton Impact Parameter}
The limit from the lepton track impact parameter distribution was obtained by CDF
using lepton plus jets events in $318\ipb$ of data~\cite{CdfNote8104}. Events
are selected requiring one isolated lepton, missing transverse energy and at
least three jets. At least one of the jets has to be identified as $b$-jet.
%Vetoes on additional leptons, events consistent with photon
%conversion and misreconstructed $Z\rightarrow\ell\ell$ are applied. 
The lepton track needs to be reconstructed with at least three $R$-$\phi$
positions in the CDF silicon tracker.

The lepton impact parameter, $d_0$, chosen as observable in this measurement is defined
as the smallest distance between the collision point and the lepton track in 
the transverse projection. The collision point is computed as the position of the beam
line in the transverse plane at the reconstructed $z$ position of
the primary vertex.

The distribution of  lepton impact parameters expected in an ideal detector 
for various top quark  lifetimes is simulated with \pythia. The resolution of the CDF detector
is measured in Drell-Yan data near the $Z$ boson resonance and used to derive
the templates for the real detector expectation.  The dominant backgrounds
like $W+$jets consist of prompt leptons, which are described by the zero
lifetime template.  But the distribution of multijet events,  backgrounds with $\tau$ leptons and
electrons from photon conversions need to be modelled. Multijets and electron
conversions are modelled from control samples in data. Backgrounds with $\tau$ leptons
are modelled using \herwig.

From these templates a likelihood as function of  $c\tau_t$ is built. The
maximal likelihood is obtained from the $0\um$ template.
Systematic uncertainties on the
signal and background systematics are computed with correspondingly varied
templates and are dominated from the uncertainty on the detector resolution
for prompt leptons. 
The Feldman-Cousins approach is used to determine the observed limit of
$c\tau_t<52.5\um$ at $95\%$~C.L.

\subsubsection{Branching fraction and  single top cross-section}
The total width of the top quark can be written as the ratio of the partial
width for the decay $t\rightarrow Wb$ and the corresponding branching
fraction: 
\beq
\Gamma_t=\Gamma(t\rightarrow Wb)/{\cal B}(t\rightarrow Wb)\quad\mbox{.}
\eeq

D\O\ measures the top quark width by relating the partial width and branching
fraction to the results of two independent analyses~\cite{Abazov:2010tm}.
The branching fraction in the denominator is equal to the branching fraction
ratio, $R_b$, measured from top quark pairs (c.f. Section~\ref{sect:Rb}). This
assumes that the top quark always decays to a $W$-boson plus a quark.  
The partial width is derived from the single top
production cross-section which is proportional to $\Gamma(t\rightarrow Wb)$. 
When considering only the $t$-channel cross-section, $\sigma_{t\mathrm{-channel}}$, this proportionality is
valid also in the presence of anomalous couplings in the $tWb$-vertex.
Thus the partial width is determined as
\beq
\Gamma(t\rightarrow Wb)=\sigma_{t\mathrm{-channel}}
  \,\frac{\Gamma^\mathrm{SM}(t\rightarrow   Wb)}{\sigma^\mathrm{SM}_{t\mathrm{-channel}}}\quad\mbox{,}
\eeq
where the superscript SM indicates the Standard Model expectations. 
As the total width is by definition larger than the partial width, a lower
bound of $\Gamma_t>1.21\GeV$ at 95\%~C.L. can be set from the $t$-channel production
cross-section~\cite{Abazov:2009pa} alone. 
In the combination of the partial width with the branching fraction
systematics are classified and treated 
as either fully correlated or uncorrelated. They
are dominated by uncertainties in the background description and the
description of the $b$-jet identification in the input analyses.

With the combination of the $t$-channel production cross-section measured in
$2.3\ifb$~\cite{Abazov:2009pa} and the branching fraction ratio measured in
$1\ifb$~\cite{Abazov:2008yn} D\O\ obtains a top quark width of
$\Gamma_t=1.99^{+0.69}_{-0.55}\GeV$ corresponding to a top quark lifetime of 
$\tau_t=(3.3^{+1.3}_{-0.9})\cdot 10^{-25}\,\mathrm{s}$~\cite{Abazov:2010tm}.

%%% Local Variables: 
%%% mode: latex
%%% TeX-master: "EPJC_TopProperties"
%%% End: 

%\subsection{Spin Correlations}
\subsection{Outlook to LHC}
Within the SM the interaction properties of the top quark are
fully defined once the top quark mass is known. The 
verification of the predicted properties establishes the top quark as the
particle expected in the SM. Huge progress has been made in the last years in
experimentally verifying the expectations. Measurements of the $W$ boson
helicity, the CKM element $V_{tb}$ and searches for FCNC confirm the expected
weak interaction properties. The electric charge has been challenged and 
% exotic topological value and gamma radiation.
the
strong interaction properties have been verified in differential cross-section
measurements, in a verification of the contribution of gluon fusion processes
and also in the forward backward charge asymmetry.  First tests of the spin
structure and determinations of the top
quark width complete the current picture. 
All current results are compatible with the SM expectations, a  single
deviation of $3.4\sigma$ 
%standard deviations 
appears in the charge asymmetry at high
$M_{t\bar t}$.

Despite the great progress, the precision of the experimental knowledge of
interaction properties of the top quark is still limited. Many results
have not yet reached a precision of $10\%$. 
As a consequence of
the experimental progress even the more precise results, like the $W$ boson
helicity and the measurement of $V_{tb}$, are statistically limited.

The near future will see updates of the results with the yet
unanalysed Tevatron data to a total $10\ifb$. In addition the LHC schedule envisions to collect
$1\ifb$ of proton-proton collisions at $\sqrt{s}=7\TeV$. % (or slightly above).
The LHC experiments will then have about twice as many top quark
pairs as the Tevatron and even three times as many single top quarks. 
For the top quark pairs the production through gluon fusion and for the
single top quark the $t$-channel production will dominate. 
When systematic uncertainties can be controlled similarly well as at the Tevatron, the LHC results on this dataset will still be limited by statistics.

Only the update of the LHC to the design centre-of-mass energy of $14\TeV$
will provide sufficiently many top quarks to enter an area of precision
measurements in the verification of top quark interactions.

An important exception to these statements is the measurement of the
forward-backward asymmetry, which cannot be measured in $pp$ collisions. 
Instead related asymmetries need to be
studied~\cite{Kuhn:1998jr,Kuhn:1998kw,Antunano:2007da,Wang:2010du,Xiao:2011kp}. 
These measurements will probably require large luminosity as only quark
annihilation diagrams contribute to the asymmetries.

%
% ATLAS has investigated the prospects for collisions at $\sqrt{s}\!=\!14\TeV$ and a
% luminosity of $1\ifb$ for various measurements~\cite{Aad:2009wy}. 
% For this reference scenario the expected results surpass the current \tevatron\ 
% results significantly. The $W$ boson helicity reaches uncertainties of 0.03 and 0.05 for $f_+$ and $f_0$,
% respectively; the branching fractions
% of FCNC can be limited to below $10^{-2}$ and $10^{-3}$ for the $tZq$ and the
% $t\gamma q$ processes, respectively.
% CMS has investigated a scenario of $10\ifb$ and
% $\sqrt{s}=14\TeV$~\cite{Ball:2007zza}. At this reference point the top quark
% spin correlations are accessible and can be extracted with a precision of $20-30\%$.

%%% Local Variables: 
%%% mode: latex
%%% TeX-master: "EPJC_TopProperties"
%%% End: 

%\cleardoublepage
\section[New Particles in Top Quark Events]{New Particles in Top Quark Events}
\label{sect:BSM}\label{sec:newPhysics}

The phenomenology of the top quark may also be altered by particles that are
not expected within the Standard Model, but in one of 
the many models of new physics. Such particles beyond the Standard Model may occur in the top
quark production or its decay, depending on the specific model or its
parameters. Some models of new physics also contain new particles with signatures that are very
similar to the Standard Model top quark. The Tevatron experiments have checked for all
these different extensions of the Standard Model in the top quark sector.

This subsection will actually start with a process that is expected in the
Standard Model though at very low rate: associated Higgs production. In some models
of new physics this process is expected to be enhanced.
Then particles beyond the Standard Model in the top quark decay 
will be discussed.
Finally, searches for the production of particles that look like the top quark
but are not are described.

\subsection[Associated  Higgs Boson Production, $ttH$]{\boldmath Associated  Higgs Boson Production, $ttH$}
Top quark pair production may be associated by the production of a Higgs boson. For
parameters where Higgs bosons dominantly decay to bottom quark pairs, i.e. low
Higgs masses, this associated production is a possibility to measure the top
quark Yukawa coupling. 
While the corresponding cross-section in the Standard Model is too low to allow a Higgs
discovery in this channel alone, it  still contributes to the combination of
the Standard Model Higgs searches. In  some models including new physics an enhancement of
$t\bar t H$ production is expected~\cite{Stange:1993td,Feng:2003uv,AguilarSaavedra:2006gw}.

\subsubsection*{D\O}
D\O\ performed an analysis searching for associated Higgs production in events
with a lepton ($e$ or $\mu$) missing transverse energy and at least four jets~\cite{d0note5739}. 
The analysis uses the scalar sum of transverse momenta, $H_T$, the number of
jets and the number of jets identified as $b$-jets to discriminate the Standard Model
backgrounds and top pair production containing no Higgs from the signal.

Signal events are simulated using \pythia. For $t\bar t$ production 
pure \pythia\ simulation was compared to  \alpgen+\pythia\
simulation. Due to the difference between the
two simulations a $50\%$ uncertainty was assigned to the contribution
of $t\bar t b\bar b$ through QCD processes. Background from $W+$jets events is
simulated with  \alpgen+\pythia\
and normalised to data. Multijet background was completely estimated from data.
Smaller backgrounds are taken from simulation normalised to NLO
cross-sections. 

In the investigation of $2.1\ifb$ 
the observed data agree with the Standard Model expectations  within statistical
and systematic uncertainties~\cite{d0note5739}. To compute limits signal and background
contributions are fitted to the data for a background only assumption 
and for a signal plus background assumption. 
Limits on $\sigma(t\bar tH)\cdot{\cal B}(H\rightarrow b\bar b)$ are then derived
using the CL$_s$ method~\cite{Junk:1999kv,Read-in-James:2000et} for Higgs masses between $105$ and $155\GeV$.

For $M_H=115\GeV$ the cross-section limit corresponds to about 60 times the Standard Model value.
While this allows to exclude unexpectedly large Higgs boson production in association with the top quark, 
its contribution to the Standard Model Higgs search remains small.

%%% Local Variables: 
%%% mode: latex
%%% TeX-master: "EPJC_TopProperties"
%%% End: 

\subsection{Charged Higgs Boson}
\label{sec:ChargedHiggs}
\begin{figure}[b]
  \centering
  \includegraphics[width=0.49\textwidth]{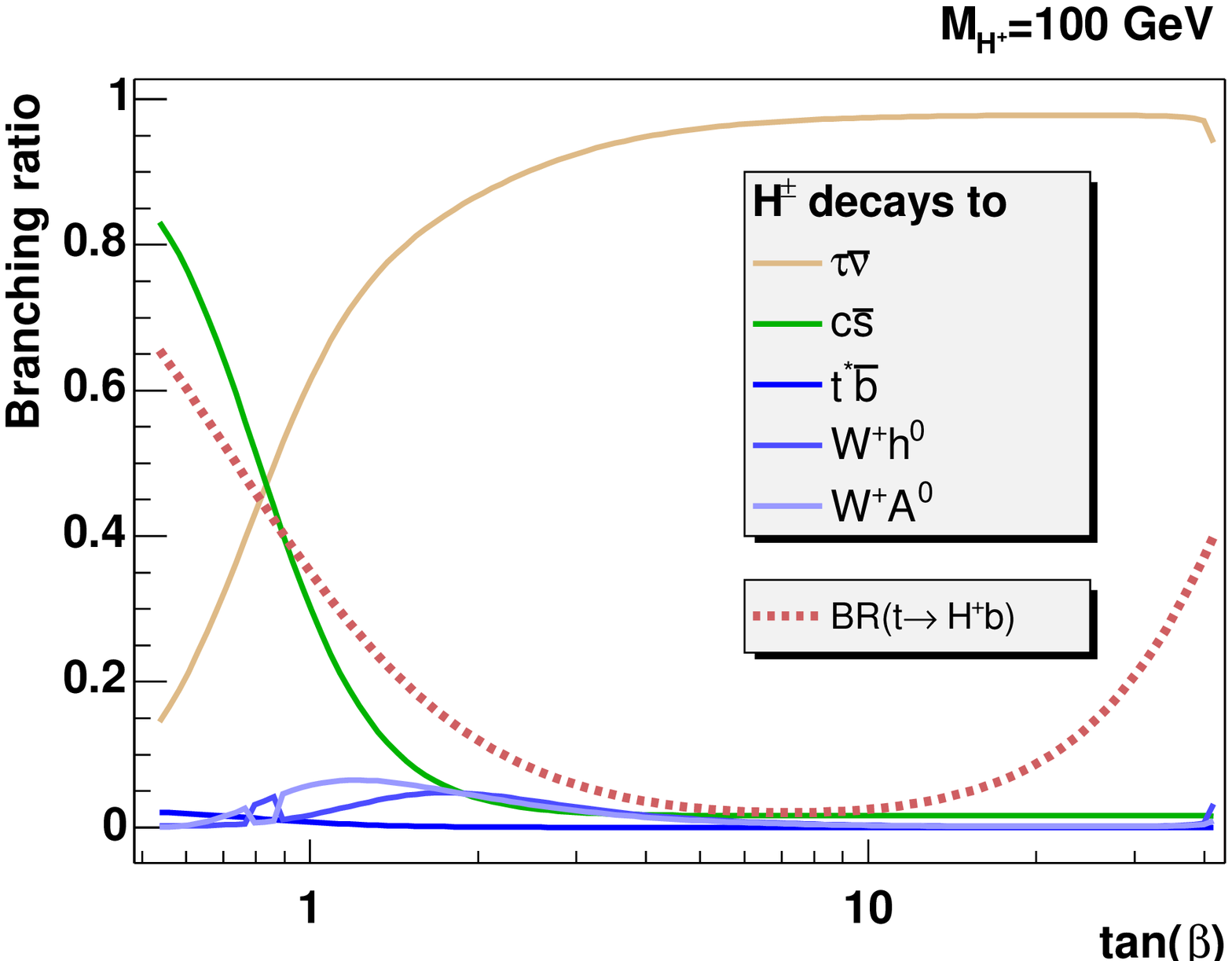}
  \includegraphics[width=0.49\textwidth]{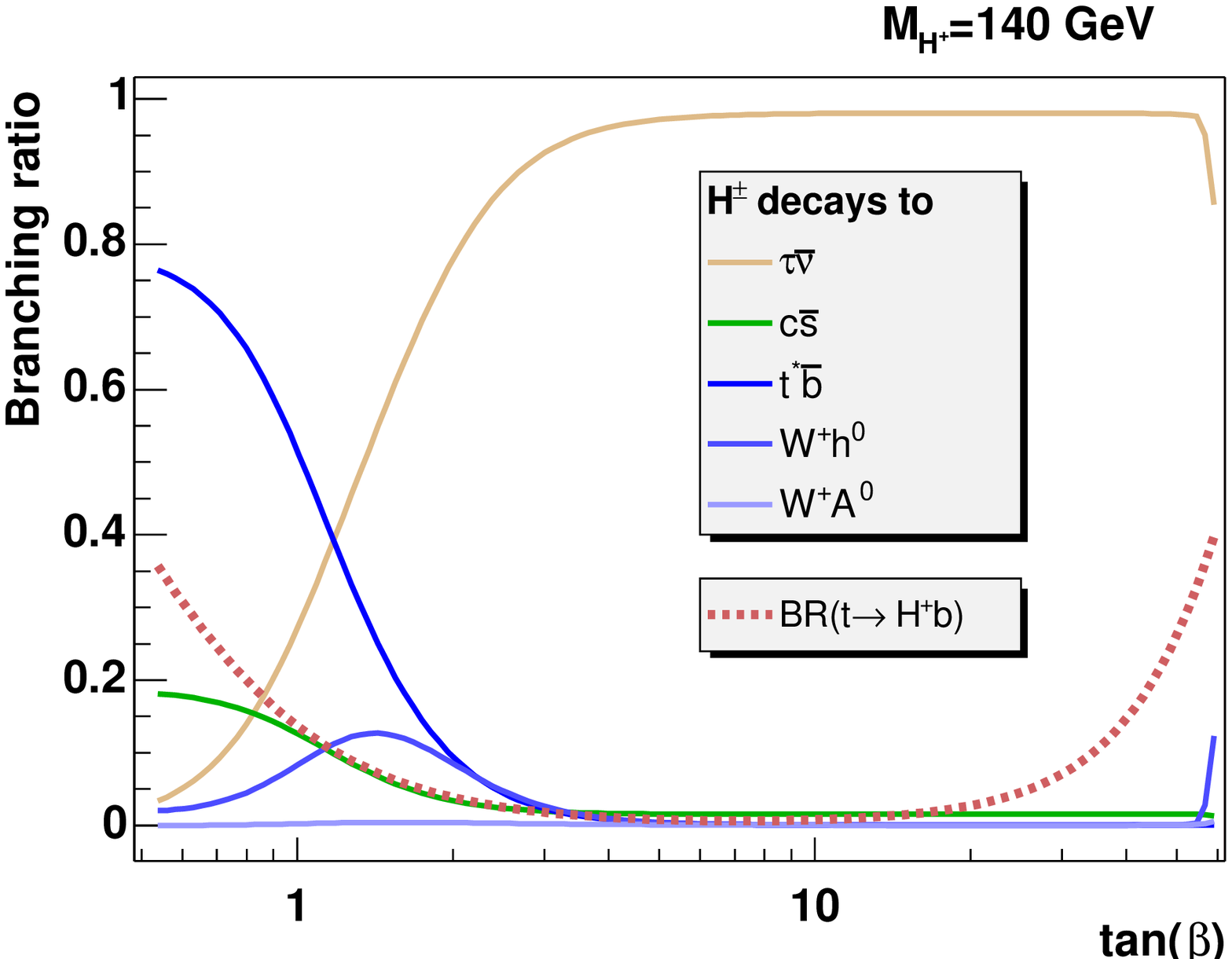}
  \caption{Charged Higgs boson branching fraction in the MSSM as function of $\tan\beta$~\cite{Abulencia:2005jd}.}
  \label{figs:ch_brs}
\end{figure}
Particles beyond the Standard Model in the final state of top quark  pair events may alter the
branching fractions of the various top quark decay channels and modify 
the kinematic properties of the final state. 

Charged Higgs bosons appear in many extentions of the Standard Model due to the need 
for an additional Higgs doublet with a separate vacuum expectation
value. These models are characterised by the ratio of the vacuum
expectation values of the two Higgs doublets, $\tan\beta$.
A charged Higgs boson can replace the $W$ boson in top quark 
decays. Because charged Higgs bosons have different branching fractions than
$W$ bosons this alters the branching fractions to the various top quark  pair decay channels.
If its mass is different from the $M_W$ it also modifies  the kinematic properties
of the top quark  pair final state. 

\begin{figure*}
  \centering
  \includegraphics[height=6.3cm]{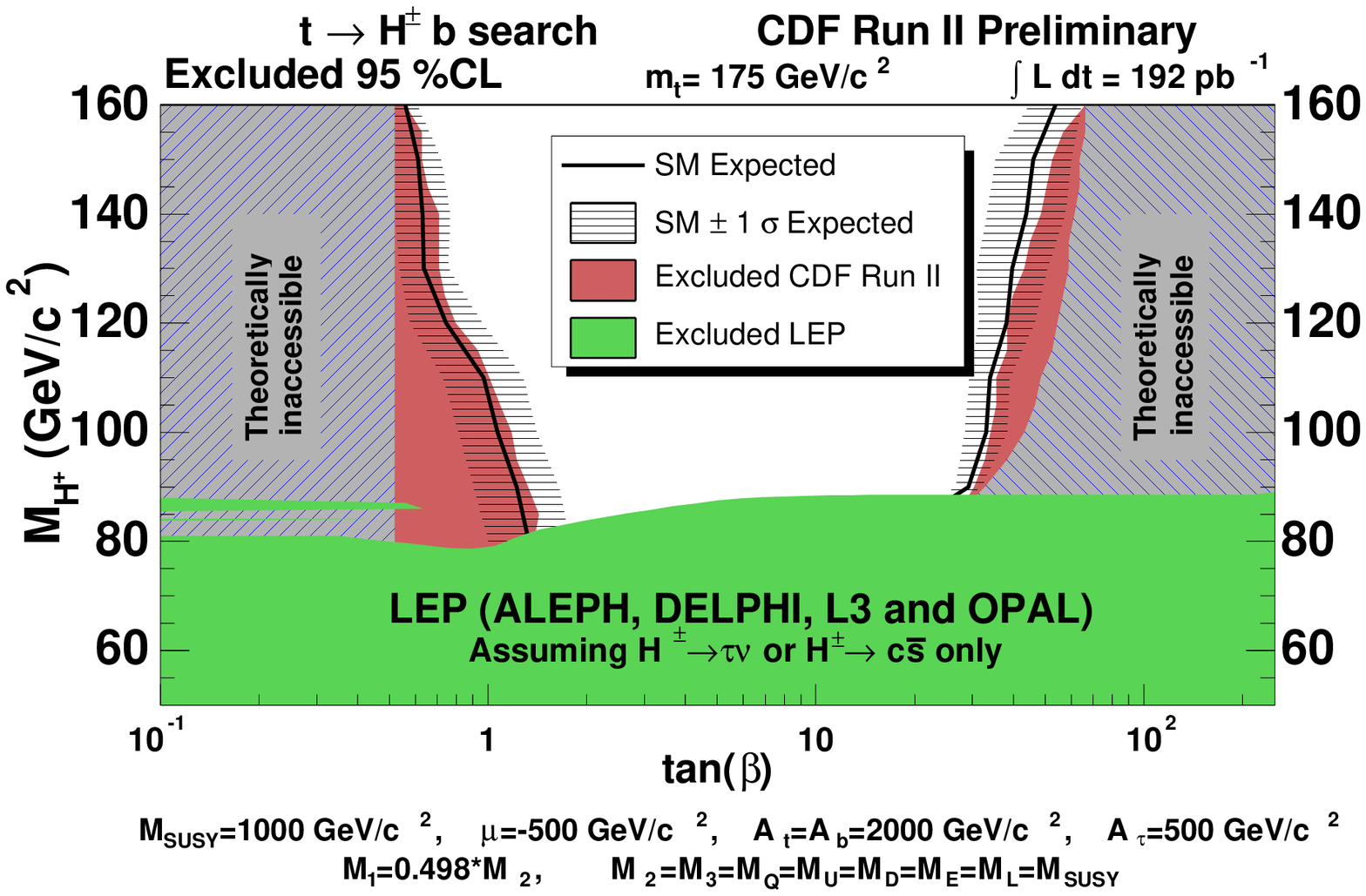}
  \includegraphics[height=6.3cm,trim=0mm -30mm 0mm 0mm]{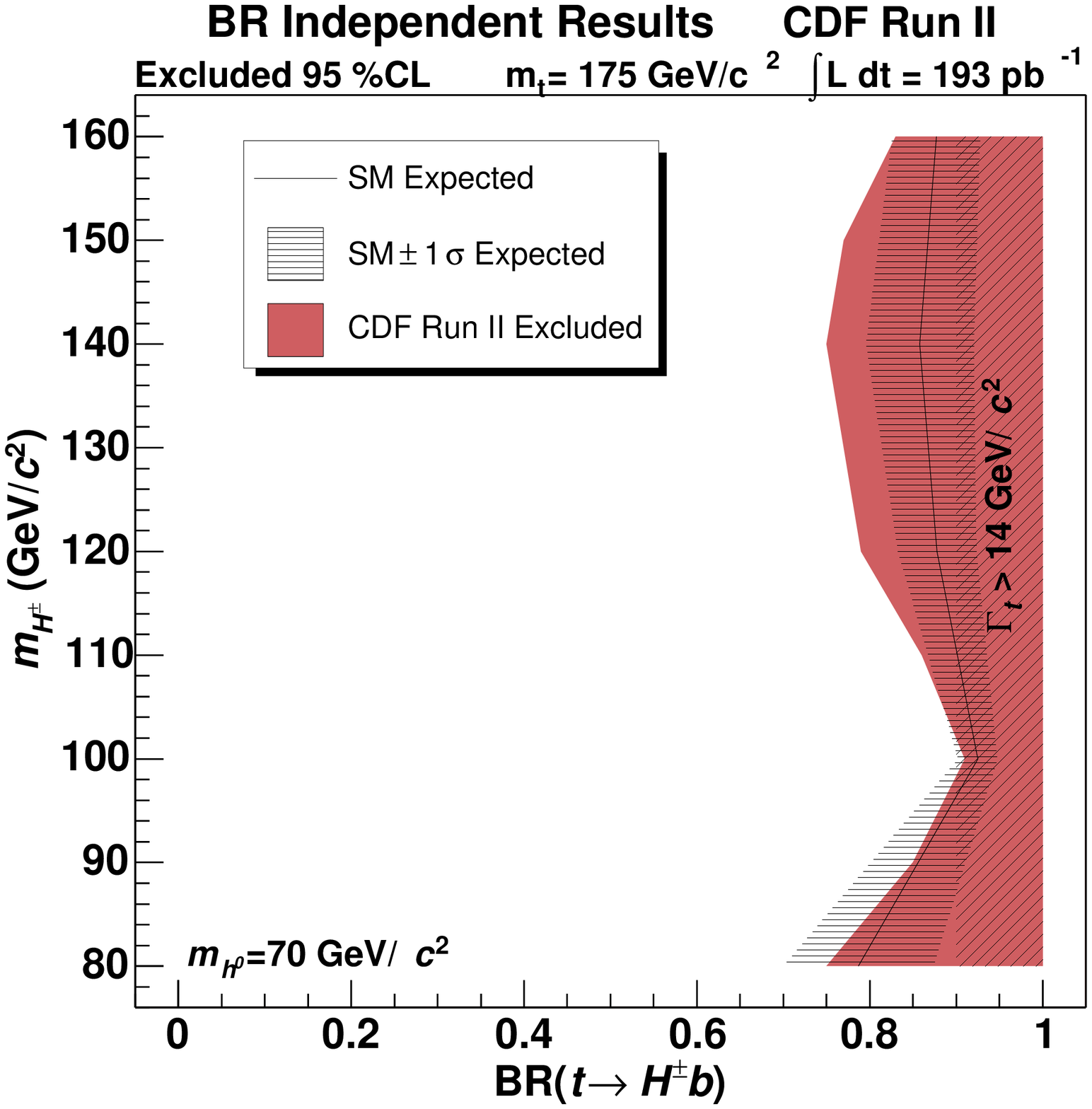}
  \caption{Results from recasting the CDF top quark  cross-section
    measurements~\cite{Abulencia:2005jd}. 
   Left: 
    Exclusion region in the MSSM $m_H$-$\tan\beta$-plane for an example benchmark scenario
    corresponding to the parameters indicated below the plot. 
    Right: Upper limits on ${\cal B}(t\rightarrow H^+b)$ derived without
    assumptions on the charged Higgs boson branching fraction.}
  \label{fig:cdf:h+tanbeta}
\end{figure*}

In the Minimal Supersymmetric Standard Model (MSSM)~\cite{Martin:1997ns} the decay at low $\tan\beta$ is dominated by hadronic decay to
$c\bar s$  at low Higgs boson masses and to  $t^*\bar b$ for Higgs boson masses above
about $130\GeV$. 
For $\tan\beta$ larger than about $1$ a leptonic decay to
$\tau\bar\nu$ dominates, c.f.~\fig{figs:ch_brs}.
The figure also shows the expected branching fraction of 
$t\rightarrow H^\pm b$ which is especially large for very low and very high
$\tan\beta$ and rather small in the intermediate range. 

\subsubsection*{CDF}
CDF has performed two analyses with different approaches. An analysis based on
$0.2\ifb$ uses the CDF $t\bar t$ cross-section measurements in various channels
and recasts the interpretation to obtain limits on the charged Higgs boson production.
A second more recent analysis on $2.2\ifb$  investigates the kinematic
differences between lepton plus jets events from top quark  pair production with Standard Model decay
and those including charged Higgs boson decays.

\subsubsubsection{Recast of Top Quark  Pair Cross-Section}

To obtain limits on a possible charged Higgs boson contribution in top quark  decay CDF
utilises cross-section measurements performed in the lepton plus jets channel (with
exactly one $b$-tag or two or more $b$ tags), the dilepton channel and the 
$\tau$ plus lepton channel~\cite{Abulencia:2005jd}.
Care is taken to avoid overlap between the various channels.
Beside the Standard Model decays of a top quark through a $W$ boson, four  decay modes through the charged
Higgs boson are considered: $H^+\rightarrow \bar\tau\nu$, $H^+\rightarrow c\bar s$,
$H^+\rightarrow t^*\bar b$ and $H^+\rightarrow W^+h^0\rightarrow W^+b\bar b$.
The latter has a non-negligible contribution at intermediate values of $\tan\beta$.

Selection efficiencies are taken from simulation of top quark  pair events for various
masses of the  top quark, the charged and neutral Higgs boson, $h^0$. The simulation
takes the dependence of the width of the top quark and the charged Higgs boson
into
account. The production cross-section is kept at its Standard Model value for
$m_t=175\GeV$: $\sigma_{t\bar t}=6.7\pm0.9\pb$.

The event counts observed in data in the four channels are compared to the
expectations in three different ways. For specific benchmarks of the MSSM a Bayesian approach is
used to set limits on $\tan\beta$. This analysis uses a flat prior on
$\log\tan\beta$ within the theoretically allowed range. These limits are
computed for various values of the charged Higgs boson mass and five different
parameter benchmarks. Figure~\ref{fig:cdf:h+tanbeta}~(left) shows the results
for one specific benchmark.

For the high $\tan\beta$ region $H^+\rightarrow \bar\tau\nu$ dominates in a
large fraction of the MSSM parameter space. Setting the branching fraction of
$H^+\rightarrow \bar\tau\nu$  to $100\%$, limits on the charged Higgs
contribution to top quark  decays are set using Bayesian statistics. A flat prior for
${\cal B}(t\rightarrow H^+b)$ between 0 and 1 is used. For  charged Higgs
boson masses
between $80\GeV$ and $160\GeV$ CDF can exclude ${\cal B}(t\rightarrow
H^+b)>0.4$ at 95\%\,C.L.

Finally, a more model independent limit is computed by scanning the full range
of possible charged Higgs boson decays. For all five $H^\pm$ decay modes considered
the branching fraction is scanned in 21 steps, assuring that the sum of
branching fractions adds to one. Limits on ${\cal B}(t\rightarrow H^+b)$ are
computed for each combination. The least restrictive limit is quoted.
Also this analysis is repeated for various charged Higgs boson masses. The limits
obtained in this more general approach,
shown in \fig{fig:cdf:h+tanbeta}~(right), exclude only very high
contributions of charged Higgs bosons to top quark  decays of above approximately $0.8$ to $0.9$, depending
on the charged Higgs boson  mass.

\subsubsubsection{Investigation of Kinematic Differences}
At low $\tan\beta$, where the charged Higgs boson can also decay to $c\bar s$, CDF used
the invariant dijet mass to search for a possible $H^\pm$ contribution in top quark 
pair events~\cite{Aaltonen:2009ke}. 
Lepton plus jet events are selected requiring at least two of
the four leading jets in $p_T$ to be $b$-tagged. The four leading jets are used
in a kinematic fit that  requires consistency of the fitted
lepton and neutrino momenta with the $W$ boson mass and reconstructed top quark 
masses to be $175\GeV$. The dijet mass of the hadronic $W$ decay remains
unconstrained.
The jet parton assignment with the best $\chi^2$ is
used and the charged Higgs boson mass is reconstructed from the non-$b$-tagged
of the 4 leading jets. 
For events with more than $4$ jets the $5$th jet is added to its closest
neighbour if their $\Delta R<1.0$ to improve the dijet mass resolution.

Background events are dominated by top quark  pair production with Standard
Model decay. Further processes
included are $W+$jets, $Z+$jets, diboson, single top quark  and multijet
events. Except for multijet events the backgrounds are estimated from simulation. The  normalisation for $W+$jets taken
from data, for the others it is taken from theory. The multijet background is fully determined from data. 
\begin{figure}[t]
  \centering
 \includegraphics[height=5.7cm,clip,trim=0mm 0mm 0mm 6mm]{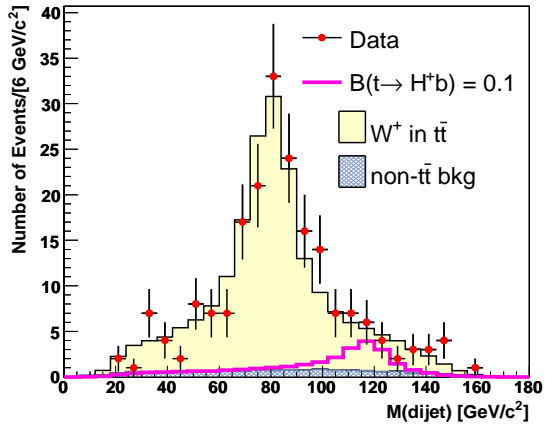}
  \caption{CDF dijet mass distribution with $120\GeV$ Higgs boson events assuming 
      ${\cal B}(t\rightarrow H^+b) = 0.1$. 
      The size of Higgs boson signal corresponds to the expected 
      upper limit branching ratio at $95\%$ C.L. for 120\GeV~\cite{Aaltonen:2009ke}.}
  \label{fig:cdf:h+mjj}
\end{figure}
\begin{figure}[t]
  \centering
 \includegraphics[height=5.7cm,clip,trim=0mm 0mm 0mm 6mm]{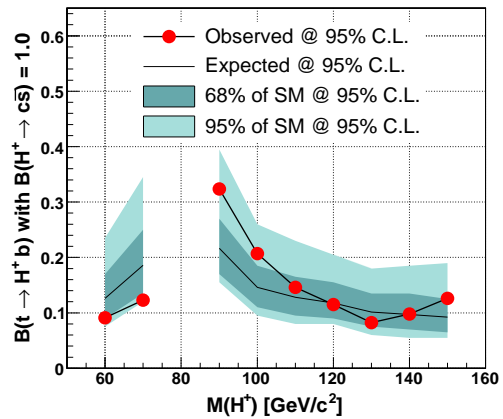}
  \caption{CDF  limits  ${\cal B}(t\rightarrow H^+b)$  observed in $2.2\ifb$ data
      (red dots) compared with the expected limit assuming the Standard Model
      (black line with uncertainty bands)~\cite{Aaltonen:2009ke}.}
  \label{fig:cdf:h+mjj-limit}
\end{figure}

To determine a possible contribution of charged Higgs boson in the decay of top quark  pair
production  a binned likelihood fit is performed. The likelihood is
constructed with templates for the backgrounds and using the branching
fraction of top quark  to charged Higgs bosons, the number of top quark  pair events and the number
of background events as parameters. The number of background events is
constrained  within the uncertainty to the expectation.
The observed dijet mass distribution and the fitted background composition is
shown in \fig{fig:cdf:h+mjj} including a charged Higgs boson contribution of
10\%.

Systematic uncertainties are computed by fitting  pseudo data created from
systematically varied templates with the standard unshifted templates. The
change of the branching ratio due to the systematic variation is taken as
systematic uncertainty for each variation considered. 
These uncertainties are included to a final likelihood by 
convoluting a Gaussian distribution to the original likelihood.
Systematic uncertainties
are dominated by the signal modeling that is derived from 
replacing the default Pythia sample of $t\bar t$ events by a Herwig sample.
For Higgs boson masses close to  the $W$ boson mass the 
jet energy scale uncertainty becomes dominant.

For various assumed Higgs boson masses 95\%CL limits on the branching fraction are
determined by integrating the likelihood distribution to 95\% of its total.
As shown in \fig{fig:cdf:h+mjj-limit} limits between about 10\% and  30\% can be
set depending on the mass of the charged Higgs, consistent with the expected
limits for pure Standard Model top quark  decays. This result is less model dependent than the
above CDF limits, but not as strict within the models used above.

\subsubsection*{D\O}
The D\O\ collaboration searched for a light charged Higgs boson contribution in top quark  decay by
reinterpreting the cross-section measurements in various decay channels 
and for heavy charged Higgs boson contributing to single top quark  production.

\subsubsubsection{Charged Higgs Boson in Top Quark  Decay}

\begin{figure*}
  \centering
\includegraphics[width=0.44\textwidth,clip,trim=0mm 0mm 0mm 3mm]{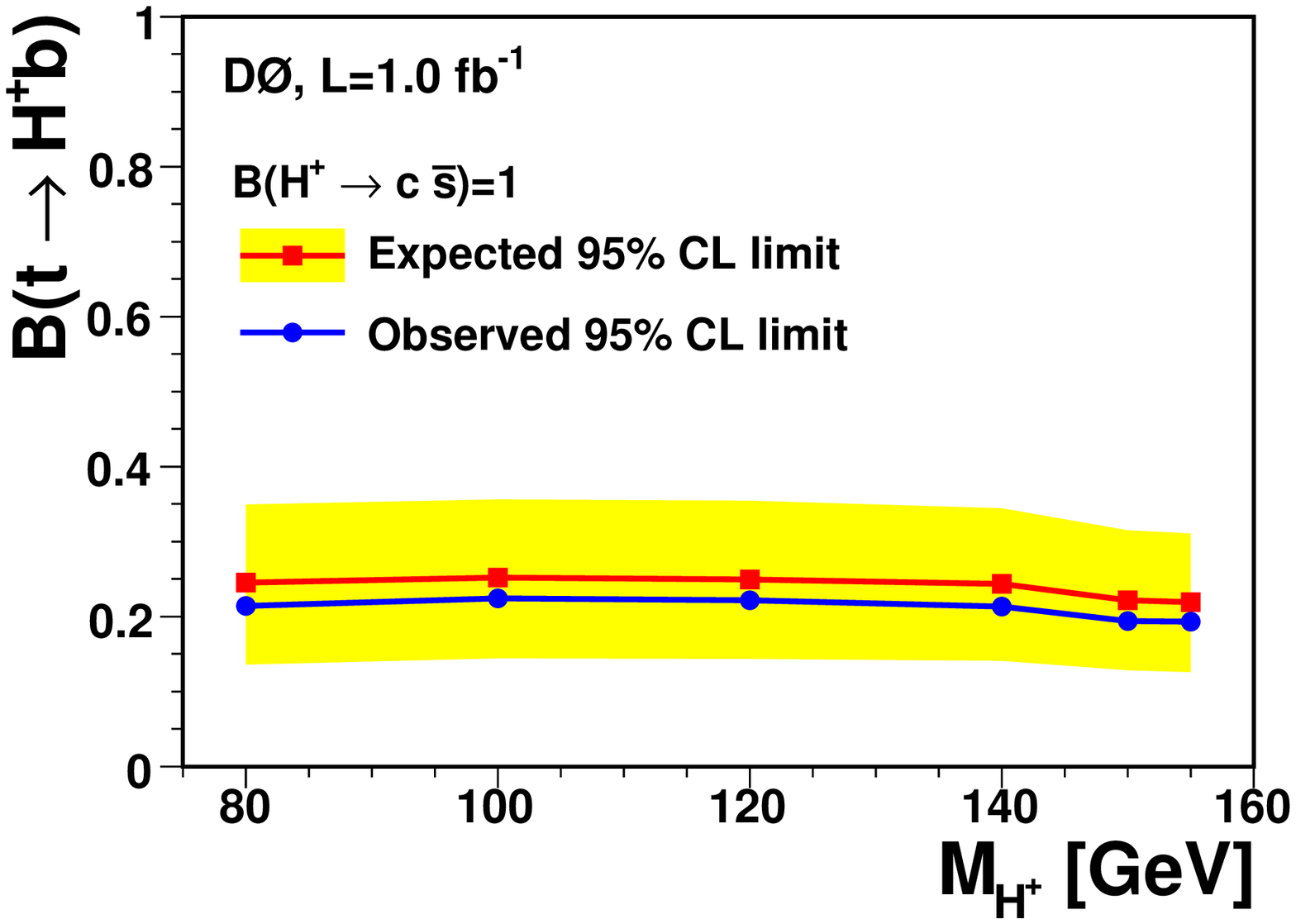}
\includegraphics[width=0.44\textwidth,clip,trim=0mm 0mm 0mm 3mm]{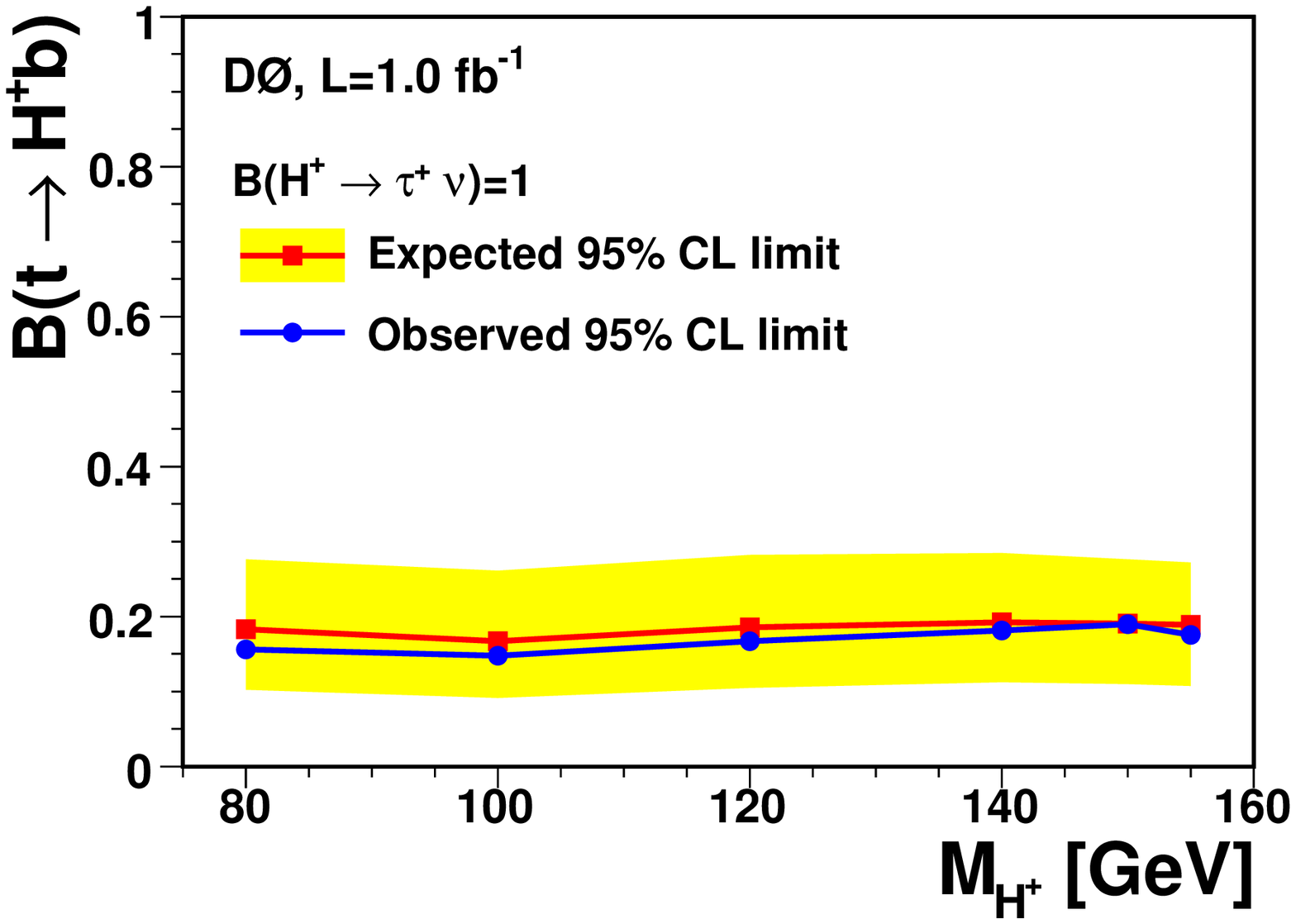}
  \caption{Limits on the contribution of a charged Higgs boson in top quark  decays for a
    leptophobic model (left) and a tauonic model (right) obtained in $1\ifb$
    of D\O\ data assuming the SM $t\bar t$ production cross-section~\cite{:2009zh}.}
  \label{fig:d0:h+limits}
\end{figure*}
In the search for a light charged Higgs boson in top quark decay D\O\ uses the
cross-section analyses for lepton plus jets, dilepton and lepton plus tau decay
channels, with lepton refering to $e$ and $\mu$ only~\cite{Abazov:2009ae,Abazov:2009wy,:2009zh}. The
channels are kept disjoint and further separated into subsamples depending on
the number of jets, the number of $b$ tags and the lepton type. 
The number of expected events for each of the subsamples is computed from
$t\bar t$ simulation with \pythia\ including  
the Standard Model decays and  decays of the top quark to a  leptophobic or a tauonic charged
Higgs boson %for a variety of branching fractions, ${\cal B}(t\rightarrow H^\pm b)$,
and charged Higgs boson masses between $80$ and $155\GeV$.

A likelihood for the observed data is built each of the two models
given the number of expected events as function of 
the branching fraction 
${\cal B}(t\rightarrow H^\pm b)$.
The observed  ${\cal B}(t\rightarrow H^\pm b)$ is extracted by maximising the
likelihood. 

In a first iteration the production cross-section is fixed at a value of
$\sigma_{t\bar t}=7.48\pb$.
%corresponding to the world
%average top quark mass at the time of the analysis, $m_t=173.1\GeV$.
Limits are set according to the Feldman-Cousins procedure
including systematic uncertainties. 
The systematic uncertainties in this method are dominated
by uncertainties due to the assumed top quark  pair cross-section, the luminosity and
$b$-jet identification. The resulting limits obtained with $1\ifb$ of data
exclude a branching fraction above
around $20\%$ for the pure leptophobic model and above  $15-20\%$ for the
tauonic model, c.f.~\fig{fig:d0:h+limits}~\cite{:2009zh}.

\begin{figure}[b]
   \centering
\includegraphics[width=0.44\textwidth,clip,trim=0mm 0mm 0mm 3mm]{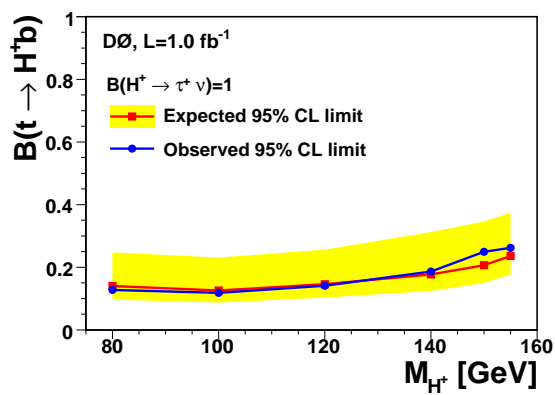}
   \caption{Limits on the contribution of a charged Higgs boson in top quark
     decays for the tauonic model obtained in $1\ifb$
    of D\O\ data in a simultaneous fit of the SM $t\bar t$ production
    cross-section~\cite{:2009zh}.}
  \label{fig:d0:h++xsec}
\end{figure}
In a second iteration the $t\bar t$ production cross-section is treated as a
free parameter and determined simultaneously with the limits on charged Higgs
production, see~\fig{fig:d0:h+limits}. This reduces the assumptions made in the determination of the
limit. In addition the result is much less sensitive to the luminosity.
In this method the description of multijet background becomes the largest
systematic uncertainty.
Such a two dimensional fit is only possible for the tauonic decay model where
this analysis includes channels that get enhanced in the presence of a charged
Higgs and others that get depleted.
In this channel the sensitivity enhances by more than $20\%$ for low Higgs boson masses~\cite{:2009zh}.

D\O\ further discusses the implication for various supersymmetric models
and sets exclusion limits on these models in the plane of model parameters
$\tan \beta$ and $M_{H^\pm}$. 

\subsubsubsection{Charged Higgs Boson in Single Top Quark  Production}
\label{sec:ChargedHiggs:SingleTop}
Heavy charged Higgs bosons can not only occur in the decay of top quarks but may
contribute to single top quark  production. Their signature is identical to Standard Model
$s$-channel single top quark  production, but may have a resonant structure in the
invariant mass distribution of its decay products, the top and the bottom
quarks.

\begin{figure}[b]
  \centering
  \includegraphics[width=0.425\textwidth]{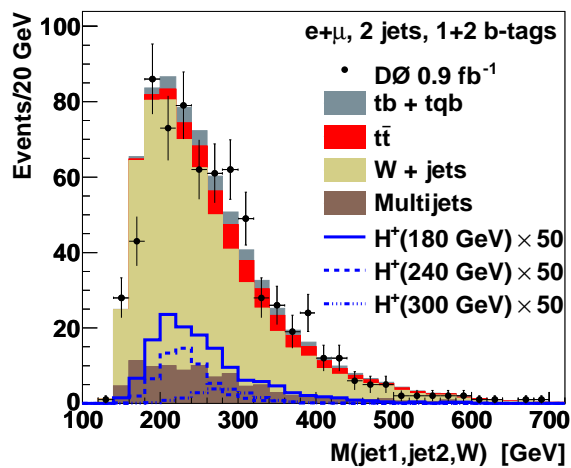}
  \caption{Observed and expected invariant mass of the $W$ boson and the two
    leading jets in $0.9\ifb$ of D\O\ data~\cite{Abazov:2008rn}.}
  \label{fig:d0:h+singletop}
\end{figure}
Following their single top quark  analysis, D\O\ selects events with an isolated
lepton, missing transverse energy and exactly two jets, one of which is
required to be identified as $b$-jet~\cite{Abazov:2008rn}. 
Background estimation for $W+$jets and $t\bar t$ production is simulated using
\alpgen. Standard Model single top quark  production is modelled using \singletop~\cite{Boos:2006af,Boos:2004kh}.
Charged Higgs boson signal events are simulated with a narrow width for the charged
Higgs boson using \comphep. Three types of two Higgs doublet models (2HDM) are
considered.
In the Type~I 2HDM one doublet gives mass to all
fermions; in the Type~II model one doublet gives mass to the $u$-type quarks
(and neutrinos), the other to the $d$-type quarks and charged leptons. This
model is realised in the MSSM. In the Type~III 2HDM both doublets contribute
to the masses of all fermions. Due to the different couplings the
cross-section of single top quark  production in these 3 models is quite different. 

\begin{figure}[t]
  \centering
  \includegraphics[width=0.425\textwidth,clip,trim=0mm 0mm 0mm 7mm]{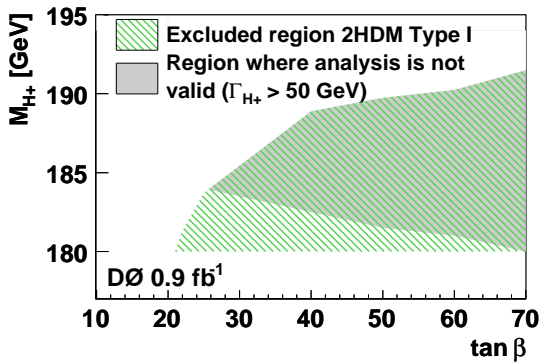}
  \caption{%
    Exclusion areas derived from  $0.9\ifb$ of D\O\ data
    for Type I two Higgs Doublet Model (2HDM)~\cite{Abazov:2008rn}.}
  \label{fig:d0:h+singletop+excl}
\end{figure}

Standard Model and charged Higgs boson production of single top quarks is separated by reconstructing
the invariant mass of the two jets and the $W$ boson. This distribution shows
good agreement between data and the Standard Model expectation, see \fig{fig:d0:h+singletop}. 
Bayesian statistics is used to set limits on the allowed cross-section for
single top quark  production through a charged  Higgs boson. 
For the Type I 2HDM  some region in $\tan\beta$ vs. $M_{H^\pm}$ can be excluded,
c.f. \fig{fig:d0:h+singletop+excl}, a significant fraction of phase space is
not accessible by the analysis in its current form due to the restriction to
small $H^\pm$ decay widths~\cite{Abazov:2008rn}.

%%% Local Variables: 
%%% mode: latex
%%% TeX-master: "EPJC_TopProperties"
%%% End: 

\subsection[Heavy Charged Vector Boson, $W'$]{\boldmath Heavy Charged Vector Boson, $W'$}

\label{sec:Wprime}
New charged gauge bosons, $W'$, are expected in extensions of the Standard Model with
additional gauge symmetries and in supersymmetric models, see e.g.~\cite{\refpdg,Martin:1997ns}. 
Its couplings may be to left-handed fermions, like for the Standard Model $W$ boson, or
include right-handed fermions.  In general a mixture of these two options is possible.
If the  $W'$ boson has left handed couplings, it will have  
a sizeable interference with the SM  $W^\pm$ boson~\cite{Boos:2006xe}.
For purely right-handed couplings, a leptonic decay may only occur when the
right-handed neutrinos are lighter than the $W'$ boson. In this case the decay
to a top and bottom quark is an interesting channel to perform direct searches for such $W'$ bosons.

Both CDF and {D\O} search for various types of $W'$ bosons decaying to $tb$ pairs
in conjunction with their single top quark  analyses. The main discriminating
observable is the reconstructed invariant mass of the decay products, which
was also utilised to search for a heavy charged Higgs boson, 
c.f.~Section~\ref{sec:ChargedHiggs:SingleTop}.

\subsubsection*{D\O}

D\O\ has published a search for a heavy $W'$ boson with decay to top and bottom quarks using
$0.9\ifb$~\cite{Abazov:2008vj}. The event selection follows the single top
quark analysis and requires one isolated lepton,
missing transverse momentum and two or three jets, one of which must be
identified as $b$-jet.

\begin{figure}[b]
  \centering
  \includegraphics[width=0.44\textwidth,clip,trim=0mm 0mm 0mm 10mm]{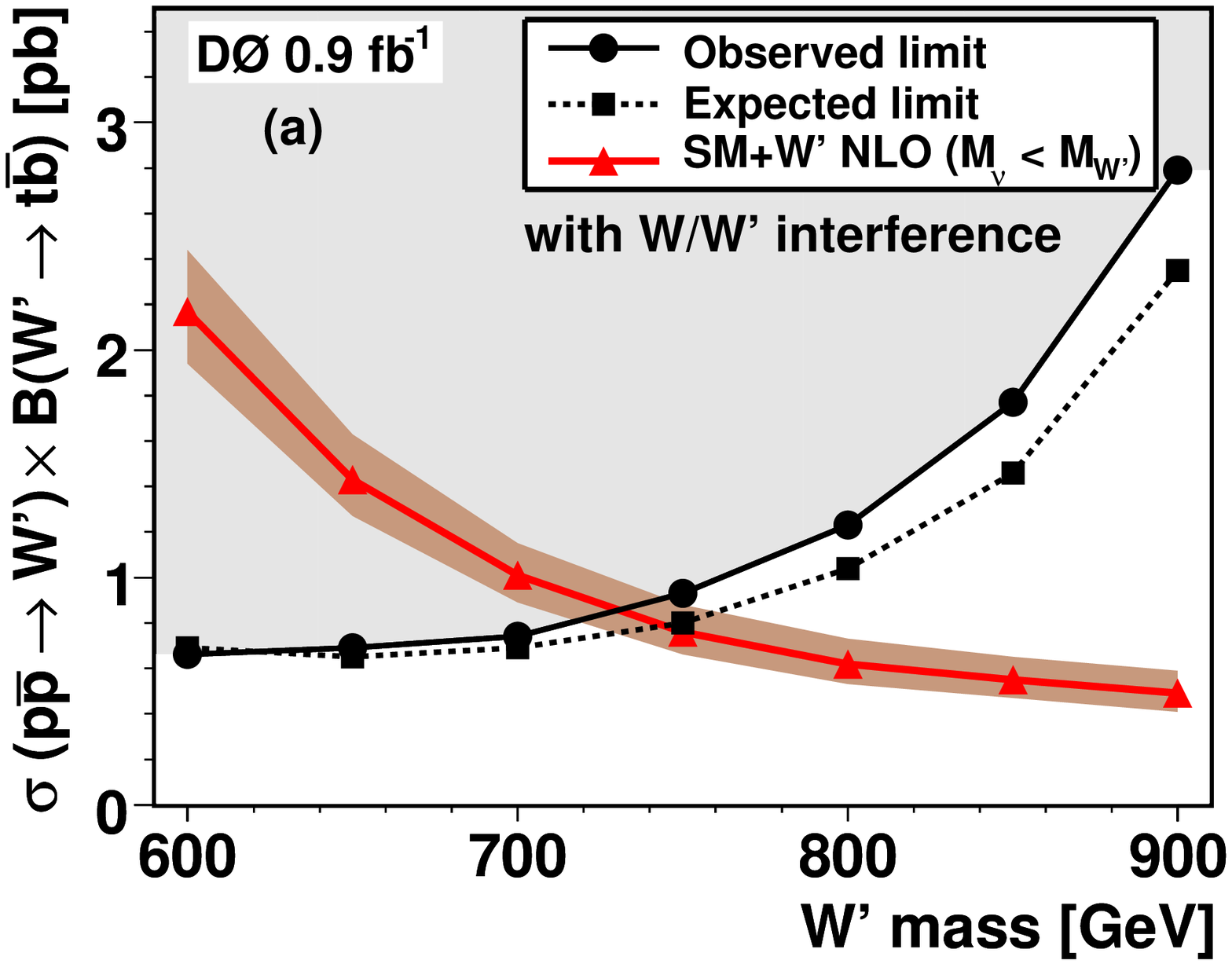}\\
  \includegraphics[width=0.44\textwidth,clip,trim=0mm 0mm 0mm 10mm]{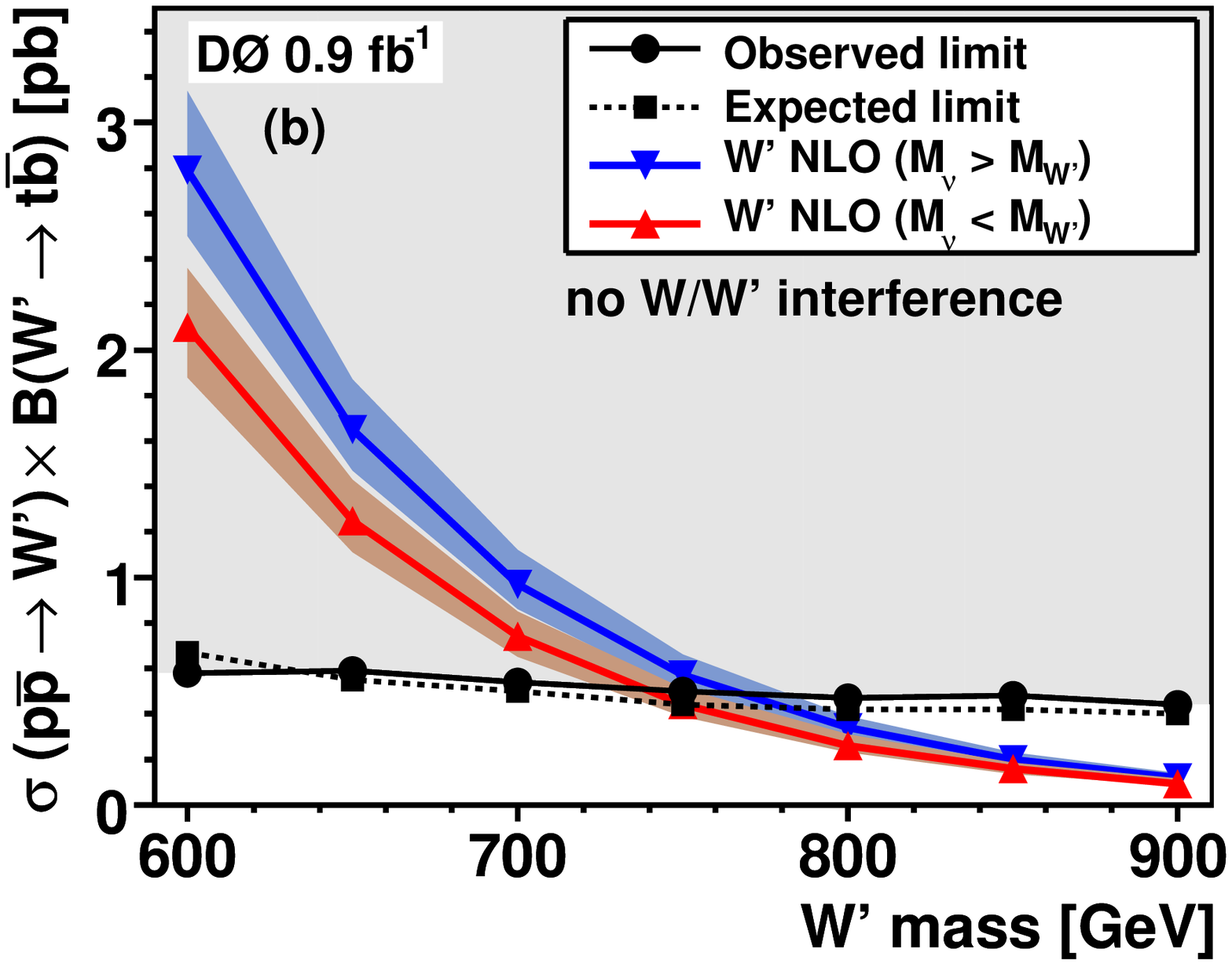}
  \caption{D\O\ results on a search for $W'$ boson decaying to top and
    bottom quark using $0.9\ifb$ of data. 
    Top: Expected and observed limits on a left-handed $W'$ production cross-section times branching
    fraction of the decay to top and bottom quark as function of $M_{W'}$ compared to the theory
    prediction. Bottom: Same but for right-handed  $W'$ production~\cite{Abazov:2008vj}.
  }
  \label{fig:d0Wprime-xsec}
\end{figure}
The invariant mass, $\sqrt{\hat s}$, of the bottom and the top quark  decay products is computed from
the measured four-momenta of the leading two jets, the charged lepton and the
neutrino. The transverse momentum of the neutrino is identified with the
transverse missing momentum, its $z$-component infered by solving
$M_W^2=\left(p_\ell+p_\nu\right)^2$ choosing the solution with the smaller
$\left|p_\nu^z\right|$.

The distributions expected within the Standard Model from a
combination of simulation and data. 
Single top quark  and top quark  pair production is generated with
\singletop\ and \alpgen+\pythia\ normalised to their
theoretical cross-sections. $W$+jets background is generated with
\alpgen+\pythia\ and normalised to data before $b$-tagging in a way that it includes
diboson backgrounds. Also the  $W$+heavy flavour fraction is derived from
data. Multijet background is fully taken from data.
Samples of $W'$ boson events with masses up to $900\GeV$ are generated in conjunction 
with the single top quark  samples taking interferences with the $W$ boson  into
account that are present for the left-handed $W'$ bosons. Because of this
interference the Standard Model single top quark   production in the $s$-channel is
treated as part of the signal in the search for $W'_{\!L}$ bosons.

The distribution of reconstructed $\sqrt{\hat s}$ measured by D\O\ agrees
with the expectation from the Standard Model. Limits on a possible
contribution from $W'$ bosons decaying to top and bottom quarks are derived as a function of
$M_{W'}$ assuming couplings like in the Standard
Model, though possibly to right handed fermions.
D\O\ uses the Bayesian approach with a flat non-negative prior on the
cross-section times branching fraction.
Expected and observed results are shown in \fig{fig:d0Wprime-xsec}. 
%and compared to the expectation.
\begin{figure}[b]
  \centering
  \includegraphics[width=0.44\textwidth]{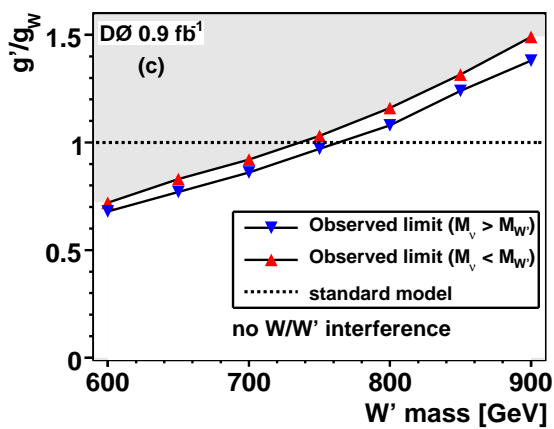}
  \caption{D\O\ results on a search for $W'$ boson decaying to top and
    bottom quark using $0.9\ifb$ of data. 
    Limits on the $W'$ boson coupling relative to the Standard Model $W$-boson coupling~\cite{Abazov:2008vj}.
  }
  \label{fig:d0Wprime-coupl}
\end{figure}
Comparing upper limits on the $W'$ boson cross-section times branching fraction to top and
bottom quark to the NLO theory predictions~\cite{Sullivan:2002jt} excludes left-handed $W'$ bosons with
$M_{W'_{\!L}}<731\GeV$. If only hadronic decays are allowed the  right handed
$W'_{\!R}$ boson is excluded for $M_{W'_{\!R}}<768\GeV$, when leptonic decays
are also possible the limit is $739\GeV$.

Without assuming the coupling strength the Bayesian approach is used to
determine a limit on the size of this coupling relative to the Standard
Model, see \fig{fig:d0Wprime-coupl}. These limits assume no interference
between the Standard Model $W$ and the $W'$ bosons.

In computing the above limits systematic uncertainties  are included.
They include effects due to uncertainties on the integrated luminosity, the theoretical
cross-sections, branchings fraction, object identification efficiencies,
trigger efficiencies, fragmentation models, jet energy scale and heavy flavour
simulation.

\subsubsection*{CDF} %CdfNote9150,Aaltonen:2009qu

In an investigation of $1.9\ifb$ of data~\cite{Aaltonen:2009qu} CDF selects  
$W+$jets events requiring one lepton ($e$, $\mu$) isolated from
jets, missing transverse energy and two or three energetic jets. At least one
of the jets must be tagged as $b$-jet. 

In these events the neutrino momentum, $p_\nu$, is infered from the missing transverse
momentum and by solving $M_W^2=\left(p_\ell+p_\nu\right)^2$ for the
longitudinal component of the neutrino. The $W$ boson mass, $M_W$, is set to its
nominal value, $p_\ell$ is the measured lepton momentum.
In case of complex solutions CDF assigns the real part of the solution to the
longitudinal neutrino momentum. The invariant mass of the lepton, the neutrino
and the two leading jets, $M_{Wjj}$, is then used as a discriminating
observable.

The distribution expected within the Standard Model is computed from a
combination of simulation  and data. The contribution of events containing a
real $W$ boson is
taken from simulation. $W+$jets samples are normalised to data before
$b$-tagging using a scale factor to correct the heavy flavour contribution to
fit the observation in $W+$1jet data. The other samples are normalised to their theoretical
expectation. The identification of heavy flavour jets is estimated from
simulation and corrected with a scale factor. Misidentification of light
flavour jets is computed from mis-tag rate functions.
The contribution from events without real $W$ bosons are
estimated from events with electrons that pass only a subset of the full electron
identification and thus are enriched with jets  misidentified as electrons.
\begin{figure}[b]
  \centering
  \includegraphics[width=0.44\textwidth]{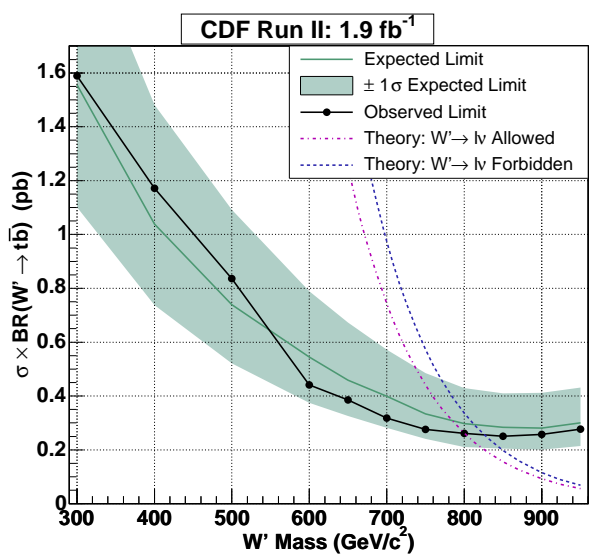}
  \caption{
   Limits on $W'_{\!R}$ cross-section times branching
    fraction to $tb$ as function $M_{W'_{\!R}}$ compared to theory obtained in
    the CDF search for $W'_{\!R}$  using  $1.9\ifb$ of
    data~\cite{Aaltonen:2009qu}.
}  
  \label{fig:CdfWprime}
\end{figure}

$W'$ boson signal events are simulated using Pythia for $W'$ boson masses between $300$
and $950\GeV$ with fermion couplings identical to the Standard Model $W$ boson. 
%Interference
%between $W$ and $W'$ is expected to be small for $M_{W'}\gg M_W$ and is
%neglected. 
When the right-handed $W'$ boson  is heavier than the right-handed neutrinos,
the branching fraction to $\ell\nu$ is corrected according to the additional
decay modes.

Limits are constructed according to the CL$_s$ method~\cite{Junk:1999kv,Read-in-James:2000et}. Probabilities are
computed from pseudo experiments which are generated including variations 
due to systematic uncertainties. The dominating systematic uncertainties are the jet energy scale and the scale
factor used to account for differences between simulation and data in the $b$-tagging
algorithm of CDF.

Limits on the $W'_{\!R}$ boson production cross-section are set  as a function of $M_{W'_{\!R}}$
assuming the Standard Model coupling strength. These are converted to mass
limits by comparison to the corresponding theoretical expectation and yield
$M_{W'_{\!R}}>800\GeV$ for $W'_{\!R}$ bosons which decay leptonically 
and $M_{W'_{\!R}}>825\GeV$ for  $M_{\nu_R}>M_{W'_{\!R}}$. 
For the more general case that the $W'_{\!R}$ coupling is a priori unknown
the $W'_{\!R}$ coupling strength, $g'$, 
relative to the Standard Model coupling, $g_W$,  is constrained. Limits
are computed from the above analysis as function of the
assumed $M_{W'}$.
The observed and expected limits derived by CDF for  $M_{W'_{\!R}}$ and
$g'/g_W$  are shown in \fig{fig:CdfWprime} and \ref{fig:CdfWprime-Coupl}.

\begin{figure}
  \centering
  \includegraphics[width=0.44\textwidth]{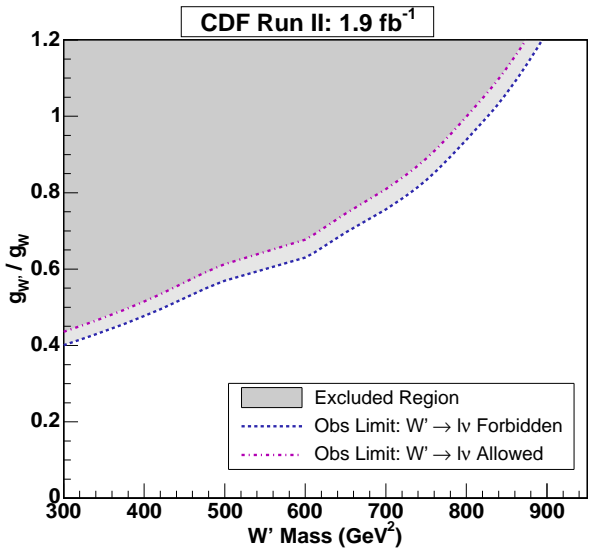}
  \caption{
    Limits on the $W'_{\!R}$ coupling strength
    relative the Standard Model coupling as function $M_{W_{\!R}'}$.
    obtained in the CDF search for $W'_{\!R}$  using  $1.9\ifb$ of data~\cite{Aaltonen:2009qu}.
}
  \label{fig:CdfWprime-Coupl}
\end{figure}

%%% Local Variables: 
%%% mode: latex
%%% TeX-master: "EPJC_TopProperties"
%%% End: 

\subsection{Resonant  Top Quark Pair Production}
Due to the fast decay of the top quark, no  resonant production of top quark 
pairs is expected within the Standard Model. However, unknown heavy resonances decaying to top quark 
pairs may add a resonant part to the Standard Model production mechanism. 
Resonant production is possible for massive $Z$-like bosons in
extended gauge theories \cite{Leike:1998wr}, Kaluza-Klein states of the gluon
or $Z$ boson~\cite{Lillie:2007yh,Rizzo:1999en}, axigluons
\cite{Sehgal:1987wi}, Topcolor \cite{Hill:1993hs,Harris:1999ya}, and other
theories beyond the Standard Model.  
Independent of the exact model, such resonant production could
be visible in the reconstructed $t\bar t$ invariant mass. 
\label{sect:BSM:Mtt}

\subsubsection*{CDF}
CDF has employed several different techniques to search for resonances in the
$t\bar t$ invariant mass distribution. All analyses use a very similar event
selection: an isolated lepton, missing transverse energy and
four or more jets. One analysis (the Matrix Element plus
Template method) is also applied to the full hadronic channel selecting six or
seven jets. Their main difference is the method to reconstruct the
$t\bar t$ invariant mass distribution. 

\subsubsubsection{Constrained Fit plus Template}
The method that uses the least assumptions reconstructs the invariant mass
using a constrained fit~\cite{cdf:2007dia} and is performed requiring at least
one identified $b$ jet. In the fit
the final state lepton and quark momenta are determined from the measured
lepton momentum, the missing transverse energy (which is assumed to stem from
the unseen neutrino) and the measured jet momenta. Constraints are imposed that
require the sum of  neutrino and
lepton momenta as well as the two light quark momenta to be consistent with
the $W$ boson mass. In addition these pairs in combination with one $b$ quark
need be  consistent with the top quark  mass of 175\GeV. The fitted momenta may
vary within the experimental resolution of their assigned measurement and the
mass constraints are varied within the natural widths of the $W$ boson and
the top quark, respectively.
The fit thus assumes the lepton plus jets decay topology of top quark  pair events.

\begin{figure}[b]
  \centering
  \includegraphics[width=0.4\textwidth]{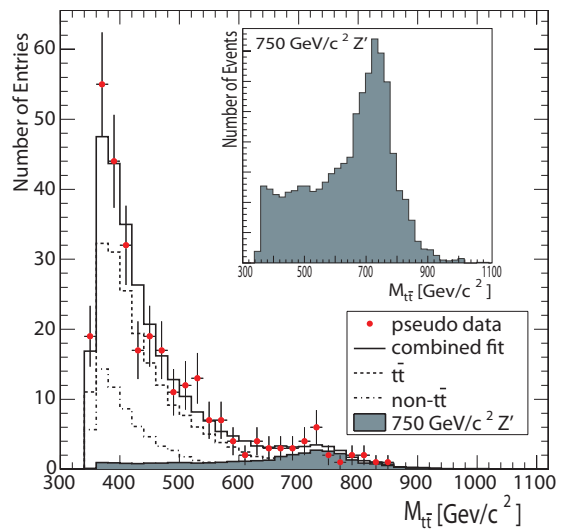}
  \caption{Distribution of the invariant top quark  pair mass reconstructed with a
    constrained fit in $1\ifb$ of CDF data~\cite{cdf:2007dia}. 
   }
  \label{fig:cdf-ttresonance-1fb-1-mtt}
\end{figure}
Of the multiple jet parton assignments the one with the best $\chi^2$
from the fit is used to compute the top quark  pair invariant mass for each event.
The expected distribution of this observable is dominated by 
Standard Model top quark  pair events which are simulated using \herwig. Further
backgrounds like $W+$jets, misidentified multijet events, diboson and single
top quark  events are modelled with combination of simulation and control data.

Templates for a resonant production of top quark pairs are simulated using \pythia\
for resonance masses between $450$ and $900\GeV$. The resonance couplings are
proportional to 
those of a Standard Model $Z$ boson. The width of this $Z'$ boson was kept at
$0.012M_{Z'}$. The observed invariant
top quark pair mass distribution is compared to the distribution expected from the
Standard Model in \fig{fig:cdf-ttresonance-1fb-1-mtt}.

From the expected distributions 
CDF constructs a likelihood for the expected bin content of the distribution
of the invariant top quark pair mass as function of the resonant production
cross-section time branching ratio, $\sigma_X{\cal B}$, the number of Standard
Model top quark pairs and the number of non-$t\bar t$ events. Nuisance parameters
with Gaussian constraints are used to implement the effect of systematic
uncertainties. These include uncertainties that affect
the relative background normalisation and the luminosity, the uncertainty of the jet
energy scale and the shape change due to the top quark mass uncertainty. Minor
contributions come from varying the PDFs between 
CTEQ6M~\cite{Pumplin:2002vw} and MRST~\cite{Martin:1999ww}
parametrisations and the uncertainty on the strength of initial and final
state radiation. 
\begin{figure}
  \centering
  \includegraphics[width=0.44\textwidth]{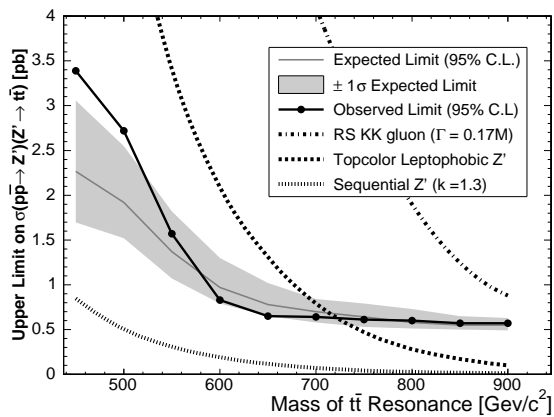}
  \caption{Limits on the
    cross-section times branching fraction for resonant top quark  pair production
     observed in $1\ifb$ of CDF data~\cite{cdf:2007dia}. Theoretical curves are shown for various model
    and used to set mass limit on the corresponding resonance. 
   }
  \label{fig:cdf-ttresonance-1fb-1-limit}
\end{figure}

To find the upper limits the maxima of the likelihood as function of
$\sigma_X{\cal B}$ is integrated to the point where the integral reaches 95\% of its
area. This is done for each assumed resonance mass between $450$ and $900\GeV$
in $50\GeV$ steps. 
Expected and observed limits are shown in \fig{fig:cdf-ttresonance-1fb-1-limit}.
At high resonance masses the observed limits exclude a resonant top quark  pair
production with  $\sigma_X{\cal B}>0.55\pb$ at 95\%~C.L. A production through leptophobic
Topcolor assisted Technicolor is excluded for resonance masses up to $720\GeV$.

\subsubsubsection{Matrix Element plus Template}

The resolution of the reconstructed invariant top quark  pair mass can be improved by
assuming additional information. In an analysis of $680\ipb$ CDF employed the
Matrix Element technique to reconstruct the invariant mass distribution that is
used to search for resonant production in lepton plus jets events~\cite{cdf:2007dz}. 
Following the mass analysis~\cite{Abulencia:2005pe} described in Section~\ref{sect:dlm}, 
for each event a probability density,
$P(\{\vec{p}\}|\{\vec{j}\})$, is computed to find the momenta of the
top quark  pair decay products (4 quarks, charged lepton and neutrino, $\{\vec{p}\}$) 
given the observed quantities, $\{\vec{j}\}$. This
probability is computed from the parton density functions, the theoretical
Matrix Element for Standard Model top quark  pair production and decay and
jet transfer functions that fold in the detector resolution. It is 
converted to a probability density for the top quark  pair invariant mass using
\beq
P_f(M_{t\bar t}|\{\vec{j}\})=\int\!\mathrm{d}\{\vec{p}\}\,\,P(\{\vec{p}\}|\{\vec{j}\})
\,\delta\!\left(M_{t\bar t}-m(\{\vec{p}\})\right)\quad\mbox{.}
\label{eq:mttFromDLM}
\eeq
The result for  all possible jet-parton assignments is summed,
before the mean value is computed and taken as the reconstructed invariant top quark 
pair mass for the event under consideration.
Here the $b$-tagging information is used to reduce the number of allowed jet parton
assignments. The events are not required to contain $b$-tagged jets in the
event selection.

\begin{figure}[b]
  \centering
  \includegraphics[width=0.42\textwidth,clip,trim=0mm 0mm 0mm 17mm]{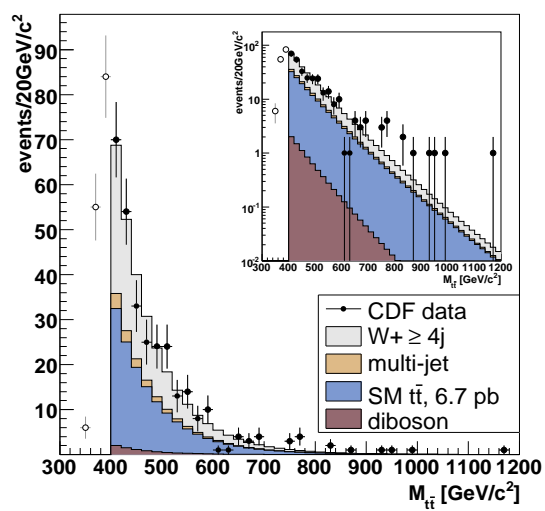}
  \caption{Distribution of the invariant top quark  pair mass reconstructed with
    the Matrix Element technique in $680\ipb$ of CDF data using semileptonic
    events~\cite{cdf:2007dz}.
  }
  \label{fig:cdf-680ipb-mtt.mtt} 
\end{figure}
Template distributions for Standard Model and resonant $t\bar t$ processes are
derived from \pythia, $W+$jets events from \alpgen+\herwig\ including full
detector simulation. 
As a resonance signal a $Z'$ boson with a width of $1.2\% M_{Z'}$ was generated.
The multijet background template was taken from data.
The Standard Model $t\bar t$ samples and diboson samples
are normalised to the theoretical cross-section, while the sum of multijet and
$W+$jet samples are scaled to fit the observed data and depend on the assumed
signal contribution.
The resulting expected distribution compared to the observed data is shown
in~\fig{fig:cdf-680ipb-mtt.mtt}. 

\begin{figure}
  \centering
  \includegraphics[width=0.4\textwidth]{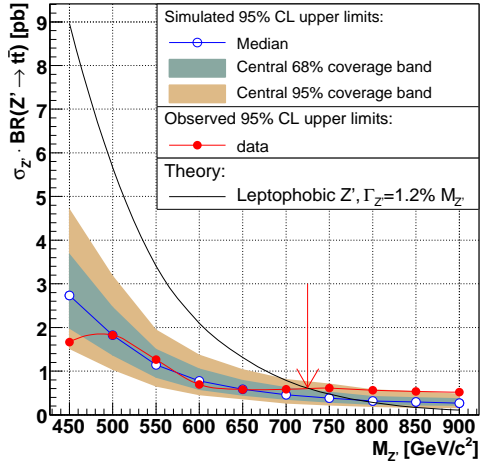}%
  \caption{
    Expected and observed limits on $\sigma_{X}\cdot {\cal B}(X\rightarrow t\bar t)$
      obtained by CDF in $680\ipb$ using the Matrix Element technique in
      lepton plus jets events~\cite{cdf:2007dz}.
  }
  \label{fig:cdf-680ipb-mtt.limit} 
\end{figure}

The possible contribution from a resonant top quark  pair production is computed
using Bayesian statistics. The posterior probability density is build from
a likelihood that implements Poissonian expectations in each bin. The
parameters of the Poisson distribution are smeared with Gaussians according to
the systematic uncertainties. Finally, the posterior  probability density is
convoluted with a flat prior in $\sigma_{Z'}\cdot {\cal B}(Z'\rightarrow t\bar t)$.
The uncertainties of the Standard Model top quark  pair production cross-section, the
jet energy scale and the variation of initial and final state gluon radiation
have the largest impact on the resulting limits. 

The events observed by CDF in $680\ipb$ of data show no evidence for resonant
top quark  pair production and  upper limits are derived for $\sigma_{Z'}\cdot {\cal
  B}(Z'\rightarrow t\bar t)$ as shown in~\fig{fig:cdf-680ipb-mtt.limit}.
A comparison to the leptophobic Topcolor assisted Technicolor model yield an
exclusion of this model for $M_{Z'}<725\GeV$ at 95\%C.L. Using additional
assumptions about the kinematics through the matrix element thus allows to
exclude slightly higher $Z'$ boson masses despite using less data.
\bigskip

\begin{figure}[t]
  \centering
  \includegraphics[width=0.4\textwidth]{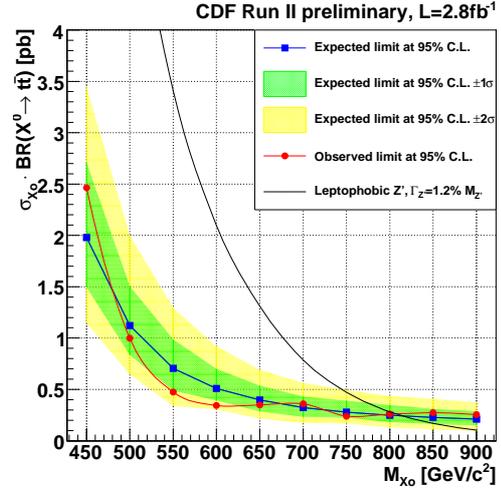}
  \caption{Expected and observed upper limits on $\sigma_{X}\cdot {\cal
      B}(X\rightarrow t\bar t)$ obtained by CDF with a Matrix Element technique in
    $2.8\ifb$ of data using all hadronic events~\cite{CdfNote9844}.}
  \label{fig:cdf-alljets-mtt}
\end{figure}
The Matrix Element plus Template method was also applied in the all hadronic decay
channel using $2.8\ifb$ of CDF data with six or seven
jets~\cite{CdfNote9844}. 
Top quark pair events are enriched with a neural net event selection
and $b$ jet identification. The dominant multijet background is described with
a data driven method from events without the $b$ identification requirement. 

No evidence for resonant top quark pair production is found. CDF computed
upper limits on the  resonant production cross-section time branching
fraction, $\sigma_{X}\cdot {\cal  B}(X\rightarrow t\bar t)$, as shown in \fig{fig:cdf-alljets-mtt}.
In the leptophobic Topcolor assisted Technicolor model resonance masses of
$M_{Z'}<805\GeV$ are excluded at 95\%C.L.

%\subsubsubsection{Massive Gluon DLM interpretation of ME+template}
\subsubsubsection{Dynamical Likelihood Method with Massive Gluon interpretation}

Another analysis of CDF is based on the dynamical likelihood
method (DLM, see also Section~\ref{sect:dlm}). It 
investigates $1.9\ifb$ of data with an isolated lepton, missing
transverse energy and exactly four reconstructed jets~\cite{Aaltonen:2009tx}.
In contrast to the previously described analysis
the invariant top quark  pair mass is reconstructed without using the production part
of the matrix element in the  construction of the probability densities in
\eq{eq:mttFromDLM}. This avoids a bias towards the Standard Model production
mechanism. 
The resulting distribution and the corresponding Standard Model expectation
is shown in \fig{fig:mtt-massivegluon.mtt}.
\begin{figure}[b]
  \centering
  \includegraphics[width=0.4\textwidth]{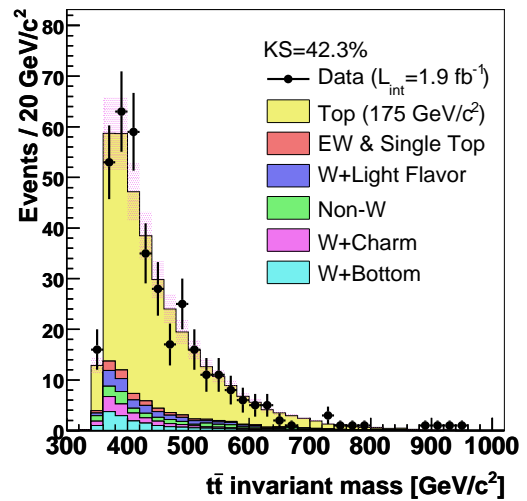}  
 \caption{Invariant top quark pair mass reconstructed with the dynamical likelihood
   method in $1.9\ifb$ of CDF data.~\cite{Aaltonen:2009tx}.}
  \label{fig:mtt-massivegluon.mtt}
\end{figure}
\begin{figure}[t]
  \centering
  \includegraphics[width=0.48\linewidth]{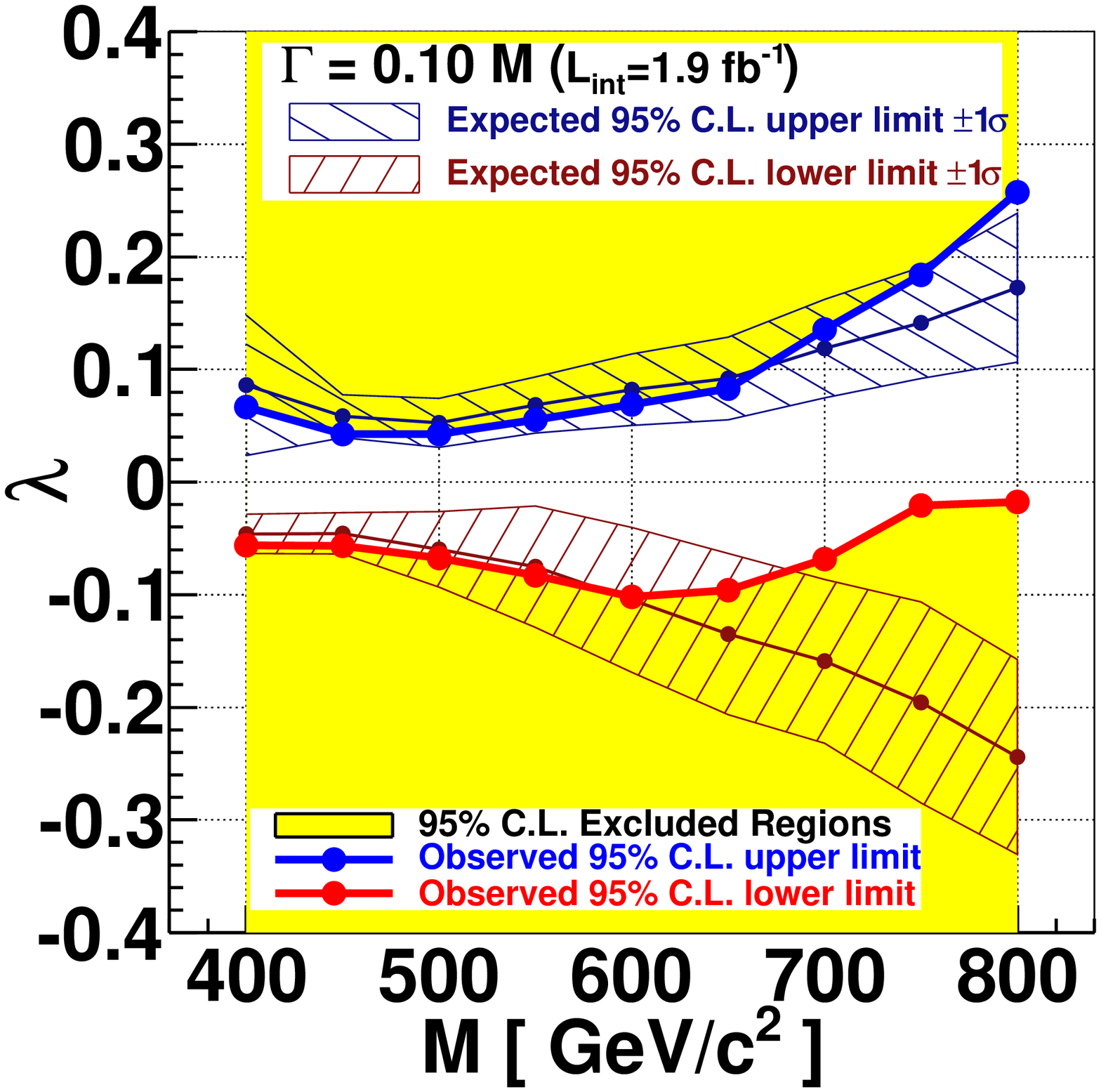}
  \includegraphics[width=0.48\linewidth]{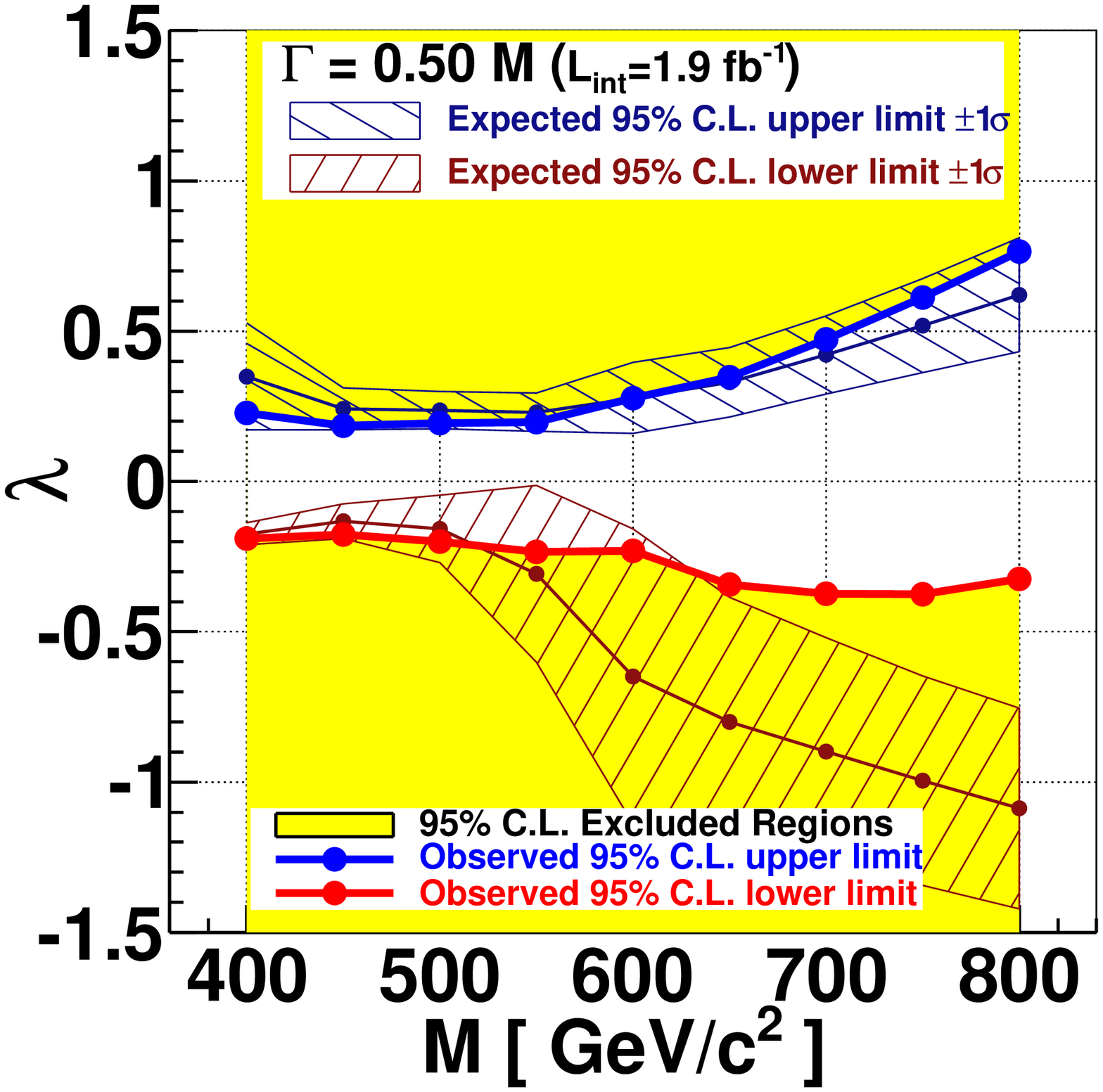}
 \caption{Limits on the coupling strength
   of a massive gluon, $G$, contribution deduced for various masses and two
   widths~\cite{Aaltonen:2009tx}.}
  \label{fig:mtt-massivegluon.limit}
\end{figure}

The distribution is then used to search for a new colour-octet particle, called
a massive gluon. 
An unbinned likelihood fit based on the production matrix elements with
and without massive gluon contribution  is used to extract the possible
coupling strengths of such a massive gluon contributing to the top quark  pair
production. The likelihood is computed for various masses and widths 
of the massive gluon.
The systematic uncertainties are incorporated in the likelihood
calculation. Jet energy scale and top mass uncertainties are the largest
contribution to the total uncertainty on the fitted coupling, $\lambda$.

The observed data agree with the Standard Model expectation within $\sim
1.7\sigma$. This agreement is cross-checked by
reconstructing the top quark $p_T$-distribution, which is also found  to be in
agreement with the Standard Model expectation.
Limits  on the possible coupling strength of a massive gluon, $G$, 
contributing to the top quark pair production are set at 95\%~C.L. for various values of the
width, $\Gamma_G$ as function of the mass,
$M_G$. Figure~\ref{fig:mtt-massivegluon.limit}
shows the expected and observed limits for two choices of the massive gluon
width. 

\subsubsection*{D\O}
%\subsubsubsection{Data Selection, Signal and Background Description}
D\O\ investigated the invariant mass distribution of top quark pairs in 
up to $3.6\ifb$ of lepton plus jets events~\cite{Abazov:2008ny,D0Note5600conf,D0Note5882conf}.  
The event selection  requires an isolated lepton, transverse missing momentum
and at least three jets. At least one of the jets needs to be identified as $b$ jet.
Signal simulation is created for various resonance masses between $350$
and $1000\GeV$. Separated resonance samples
were generated with couplings proportional to Standard Model $Z$ boson
couplings, with pure vector couplings and with pure axial-vector couplings.
The width of the resonances was chosen to be $1.2\%$
of their mass, which is much smaller than the detector resolution.

% \subsubsubsection{Top Pair Invariant Mass}
The top quark pair invariant mass, $M_{t\bar t}$, is reconstructed directly from the
reconstructed physics objects. A constrained kinematic fit
is not applied. Instead the momentum of the
neutrino is reconstructed from the transverse missing energy, $\met$,
which is identified with the transverse momentum of the neutrino and by solving
$M_W^2=(p_\ell+p_\nu)^2$ for the $z$-component of the neutrino
momentum. $p_\ell$ and $p_\nu$ are the four-momenta of the lepton and
the neutrino, respectively.

% \begin{figure}[b]
%   \centering
% \includegraphics[width=0.46\linewidth]{figs/Tagged__hmtt_nohf_alljets_3_ttbarZprimeall_comb.eps}
% \includegraphics[width=0.46\linewidth]{figs/Tagged__hmtt_nohf_4_ttbarZprimeall_comb.eps}
%   \caption{Shape comparison of the expected $t\bar t$ invariant mass distributions for 
% Standard Model top quark pair production (histogram)
% and 
% resonant  production from narrow-width resonances of mass 
% $M_X=450$, $650$, and $1000\GeV$, % (left to right), 
% for (a) $3$ jets events
% and (b)~$\ge 4$~jets events.}
%   \label{fig:mtt.mttshapes}
% \end{figure}
\begin{figure}[b]
  \centering
\includegraphics[width=.46\textwidth]{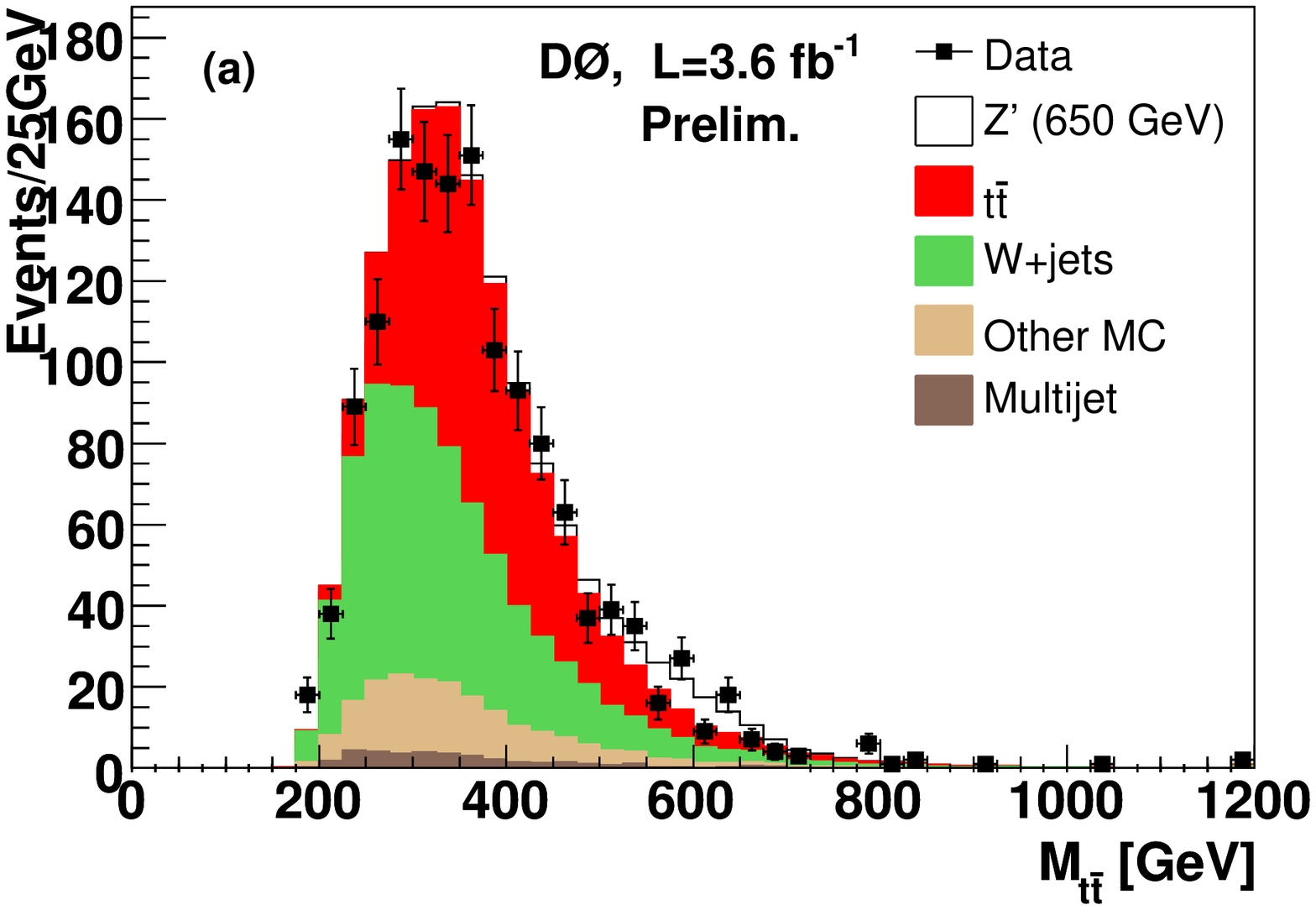}
\includegraphics[width=.46\textwidth]{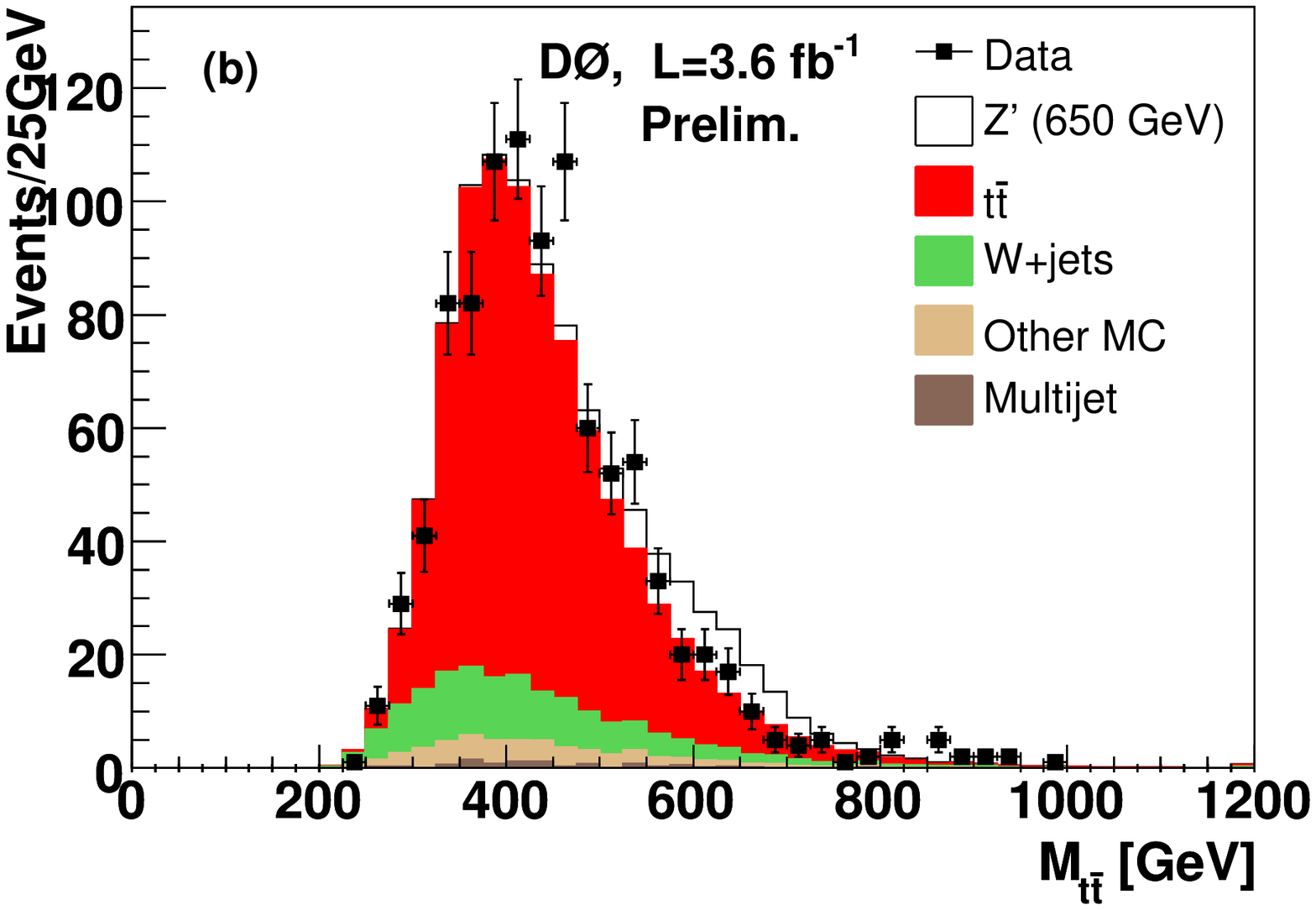}%
  \caption{Expected and observed $t\bar t$ invariant mass
  distribution for the combined (a) $\ell+3$~jets and (b) $\ell+4$ or more jets
  channels, with at least one identified $b$ jet.
Superimposed as white area is the expected signal for a
Topcolor assisted Technicolor $Z'$ boson with $M_{Z'}=650\GeV$~\cite{D0Note5882conf}.}
  \label{fig:mtt.mtt}
\end{figure}

The $t\bar t$ invariant mass can then be computed without any
assumptions about a jet-parton assignment that is needed in
constrained fits. Compared to the constrained fit reconstruction, applied
in an earlier analysis, this gives  better performance at high
resonance masses and in addition allows the inclusion of $\ell+$3 jets events.
%The expected signal shapes for various resonance
%masses are compared to the Standard Model top pair distribution in
%\fig{fig:mtt.mttshapes}.
The expected distribution of Standard Model processes and the measured data is
shown in \fig{fig:mtt.mtt}.  For comparison a resonance with a mass
of $650\GeV$ is shown at the cross-section expected in the Topcolor
assisted Technicolor model used for reference. 
%More details on the analysis procedure 
%can also be found in the {D\O} analysis notes~\cite{d0note5860,d0note5434}.

\begin{figure*}
  \centering
%  \includegraphics[width=.49\textwidth]{figs/compare_observed_wxsec_p17p20_nohf_4_nohf_alljets_3.eps}
%\includegraphics[width=.49\textwidth]{figs/T65F3.eps}%
%\\
\includegraphics[width=0.49\linewidth]{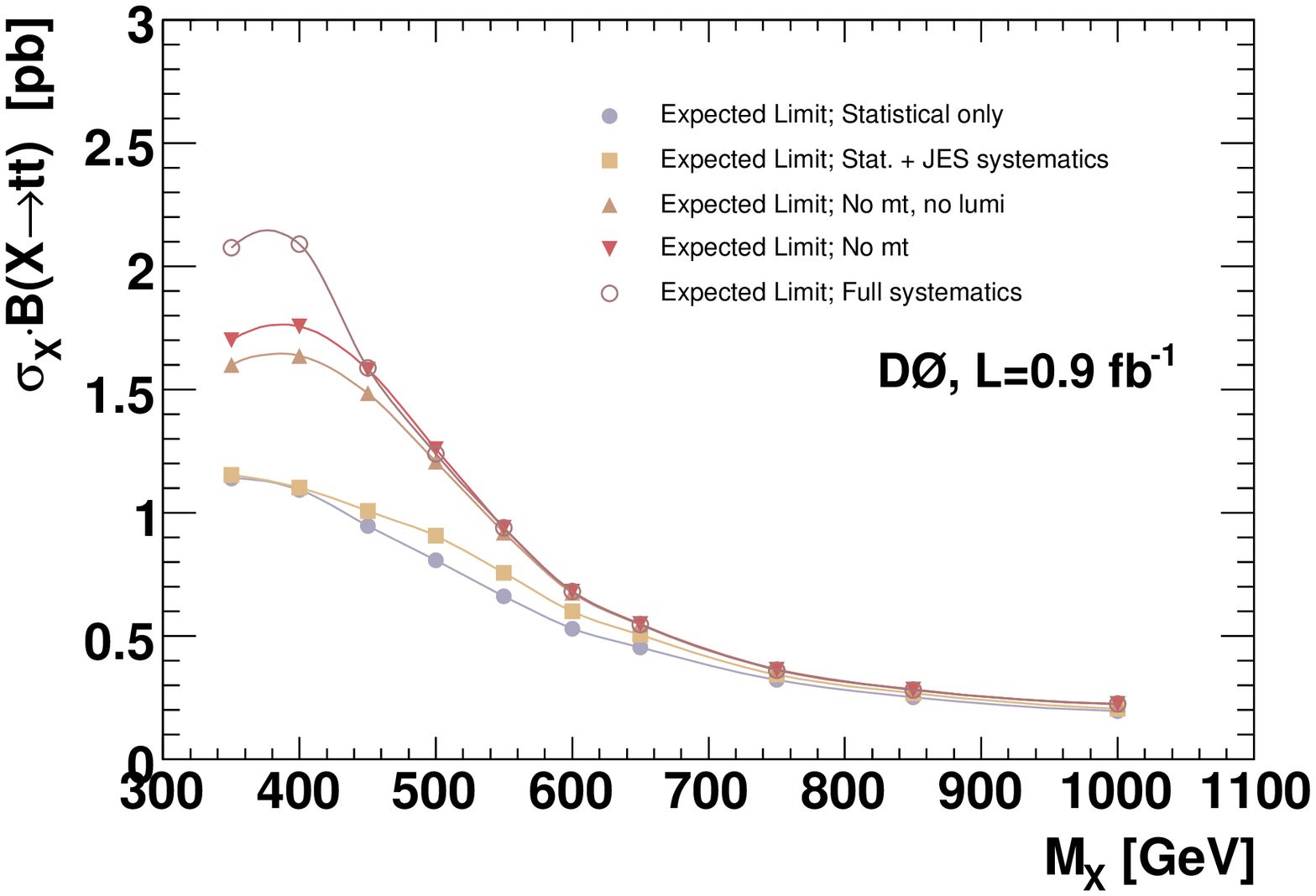}
\unitlength=0.01\linewidth
\begin{picture}(49,33)
\put(0,0){\includegraphics[width=.49\linewidth]{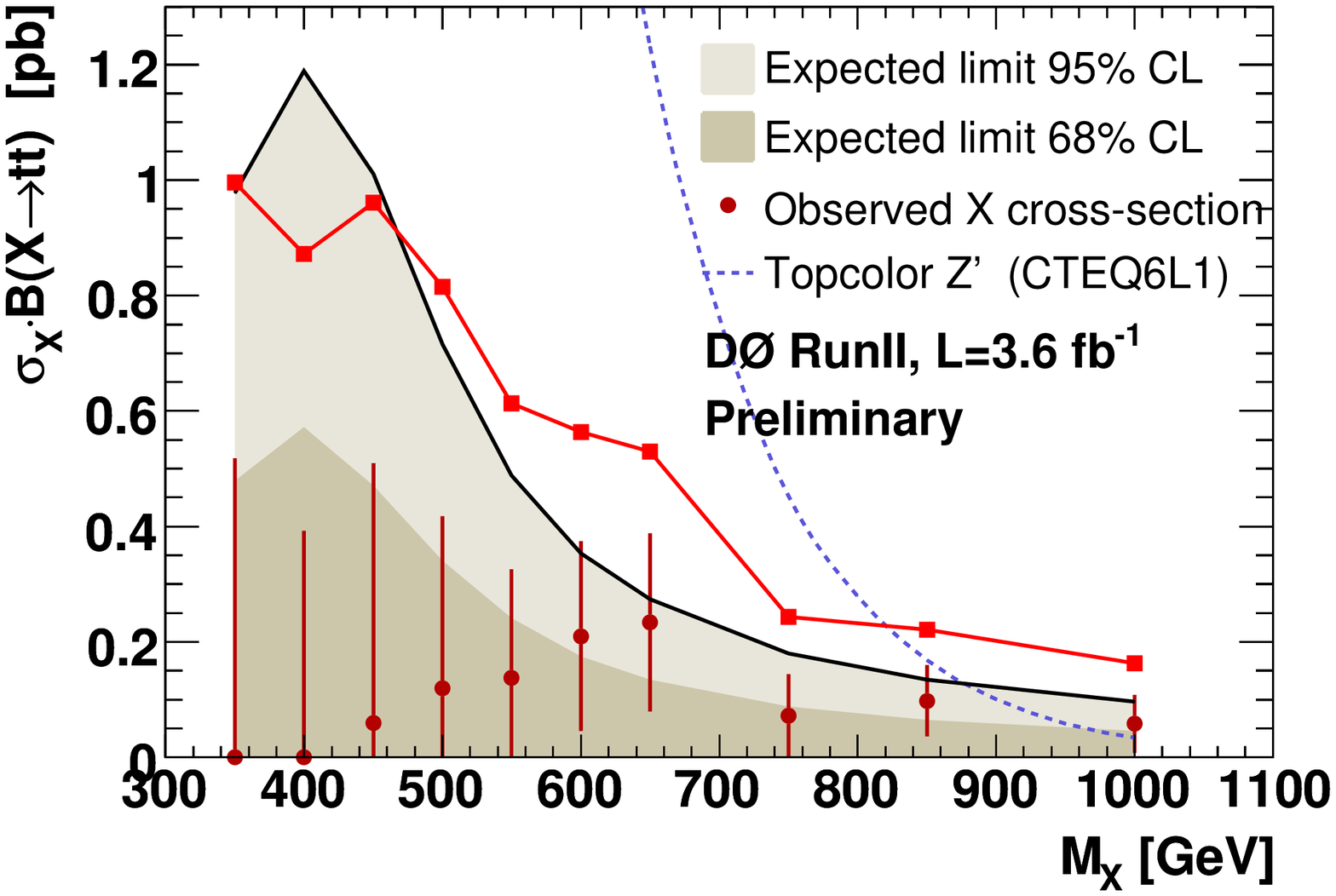}}
\put(25.5,23.5){\includegraphics[height=0.0775\linewidth,clip,trim=104mm 93mm 17mm 11mm]{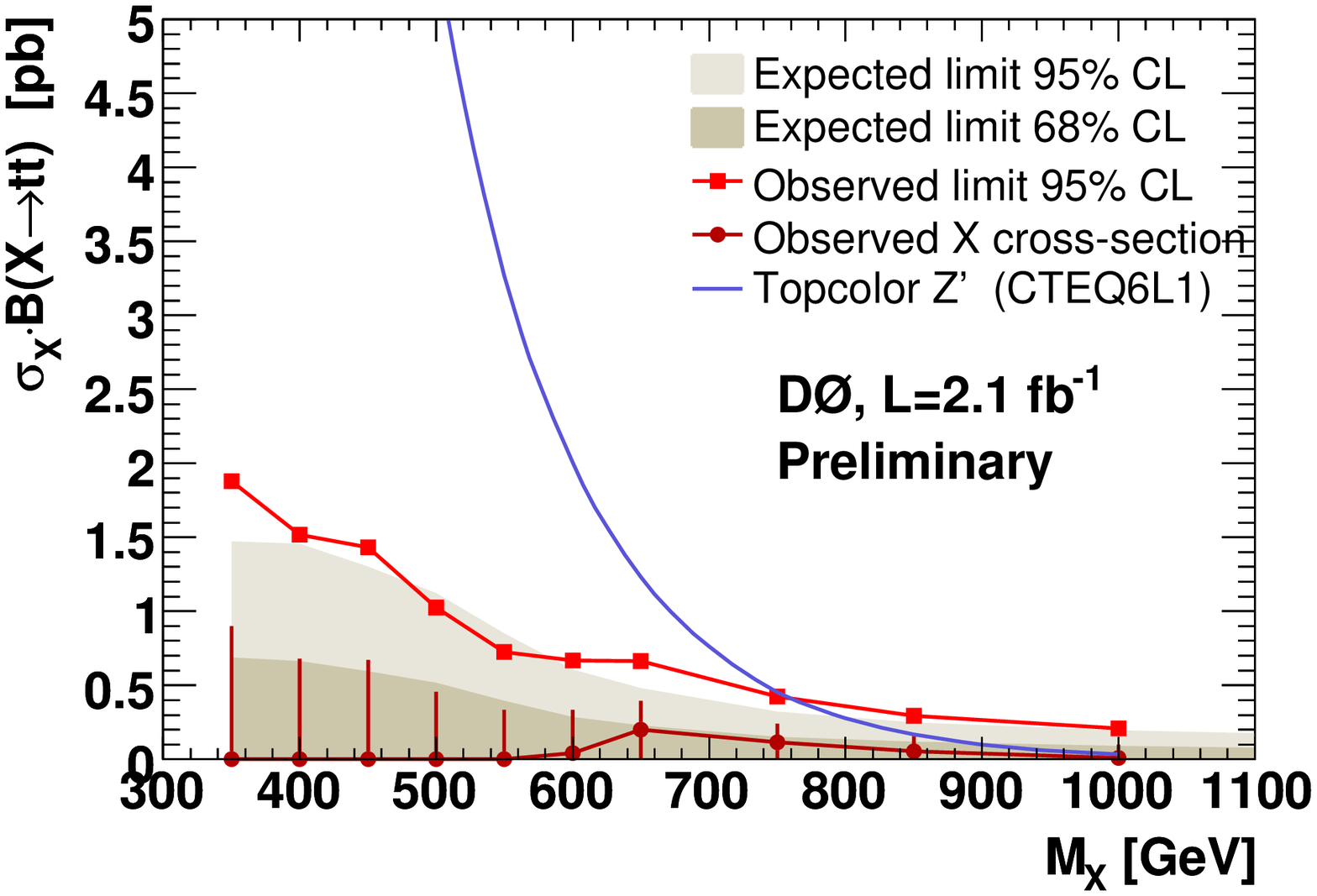}}
\end{picture}
  \caption{%
Left:
Expected limits  $\sigma_X\cdot B(X\rightarrow t\bar t)$ vs.\ 
the assumed resonance mass for $0.9\ifb$. From bottom to top the lines
represent the limit expected without systematics, including only JES
systematics, excluding selection efficiencies, $m_t$ and luminosity,
 all  except $m_t$ and complete systematics~\cite{Wicke:2008st}.
Right: 
Expected  upper limits with complete systematics in $3.6\ifb$
compared to the observed cross-section and exclusion limits at
95\%CL~\cite{D0Note5882conf}.
  }
 \label{fig:mtt.syst}
  \label{fig:mtt.result} 
\end{figure*}

%\subsubsubsection{Limit Calculation and Systematics}
Cross-sections for resonant production are evaluated with
Bayesian statistics 
using a non-zero flat prior (for positive values) of the resonant top quark
pair cross-section time branching fraction, $\sigma_X\cdot B(X\rightarrow t\bar t)$.
A Poisson distribution is assumed for the number of events observed in
each bin of the likelihood. The prior for the combined signal
acceptance and background yields is a multivariate Gaussian with
uncertainties and correlations described by a covariance matrix. 
The measured  $\sigma_X\cdot B(X\rightarrow t\bar t)$
correspond to the maximum of the Bayesian posterior probability density, limits
are set at the point where the integral of the posterior probability
density from zero reaches $95\%$ of its total. Expected limits are
obtained by applying the procedure when assuming that the observed
result corresponded to the Standard Model expectation.
The limits obtained for the $Z$-like,
vector and axial-vector resonances are found to agree with each other,
thus these limits are valid for a general (colour neutral)
narrow resonance. 

These expected limits were used to optimise major analysis cuts and the
$b$-tag working point. 
In \fig{fig:mtt.syst}~(left) the expected limits are
used to visualise the effect of the various systematics by including one
after another. The lowest curve corresponds to a purely statistical
limit. Adding the jet energy uncertainty shows that this uncertainty
mainly contributes at medium resonance masses. The various object
identification efficiencies and the luminosity are added. They
essentially scale like the background shape. Finally the effect of
the top quark mass is included and it is most important at low resonance masses~\cite{Wicke:2008st}.

%\subsubsubsection{Results}

In $3.6\ifb$ of {D\O} data the observed cross-sections are close to zero for all considered
resonance masses, as shown in~\fig{fig:mtt.result}~(right). The largest
deviation (around $650\GeV$) corresponds to a little more than $1.5$ standard deviations. Thus limits
are set on the $\sigma_X\cdot B(X\rightarrow t\bar t)$ as function of
the assumed resonance mass, $M_X$. The excluded values range from
 about $1\pb$ for low mass resonances to less then $0.2\pb$ for the highest
considered resonance mass of $1\TeV$. 
The benchmark Topcolor assisted Technicolor model can be excluded for
resonance masses of $M_{Z'}<820\GeV$ at $95\%$CL.

%%% Local Variables: 
%%% mode: latex
%%% TeX-master: "EPJC_TopProperties"
%%% End: 

\subsection{Admixture of Stop Quarks}
A final very fundamental question that may be asked in the context of top
quark physics beyond the Standard Model is, whether the events that are
considered to be top quarks  actually are 
all top quarks or whether some additional unknown new particle is hiding in
the selected data. 
The top quark's supersymmetric partners, the stop quarks $\tilde{t}_1$ and
$\tilde{t}_2$, are possible candidates in such a scenario. 

\subsubsection{D\O, Lepton plus Jets}
The stop quark  decay modes to neutralino and top quark, $\tilde\chi_1^0 t$, or through chargino and
$b$ quark, $\tilde\chi_1^\pm b$, both yield a final state with a neutralino,
a $b$ quark and a $W$ boson,
$\tilde\chi_1^0 b W$. The  neutralino is the lightest supersymmetric
particle in many models and is stable if $R$-parity is conserved. 
Then it escapes the detector and the experimental signature of  stop quark  pair
production differs from that of semileptonic top quark pair events only by the 
additional contribution to the  missing transverse energy from the
neutralino.
%\beq
% \tilde\chi^\pm_1  \rightarrow  \tilde\chi^0_1+ W^\pm\rightarrow\tilde\chi^0_1+\ell +\nu 
%\eeq

%\subsubsubsection{Data Selection, Signal and Background Description}\label{sect:stop.selection}
{D\O} has searched for a contribution of such stop quark  pair production in
the semileptonic channel in data with $0.9\ifb$~\cite{d0note5438conf,Abazov:2009ps}. Semileptonic
events were selected following the corresponding $t\bar t$
cross-section analysis by looking for isolated leptons ($e$ and $\mu$),
missing transverse energy and four or more jets. At least one of the jets
was required to be identified as $b$-jet using {D\O}'s neural network algorithm. 

To describe the Standard Model expectation a mixture of data and simulation
is employed.  The description of top quark  pair production (and of further
minor backgrounds) is taken fully from simulation normalised to the
corresponding theoretical cross-sections. For $W+$jets the kinematics
is taken from simulation, but the normalisation is taken from data. 
Multijet background is fully estimated from data.  
As signal the lighter of the two stop quarks, $\tilde{t}_1$, is considered.
Production of $\tilde{t}_1 \bar{\tilde{t}}_1$ is simulated for various
combinations of stop quark  and chargino masses, $m_{\tilde{t}_1}, m_{\tilde\chi^\pm_1}$. For the sake of this
analysis the stop quark  mass was assumed to be less or equal to the top quark  mass. The
neutralino mass $m_{\tilde\chi^0_1}=50\GeV$ was chosen to be close
to the experimental limit.

%\subsubsubsection{Signal Background Separation} 
%
\begin{figure*}
  \centering
\includegraphics[width=0.45\linewidth]{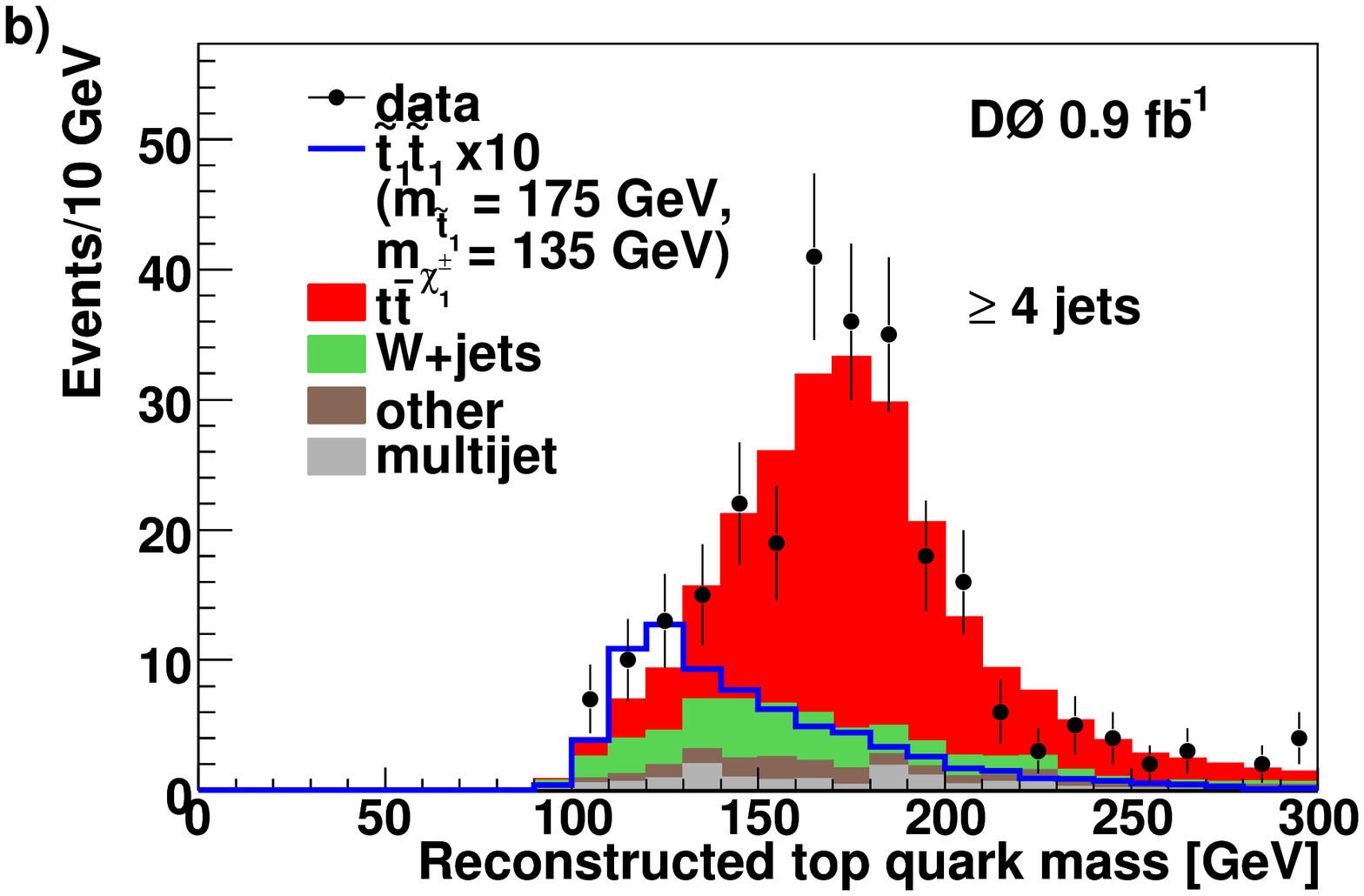}
\includegraphics[width=0.45\linewidth]{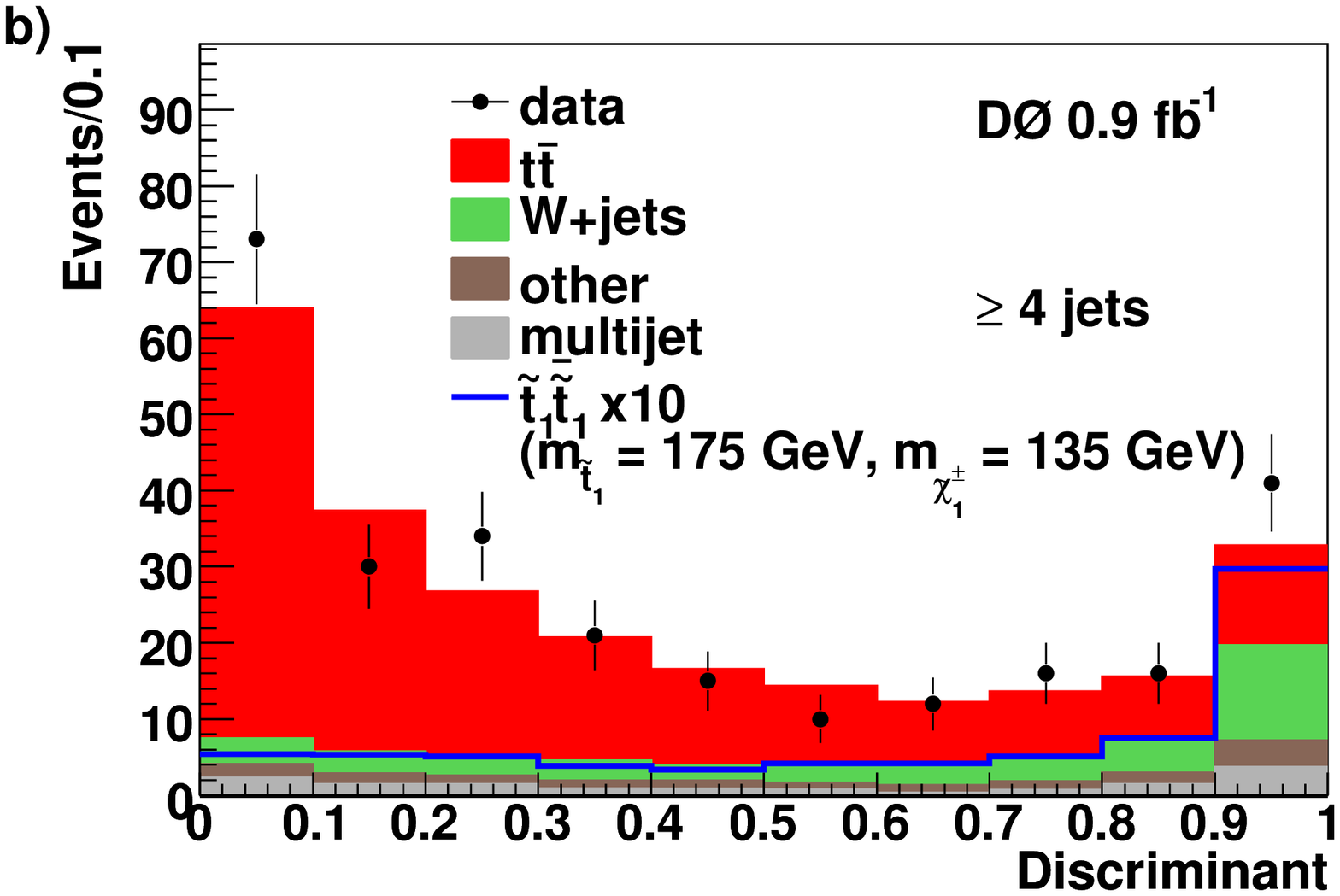}%
  \caption{\label{fig:stop.likelihood}%
Left: Expected and observed distribution for the reconstructed top quark mass.
Right: Expected likelihood distribution
for Standard Model and signal compared to data.  In both plots only events with four or
more jets are shown. The stop quark 
contribution  corresponds  to ten times the expectation  in the MSSM with $m_{\tilde{t}_1}=175\GeV$,
$m_{\tilde\chi_1^\pm}=135\GeV$~\cite{Abazov:2009ps}.
}
\end{figure*}
\begin{figure*}
  \centering
\includegraphics[width=0.32\linewidth]{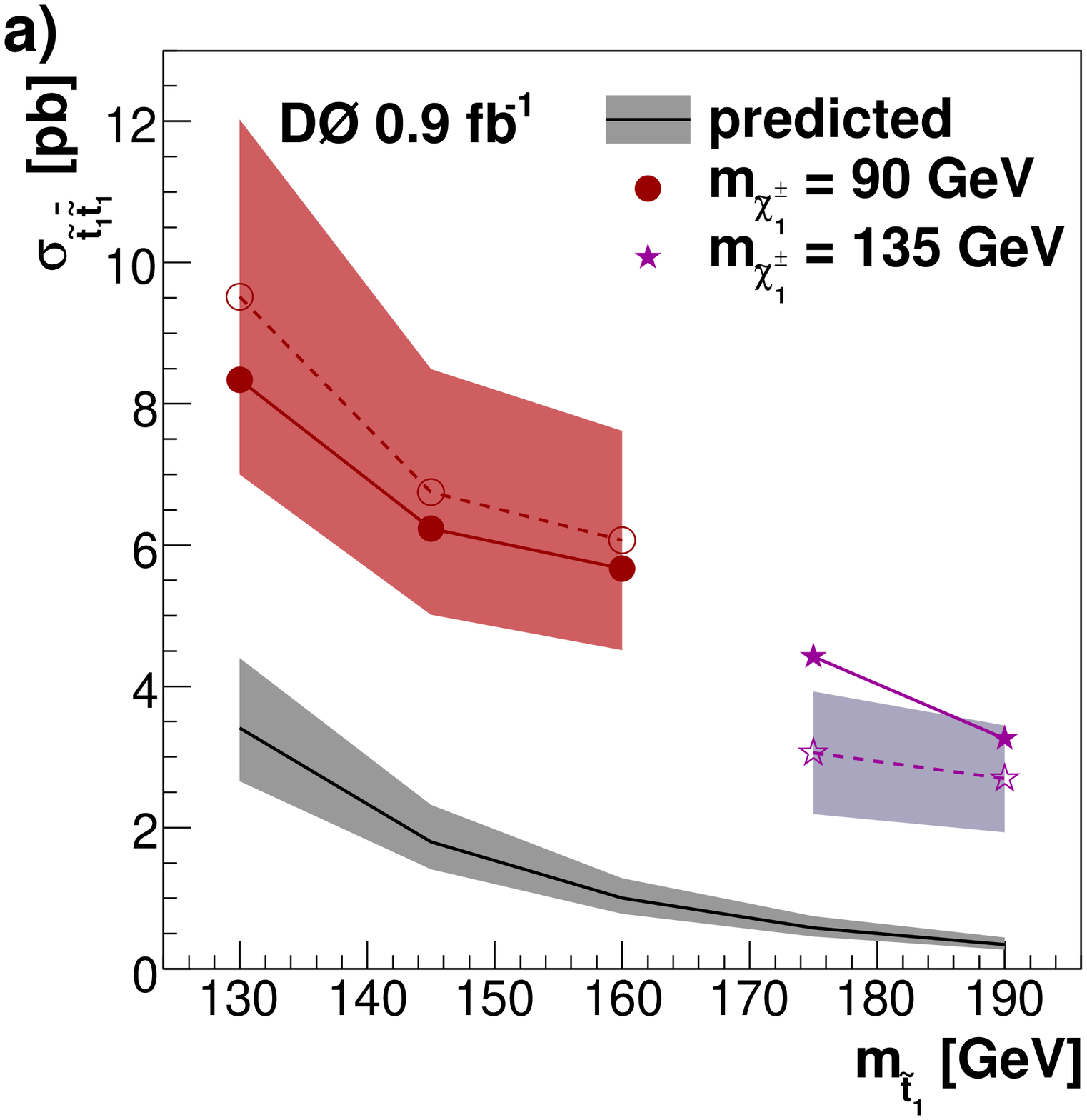}
\includegraphics[width=0.32\linewidth]{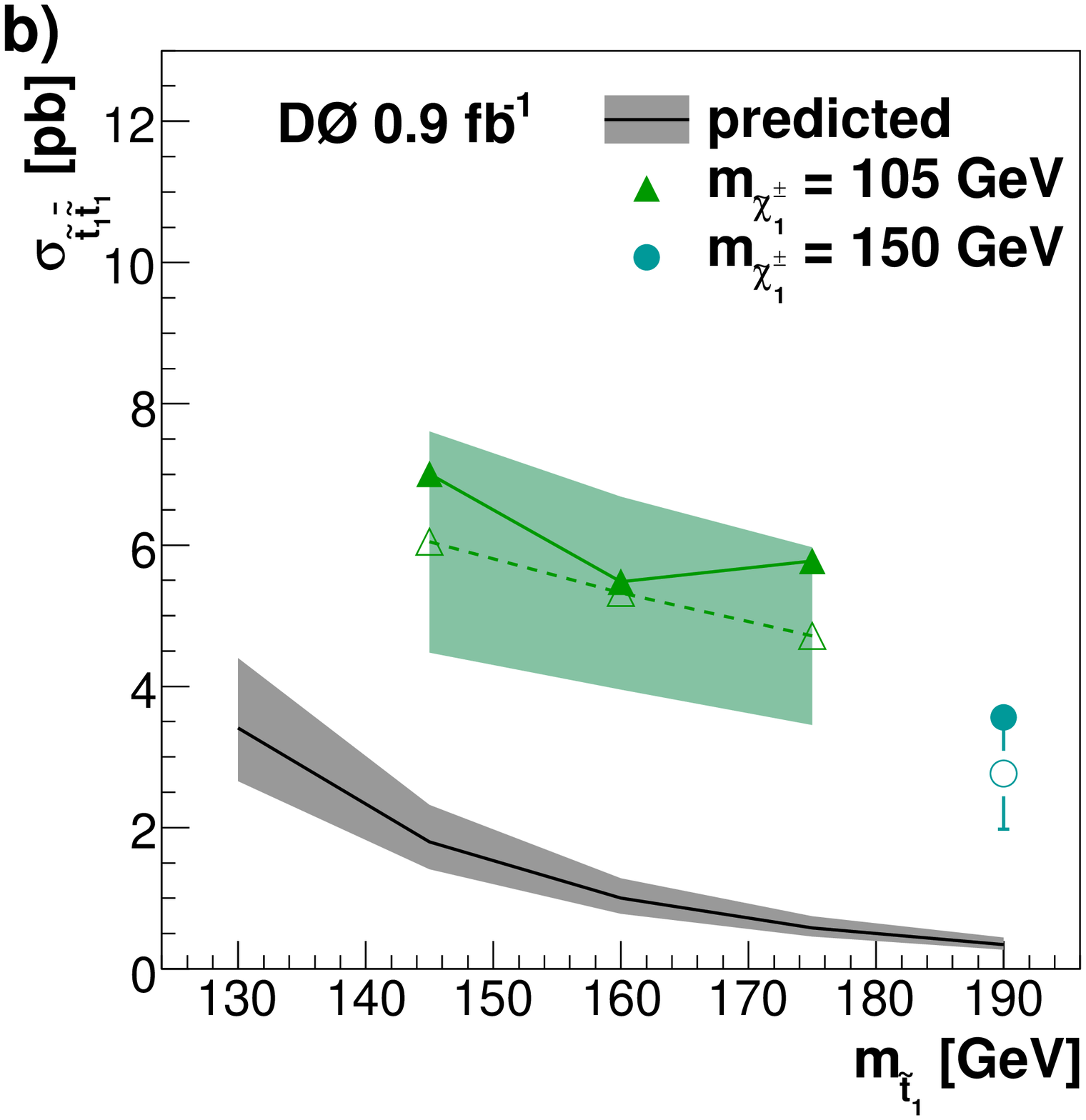}
\includegraphics[width=0.32\linewidth]{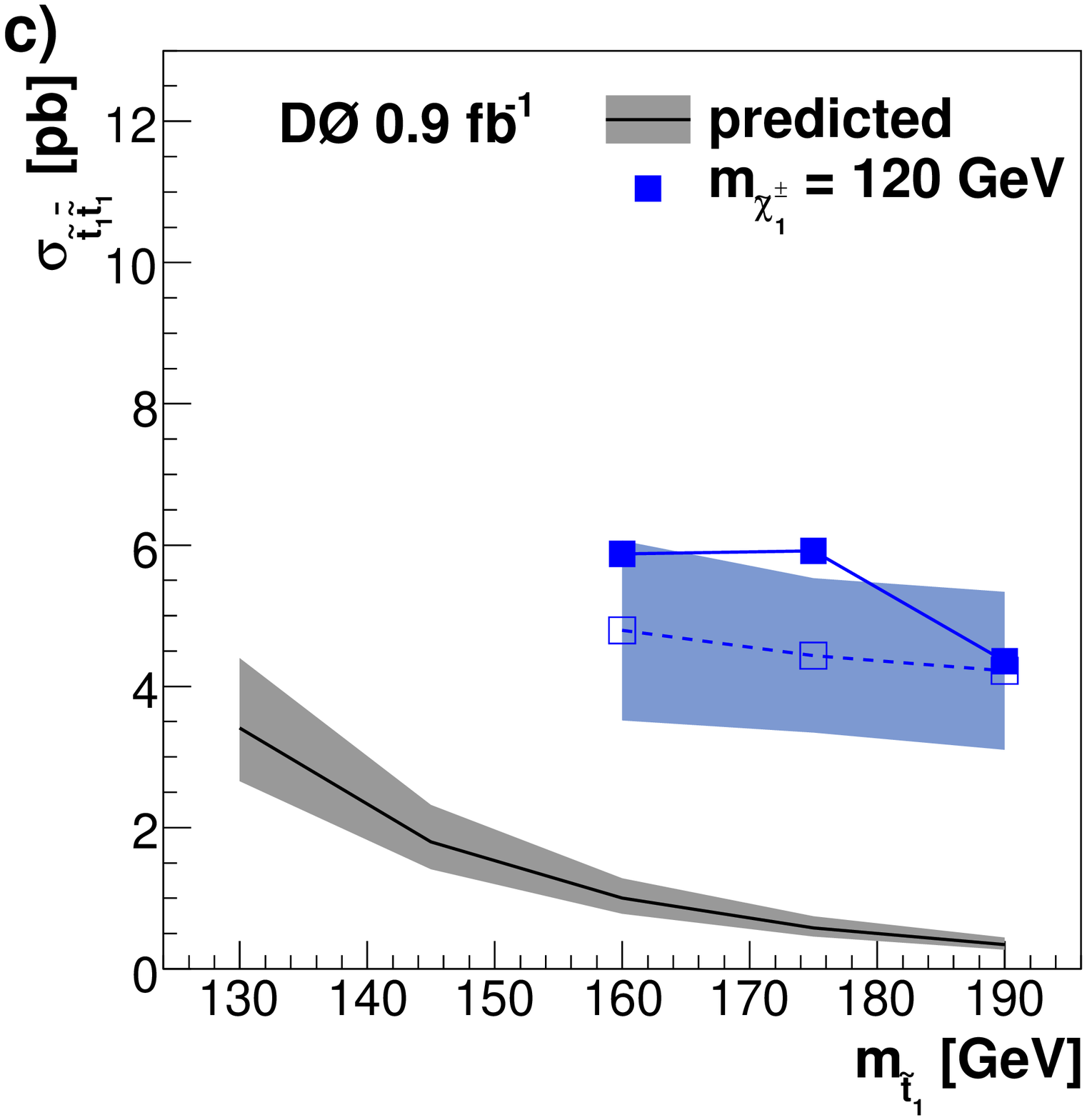}
  \caption{\label{fig:stop.limits}Expected and observed limits on the stop quark 
    pair production cross-section compared to the expectation in the
    various MSSM parameter sets~\cite{Abazov:2009ps}.}
\end{figure*}
To detect a possible contribution of
stop quark  pairs the differences between stop quark  pair events and Standard Model top quark  pair production  
kinematic variables are combined into a likelihood, 
${\cal  L}=P_\mathrm{stop}/\left(P_\mathrm{stop}+P_\mathrm{SM}\right)$.
The kinematic variables considered include the transverse momentum of
the (leading) $b$ jet, distances between leading $b$ jet and lepton or
leading other jet. Additional variables were reconstructed by applying
a constrained fit.  In this fit reconstructed physics objects (lepton,
missing transverse energy and jets) are assigned to
the decay products of an assumed semileptonic top quark  pair event and 
the measured quantities are allowed to vary within their experimental
resolution to fulfil additional constraints. It was required that the $W$
boson mass is consistent with the invariant mass
of the jets assigned to the two light quarks as well as with the
invariant mass of the lepton with the neutrino.
The masses of the reconstructed top quark  pairs were constrained to be equal.
Of the possible jet parton assignments only the one with the best
$\chi^2$ was used. 
From the constrained fits observables the angle between the $b$-quarks and the
beam axis in the $b\bar b$ rest frame, the $b\bar b$ 
invariant mass, the distances between the $b$'s and
the same-side or opposite-side $W$ bosons and the reconstructed top quark 
mass are considered. % discriminant.
The likelihood was derived  for each $m_{\tilde{t}_1},
m_{\tilde\chi^\pm_1}$ combination separately and the selection of variables used
has been optimised each time. Figure~\ref{fig:stop.likelihood}
shows the separation power of the reconstructed top quark mass and the full likelihood for the case of
$m_{\tilde{t}_1}=175\GeV$, $m_{\tilde\chi^\pm_1}=135\GeV$ and the comparison to the observed data.

To determine limits on the possible contribution of stop quark  pair
production in the selected channel  Bayesian statistics is employed
using a non-zero flat prior for positive values of the stop quark  pair cross-section.
A Poisson distribution is assumed for the number of events observed in
each bin of the likelihood. The prior for the combined signal
acceptance and background yields is a multivariate Gaussian with
uncertainties and correlations described by a covariance matrix. 
The systematic uncertainty is dominated by the uncertainties on the theoretical
cross-section of top quark  pair production, on the selection efficiencies 
and the luminosity determination. 
Figure~\ref{fig:stop.limits} shows the expected and observed limits
on the stop quark  pair production cross-section compared to the MSSM
prediction for various values of $m_{\tilde{t}_1}$ and $m_{\tilde\chi^\pm_1}$. 
%While the $e+$jets channel shows agreement between observed and
%expected limits, the observed limits in the $\mu+$jets channel are
%weaker than the expectations. 
The theoretically expected
stop quark  signal cross-section in the MSSM is smaller than the
experimental limits for all parameter points considered.

\subsubsection{CDF, Dilepton}
%\subsubsubsection{Data Selection, Signal and Background description}

CDF has searched for a contribution of stop quarks in the dilepton channel
using upto $2.7\ifb$ of data~\cite{CdfNote9439}. The dilepton event signature was chosen to cover
chargino decay modes to $\ell+\nu$ that do not involve an intermediate $W$
boson and thus may not have a corresponding hadronic decay to build
semileptonic events. 
\bea
\begin{array}{rclcl}
 \tilde\chi^\pm_1 & \rightarrow & \tilde\chi^0_1+ H^\pm&\rightarrow&\tilde\chi^0_1+\ell +\nu \\
 \tilde\chi^\pm_1 & \rightarrow &\ell+\tilde\nu_\ell &\rightarrow&\tilde\chi^0_1+\ell +\nu \\
 \tilde\chi^\pm_1 & \rightarrow & \tilde\ell+\nu &\rightarrow&\tilde\chi^0_1+\ell +\nu \\
 \tilde\chi^\pm_1 & \rightarrow & \tilde\chi^0_1+ G^\pm&\rightarrow&\tilde\chi^0_1+\ell +\nu \\
\end{array}
\eea

CDF collected data using an inclusive high-$p_T$ trigger and
selects events with 2 leptons, one of which needs to be isolated from calorimeter energies not
associated to that lepton. The events are required to have missing transverse
energy and at least 2 jets. A $Z$ boson veto is applied for $ee$ and $\mu\mu$
events. 
To suppress the leading background, Standard Model top quark  pair events, from the selected event a cut in the plane of
$H_T$ vs. $\Delta$ plane is applied, where $\Delta$
is the product of the azimuthal angles between the leading jets and the two
leptons:
$\Delta=
\Delta\phi(\mathrm{jet}_1,\mathrm{jet}_2)\Delta\phi(\ell_1,\ell_2)$.
This cut reduces top quark  pair production by a factor of 2, but reduces stop quark  by
approximately $12\%$ only.

The Standard Model background expectation is modelled using simulation and control
data.  Simulation of top quark  pair production and other minor backgrounds are 
normalised to their NLO cross-section. $Z+$jets samples are normalised to
control data of low missing transverse energy near the $Z$-pole separately for
events with and without $b$-tags. 
To model events with faked leptons that may stem from other top quark  pair decay
channels, from $W+$jets or from multijet events, parametrised lepton fake
rates are derived from a large sample of generic jets. These fake rates are
applied to events with lepton plus electron or muon like events to find the
contribution of fakes in the signal region.

To describe signal events stop quark  pair production is simulated with various
combinations of neutralino, chargino and stop quark  masses lighter than the top quark 
mass. Generated samples are interpolated through a template morphing technique
to obtain any combination of masses within the generated range.

%\subsubsubsection{Signal Background Discrimination}
The reconstructed stop quark  mass is used to distinguish a stop quark  signal from Standard Model
backgrounds including top quark  pair production. The stop quark  mass, $m_{\tilde{t}_1}$,
is determined following an extention of the dilepton neutrino weighting
technique. 

\begin{figure}
  \centering
  \includegraphics[width=0.49\textwidth]{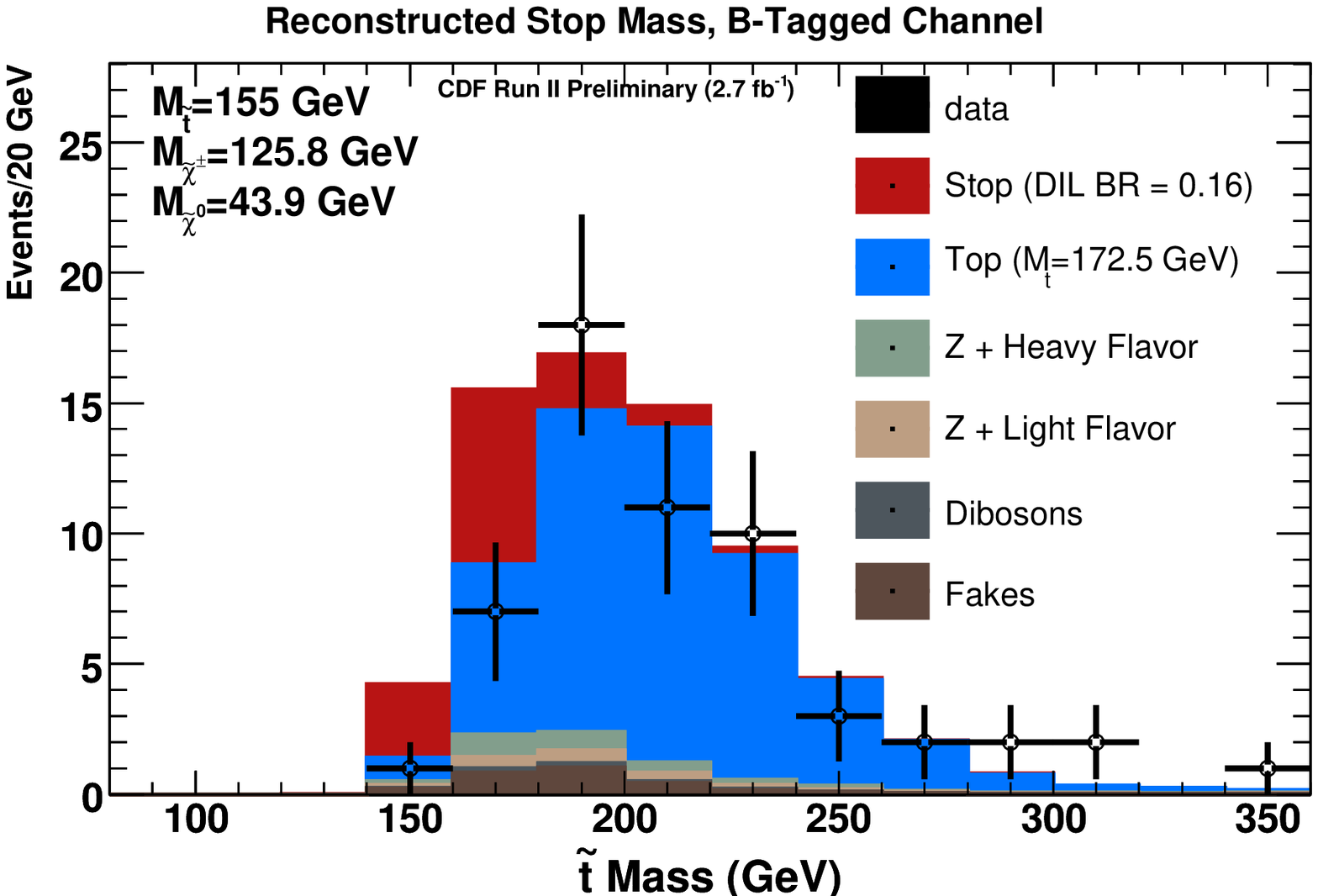}
  \includegraphics[width=0.49\textwidth]{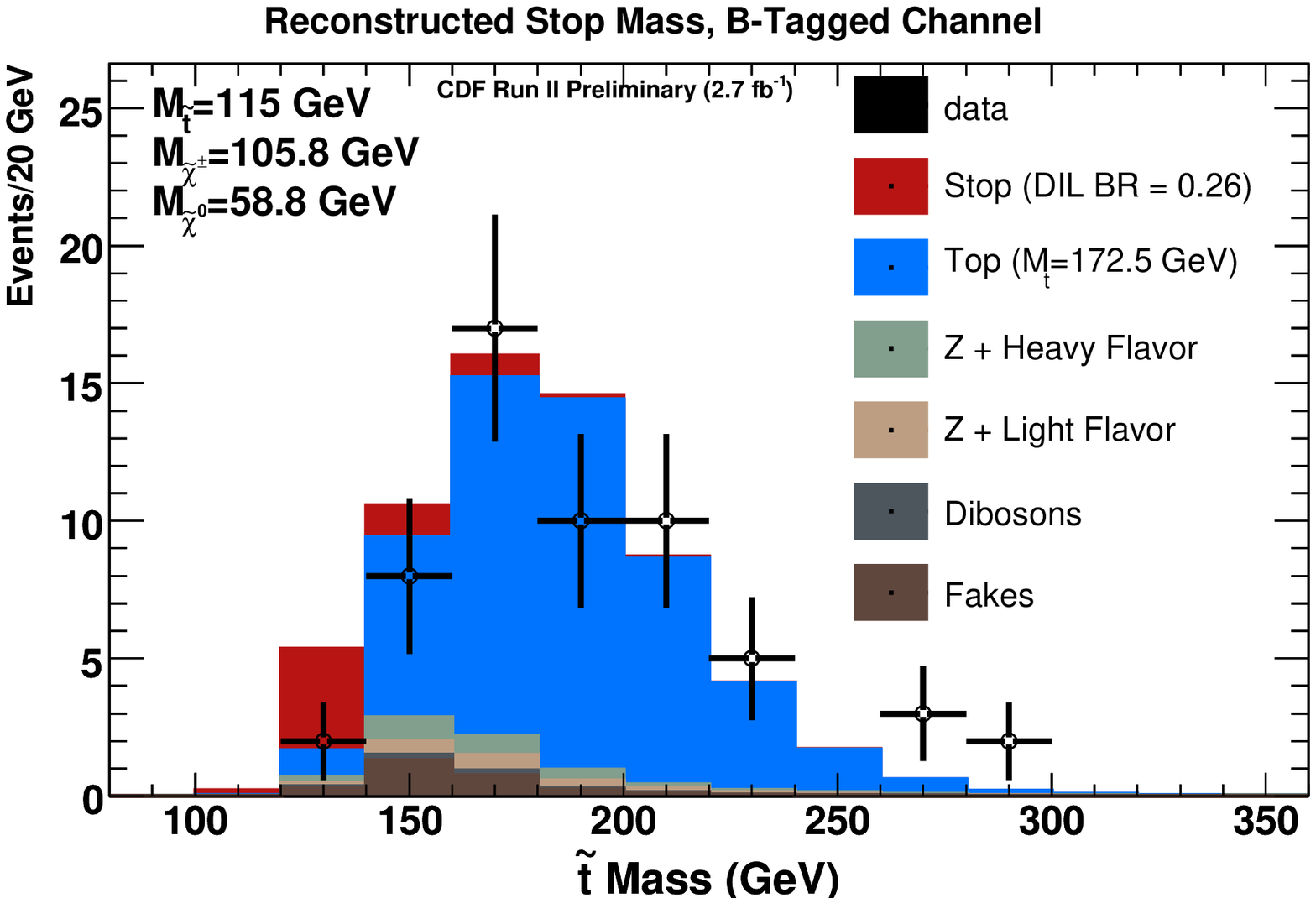}
  \caption{Distribution of reconstructed stop quark  masses in data and simulation
    for two choices of the stop quark, chargino and neutralino masses~\cite{CdfNote9439}.}
  \label{fig:cdf-stop-masss}
\end{figure}
$b$-jets are assigned to their proper lepton  based on jet-lepton 
invariant mass
quantities. 
A correct assignment is reached in $85\%$ to $95\%$ of the cases
with both $b$-jets being the leading 2 jets.
Neutralino and neutrino are considered as a single, though massive, pseudo
particle. 
For given $\phi$ directions of the pseudo particles the particle momenta are
determined with a fit to the measured quantities using constraints on
the assumed pseudo particle mass, the assumed chargino mass and the equality
of the two stop quark  masses. The reconstructed stop quark  mass is computed as
weighted average of the fitted stop quark  masses, where the average is computed over all
values of $\phi$, with weights of  $e^{-\chi^2}$.
The expected and observed distributions for two choices of 
parameters are shown in~\fig{fig:cdf-stop-masss}.

%\subsubsubsection{Limits}
The combination of reconstructed stop quark  mass templates from the signal and the
various background components is fitted to data. Systematic uncertainties
enter the fit through nuisance parameters, signal and background contributions
are allowed to vary within their rate 
uncertainties and the shape may vary according to CDFs morphing technique. 
The ratio of the likelihoods is used to do the limit-setting according to the
CL$_s$ technique~\cite{Junk:1999kv,Read-in-James:2000et}.

\begin{figure*}
  \centering
  \includegraphics[width=0.49\textwidth]{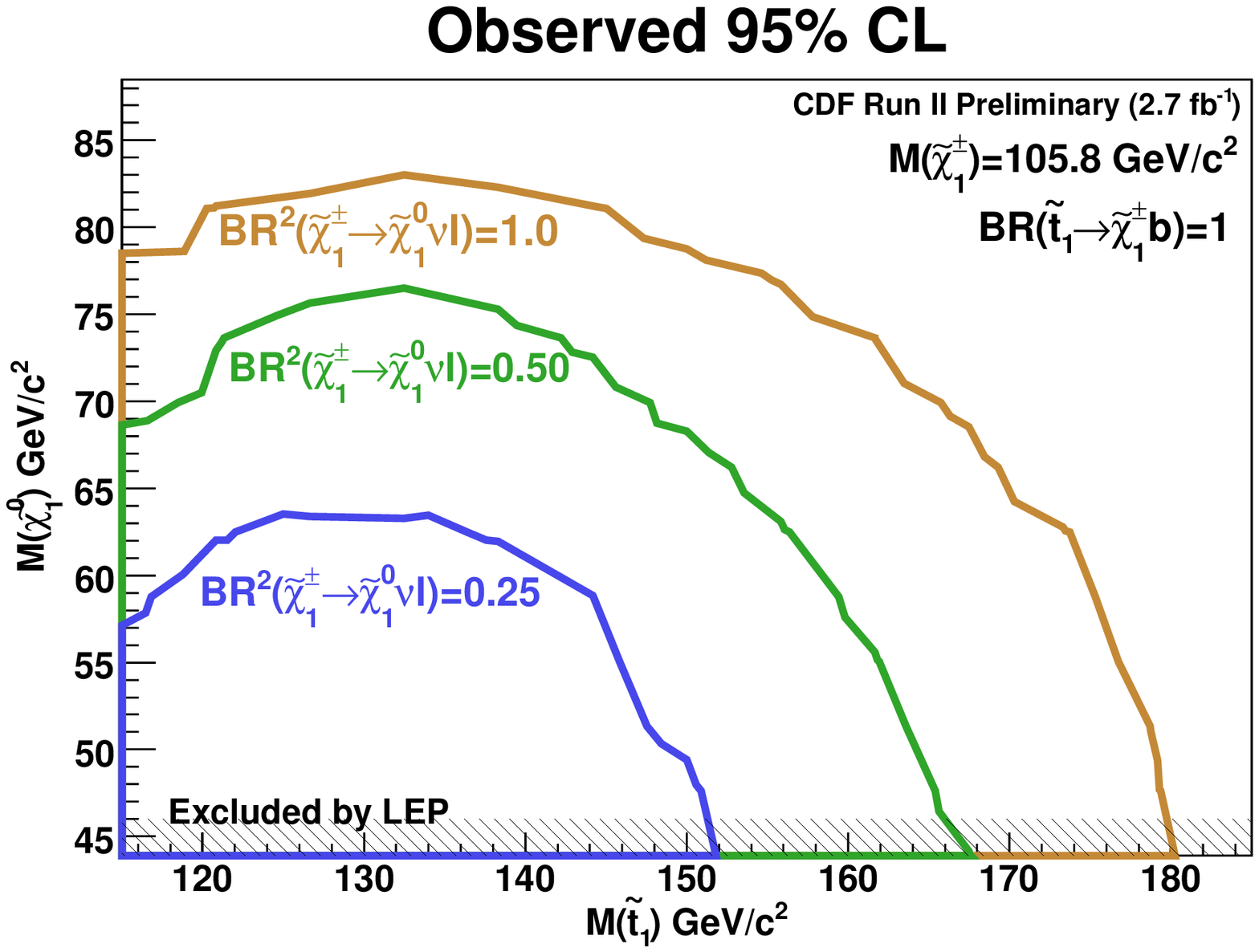}
  \includegraphics[width=0.49\textwidth]{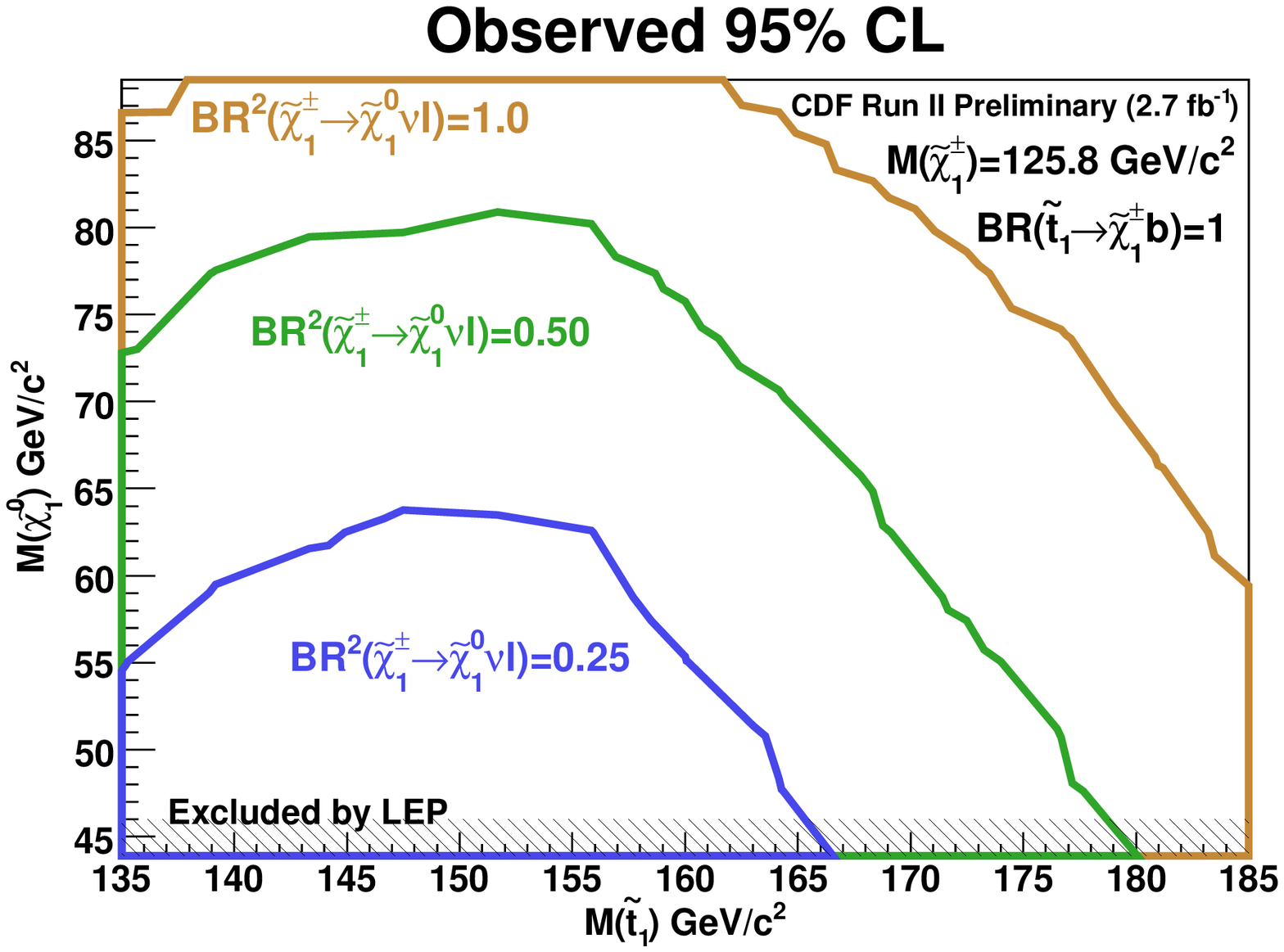}
  \caption{CDF observed 95\%CL limits in the stop mass vs. chargino mass
    plane. The left plot shows limits for a chargino mass of 105.8\GeV, the
    right for 125.8\GeV~\cite{CdfNote9439}.}
  \label{fig:cdf-stop-limits}
\end{figure*}

Depending on the dilepton branching ratio limits are set in the stop quark  vs.
neutralino mass plane. Figure~\ref{fig:cdf-stop-limits} shows the results for
two choices of parameters. These limits are
derived using only very few assumptions, these are
(a)~$\tilde\chi_1^0$ is the LSP and $\tilde q$, $\tilde\ell$ and $\tilde \nu$
  are heavy,
(b)~$m_{\tilde{t}_1}\lesssim m_t$ and
(c)~$m_{\tilde\chi_1^\pm}< m_{\tilde t_1}- m_b$.
Thus the limits are valid over a large range of SUSY parameter space.

%%% Local Variables: 
%%% mode: latex
%%% TeX-master: "EPJC_TopProperties"
%%% End: 

\subsection[Heavy Top-like Quark, $t'$]{\boldmath Heavy Top-like Quark, $t'$}
Another class of hypothetical particles that may hide in samples usually considered
as top quarks are new heavy quarks, in this context usually called $t'$~quark. 
These new particles a considered to decay to $Wq$ and thus show a
signature very similar to that of top quarks. 

Heavy top-like quarks appear in a large number of new physics models: A fourth
generation of fermions~\cite{Kribs:2007nz}, Little Higgs models~\cite{Han:2003gf} and
more named in~\cite{cdf:2008nf}. Strong bounds are placed on such models by
electroweak precision data, but 
for special parameters the effects of the fourth generation particles
on electroweak observables compensate. Among other settings a small mass
splitting between the fourth $u$-type quark, $t'$, and its isospin partner,
$b'$, is preferred, i.e. $m_{b'}+M_W>m_{t'}$~\cite{Kribs:2007nz}.
Especially, when the new top-like quark is
very heavy, it should be distinguishable from Standard Model top production
in kinematic observables. A search for such
heavy top-like quarks has been performed by both Tevatron experiments.

\subsubsection*{CDF}
The CDF collaboration has repeatedly analysed their samples of lepton plus
jets events to search for a new top-like quark. The published result uses
$760\ipb$~\cite{cdf:2008nf} and the preliminary updated
results $4.6\ifb$ of data~\cite{CdfNote10110}. 
The analyses consider pair production of a new quark 
heavier than the top quark and with a subsequent decay to $Wq$. 
The event selection requires exactly one isolated lepton,
large missing transverse energy and at least four energetic jets. 

The dominant Standard Model processes that pass this selection are top quark pair
production which is simulated using \pythia, $W+$jets events which are
simulated using \alpgen+\herwig\ or \pythia\ and normalised to data and multijet events
which are modelled from data with reversed lepton identification. Minor
backgrounds like $Z+$jets and diboson events are considered to be described
by the $W+$jets simulation. 
The simulation of $t'\bar t'$ signal is performed using \pythia.

\begin{figure*}[t]
  \centering
  \includegraphics[width=0.44\textwidth,trim=0mm 3mm 0mm 7mm]{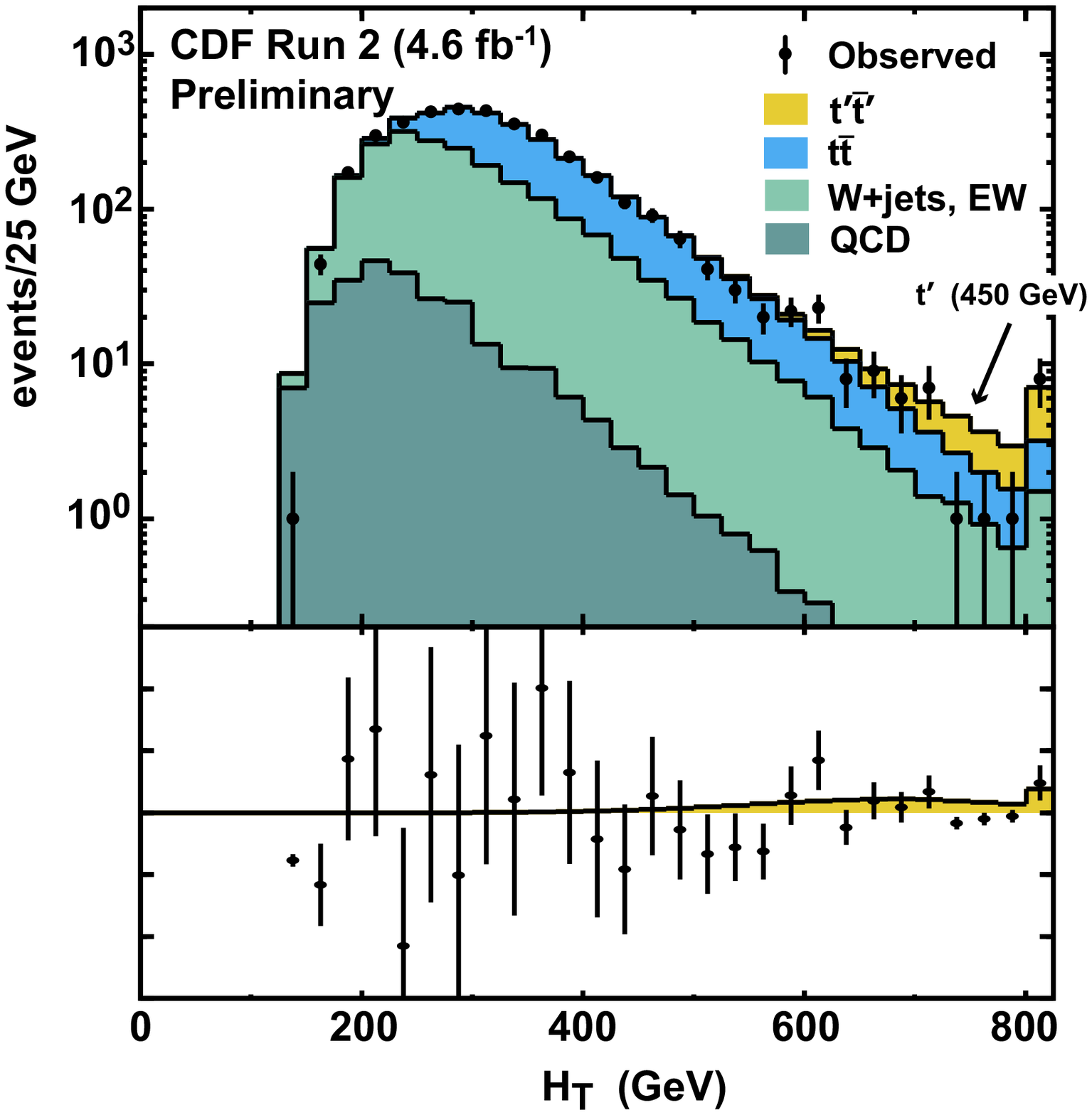}
  \includegraphics[width=0.44\textwidth,trim=0mm 3mm 0mm 7mm]{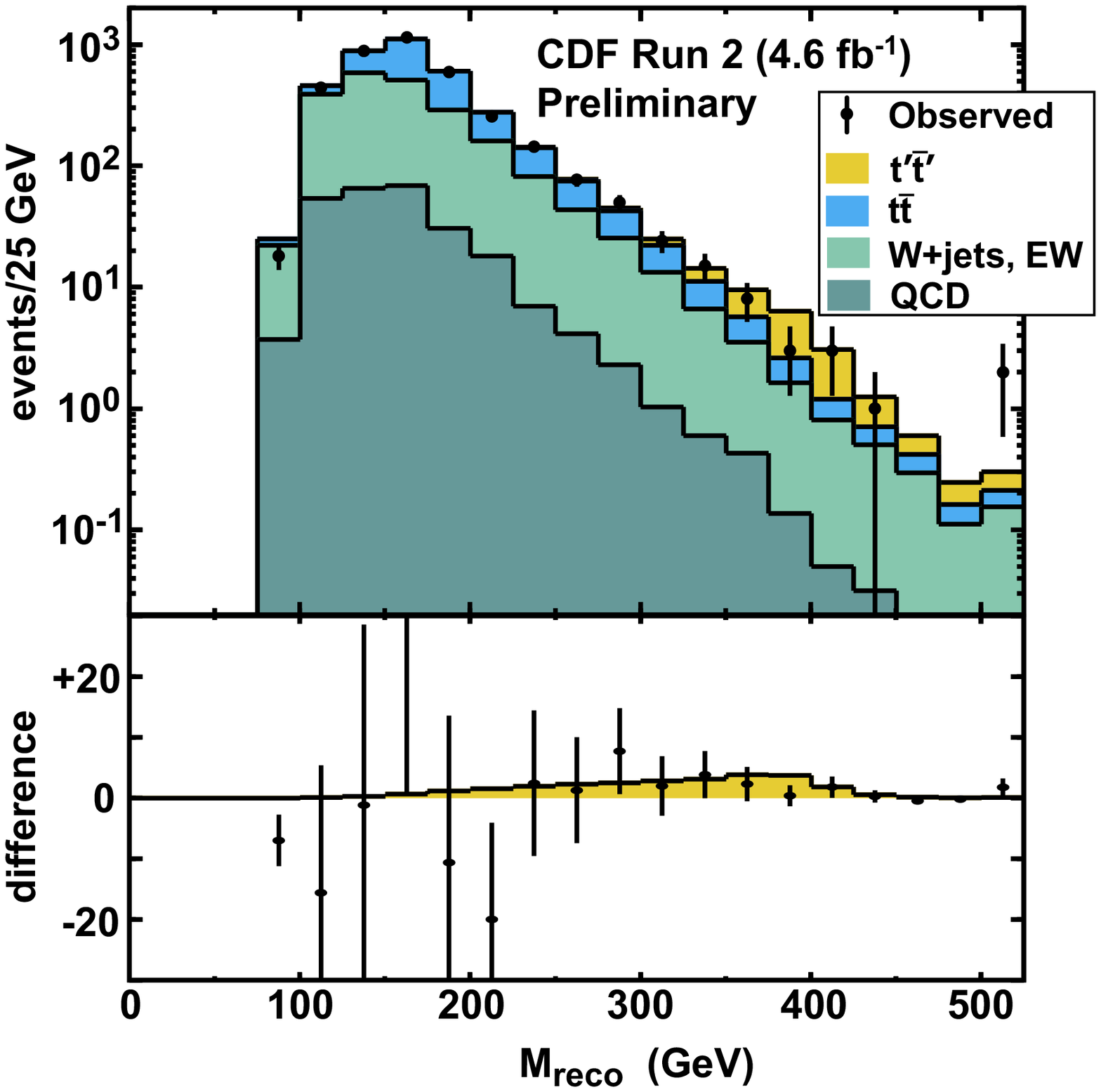}
  \caption{Expected and observed distribution of the the total transverse
    energy, $H_T$, and the reconstructed
    $t^{(\prime)}$ mass in $4.6\ifb$ of CDF data including a hypothetical
    signal at $m_{t'}=450\GeV$~\cite{CdfNote10110}.}
  \label{fig:cdf-tprime-ht-mrec}
\end{figure*}
 As the $t'$ quark decay chain is the same as the top quark decay chain a
constrained fit to the kinematic properties is performed. The 
momenta of the quarks and leptons from the $t^{(\prime)}$ quark and the following
$W$ boson decays are fitted to the observed transverse momenta.
The constraints require that the $W$ boson decay products form the nominal $W$ mass 
and that the decay products of the $t^{(\prime)}$ quark yield the same mass on 
the hadronic and the leptonic side. Of the 12 different jet-parton assignments
CDF chooses the one with the best $\chi^2$ of the fit. The corresponding
$t^{(\prime)}$ mass from the fit, $M_\mathrm{Reco}$, is used as one observable
to separate signal from background. The second observable is the total
transverse energy, $H_T$, i.e. the sum of transverse energies of the observed
jets, the lepton and the missing transverse energy.  The choices
explicitly avoid imposing $b$-quark tagging requirements.
The expected and observed
individual distributions are shown in \fig{fig:cdf-tprime-ht-mrec}.

The signal and background shapes in the two dimensional
$M_\mathrm{Reco}$-$H_T$ 
plane are used to construct a likelihood for the observed data as
function of the assumed $t'$ quark cross-section, $\sigma_{t'}$. Then Baysian
statistics is employed to compute expected and observed limits on
$\sigma_{t'}$ that are shown in \fig{fig:cdf-tprime}.
\begin{figure}
  \centering
  \includegraphics[width=0.44\textwidth]{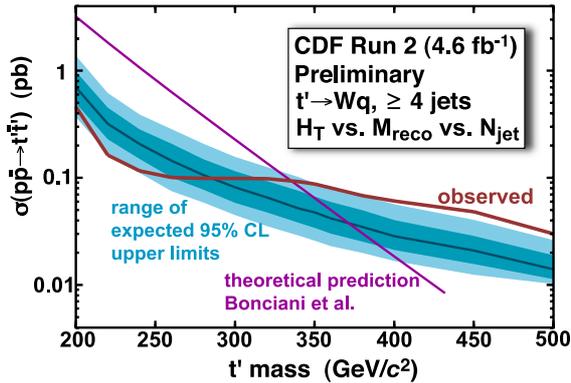}
  \caption{Expected and observed limits on the cross-section of a new top-like
    quark determined from $4.6\ifb$ of CDF data~\cite{CdfNote10110}. The shaded
   bands show the  expected one and two sigma variation on the expected upper limit.}
  \label{fig:cdf-tprime}
\end{figure}

Systematic uncertainties are implemented through nuisance parameters
that are constrained in the fit with a Gaussian function 
to their nominal value within their expected uncertainty.
The jet energy scale is named as one of the largest uncertainties. In addition
uncertainties on the $Q^2$ scale used in simulating $W+$jets,  on initial state
and final state radiation, on the multijet background determination, the
integrated luminosity, the lepton identification, the parton density
functions and the expected $t'$ quark cross-section as function of the $t'$
quark mass are
considered.

The limits on $t'$ quark pair production cross-section, $\sigma_{t'}$, determined in 
this search for a new top-like heavy quark are compared to the theoretical
prediction~\cite{Cacciari:2003fi}.  % Bonciani:1998vc ??
Assuming $t'\rightarrow Wq$ 
CDF concludes that a  $t'$ pair production
can be excluded for $m_{t'}<335\GeV$ at 95\%CL. However, for masses of
$m_{t'}\simeq 350\GeV$ the observed limit is worse than the expectation by
about two standard deviations, which indicates a surplus of data for that range.

\subsubsection*{D\O}
The D\O\ collaboration has searched for a heavy top like quark with a decay to
a $W$-boson and a quark using $4.3\ifb$ of data~\cite{d0note5892}.  
Events are selected which show  an isolated lepton, large missing transverse
energy and at least four jets. 

The dominating backgrounds from Standard Model processes are top quark pair
and $W+$jets production, which are simulated using \alpgen+\pythia. The
$t\bar t$ simulation and additional smaller backgrounds from diboson and
$Z$-boson production are normalised to the expected Standard Model
cross-sections. The normalisation of the $W+$jets contribution is determined below.
The contribution of multijet background faking a lepton is
derived from a sample with loosened lepton selection.
Signal simulation is obtained from \pythia\  for 13 different $t'$  mass
values using a fixed decay width of $10\GeV$.

Also D\O\ uses a reconstructed $t^{\prime}$-quark invariant mass and the scalar sum of
transverse momenta to distinguish the $t'$-quark signal from the Standard
Model backgrounds. The $t^{\prime}$-quark invariant mass is derived in  a
full event reconstruction assuming the expected decay 
$t'\bar t'\rightarrow \ell\nu bq\bar q\bar b$. In this reconstruction the
momenta of the assumed decay products are fitted to the reconstructed measured
objects applying constraints that enforce the mass of the intermediate
$W$-bosons and equality of the initial $t'$ and $\bar t'$ masses. All possible
associations of the four leading jets in $p_T$ to the quarks are considered.
To select the best association in addition to the fit $\chi^2$ 
a $\Delta\chi^2$ term is computed that prefers low transverse momenta for the
reconstructed $t'$-quark. The $t'$-mass reconstructed, $m_{\mathrm{fit}}$, with the
association that minimises $\chi^2+\Delta\chi^2$  is taken for further analysis.
\begin{figure*}[t]
  \centering
  \includegraphics[width=0.48\textwidth]{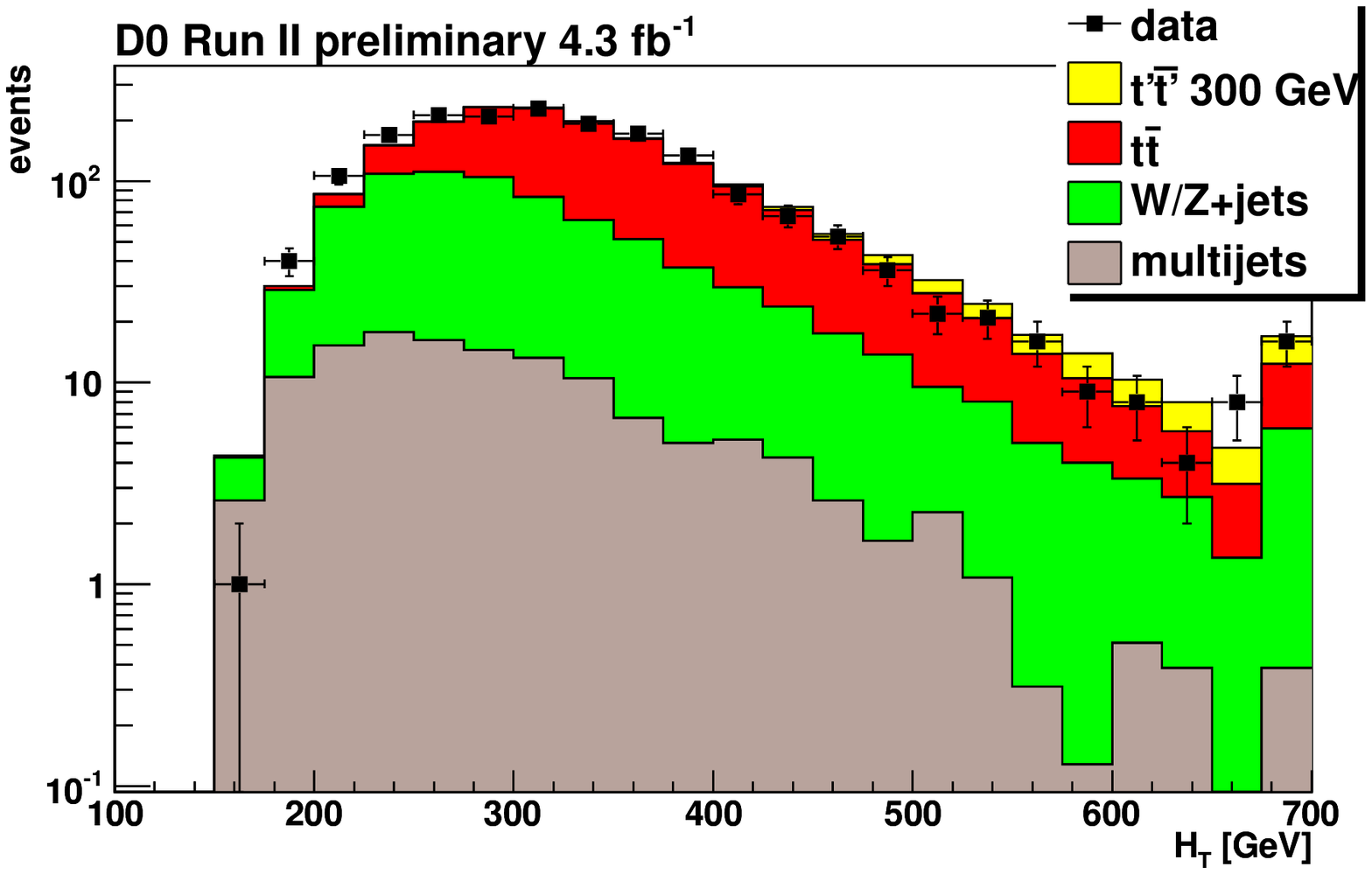}
  \includegraphics[width=0.48\textwidth]{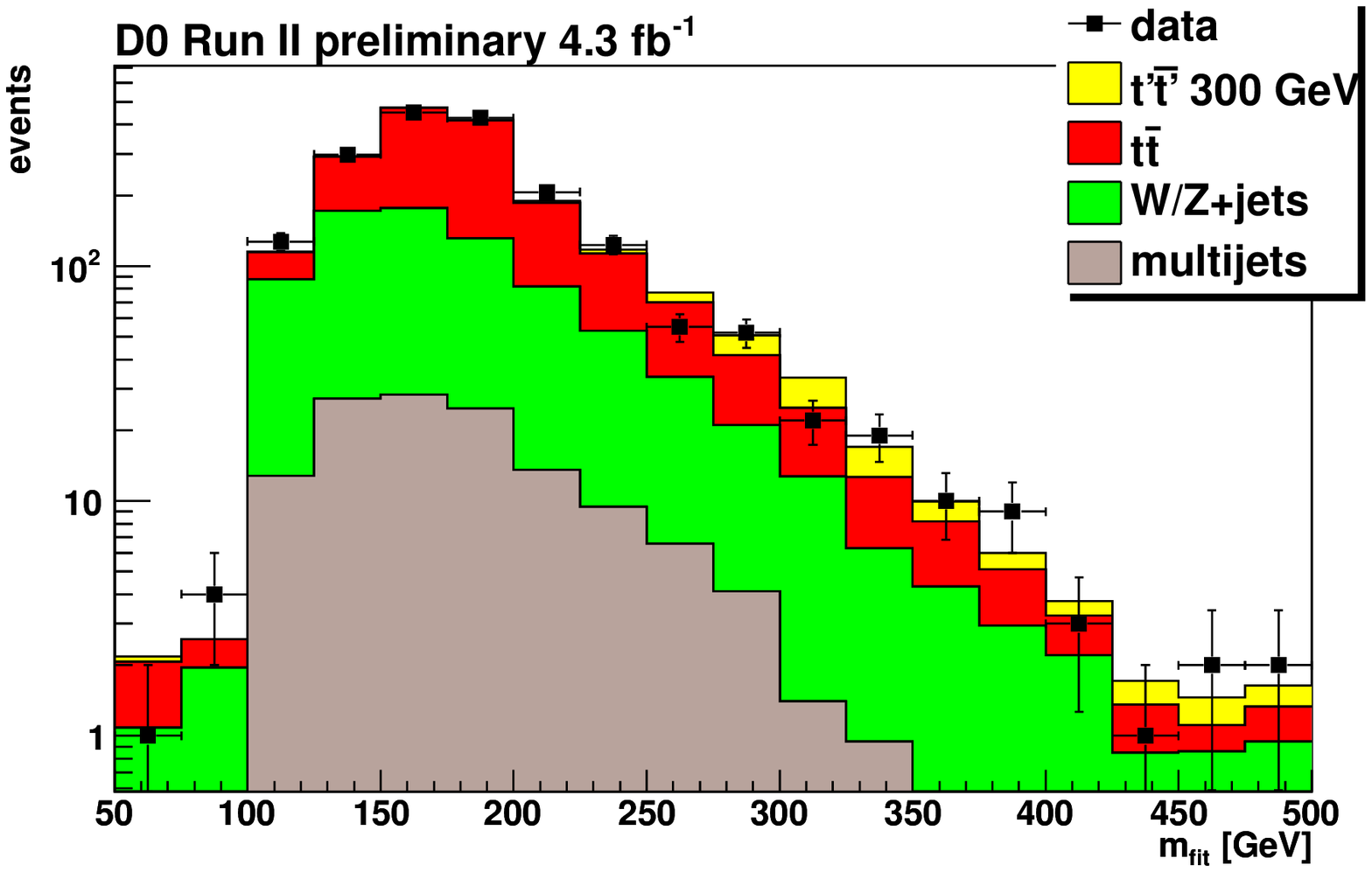}
  \caption{Expected and observed distribution of the the total transverse
    energy, $H_T$, and the reconstructed
    $t^{(\prime)}$ mass in $4.3\ifb$ of D\O\ data including a hypothetical
    signal at $m_{t'}=300\GeV$~\cite{d0note5892}.}
  \label{fig:d0-tprime-ht-mrec}
\end{figure*}

Two dimensional distributions of $H_T$ vs. $m_{\mathrm{fit}}$ are used to
fit the data compositions with free $W+$jets and signal normalisations, c.f. \fig{fig:d0-tprime-ht-mrec}. This is
done for all 13 $t'$-quark mass hypotheses and for a background only hypothesis. 
D\O\ uses the CL$_s$ method~\cite{Junk:1999kv,Read-in-James:2000et} with a
Poisson likelihood as the test statistics to determine the cross-section
limits. D\O\ includes several systematic uncertainties affecting the
normalisation of the fixed background contributions as well as shape changing
effects in the limit calculation. Of the normalisation uncertainties the
luminosity uncertainty gives the largest single contribution. 

The limits on a $t'$-quark production cross-section observed by D\O\ in
$4.3\ifb$~\cite{d0note5892} and the corresponding expectations are shown
in \fig{fig:d0-tprime-limit} as function of the hypothetical $t'$ mass. By comparing to the
theoretical prediction~\cite{Cacciari:2003fi} 
D\O\ excludes heavy top like
quarks with masses less than $296\GeV$ at 95\%~C.L. Over almost the full range
the observed limit stays behind the expected limit indicating a surplus of
data in the $t'$-signal range. Correspondingly the
expected mass exclusion of $330\GeV$ is not reached. 

\begin{figure}
  \centering
  \includegraphics[width=0.88\linewidth]{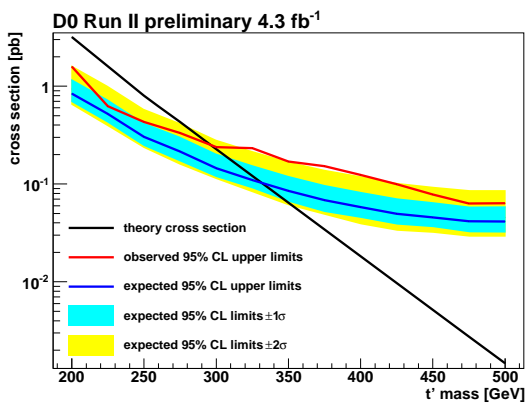}
  \caption{Expected and observed limits for the production of a $t'$-quark obtained
    in $4.3\ifb$ of D\O\ data. The shaded bands indicate the one and two sigma
    deviations from the expected limit~\cite{d0note5892}.}
  \label{fig:d0-tprime-limit}
\end{figure}
%%% Local Variables: 
%%% mode: latex
%%% TeX-master: "EPJC_TopProperties"
%%% End: 

\subsection{Outlook to LHC}
Due to the high mass of the top quark
it has been speculated that the top quark may play a special role in particle
physics. Many models of new physics involve new particles that may occur in
top quark production or decay. The Tevatron experiments have investigated a
wide range of options for such new particles in top quark events and also
looked for particles with signatures very similar to that of top quarks. So far
no significant deviation has been found. 

In the near future the LHC is scheduled to deliver at least $1\ifb$ at
$\sqrt{s}=7\TeV$. With such a dataset the LHC experiments will have roughly
twice as many top quark pairs and three times as many single top quarks 
as the Tevatron with the expected $10\ifb$. The benefit of this data,
however, depends on the process in which the new particle appears.
 
Searches for new particles that occur in the top quark decay, like e.g. a
charged Higgs boson,  will benefit directly from the increased dataset and
yield results competitive to those of the Tevatron.
Searches for new particles that occur in the production like, $Z'$ or $W'$ bosons or
a $t'$ quark, will  benefit from the larger centre-of-mass energy.
The cross-sections of such processes, however, scale with the corresponding parton luminosity,
which increases much more for the gluon than for the quark luminosity.
Particles like the $Z'$ boson that can by produced only by quark-antiquark annihilation thus 
will have less signal events in $1\ifb$ than the Tevatron will have in the
expected $10\ifb$.  Such analyses will still extend the Tevatron results to
the phase space areas opened with the higher centre-of-mass energy. 
Finally, the search for a boson produced in the single top quark
$s$-channel, like the $W'$, will suffer from the comparably moderate increase
of the corresponding cross-section and thus require more statistics at the LHC
to become competitive to the Tevatron results.

With the design LHC at $\sqrt{s}=14\TeV$ the larger production cross-section
and the enlarged phase space will again extend the results significantly.
An ATLAS study assuming collisions at $\sqrt{s}=14\TeV$ and a
luminosity of $1\ifb$~\cite{Aad:2009wy} has considered the potential for
discovering resonant top quark pair production through a narrow $Z'$ boson.
A significant  degradation of the selection efficiency is expected at high top
pair invariant masses because the top quark decay products join into the
same jet more and more often. 
Given the performance visible in the
first results shown at conferences from LHC experiments 
searches for new physics in top quark events will greatly advance at each new
energy reached by the LHC after collecting a moderate luminosity of e.g. $1\ifb$.

%%% Local Variables: 
%%% mode: latex
%%% TeX-master: "EPJC_TopProperties"
%%% End: 

\section{Conclusions}
\label{sect:conclustions}
% $Id: Conclusions.tex,v 1.12.2.8 2011-03-22 17:46:50 wicke Exp $ 
%
More than fifteen years after the discovery of the top quark the experimental
verification of the properties expected in the Standard Model has reached a
first level of maturity. With the luminosity delivered by the Tevatron
accelerator in its Run\,II to the CDF and D\O\ experiments many top quark
properties have been challenged and contributions for new particles have been
searched for. 

% Masses

The top quark mass measurements,
with the elaborate statistical methods of the Matrix Element technique, 
have now reached a statistical precision of less than $1\GeV$
and experimental systematics of $1\GeV$ making the top quark mass the most
precisely known mass of all quarks. This precision significantly exceeds
the precision goals of about $3\GeV$  of the Tevatron Run\,II programme set for 
$2\ifb$~\cite{Blair:1996kx,Amidei:1996dt}.
Despite the low branching fraction of the dileptonic channel, even results in
this channel alone now have achieved 
an experimental precision exceeding this goal. 
In order to reach these small uncertainties, it was important to constrain the
jet energy scale, leading to the dominating uncertainty, in-situ to data.
%To reach these small uncertainties
%it was important that the dominating uncertainty that is due to the overall jet energy scale
%uncertainty could be constrained in-situ from data. 
Unfortunately, the experimental precision on the top quark mass is currently not matched by a
corresponding theoretical understanding. The uncertainty of the mass
definition used in the simulations that the experiments apply to calibrate
their measurements is not known better than to the order of $1\GeV$.
This complicates the comparison of the top quark mass 
results, e.g. with other electroweak
precision data, and is a fundamental problem also for the LHC programme. 
Discussions on these issues have started between the experimental, the
generator and the theoretical community in order to collect the information
needed to overcome this issue.  The development of
methods that precisely determine the mass in well defined definitions 
is an important
problem in current top quark physics.

Many interaction properties have been challenged by the 
Tevatron experiment in the recent years. 
Production properties such as the quark and gluon induced production
rates, the electrical charge and the various decay 
properties did not show significant deviation from the expectation and
thus confirm that we observe the top quark expected in the SM. 
The $V-A$ structure of the weak top quark decay has been tested in the
$W$ boson helicity measurements to the 5\% level and weak decays through 
flavour changing neutral currents are constrained to be below the 4\% level. 
The largest deviation ($3.4\sigma$) from the SM was found in the forward
backward charge asymmetry at high $m_{t\bar t}$.
While the sheer number of tests documents a great progress the precision of
these measurements is generally limited by statistics. 
Searches for new particles have been performed in events with top quark
like signatures. Neither specific searches
for supersymmetric top quark partners, for charged Higgs bosons nor searches
for $W'$ bosons, $t'$ quarks or generic resonances found
any significant deviations from the SM expectations.
Also these studies are limited by the statistics available so far. 

The Tevatron is planned to be shutdown in autumn 2011. Until then it is
expected to increase the total integrated luminosity for the CDF and D\O\
experiments to more than $10\ifb$. The analyses of the Tevatron 
take advantage of several years of optimising the reconstruction of the
recorded events. Thus further updates 
of existing analyses, but also investigation of yet untested properties can be
expected. % in the near future.
In addition the LHC has started to operate at $\sqrt{s}=7\TeV$
well above that of the Tevatron centre-of-mass energy. It is expected to
deliver an integrated luminosity of at least $1\ifb$ throughout 2011.
Due to the increased collision energy at the LHC the top production
cross-sections are significantly higher than at the Tevatron. With $1\ifb$ the
LHC experiments will have collected about twice as many top quark pairs and
three times as many single top quarks than the Tevatron experiments at $10\ifb$.

Given the impressive performance of the LHC experiments shown at recent
conferences it is natural to assume that the systematic uncertainties can be
controlled comparably well as at the Tevatron experiments, if not better. 
We will thus see a plethora of competitive results
from the LHC experiments. Clearly some
searches profit from the increased phase space and will be dominated by the LHC.
There are also results where the Tevatron has an advantage, like the
charge asymmetry and the single top quark $W'$ search, however, most results
will be comparable. Also at the LHC many  will be statistically
limited which makes it useful to foresee a combination of the results. 

After an update to the design LHC with proton-proton collisions at
$\sqrt{s}=14\TeV$ the field will be taken over by the LHC experiments. 
With a top quark pair  cross-section which is $100$ times higher than at the
Tevatron and dominant backgrounds like $W+$jets and $Z+$jets increasing only by
factors of about $10$ the LHC will become a real top quark
factory. Again many searches for new particles will profit from the enlarged
phase space. Measurements of interaction properties and searches for new
particles in top quark decays will become limited by
systematics after only a short running at design luminosity.
At the same time with the huge expected statistics new types of
measurement methods on rare subsamples will become feasible.
For the top quark mass methods using leptonic $J/\psi$ decays or using events
with very high top quark momentum will have very different experimental and
theoretical uncertainties. Such alternative methods with different systematic
uncertainties will contribute to the combination of results. And may even help
to solve the puzzle of 
the meaning of the top quark mass.

%%% Local Variables: 
%%% mode: latex
%%% TeX-master: "EPJC_TopProperties"
%%% End: 

\section*{Acknowledgments}

The author would like to thank his colleagues in the D\O\ and CDF collaborations
for their support in the preparation of this review.
Special thanks belong to
Frank Fiedler, Lina Galtieri, Alexander Grohsjean, Klaus Hamacher, 
Michele Weber and Wolfgang Wagner
for valuable discussions about experimental details
and to Werner Bernreuther,  Andr\'e Hoang, Sven Moch and Peter Uwer 
for detailed support and suggestions on theoretical questions.
Finally, I would like to thank Stefan Tapprogge for valuable suggestions in
the preparation of this work.

%%% Local Variables: 
%%% mode: latex
%%% TeX-master: "EPJC_TopProperties"
%%% End: 

\addcontentsline{toc}{section}{References}
\bibliographystyle{utphys}
\begin{flushleft}
\bibliography{Habilitation,CDF,Dzero}
\end{flushleft}

\end{document}